\documentclass{aa}  

\usepackage{graphicx}
\usepackage{grffile}
\usepackage[varg]{txfonts}
\usepackage{subfigure}
\usepackage{amssymb}
\usepackage{cancel}
\usepackage{bm}
\usepackage[draft]{hyperref}
\usepackage{color}
\usepackage{footnote}
\usepackage{eqnarray}
\newcommand{\rev}[1]{{#1}}

\begin{document}

   \title{Different dust and gas radial extents in protoplanetary disks:\\
   consistent models of grain growth and CO emission}
   
\titlerunning{Different dust and gas radial extents in protoplanetary disks}

   \author{S. Facchini\inst{1}   
           \and
           T. Birnstiel\inst{2,}\inst{3}
           \and
           S. Bruderer\inst{1}
           \and
           E.~F. van Dishoeck\inst{1,}\inst{4}
          }

   \institute{Max-Planck-Institut f\"ur Extraterrestrische Physik, Giessenbachstrasse 1, 85748 Garching, Germany,\\\email{facchini@mpe.mpg.de}    
                         \and
                 Max-Planck-Institut f\"ur Astronomie, K\"onigstuhl, D-69117 Heidelberg, Germany
                 \and
                 University Observatory, Faculty of Physics, Ludwig-Maximilians-Universit\"at M\"unchen, Scheinerstr. 1, 81679 Munich, Germany
                 \and
                 Leiden Observatory, Leiden University, P.O. Box 9513, NL-2300 RA Leiden, The Netherlands
             }

   \date{Received; accepted}

% \abstract{}{}{}{}{} 
% 5 {} token are mandatory
 
  \abstract
   {ALMA observations of protoplanetary disks confirm earlier indications that there is a clear difference between the dust and gas radial extents. The origin of this difference is still debated, with both radial drift of the dust and optical depth effects suggested in the literature.}
   {In thermo-chemical models, the dust properties are usually prescribed by simple parametrisations. In this work, the feedback of more realistic dust particle distributions onto the gas chemistry and molecular emissivity is investigated, with a particular focus on CO isotopologues.}
   {The radial dust grain size distribution is determined using dust evolution models that include growth, fragmentation, and radial drift for a given static gas density structure. The vertical settling of dust particles is computed in steady-state. A new version of the code DALI is used to take into account how dust surface area and density influence the disk thermal structure, molecular abundances, and excitation. Synthetic images of both continuum thermal emission and low $J$ CO isotopologues lines are produced.}
   {The difference of dust and gas radial sizes is largely due to differences in the optical depth of CO lines and millimeter continuum, without the need to invoke radial drift. The effect of radial drift is primarily visible in the sharp outer edge of the continuum intensity profile. The gas outer radius probed 
by $^{12}$CO emission can easily differ by a factor of $\sim$ two between the models for a turbulent $\alpha$ ranging between $10^{-4}$ and $10^{-2}$, with the ratio of the CO and mm radius $R_{\rm CO}^{\rm out}/R_{\rm mm}^{\rm out}$ increasing with turbulence. Grain growth and settling concur in thermally decoupling the gas and dust components, due to the low collision rate with large grains. As a result, the gas can be much colder than the dust at intermediate heights, reducing the CO excitation and emission, especially for low turbulence values. Also, due to disk mid-plane shadowing, a second CO thermal desorption (rather than photodesorption) front can occur in the warmer outer mid-plane disk. The models are compared to ALMA observations of HD 163296 as a test case. In order to reproduce the observed CO snowline of the system, a binding energy for CO typical of ice \rev{mixtures}, with $E_{\rm b} \geq 1100$\,K, needs to be used rather than the lower pure CO value.}
   {The difference between observed gas and dust extent is largely due to optical depth effects, but radial drift and grain size evolution \rev{also} affect the gas and dust emission in subtle ways. In order to properly infer fundamental quantities of the gaseous component of disks, such as disk outer radii and gas surface density profiles, simultaneous modelling of both dust and gas observations including dust evolution is needed.}


   \maketitle
%
%________________________________________________________________
\section{Introduction}

\label{sec:intro}

The advent of the Acatama Large Millimeter/submillimeter Array (ALMA) is providing an unprecedented level of angular resolution and sensitivity at (sub)mm wavelengths. This technological step forward allows us to determine more precisely fundamental quantities of protoplanetary disks, the cradles of planet formation. In particular, two quantities of great interest are the outer radius of disks and the surface density radial profile $\Sigma(R)$ \citep[see][for recent reviews]{2011ARA&A..49...67W,armitage_11,2015arXiv150906382A,2016JGRE..121.1962M}. Both of them are of fundamental importance because they are directly related to the amount of mass present in the disk, which is obviously linked to the planet formation potential of a system. The surface density profile also provides one of the main parameters for any planet formation model. Moreover, in the commonly assumed disk evolution framework \citep[the so-called $\alpha$-disk scenario,][]{1952ZNatA...7...87L,shakura73,1974MNRAS.168..603L}, the gas outer radius determines the global evolution of the gaseous component, since the viscous timescale is directly related to the radial extent of the disk. In star forming regions, it can also imprint the respective importance of environmental effects, \rev{such} as star-disk encounters \citep[e.g.][]{1993MNRAS.261..190C,2006ApJ...642.1140O,2014MNRAS.441.2094R,2015MNRAS.449.1996D,2015A&A...577A.115V} and external photoevaporation \citep[e.g.][]{hollenbach_94,johnstone_98,2004ApJ...611..360A,2016MNRAS.457.3593F,2016MNRAS.463.3616H}, which can both truncate the disks radially.

Interestingly, observations have shown that the outer radius of the gaseous component of a disk, as probed by the bright $^{12}$CO emission, is generally larger than that of the dust component, probed by (sub)mm continuum emission. Historically, this difference has been considered to be due to optical depth effects, with the gas lines (in particular the $^{12}$CO line) more optically thick than the continuum emission at similar wavelengths \citep[e.g.][]{1998A&A...338L..63D,1998A&A...339..467G,2009A&A...501..269P}. This assumption motivated observers to try fitting both the gas tracers and continuum emission with the same surface density profiles, leading \citet{2008ApJ...678.1119H} and \citet{2009ApJ...700.1502A} to propose a tapered power law profile, theoretically justified by the self-similar solutions by \citet{1974MNRAS.168..603L}. However, recent observations with higher sensitivity and angular resolution have shown that the apparent friction between the dust (continuum) and gas (molecular) intensity profiles in disks cannot be reconciled even with such a surface density profile. In well resolved systems, in particular those imaged with ALMA, the dust outer edge decreases too sharply with radius, and cannot be reproduced by a tapered outer disk \citep[e.g.][]{2012ApJ...744..162A,2016ApJ...820L..40A,2013A&A...557A.133D,2014A&A...564A..95P,2016ApJ...832..110C}.

The natural explanation that has been proposed to interpret such observations is a combination of grain growth and consequent radial drift of large particles from the outer to the inner disk. The fact that dust particles can grow to at least mm/cm sizes in protoplanetary disks has now been proven by many observations at different radio wavelengths \citep[e.g.][]{2003A&A...403..323T,2004A&A...416..179N,2007A&A...462..211L,2010A&A...515A..77L,2005ApJ...631.1134A,2007ApJ...671.1800A,2010A&A...521A..66R,2010A&A...512A..15R}. Recent reviews on the topic are \citet{2014prpl.conf..339T,2015PASP..127..961A,2016SSRv..tmp...32B}. Moreover, there is increasing evidence that the maximum grain size attained in a disk is a function of distance from the star, as expected from grain growth models \citep[e.g.][]{2011A&A...529A.105G,2012ApJ...760L..17P,2015ApJ...813...41P,2014A&A...564A..93M,2016A&A...588A..53T}. These observational constraints have been accompanied by further theoretical modelling of both grain growth and radial drift processes \citep[e.g.][]{2010A&A...513A..79B,2012A&A...539A.148B}. In particular, \citet{birnstiel_14} have shown that the sharp edge observed in the (sub-)mm continuum is predicted by dust evolution models.

In order to investigate this situation further, simultaneous modelling of both dust and gas is needed. The size distribution evolution of the dust particles does affect the chemistry occurring in protoplanetary disks, in particular the CO abundance and excitation, via multiple physical-chemical processes some of which are directly linked to the total dust surface area available at any point of the disk. A first important effect is that grain growth, vertical settling, and radial drift affect the penetration depth of the UV (ultraviolet) photons into the disk; these photons can originate both from the central star or from the surrounding environment. Some studies have started addressing this topic in static disks, for example \citet{2004A&A...428..511J,2007A&A...463..203J,2007ApJ...661..334N,2011ApJ...727...76V,2011ApJ...726...29F,2016A&A...586A.103W,2016ApJ...816L..21C}. Enhanced UV penetration affects both the chemistry and the gas temperature, which both determine the molecular (and atomic) emission. Moreover, the relation between UV penetration depth and grain growth can also have an important dynamical effect, making the disk more or less prone to photoevaporate \citep{2015ApJ...804...29G,2016MNRAS.457.3593F}. Another important connection between dust evolution and chemistry is that the amount of dust surface area available determines the level of thermal coupling between the gas and solid phases. Finally, the dust surface area affects the balance between freeze-out and desorption, since they both depend on the total surface area. The non linear combination of all these effects make the modelling of CO emission lines (and of other molecules) less straightforward than simply assumed.

In this paper, we aim to explore how the properties of physically justified dust grain size distributions affect the thermal and chemical structure of a protoplanetary disk. In particular, we focus on how dust evolution (grain growth, radial drift, and vertical settling) affects the observability of both dust and gas outer radius and surface densities, as probed by (sub-)mm continuum observations and low-$J$ rotational lines of different CO isotopologues. To do so, we combine the dust evolution models by \citet{2015ApJ...813L..14B} with the thermo-chemical code DALI \citep[Dust And LInes,][]{2012A&A...541A..91B,2013A&A...559A..46B}. The method used in the paper is detailed in Sec.~\ref{sec:method} and the setup parameters for the models investigated are given in Sec.~\ref{sec:setup}. In Sec.~\ref{sec:results} we describe our findings, which are then discussed in Sec.~\ref{sec:discussion}. In Sec.~\ref{sec:conclusion} we summarise the results and draw our conclusions. 

\section{Method}
\label{sec:method}

\begin{figure*}
\center
\includegraphics[width=.95\textwidth]{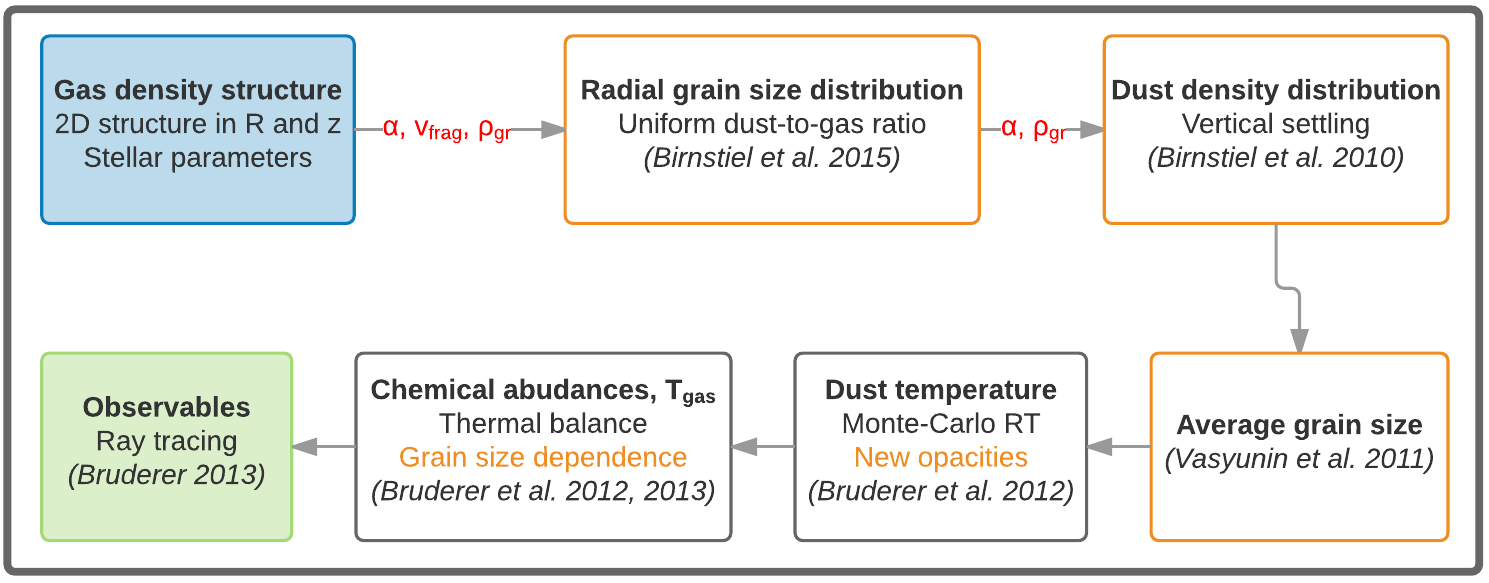}
\caption{Diagram of the method used in this paper. The blue box indicates the input parameters and the green box indicates the output observables. Boxes with orange borders show new modules, compared to the standard DALI code. Orange text underlines modifications in previous DALI modules. The red text lists input parameters needed to compute the dust distribution.}
\label{fig:diagram}
\end{figure*}

In this Section we present the method used in this paper to couple dust evolution with the thermo-chemical processes in a protoplanetary disk. The method consists in merging the thermo-chemical code DALI \citep{2009ApJS..183..179B,2012A&A...541A..91B,2013A&A...559A..46B}, which has been widely used to model gas emission in protoplanetary disks \citep[e.g.][]{2014A&A...572A..96M,2016A&A...594A..85M,2014A&A...562A..26B,2015A&A...575A..94B,2015A&A...579A.106V,2016A&A...585A..58V,2016A&A...592A..83K,2016A&A...591A..95F}, with the grain size distribution reconstruction routine by \citet{2015ApJ...813L..14B}. More precisely, given an initial gas density distribution, the radial dependence of the grain size distribution is computed with the grain size reconstruction routine. Subsequently, the vertical distribution of the dust is calculated solving the advection-diffusion equation of vertical settling in steady-state at every radius for every dust bin considered. The local grain size distribution is then used to compute the opacities at every grid point of the disk (radius and height), which are then used in the continuum radiative transfer module. Moreover, the local particle distribution is used to compute the total dust surface area available for the thermo-chemical processes, in particular gas-grain collisions, H$_2$ formation rate, freeze-out, thermal and non-thermal desorption, and hydrogenation. With these new rates, both the gas temperature and the chemical abundances are computed self-consistently. A schematic of the method is illustrated in Fig.~\ref{fig:diagram}.

\subsection{Gas density distribution and stellar parameters}
\label{sec:method_input}

The main input physical quantity for the used method is the gas density structure. We use the following gas surface density $\Sigma_{\rm gas}$ dependence on cylindrical radius $R$ \citep{2009ApJ...700.1502A,2013A&A...559A..46B}:

\begin{equation}
\Sigma_{\rm gas} (R)= \Sigma_{\rm c} \left(\frac{R}{R_{\rm c}}\right)^{-\gamma} \exp{ \left[  -\left( \frac{R}{R_{\rm c}} \right)^{2-\gamma} \right] },
\label{eq:surf_dens}
\end{equation}
where $R_{\rm c}$ is the characteristic radius determining the radial length scale of the disk, and $\Sigma_{\rm c}$ determines the total mass of the disk. Assuming hydrostatic equilibrium, isothermality in the vertical direction, \rev{and $z\ll R$} (where $z$ is the vertical coordinate), we obtain the gas mass density $\rho_{\rm gas}$:

\begin{equation}
\rho_{\rm gas} (R,z)= \frac{\Sigma_{\rm gas}(R)}{ \sqrt{2\pi} R h } \exp { \left[ -\frac{1}{2} \left( \frac{z}{Rh} \right)^2 \right] },
\label{eq:gas_dens}
\end{equation}
where $h=h_{\rm c}(R/R_{\rm c})^\psi$, $\psi$ is the flaring index, and $h_{\rm c}$ is the aspect ratio of the disk at the characteristic radius. The expression in Eq.~\ref{eq:surf_dens} is justified by the self-similar solutions by \citet{1974MNRAS.168..603L}. This surface density profile would require the system to be older than a viscous timescale, which is not always the case. Thus the outer regions of protoplanetary disks might have surface densities that do not follow such profiles and instead reflect the initial conditions. However, we choose to use it throughout this paper since it can describe the surface densities with very few parameters, and to compare our models with those present in the literature exploiting the same parametrisation.

The other input parameters describe the stellar properties, in particular the stellar mass $M_*$ and radiation parameters. The stellar spectrum is modelled as a black body, determined by its total luminosity $L_*$ and effective temperature $T_{\rm eff}$. Accretion onto the star is also considered, which can be important in setting the total high energy radiation. The accretion luminosity is modelled as a black body emitting at $10000$\,K, with total luminosity equalling the total potential energy per time unit emitted on the stellar surface by a mass accretion rate $\dot{M}$ \citep{2016A&A...592A..83K}. This prescription is used to estimate the UV excess emitted in the accretion shock. We note, however, that the disk evolution is not modelled self-consistently with this mass accretion rate.

\subsection{Radial grain size distribution}

In order to determine the grain size distribution at every radius, given the gas density structure defined as above, the semi-analytical prescription described in \citet{2015ApJ...813L..14B} is used, which has proven to be a good representation of the more complete numerical models by \citet{2010A&A...513A..79B}. Here we briefly summarise the main physical mechanisms determining the radial distribution of particles sizes. The analytical details are given in \citet{2015ApJ...813L..14B}.

Grain growth in disks occurs due to the mutual collisions of dust particles. Depending on the relative velocity $\Delta v$, such collisions lead either to sticking of the particles or to fragmentation or erosion (or various other outcomes that are not modelled here). The boundary between these two regimes is set by the fragmentation velocity $v_{\rm frag}$, where we assume that perfect sticking occurs when $\Delta v < v_{\rm frag}$. The relative motions of dust particles and their transport within the disks are driven by four processes: turbulent mixing, Brownian motions, gas drag, and radial drift \citep{1977MNRAS.180...57W}. The combination of such processes can lead to two different regimes: 1) dust grains grow until the relative velocities are high enough that fragmentation occurs (fragmentation barrier); 2) the drifting timescale becomes shorter than the collisional timescale, and the maximum particle size is determined by radial drift (drift barrier).

The physical conditions of protoplanetary disks are such that the fragmentation barrier is unlikely to dominate in the entire disk. It is possible to assign a fragmentation radius $R_{\rm frag}$, inside which the maximum grain size is set by fragmentation. In the outer regions of the disk, the maximum grain size is determined by radial drift. The actual boundary between the two regimes is smoothed by diffusion. As in \citet{2015ApJ...813L..14B}, in this paper the grain size distribution inside $R_{\rm frag}$ is determined by the analytical fits of coagulation and fragmentation equilibrium by \citet{2011A&A...525A..11B}. Outside the fragmentation radius, semi-analytical treatments including diffusion and radial drift are exploited \citep{2007A&A...471..833P,2015ApJ...813L..14B}. In this paper, a $p_{\rm d}$ parameter equalling $2.5$ is assumed, where $p_{\rm d}$ is the power law index of the radial dependence of a given grain size bin in the outer disk \citep[see Sec. 2.3 in][for details]{2015ApJ...813L..14B}.

The final radial distribution of grain sizes will depend on many input parameters, in particular the gas surface density of the disk, the dust-to-gas mass ratio $\Delta_{\rm dg}$, the mass of the star, and the dust temperature profile, which is implicitly set by the flaring angle $\psi$. Besides these quantities, which are all implemented as initial input parameters (see Sec. \ref{sec:method_input}), the output of the grain size reconstruction routine depends on the fragmentation velocity, the disk turbulence, and the grain mass density $\rho_{\rm gr}$. Laboratory experiments have shown that fragmentation velocity can assume values of a few $1-10$\,m\,s$^{-1}$ \citep[e.g.][]{2000Icar..143..138B,2015ApJ...798...34G}. Throughout this paper we assume $v_{\rm frag}=10$\,m\,s$^{-1}$. We parametrise turbulence via the dimensionless parameter $\alpha$ \citep{shakura73}, and we will vary this parameter in the simulations presented below. Finally, we assume compact dust grains, with $\rho_{\rm gr}=2.5\,$g\,cm$^{-3}$, and assume the dust-to-gas ratio to be uniform in the whole disk, with $\Sigma_{\rm dust}(R)/\Sigma_{\rm gas}(R)=\Delta_{\rm dg}=0.01$. In future work, we will address how more complex models, with a radially varying dust-to-gas ratio, will affect the final results. A combination of some of these parameters can easily lead to an analytical estimate of the fragmentation radius \citep{2015ApJ...813L..14B}:

\begin{equation}
\frac{R_{\rm frag}}{100\,{\rm AU}} \sim \frac{M_*}{M_\odot} \frac{\alpha}{10^{-3}} \frac{\Delta_{\rm dg}}{0.01} \Bigg( \frac{|\gamma|}{2.75} \Bigg)^{-1} \Bigg( \frac{v_{\rm frag}}{10\,{\rm m\,s}^{-1}} \Bigg).
\label{eq:r_frag}
\end{equation}

Given the input parameters defined in Sec.~\ref{sec:method_input} and the additional ones listed in this Section, a radially dependent grain size distribution $\Sigma_{\rm dust}(R,a)$ can be retrieved, which is defined as the dust surface density at a given radius $R$, with particle sizes between $a$ and $a+da$ (in radius). For numerical reasons, a discrete grain size grid of $250$ size bins logarithmically spaced is used, with a minimum grain size $a_{\rm min}=50\,\AA$, and a maximum grain size $a_{\rm max}=1$\,m.

\begin{figure}
\center
\includegraphics[width=\columnwidth]{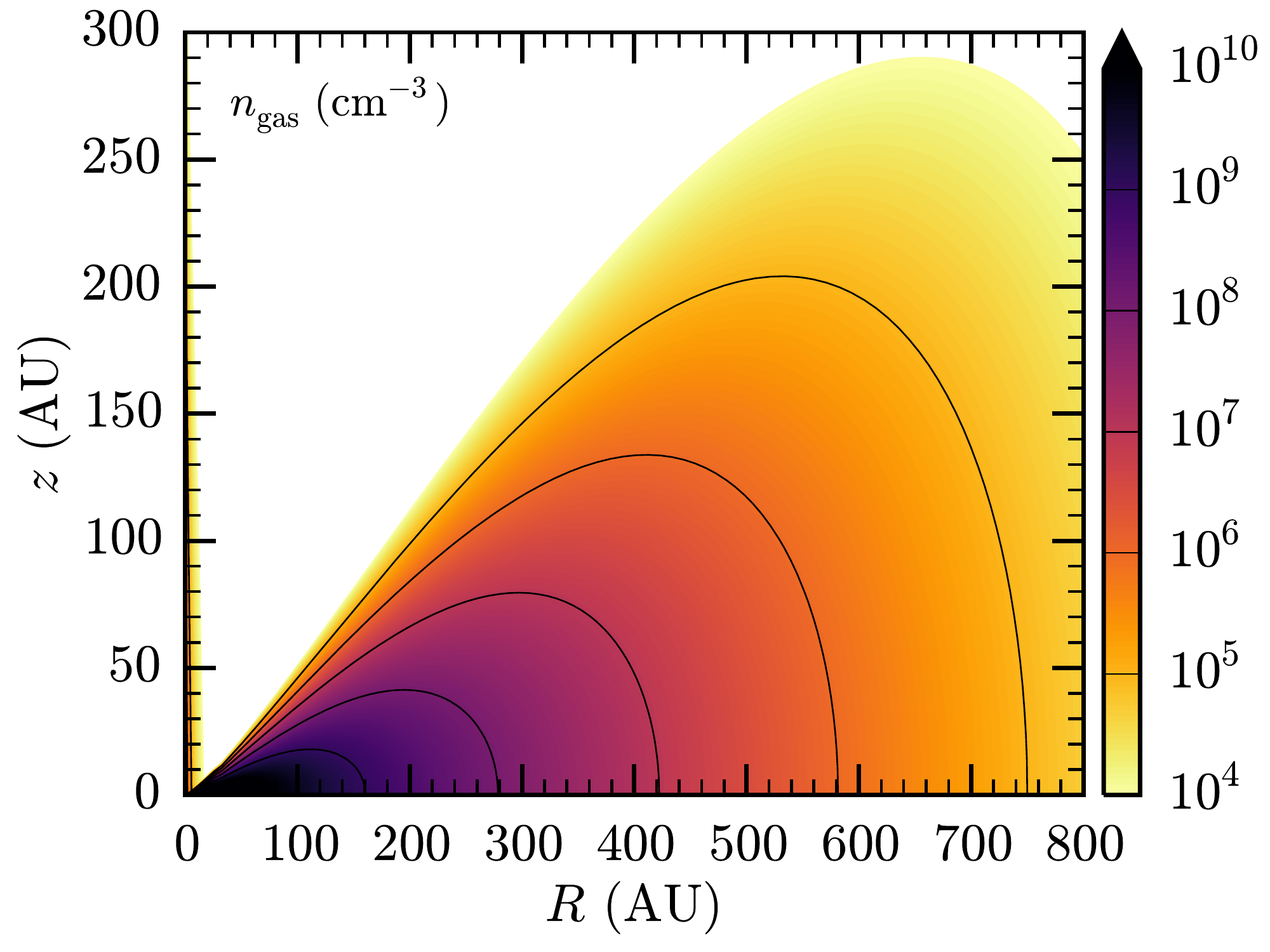}
\caption{Gas density structure used in all models.}
\label{fig:n_gas}
\end{figure}

\subsection{Dust vertical settling}

Once the radial dependence of the grain size distribution is computed, the next step is to determine the vertical distribution of dust particles. This can be done by taking into account the physical processes responsible for dust settling, with the assumption that the turbulence acting in the vertical direction is the same as that responsible for the angular momentum transport and for the turbulent motions leading to particle collisions, thus with a gas diffusion coefficient:

\begin{equation}
D_{\rm gas} \sim \nu = \frac{\alpha c_{\rm s}^2}{\Omega},
\end{equation}
where $\nu$ is the kinematic viscosity \citep{shakura73}, and $\Omega=\sqrt{GM_*/R^3}$. 

Following \citet{2010A&A...513A..79B}, we use the results derived by \citet{2007Icar..192..588Y} and we relate the dust diffusivity to the gas diffusivity through the Schmidt number:

\begin{equation}
{\rm Sc} \equiv \frac{D_{\rm gas}}{D_{\rm dust}} = 1+{\rm St}^2,
\end{equation} 
where the Stokes number ${\rm St}$ in the Epstein regime in the disk mid-plane can be computed as \citep[e.g.][]{2011A&A...525A..11B}:

\begin{equation}
\label{eq:stokes}
{\rm St} = \frac{a \rho_{\rm gr}}{\Sigma_{\rm gas}} \frac{\pi}{2}.
\end{equation}

The equation regulating the vertical motions of dust particles can be written as \citep{1995Icar..114..237D,2004A&A...421.1075D,2006A&A...452..751F,armitage_book_10}:

\begin{equation}
\frac{\partial \rho_{\rm dust}(z,a)}{\partial t} = D_{\rm dust} \left[ \rho_{\rm gas} \frac{\partial}{\partial z} \left( \frac{\rho_{\rm dust}}{\rho_{\rm gas}} \right) \right] + \frac{\partial}{\partial z} ( \Omega^2 t_{\rm fric} \rho_{\rm dust} z ),
\label{eq:adv_diff}
\end{equation}
where the friction timescale of a particle of size $a$ in the Epstein regime is:

\begin{equation}
t_{\rm fric} = \frac{\rho_{\rm gr}}{\rho_{\rm gas}} \frac{a}{v_{\rm th}},
\end{equation}
and the thermal velocity of the gas $v_{\rm th}$ is:

\begin{equation}
v_{\rm th} = \sqrt{\frac{ 8k_{\rm B} T }{ \pi \mu m_{\rm H} }} = \sqrt{\frac{8}{\pi}} c_{\rm s}.
\label{eq:v_th}
\end{equation}
The temperature $T$ (or better, the sound speed $c_{\rm s}$) used in Eq.~\ref{eq:v_th} is derived from the flaring angle of the disk, which is an input parameter of the models. In particular, $c_{\rm s}=hR/\Omega$ (from the ideal gas law).

In this work, we assume that the dust structure has reached a steady-state in the vertical direction, that is when the gravitational force acting along the vertical direction and the aerodynamical drag caused by turbulent motions balance out \citep{2009A&A...496..597F}. We can therefore neglect the term on the left hand side of Equation \ref{eq:adv_diff}, and solve the advection-diffusion equation for $\rho_{\rm d}(R,z,a)$ with a 1D integration in $z$. We normalise each bin by requiring that:

\begin{equation}
\int_{\infty}^{\infty}{ \rho_{\rm dust}(R,z,a)dz } = \Sigma_{\rm dust}(R,a).
\end{equation}
We note that the ``canonical'' grain size distribution $f(R,z,a)$ is related to $\rho_{\rm dust}(R,z,a)$ by:
\begin{equation}
 \rho_{\rm dust}(R,z,a) = \frac{4}{3}\pi \rho_{\rm gr} a^3 f(R,z,a)da,
\end{equation}
where $f(R,z,a)da$ is the number of grains per unit volume, with size between $a$ and $a+da$. At this stage, we have obtained the dust density distribution for every size bin in every grid point of the disk. 

\rev{A limitation of our model is the underlying vertical gas density distribution (see Eq.~\ref{eq:gas_dens}), which implicitly assumes isothermality along the $z$-direction. As shown in this paper, and in all the literature computing the thermal balance of the gas phase material, the gas temperature is an increasing function of $z$, with the upper layers being more heavily irradiated by the central star. Thus, the disk upper layers are expected to ``puff up'' and be less dense than what we assume here, in order to reach hydrostatic equilibrium in the vertical direction. Moreover, smaller particles are expected to be stirred up more easily than what we assume, since Eq.~(\ref{eq:v_th}) considers the temperature derived from the flaring index (i.e. assuming vertical isothermality again), and does not take into account a vertical thermal gradient. In order to obtain a completely consistent density profile, one should} iterate \rev{between steps 1 and 6 of the diagram shown in Fig.~\ref{fig:diagram}} to obtain a vertical hydrostatic solution for \rev{both} the gas \rev{and} the dust density structure. \rev{At the moment, this is} computationally too expensive for these complex models. \rev{Moreover, such iterations may result in disks that are too strongly flared and/or unstable in the very inner regions \citep[e.g.][]{2009A&A...501..383W}}.

\subsection{Average grain size}
\label{sec:method_aver_a}

The grain size distribution affects both the opacities and the thermo-chemistry of the disk. To reduce the computational cost of such calculations, it is possible to define average quantities of the grain size distribution, where an ensemble of dust defined by these average properties has the same mass and total surface area as the original ensemble. In particular, following \citet{2011ApJ...727...76V}, the second and third moment of the grain size distribution can be defined as:
\begin{equation}
\langle a^2\rangle (R,z)= \int{a^2f(R,z,a)da};
\end{equation}
\begin{equation}
\langle a^3\rangle (R,z)= \int{a^3f(R,z,a)da}.
\end{equation}
The average grain size is then given by:
\begin{equation}
\bar{a} (R,z)= \frac{\langle a^3\rangle(R,z)}{\langle a^2\rangle(R,z)}.
\end{equation}
This approach employs an average size $\bar{a}$ that yields the same mass and total surface area of the original ensemble \citep{2011ApJ...727...76V}.

Dividing the total dust mass per cm$^{-3}$ $\rho_{\rm dust}(R,z)$ by the mean mass of one dust particle $(\bar{m}_{\rm gr}=\rho_{\rm gr}4\pi/3\times\bar{a}^3),$ we can obtain the number of grains per volume $n_{\rm gr}$. Defining the mean geometrical cross section as $\langle \sigma \rangle = \pi \bar{a}^2$, the total grain geometrical cross section per volume at an $(R,z)$ location in the disk is given by:

\begin{equation}
n_{\rm gr}\langle \sigma \rangle (R,z) = \frac{\rho_{\rm dust}(R,z)}{\rho_{\rm gr}} \frac{3}{4\bar{a}(R,z)},
\label{eq:n_gr_sigma}
\end{equation}
where
\begin{equation}
\rho_{\rm dust}(R,z) = \int_{a_{\rm min}}^{a_{\rm max}}{\frac{4}{3}\pi \rho_{\rm gr} a^3 f(R,z,a)da}.
\end{equation}

\subsection{Dust opacities and PAHs abundance}
\label{sec:method_opacities}

The grain size distribution is used to compute the dust opacities at every position in the disk. We have first computed a library of dust opacities containing the opacities $\kappa_\nu(a_i)$ for every bin size $a_i$. These opacities are computed using a standard ISM (interstellar medium) dust composition following \citet{2001ApJ...548..296W}, with an MNR \citep{1977ApJ...217..425M} grain size distribution between $a_i$ and $a_i + \Delta a_i$. The mass extinction coefficients are calculated using Mie theory with the \verb!miex! code \citep{2004CoPhC.162..113W} and optical constants by \citet{2003ApJ...598.1017D} for graphite and \citet{2001ApJ...548..296W} for silicates. This opacity library is then used to obtain the dust opacities in every grid point of the disk by mass averaging:

\begin{equation}
\kappa_\nu(R,z) = \frac{ \displaystyle\sum_i{\kappa_\nu(a_i) \rho_{\rm dust}(R,z,a_i) } }{\displaystyle\sum_i{\rho_{\rm dust}(R,z,a_i)}}.
\end{equation}
These opacities are implemented in the Monte-Carlo continuum radiative transfer module of DALI, as described in the Appendix of \citet{2012A&A...541A..91B,2013A&A...559A..46B}. A first stage of the Monte-Carlo method computes the dust temperature $T_{\rm dust}$, whereas a second stage computes the mean intensities over the entire spectrum, from UV to radio frequencies. 

In the radiative transfer, we also consider the opacity of PAHs (Polycyclic Aromatic Hydrocarbons). PAHs are also taken into account in the thermo-chemical modelling, where they can be important heating sources via photoelectric effect. In this work, the PAHs abundance is assumed to be $0.1$ of the typical ISM abundance in the whole disk \citep[see][]{2012A&A...541A..91B}, following observations suggesting a PAHs deficit in protoplanetary disks \citep[e.g.][]{2006A&A...459..545G,2010ApJ...714..778O}.

\subsection{Thermo-chemistry}

The thermo-chemical module of DALI is used to determine the chemical abundances, molecular and atomic excitations, and gas temperature $T_{\rm gas}$ with non-LTE (local thermodynamical equilibrium) calculations. The details are described in \citet{2012A&A...541A..91B,2013A&A...559A..46B}. Since some reaction rates in the thermal-chemical module do depend on the total dust surface area available, we apply small changes that take into account such dependence. The details of such rates, and their dependence on the total dust surface area, are explained below. After this final step, the synthetic emission maps of both continuum, and of specific molecular lines, can be obtained by using the ray-tracer described in \citet{2012A&A...541A..91B,2013A&A...559A..46B}.

\subsubsection{Gas-grain collisions}

Gas-grain collisions are important in setting the thermal coupling between dust and gas. We follow the prescription by \citet[][see their Appendix]{2004ApJ...614..252Y}, where the dust-gas energy transfer is expressed as:

\begin{eqnarray}
\nonumber \Lambda_{\rm gd} = 
& \displaystyle 2.85\times10^{-29}n_{\rm gas}^2\sqrt{\frac{T_{\rm gas}}{1000\,{\rm K}}} \Bigg[ 1-0.8\exp{\Bigg( -\frac{75\,{\rm K}}{T_{\rm gas}}} \Bigg) \Bigg] \times \\
& \displaystyle \frac{(T_{\rm dust}-T_{\rm gas})}{1000\,{\rm K}} \Bigg( \frac{S_{\rm d}}{6.09\times10^{-22}\,{\rm cm}^2} \Bigg)\ \ {\rm erg\,cm}^{-3}\,{\rm s}^{-1},
\end{eqnarray}
where $S_{\rm d}$ is the dust geometrical cross section per baryon, that is:

\begin{equation}
S_{\rm d} = \frac{n_{\rm gr}\langle\sigma\rangle}{n_{\rm gas}} \propto \frac{\delta_{\rm dg}}{\bar{a}}.
\label{eq:sigma_d}
\end{equation}
The direct proportionality to the local dust-to-gas ratio $\delta_{\rm dg}=\rho_{\rm dust}/\rho_{\rm gas}$ shows that the total dust cross section is higher for a larger amount of dust, whereas the inverse dependence on $\bar{a}$ indicates that a smaller average grain size increases the surface-to-mass ratio of the grain size distribution. Thus, a more top-heavy grain size distribution leads to a less effective thermal coupling between the gas and dust components. We note that usually DALI assumes $S_{\rm d} = \delta_{\rm dg}\times6.09\times10^{-20}\,{\rm cm}^2$ \citep{2004ApJ...614..252Y}.

\subsubsection{H$_2$ formation rate}

The H$_2$ formation rate on the surface of dust grains depends on the total dust surface area available. Following the prescription by \citet{2002ApJ...575L..29C,2002ApJ...577L.127C}, we use a recombination rate for H$_2$ equal to \citep[see also][]{1971ApJ...163..165H}:

\begin{equation}
k({\rm H}_2)=\frac{1}{2} n_{\rm H}^{-1} v_{\rm th} n_{\rm gr} \langle \sigma \rangle \epsilon_{{\rm H}_2} S_{\rm H}(T_{\rm dust}),
\end{equation}
where $n_{\rm H}$ is the abundance of H atoms in the gas phase, $\epsilon_{{\rm H}_2}$ is the H$_2$ recombination efficiency, and $S_{\rm H}(T_{\rm dust})$ is the sticking coefficient of H atoms on the grain surface, which depends on the dust temperature.

\subsubsection{Freeze-out, desorption, and hydrogenation}
\label{sec:freeze_out}

The other chemical mechanisms that depend on the dust surface area in the chemical network used in this paper are freeze-out (or adsorption), evaporation (or desorption), and grain-surface hydrogenation reactions.
For freeze-out, we follow \citet{2001A&A...378.1024C} and \citet{2009A&A...495..881V}, where the adsorption rate coefficient for a molecule $X$ can be expressed as:

\begin{equation}
k_{\rm ads}(X) = n_{\rm gr} \langle \sigma \rangle  \frac{v_{\rm th}}{\sqrt{M(X)}},
\label{eq:ads}
\end{equation}
where $M(X)$ is the molecular mass of species $X$.

Thermal desorption is treated similarly. Following \citet{2011A&A...534A.132V}, the thermal evaporation rate is prescribed as:

\begin{equation}
k_{\rm thdes} = 4 n_{\rm gr} \langle \sigma \rangle A(X) f(X) \exp{ \Bigg[ -\frac{E_{\rm b}(X)}{k_{\rm B}T_{\rm dust}} \Bigg] },
\end{equation}
where $E_{\rm b}(X)$ is the binding energy of species $X$. The most relevant molecule in this work is CO. The binding energy for pure CO ice is $\sim855\,$K \citep{1993ApJ...417..815S,2003ApJ...583.1058C,2005ApJ...621L..33O,2006A&A...449.1297B}, but it can be as high as $\sim1100-1500\,$K on H$_2$O ice or mixed with CO$_2$ ice \citep{2014A&A...564A...8M,2016ApJ...816L..28F}. We assume a binding energy for pure CO ice, $855\,$K, unless stated otherwise. The values of the pre-exponential factor $A(X)$ can be found in \citet{2011A&A...534A.132V}. The factor $f(X)$ sets the boundary between zeroth to first order desorption, where only one monolayer of ice is considered active in the evaporation process:

\begin{equation}
f(X) = \frac{1}{\max{(n_{\rm ice},\,N_{\rm ss}n_{\rm gr} \langle \sigma \rangle)}},
\end{equation}
where the number of binding sites per unit grain surface $N_{\rm ss}=8\times10^{14}$\,cm$^{-2}$ \citep[e.g.][]{2009ApJ...690.1497H} and $n_{\rm ice}$ is the sum of number densities of all individual ice species.

Photodesorption is modelled following again \citet[][see their eq. 6]{2011A&A...534A.132V}:

\begin{equation}
k_{\rm phdes} = n_{\rm gr} \langle \sigma \rangle f(X) Y(X) F_{\rm UV},
\end{equation}
where $F_{\rm UV}$ is the local value of UV flux (obtained from the radiative transfer module, see Sec.~\ref{sec:method_opacities}) and $Y(X)$ is the number of photodesorbed molecules per grain per incident UV photon. The values used in this paper are the same as in \citet{2011A&A...534A.132V}, and are taken from laboratory experiments \citep{2007ApJ...662L..23O,2009ApJ...693.1209O,2009A&A...496..281O}. More recent works on the subject can be found in \citet[][and references therein]{2016A&A...592A..67P}. To mimic photodesorption induced by cosmic rays in the dense optically thick regions of the disk, we assume a floor value for the local UV flux of $10^4\,$cm$^{-2}$\,s$^{-1}$ \citep{2004A&A...415..203S}. We do not include cosmic ray spot heating.

Finally, DALI includes a set of simplified grain-surface hydrogenation reactions, in particular to turn C, N, and O into CH$_4$, NH$_3$ , and H$_2$O. The simple prescription implemented is as follows:

\begin{equation}
k_{\rm hydro}(X) = n_{\rm gr} \langle \sigma \rangle \frac{v_{\rm th}}{\max{(n_{\rm hydro},\,N_{\rm ss}n_{\rm gr} \langle \sigma \rangle)}},
\end{equation}
where again the reactions are allowed only in the top monolayer and $n_{\rm hydro}$ represents the sum of the number densities of all ice species that can be hydrogenated; in this paper, C, CH, CH$_2$, CH$_3$, N, NH, NH$_2$, O, and OH. In freeze-out, evaporation, and hydrogenation rates, DALI usually assumes a typical grain size of $0.1\,\mu$m and a grain number density $n_{\rm gr}=10^{-12}n_{\rm H}$. Instead, in this paper $n_{\rm gr}$ and $\bar{a}$ are computed directly from the local grain size distribution.

With these new rates, the molecular abundances and excitations and gas temperature are computed. Using the ray-tracing module, the synthetic emission maps in both continuum and molecular and atomic lines can be obtained.

\subsection{Standard DALI models}
\label{sec:stn_models}

\begin{figure*}
\center
\includegraphics[width=.323\textwidth]{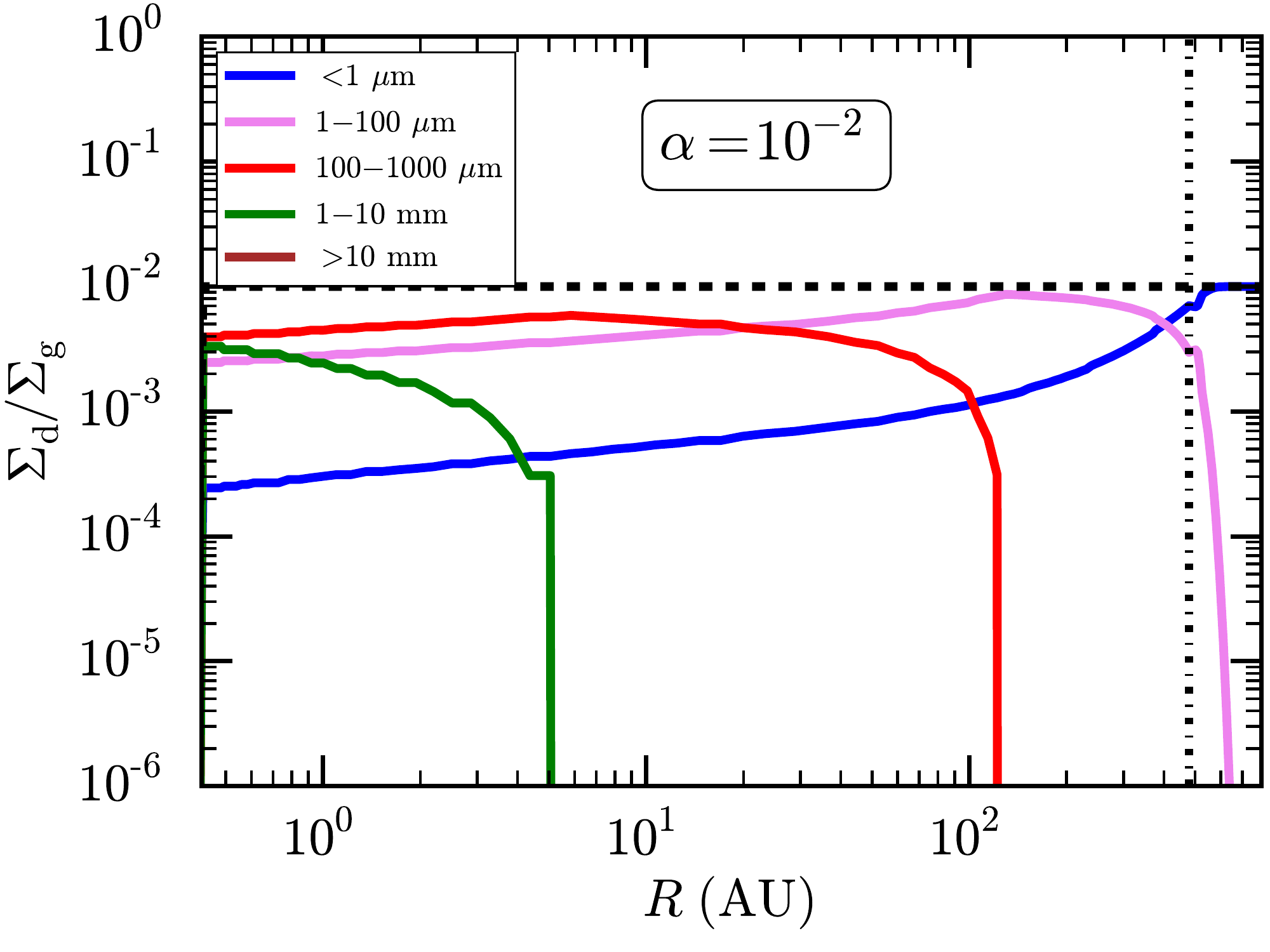}
\includegraphics[width=.323\textwidth]{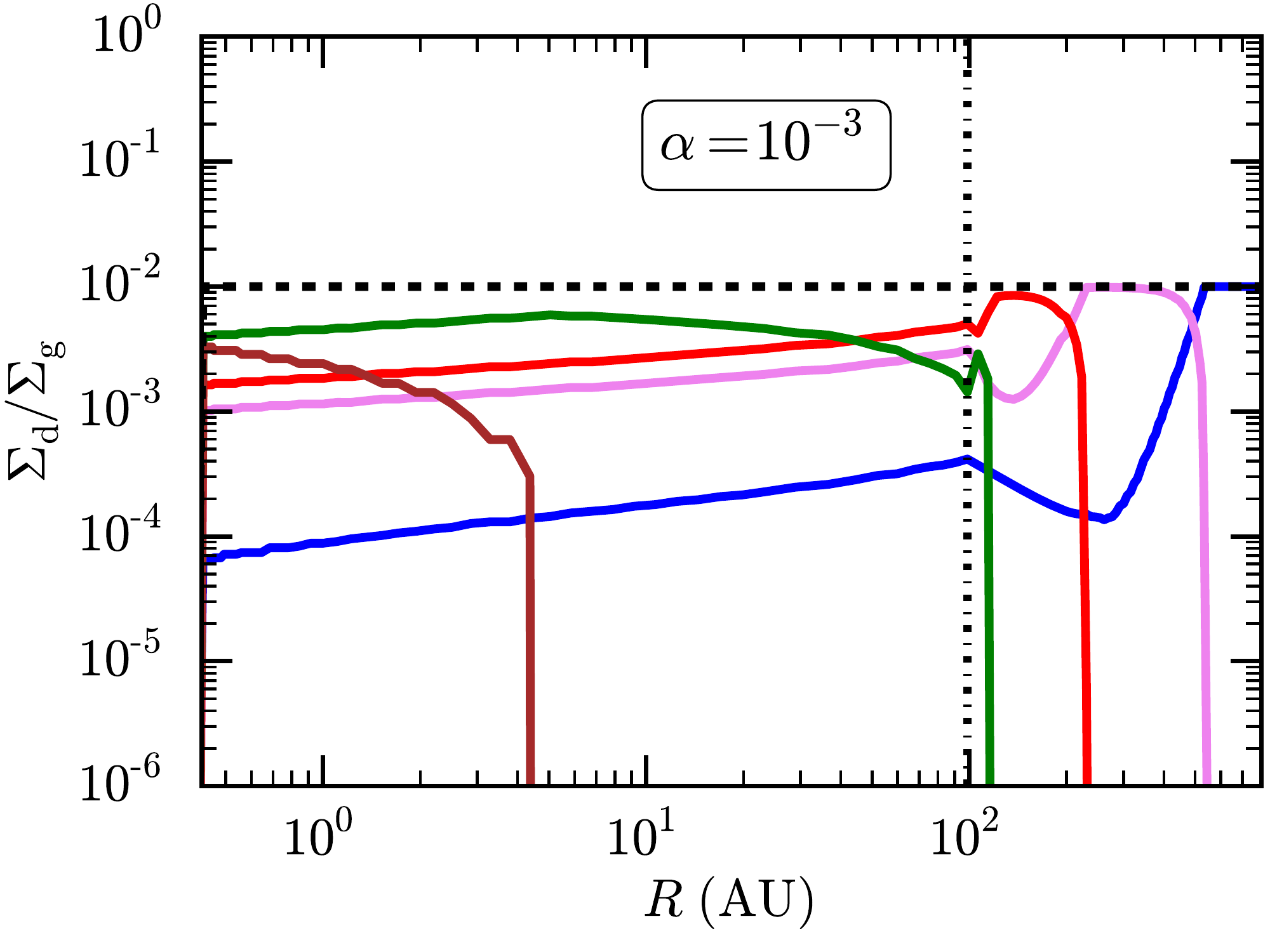}
\includegraphics[width=.323\textwidth]{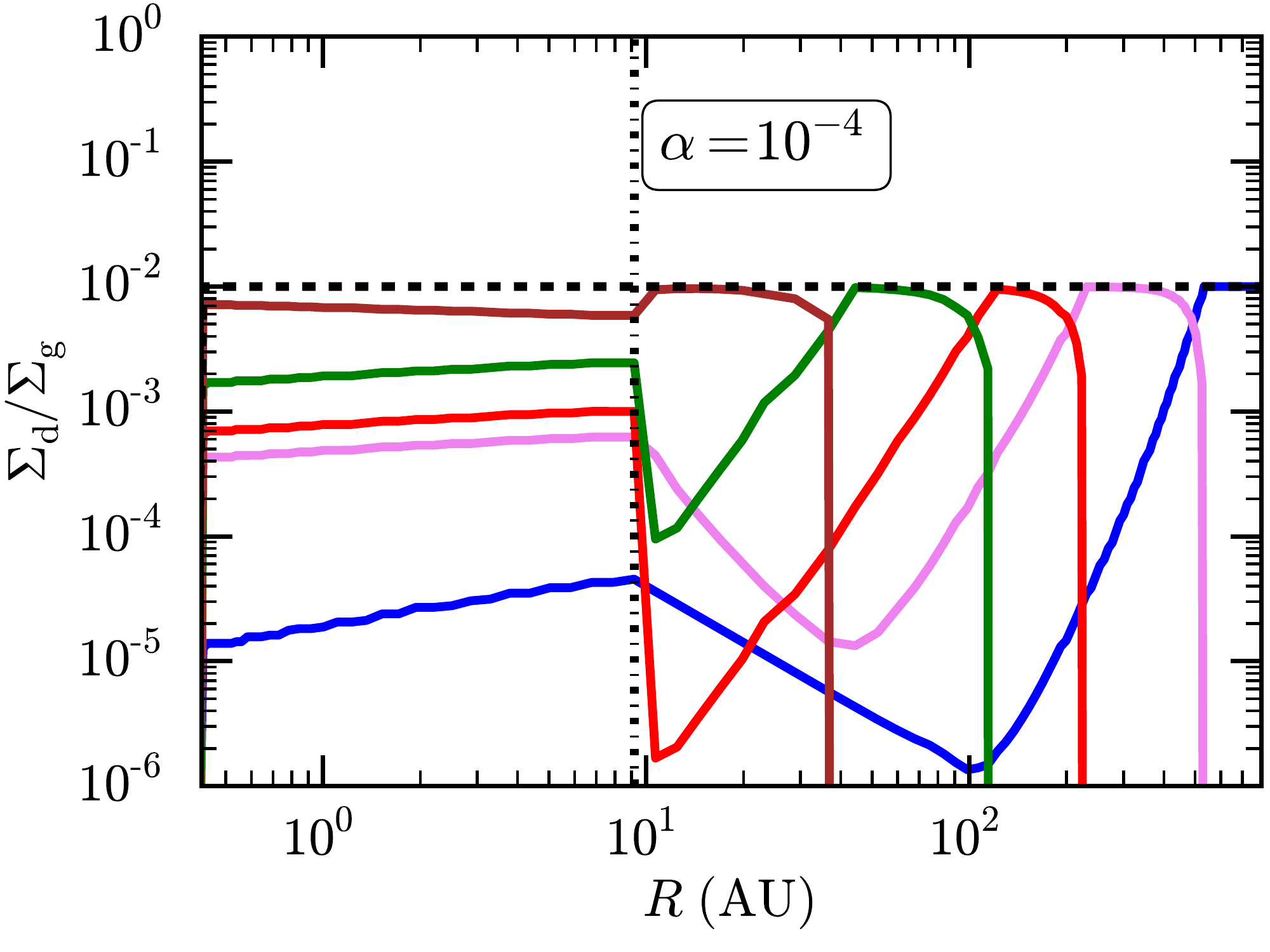}\\
\includegraphics[width=.33\textwidth]{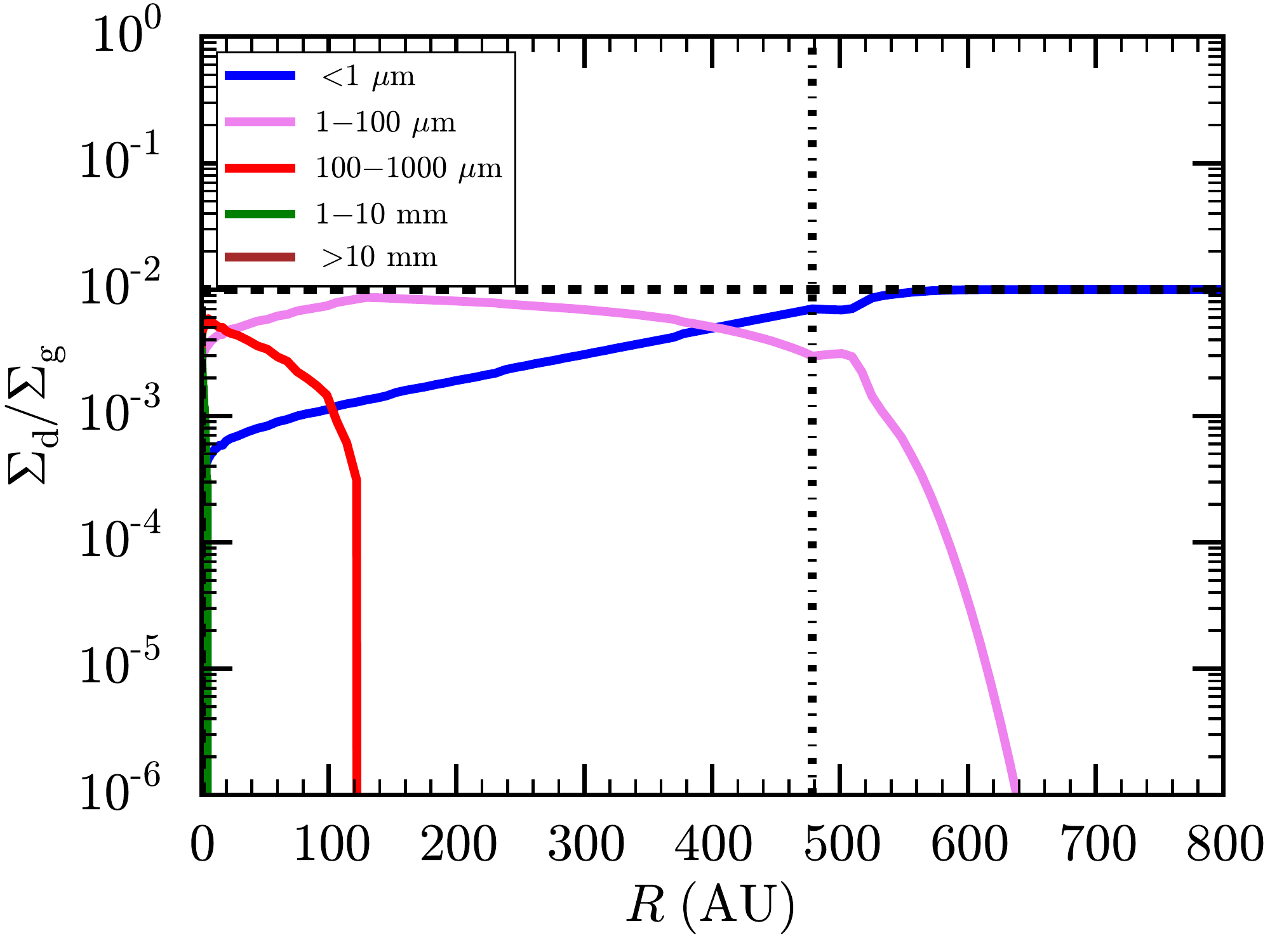}
\includegraphics[width=.33\textwidth]{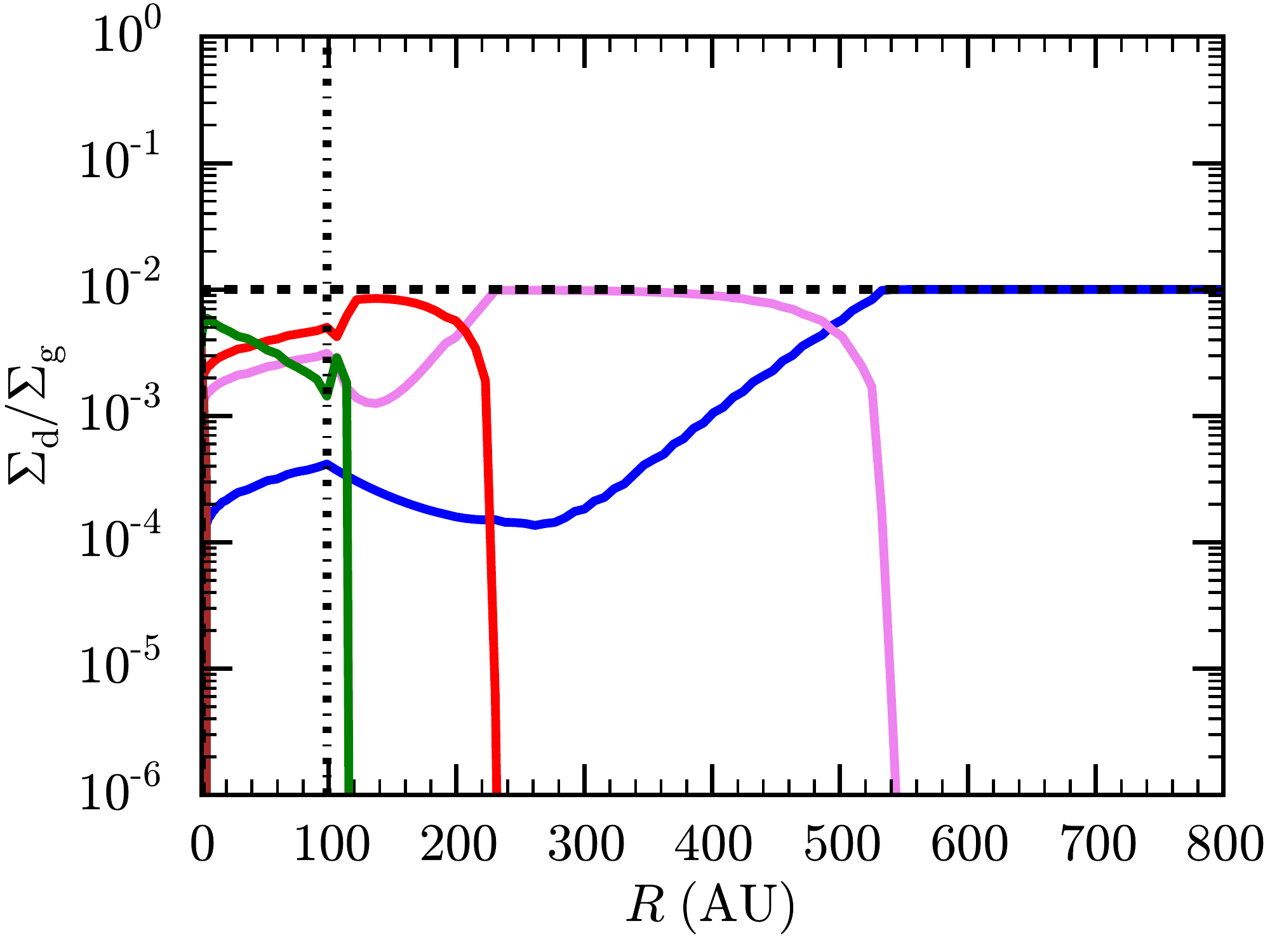}
\includegraphics[width=.33\textwidth]{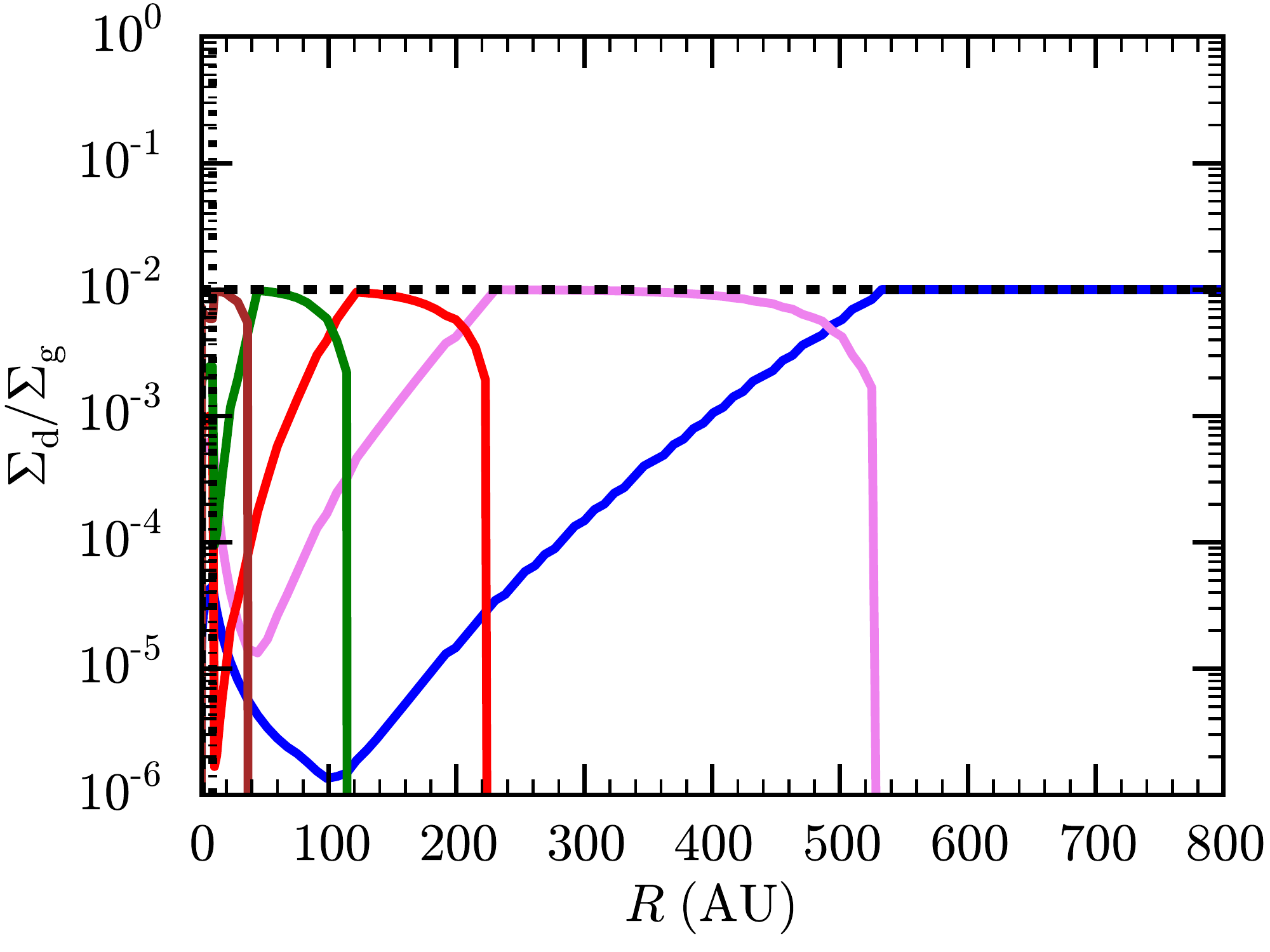}\\
\caption{Grain size distribution for $\alpha=$$10^{-2}$, $10^{-3}$ and $10^{-4}$, from left to right, respectively. The two rows show the same plots in two different radial scales. The dashed horizontal line is the total dust-to-gas ratio $\Delta_{\rm dg}=0.01$, whereas the dashed-dotted vertical line indicates the fragmentation radius $R_{\rm frag}$.}
\label{fig:d2g_vs_r}
\end{figure*}

Since the implementation of a more realistic grain size distribution into DALI requires a few changes, we will compare the results of this new methodology with similar results from the standard DALI code to better estimate the effects of grain growth and radial drift onto the chemistry and emission of a few molecular lines. To mimic grain growth and settling, DALI usually considers two populations of dust grains, with sizes ranging between $50\,\AA$ and $1\,\mu$m for small grains and ranging between $50\,\AA$ and $1\,$mm for large grains \citep[as in][]{2006ApJ...638..314D}. The scale height for the small population is the same as for the gas, whereas the scale height for the large population is reduced to $\chi h$, where $\chi<1$. The mass ratio between the large and small populations is defined as $f$.

To be more specific, these standard parametric models have a gas density distribution as defined in Eq.~\ref{eq:gas_dens}, but a dust density distribution of:

\begin{equation}
\rho_{\rm dust,small} (R,z)= \frac{(1-f)\Delta_{\rm dg}\Sigma_{\rm gas}(R)}{ \sqrt{2\pi} R h } \exp { \left[ -\frac{1}{2} \left( \frac{z}{Rh} \right)^2 \right] };
\end{equation}

\begin{equation}
\rho_{\rm dust,large} (R,z)= \frac{f\Delta_{\rm dg}\Sigma_{\rm gas}(R)}{ \sqrt{2\pi} R \chi h } \exp { \left[ -\frac{1}{2} \left( \frac{z}{R\chi h} \right)^2 \right] }.
\end{equation}
We will compare these standard models with the more complete ones including dust evolution.

\section{Models setup}
\label{sec:setup}

As our representative model, a model using a gas density distribution that resembles early ALMA observations of the HD 163296 system is considered. We stress here that this paper is not aimed to model specifically the HD 163296 system. However, we consider it as a good representative case, since \citet{2013A&A...557A.133D} have shown that the dust and gas emission cannot be modelled simultaneously by a tapered outer disk surface density profile. Moreover, the disk is bright and large enough that resolved intensity profiles have been obtained for both continuum and CO isotopologues. We thus take disk parameters that have been used by \citet{2013A&A...557A.133D} to fit the Science Verification ALMA data. Other parametrisations for the disk structure have been used \citep[e.g.][]{2013ApJ...774...16R,2016ApJ...830...32W}, but we will discuss them more thoroughly in Sec.~\ref{sec:results_hd16}. In particular, we use $R_{\rm c}=125\,$AU, $\gamma=0.9$, $\Sigma_{\rm c}=7.03\,$g\,cm$^{-2}$, $h_{\rm c}=0.11$, and $\psi=0.25$. These parameters yield a disk gas mass $M_{\rm disk}\sim7\times10^{-2}\,M_\odot$. The gas density structure used in all models is shown in Fig.~\ref{fig:n_gas}. The total dust-to-gas mass ratio $\Delta_{\rm dg}$ is set to $0.01$. We use a stellar mass of $2.47\,M_\odot$, with a bolometric luminosity $L_*=38\,L_\odot$, and an effective temperature of $10^4\,$K \citep{2012A&A...538A..20T}. The X-ray luminosity is set to $L_X=6\times10^{29}\,$erg\,s$^{-1}$ \citep{2009A&A...494.1041G}. Finally, we use an accretion rate of $4.5\times10^{-7}\,M_\odot$\,yr$^{-1}$ \citep{2013ApJ...776...44M}.

All disk models are represented with a 2D $(R,z)$ grid, with $150$ grid points in the radial direction and $80$ points in the vertical direction. The first $50$ radial points are logarithmically sampled between $R_{\rm in}$ and $25$\,AU. We set an inner disk radius $R_{\rm in}=0.42\,$AU, since the dust sublimation radius is $\sim 0.07\sqrt{L_*/L_\odot}$, if a dust sublimation temperature of $\sim 1500\,$K \citep{2001ApJ...560..957D} is assumed. The grid is then sampled linearly between $25$ and $800\,$AU. The vertical grid is linear and samples the disk between $0$ and $6$ local scale heights.

As standard reference models using standard DALI, we define two models, which we label STN (which stands for ``standard'') and STN-SM, where the main difference between the two is that the second one does not have grains larger than $1\,\mu$m in the disk. For model STN, we consider $f=0.85$ and a settling parameter $\chi=0.2$ (see Section~\ref{sec:stn_models}). Instead, for model STN-SM, we set $f=0$, that is, all the dust is in the population of small grains. In these reference models, we compute the dust opacities as described in Sec~\ref{sec:method_opacities}, with just two large size bins. The rather unrealistic STN-SM model is used mostly to determine the thermal structure of a hypothetical disk that did not witness grain growth yet, in order to compare chemical timescales with grain growth timescales in Section~\ref{sec:discussion_timescales}.

\begin{figure*}
\center
\includegraphics[width=.33\textwidth]{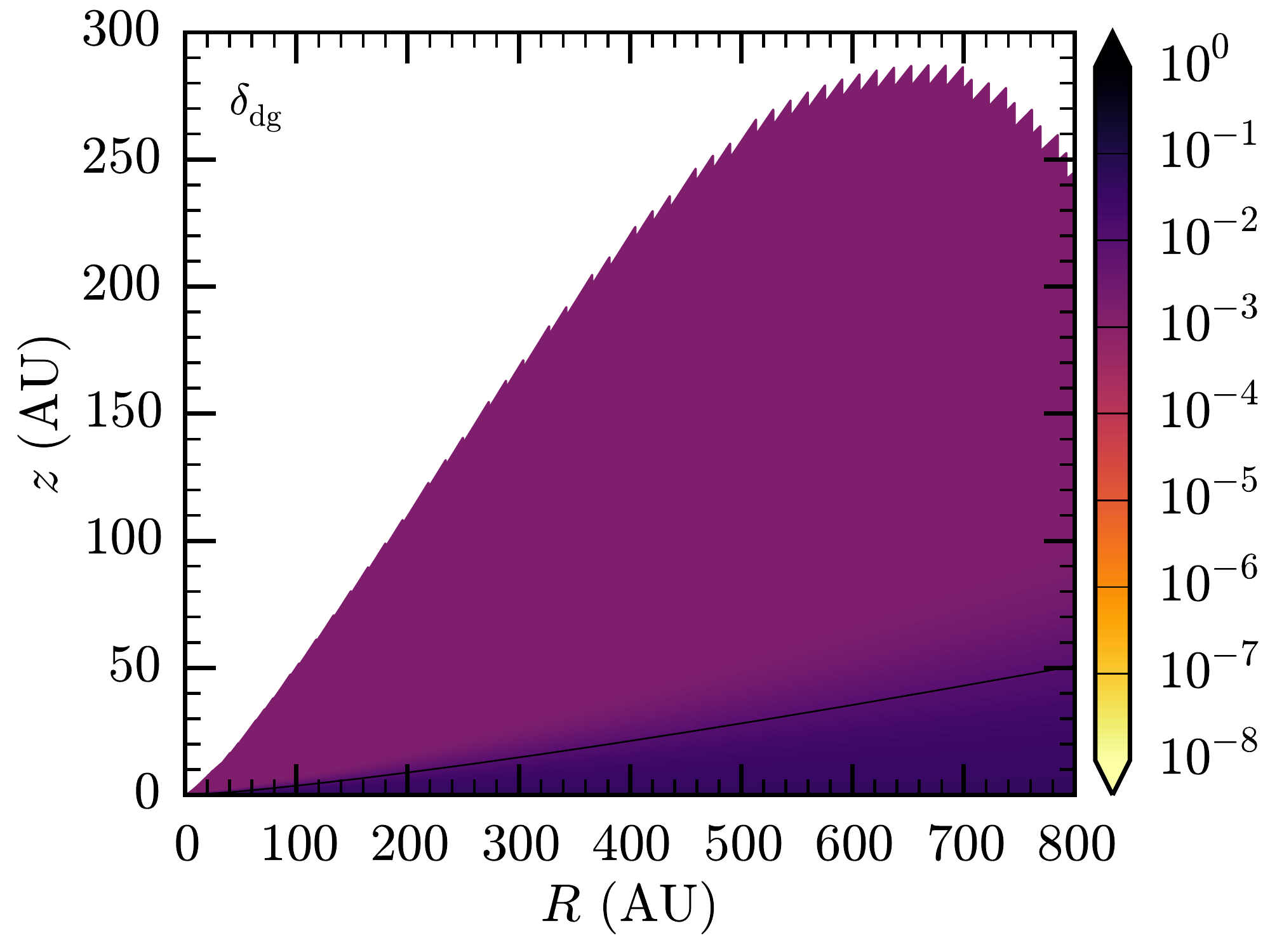}
\includegraphics[width=.33\textwidth]{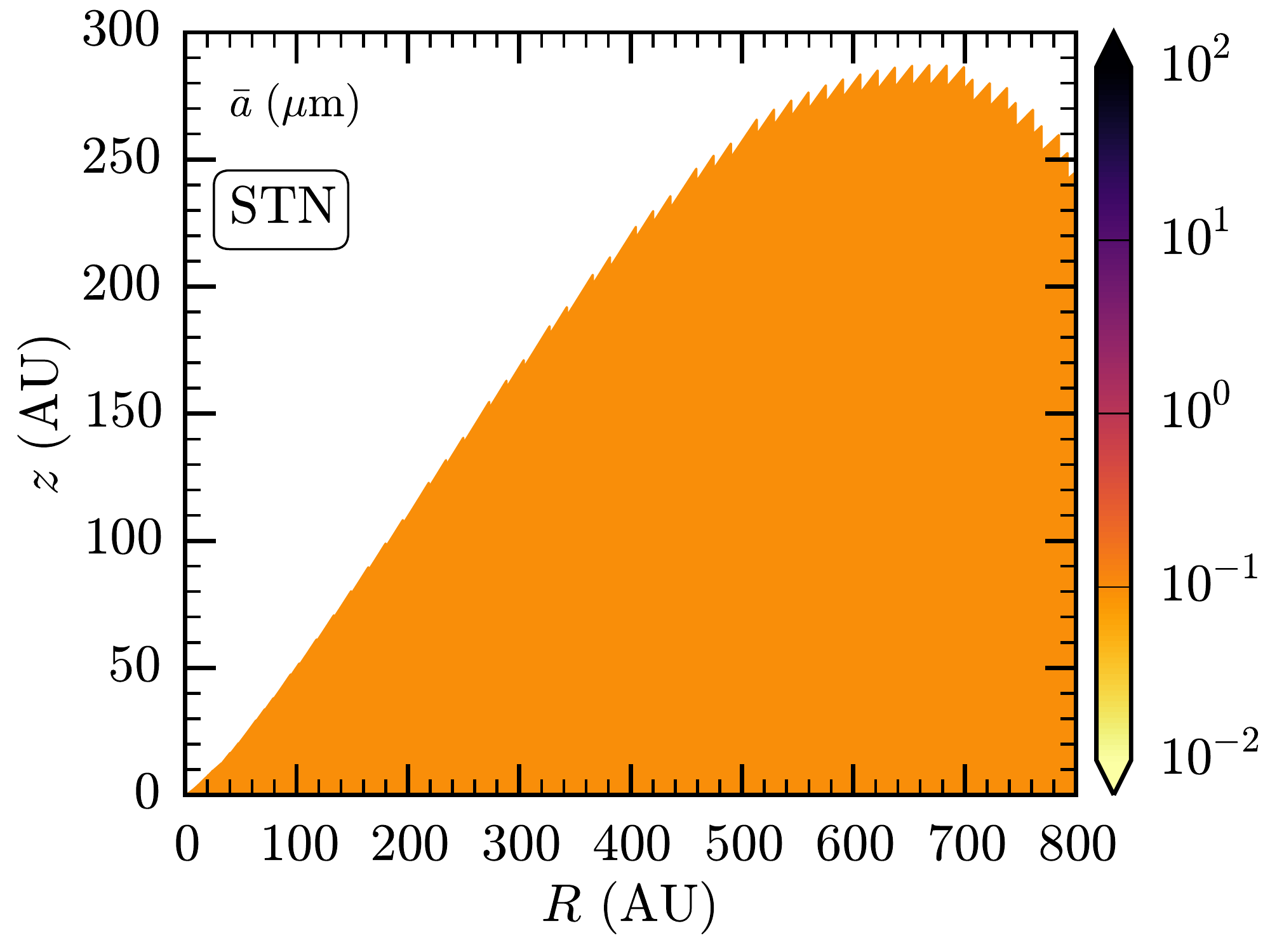}
\includegraphics[width=.33\textwidth]{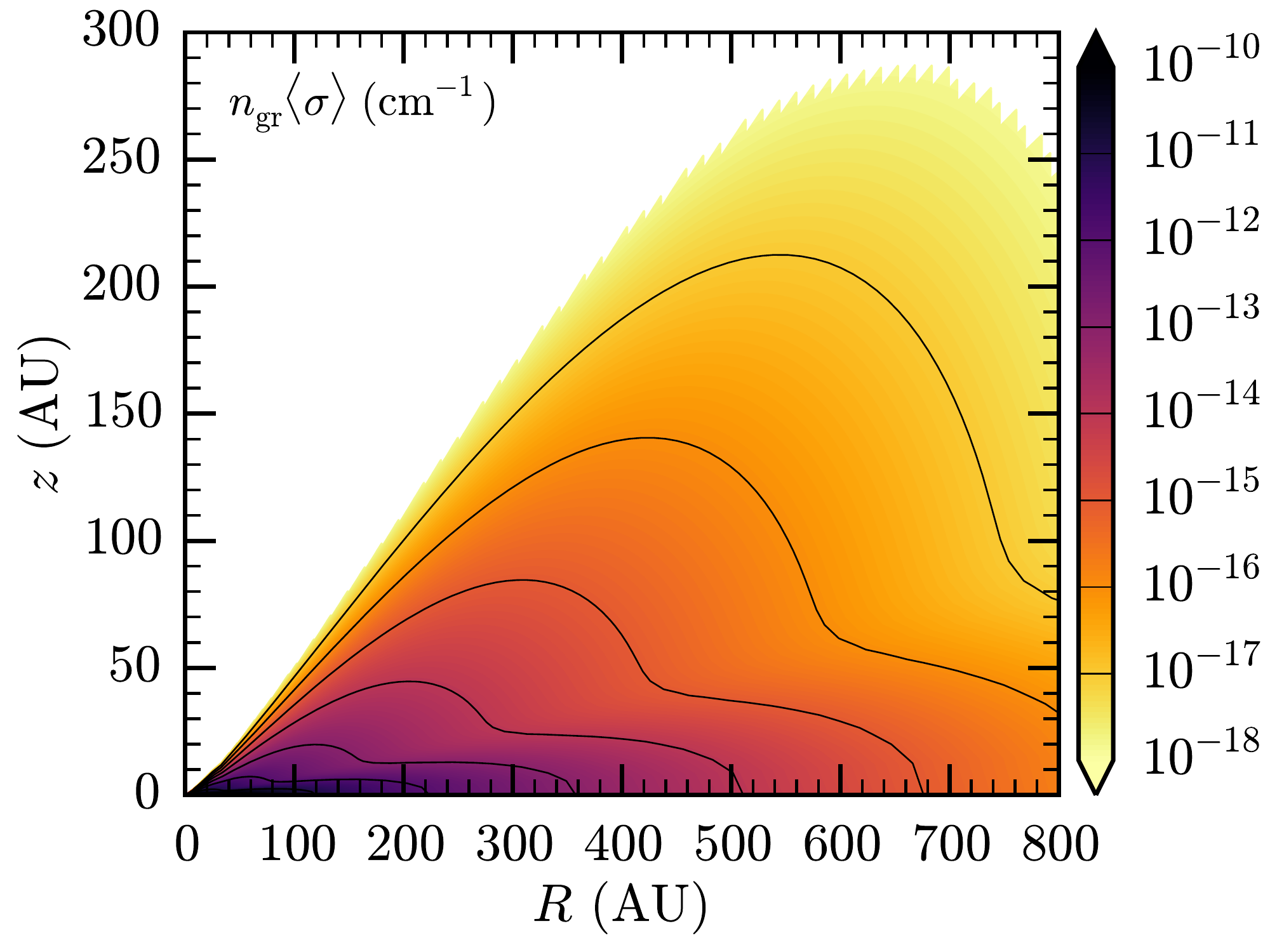}\\
\includegraphics[width=.33\textwidth]{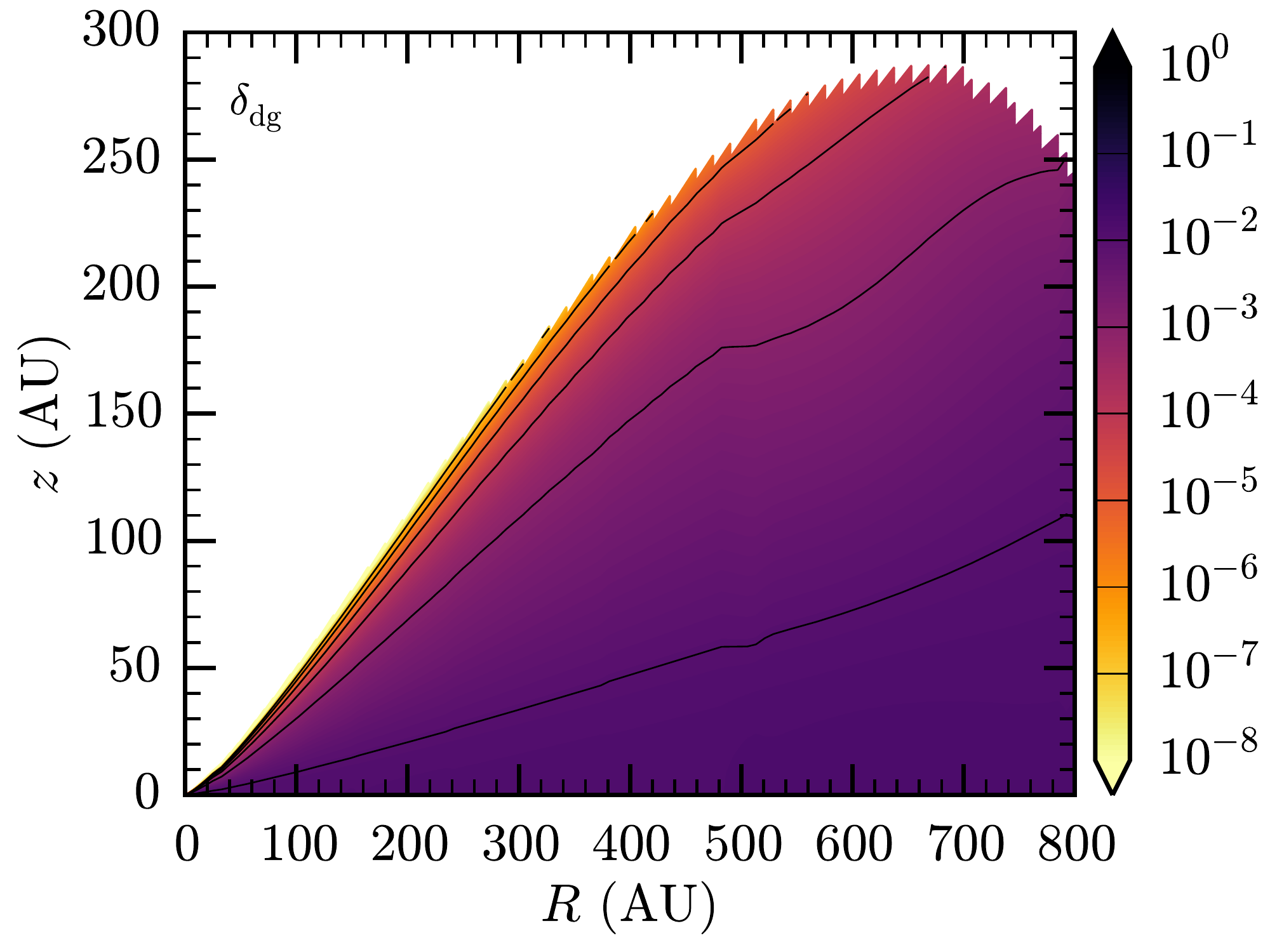}
\includegraphics[width=.33\textwidth]{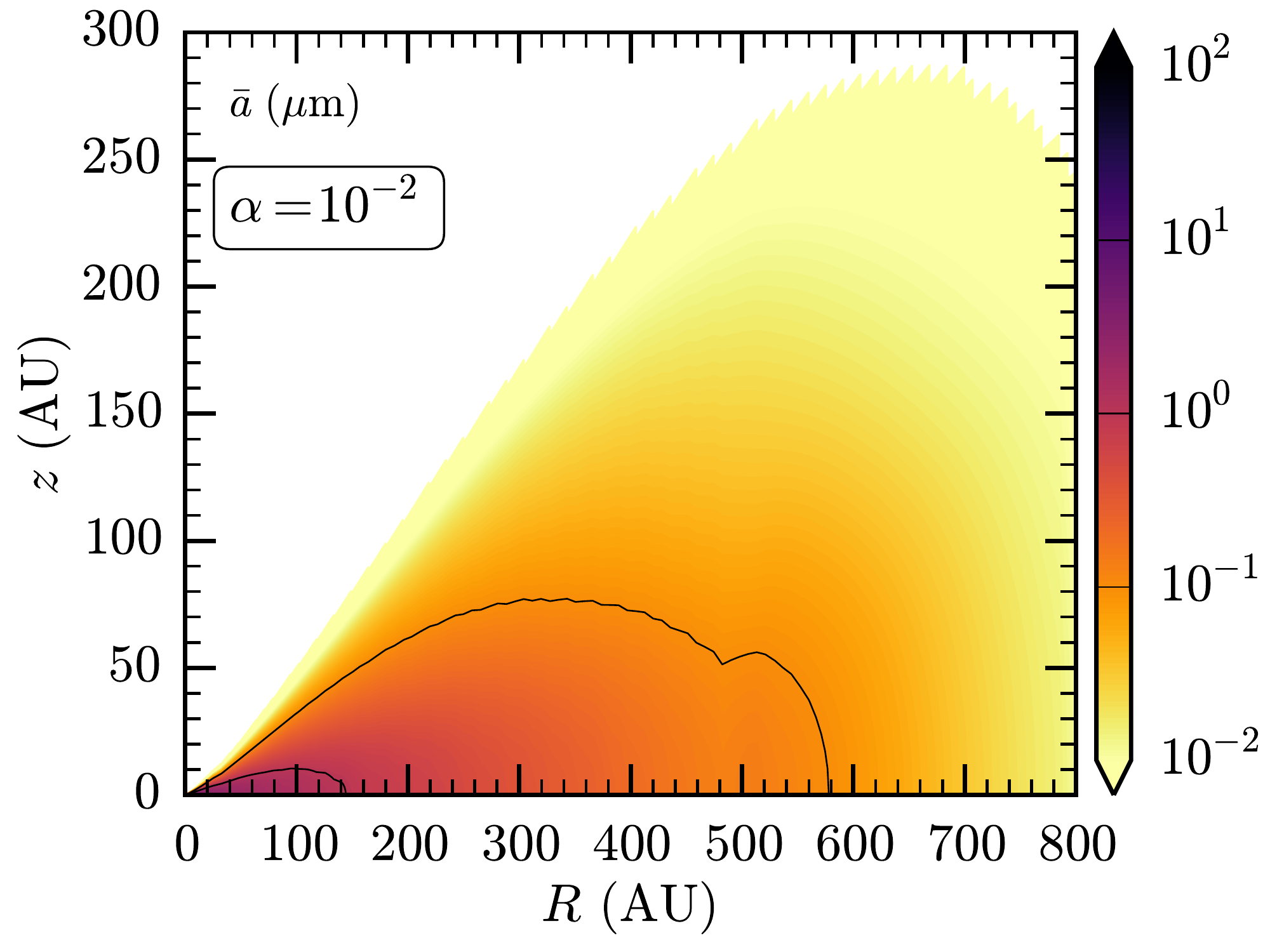}
\includegraphics[width=.33\textwidth]{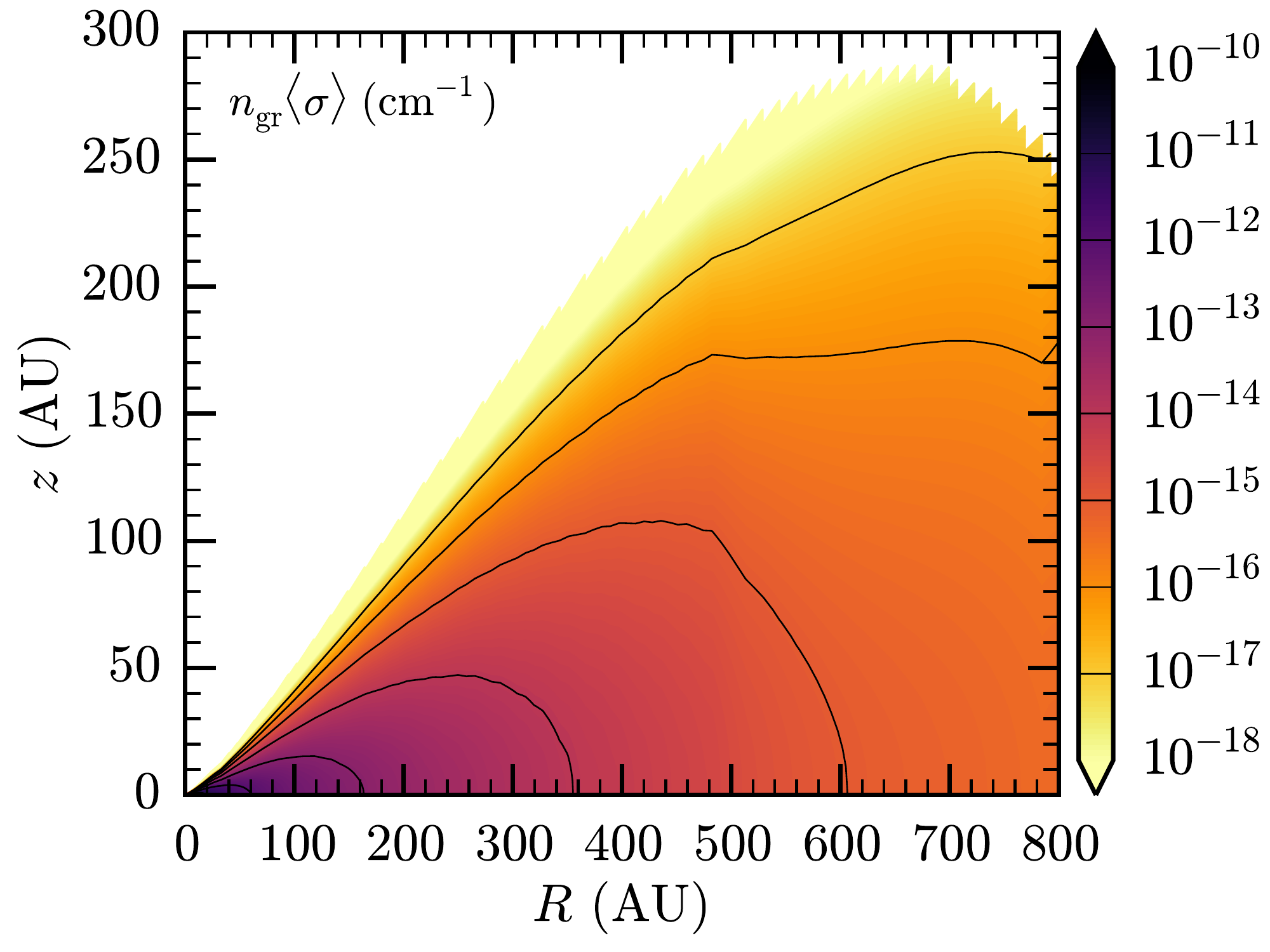}\\
\includegraphics[width=.33\textwidth]{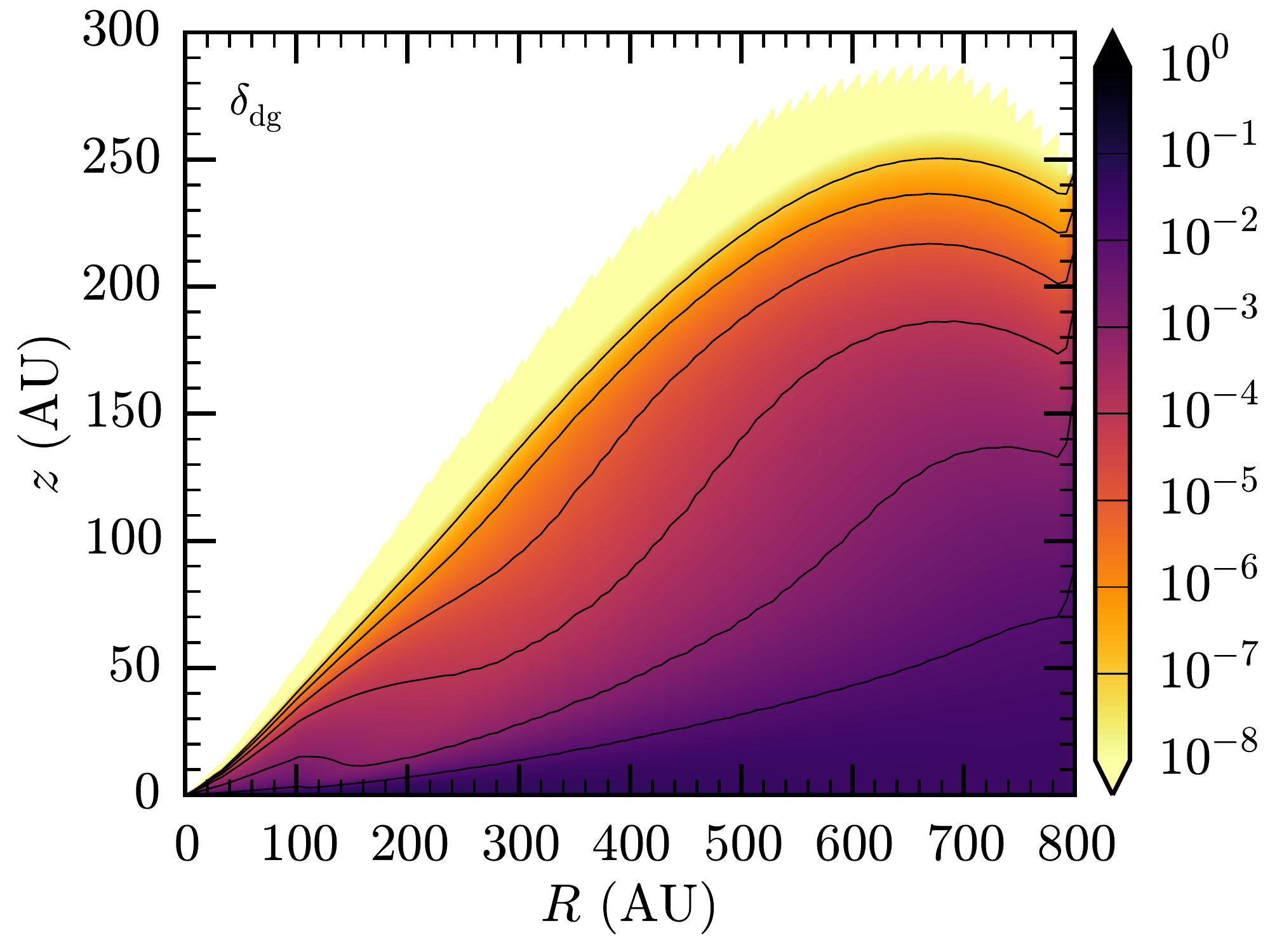}
\includegraphics[width=.33\textwidth]{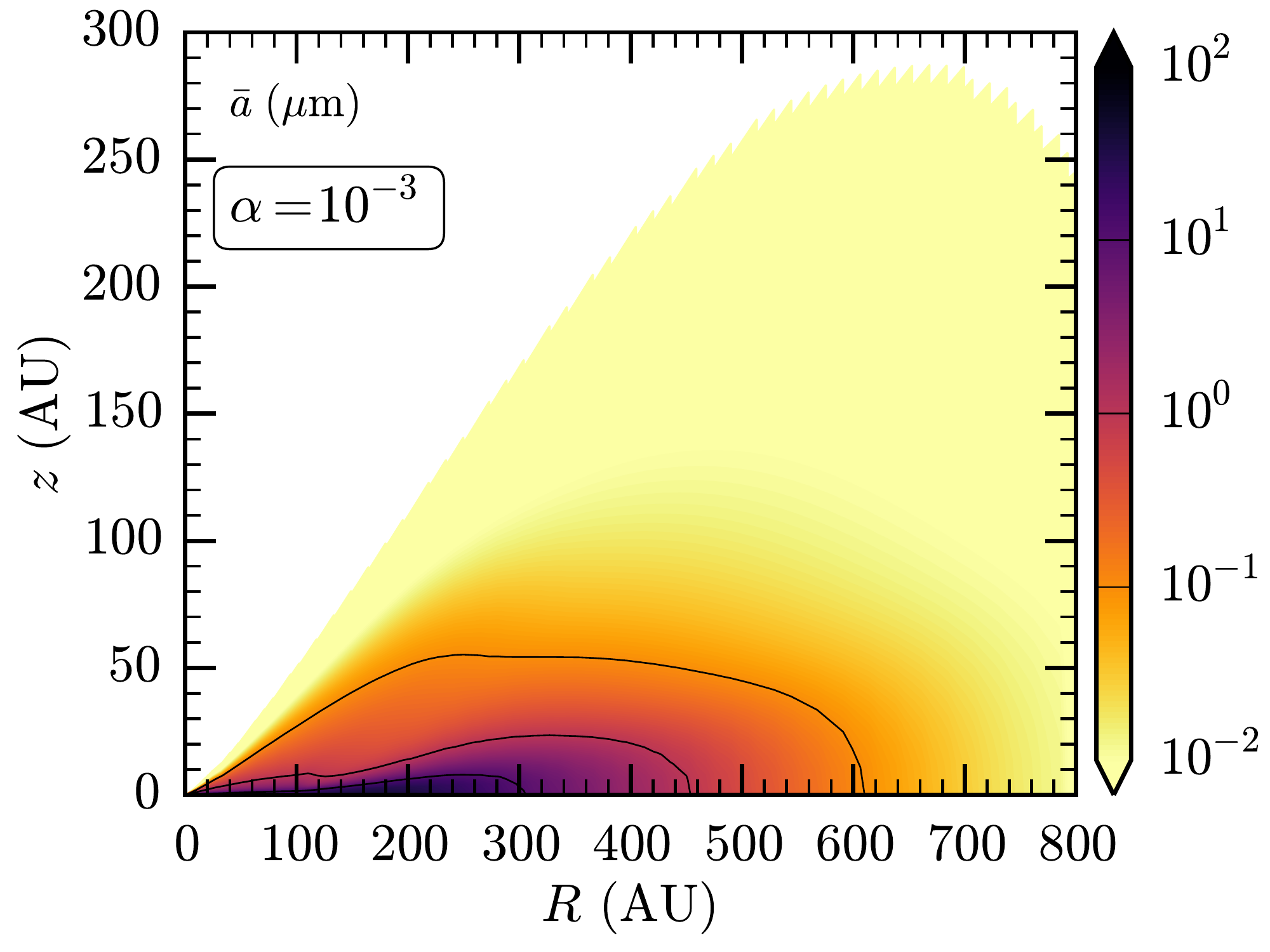}
\includegraphics[width=.33\textwidth]{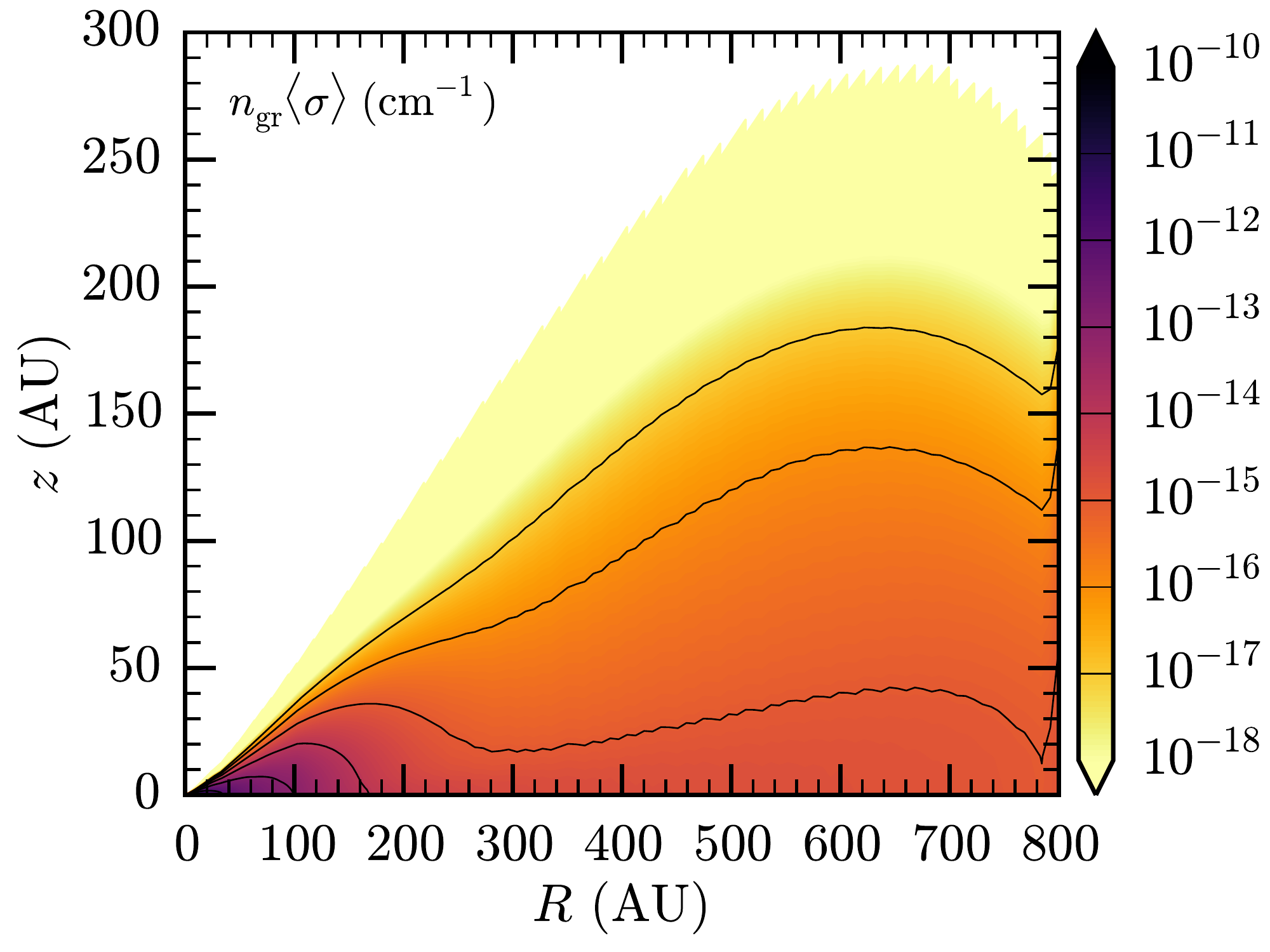}\\
\includegraphics[width=.33\textwidth]{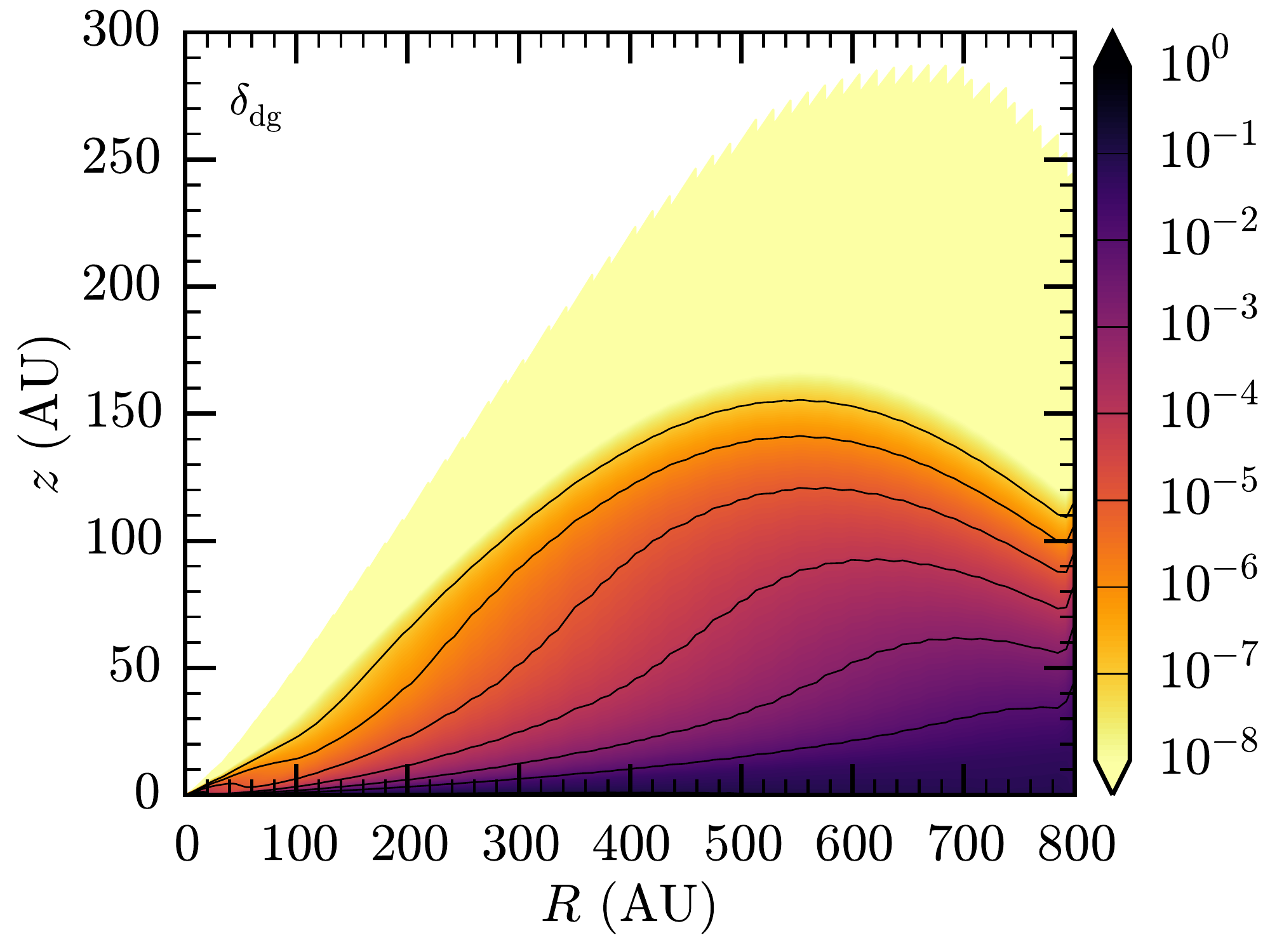}
\includegraphics[width=.33\textwidth]{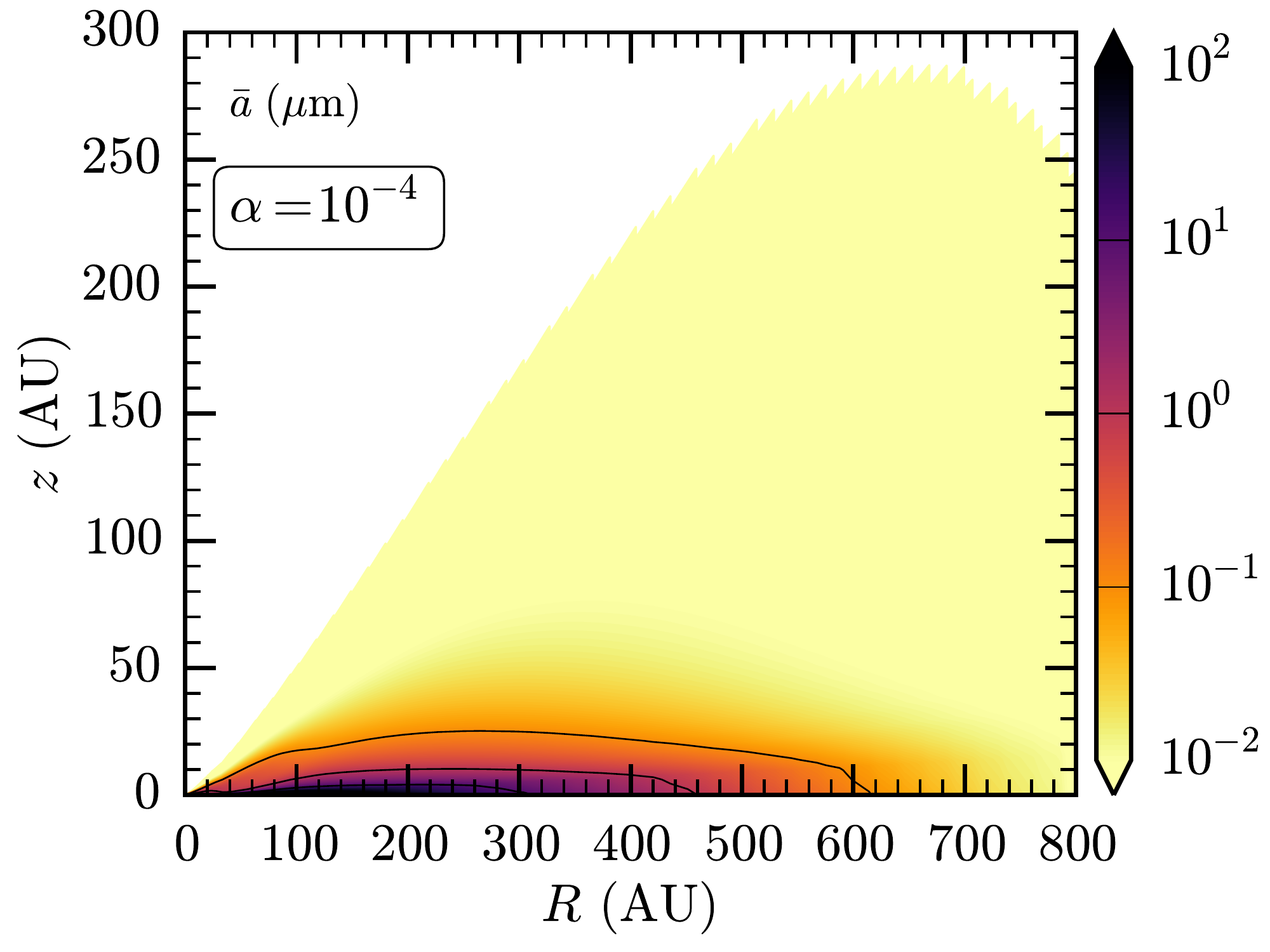}
\includegraphics[width=.33\textwidth]{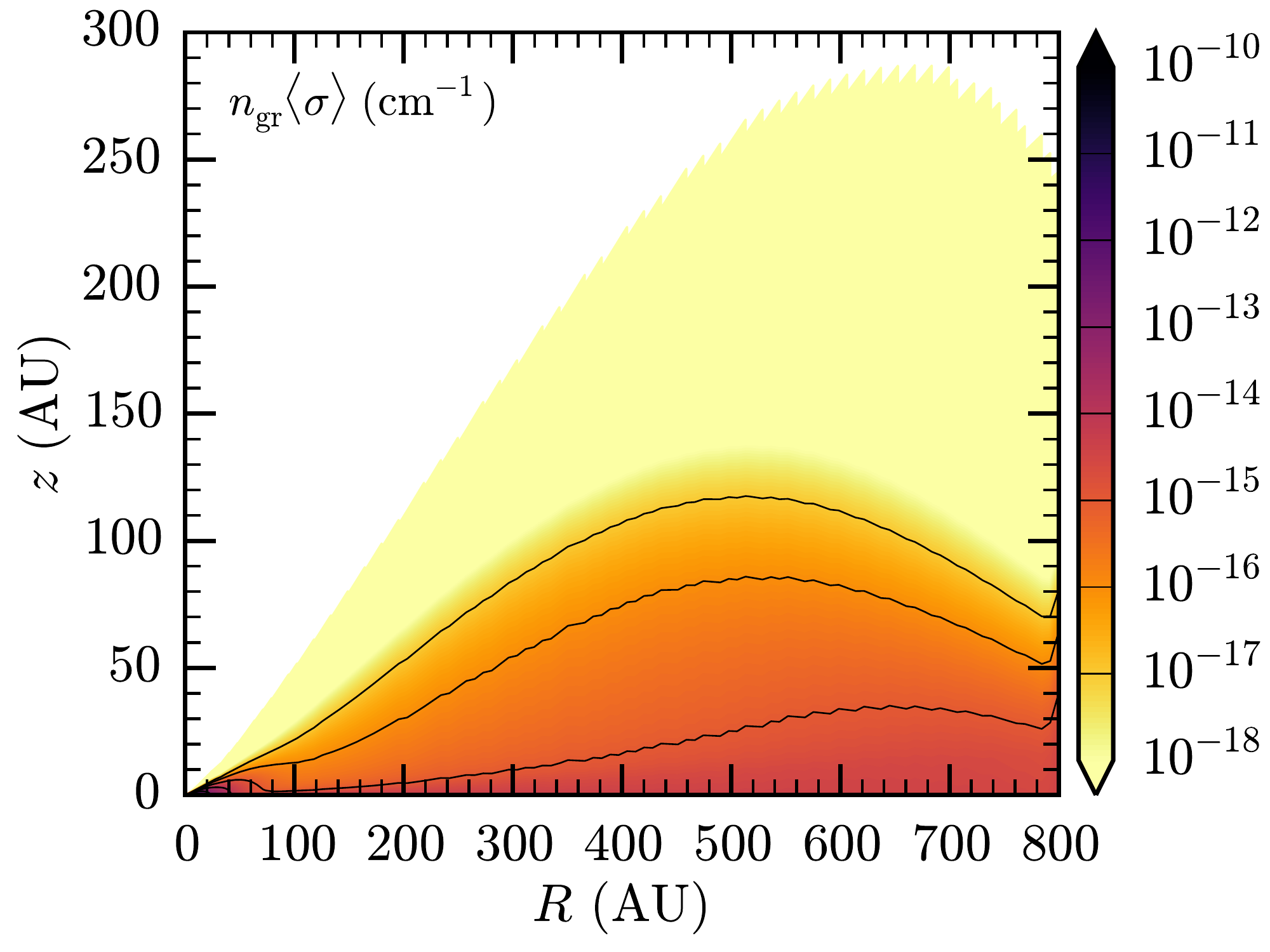}\\
\caption{From top to bottom: STN, $\alpha=10^{-2}$, $10^{-3}$ and $10^{-4}$ models. From left to right: local dust-to-gas ratio $\delta_{\rm dg}$, average grain size $\bar{a,}$ and total dust surface area per volume $n_{\rm gr}\langle \sigma \rangle$ used in the thermo-chemical module. Contour levels are indicated in the colour bar.}
\label{fig:vert_settling}
\end{figure*}

For the models including dust evolution, the three new parameters defining the dust distribution are $\alpha$, $v_{\rm frag}$ , and $\rho_{\rm gr}$. As noted above, $v_{\rm frag}=10\,$m\,s$^{-1}$ and $\rho_{\rm gr}=2.5\,$g\,cm$^{-3}$. Here we focus on exploring the dependence of the dust and gas properties on turbulence, by assigning $\alpha$ the values of $10^{-2}$, $10^{-3}$ and $10^{-4}$, that is, ranging between high and low values of disk viscosity \citep[e.g.][where the latter is a recent theoretical review]{1998ApJ...495..385H,2014prpl.conf..411T}. We note that the first tentative measurement of disk turbulence by \citet{2011ApJ...727...85H}, \citet{2015ApJ...813...99F} and \citet{2016A&A...592A..49T} seem to indicate that turbulence is quite low ($\alpha\lesssim10^{-3}$), at least in the outer disks.

In all the models we assume other parameters that are relevant for the calculations performed for this paper. In particular, in the radiative transfer module in the first stage we use $3\times10^7$ photon packages for both the star and the environment radiation, which is assumed to be $1\,G_0$, where $G_0\sim2.7\times10^{-3}\,$erg\,s$^{-1}$\,cm$^{-2}$ is the UV interstellar radiation field between $911\,\AA$ and $2067\,\AA$ \citep{1978ApJS...36..595D}. The number of photon packages used in stage 2 is $3\times10^6$ in every wavelength bin \citep[see][for more details]{2013A&A...559A..46B}. The number of photon packages in both stages has been chosen such that the dust temperature and the intensity field are smooth functions in the spatial grid specified above.

In the thermo-chemical module, we assume initial ISM-like abundances, with $[{\rm C}]/[{\rm H}]=1.35\times10^{-4}$ and $[{\rm O}]/[{\rm H}]=2.88\times10^{-4}$, where notation $[X]$ indicates element $X$ in all its volatile forms (i.e. not locked up in refractory dust). We do not consider CO isotope selective photodissociation \citep[e.g.][]{1988ApJ...334..771V}, since the mass of the disk and the stellar mass used in these models are high enough that this effect should not be too significant \citep{2014A&A...572A..96M,2016A&A...594A..85M}. The assumed cosmic ray ionisation rate is $\zeta_{\rm CR}=5\times10^{-17}\,$s$^{-1}$. The calculations are performed in time-dependent mode, for a timescale of $1\,$Myr. The upper CO emitting layers of disks reach chemical equilibrium in $<1\,$Myr. The same timescale is used in the dust evolution calculations, where, however, the grain size distribution reaches a steady-state in $\sim2\times10^5\,$yr.

Finally, the ray tracing assumes a distance that is comparable to that of the HD 163296 system, $122$\,pc \citep{1998A&A...330..145V}. The synthetic observables are ray traced assuming a disk inclination of $45^\circ$, with a beam convolution of $0.52\arcsec\times0.38\arcsec$, where the position angle of the beam is $82^\circ$ \citep{2013A&A...557A.133D}. The position angle of the major axis of the disk is taken as $137^\circ$. 

\section{Results}
\label{sec:results}

In this Section we show the results of the models, focusing in particular on the dust properties and on the effects that these have on the chemical abundances and excitation of the main CO isotopologues, $^{12}$CO, $^{13}$CO, and C$^{18}$O.

\subsection{Dust density structure and average grain size}
\label{sec:results4_1}

The three values of turbulence induce important differences in the radial dependence of the grain size distribution (see Fig.~\ref{fig:d2g_vs_r}). In the inner regions, lower viscosities lead to larger grain sizes since turbulent velocity decreases with $\alpha$. Moreover, for low viscosities, most of the grain size distribution is limited by radial drift and not by fragmentation. This is apparent by looking at the location of the fragmentation radius $R_{\rm frag}$ for the three different cases, where $R_{\rm frag}$ varies between $\sim10$\,--\,$500$\,AU with $\alpha$ ranging between $10^{-4}$ and $10^{-2}$ (in fact $R_{\rm frag}$ is expected to be roughly linear with turbulence parameter, see Eq.~\ref{eq:r_frag}). In the outer regions, where the gas surface density exponentially decays, higher viscosities lead to grain size distributions which are less bottom-heavy. \rev{The reason is that in the outer exponential tail of the surface density profile, even the smallest grains have a Stokes number $\lesssim1$, and thus all grains reside in the smallest size bin, being the maximum grain size set by the radial drift limit. However, as viscosity gets higher, some larger grains from smaller radii are diffused outwards, leading to a more top-heavy grain size distribution in the outer regions of the disk when compared to the one resulting from lower viscosities.} As we will show, this is important in setting the penetration depth of UV photons into the outer disk. All three models show a quite sharp transition between mm-size particles and micron-size grains at $\sim200\,$AU. Such a transition becomes sharper for lower viscosities, since the maximum grain size attained at every radius becomes a steeper function of the distance from the star. Interestingly, in all models there is a significant depletion of small grains just outside the fragmentation radius. As explained in \citet{2015ApJ...813L..14B}, this occurs because the relative motions between dust particles are not high enough to induce fragmentation. Thus, particles grow to large enough sizes that they radially drift inwards and the small particles are not replenished efficiently. This is in fact a possible origin of gaps in scattered light images, as detailed in \citet{2015ApJ...813L..14B}. Such a gap in scattered light should be correlated to a possibly small enhancement of the opacity at (sub-)mm wavelengths.

\begin{figure}
\center
\includegraphics[width=\columnwidth]{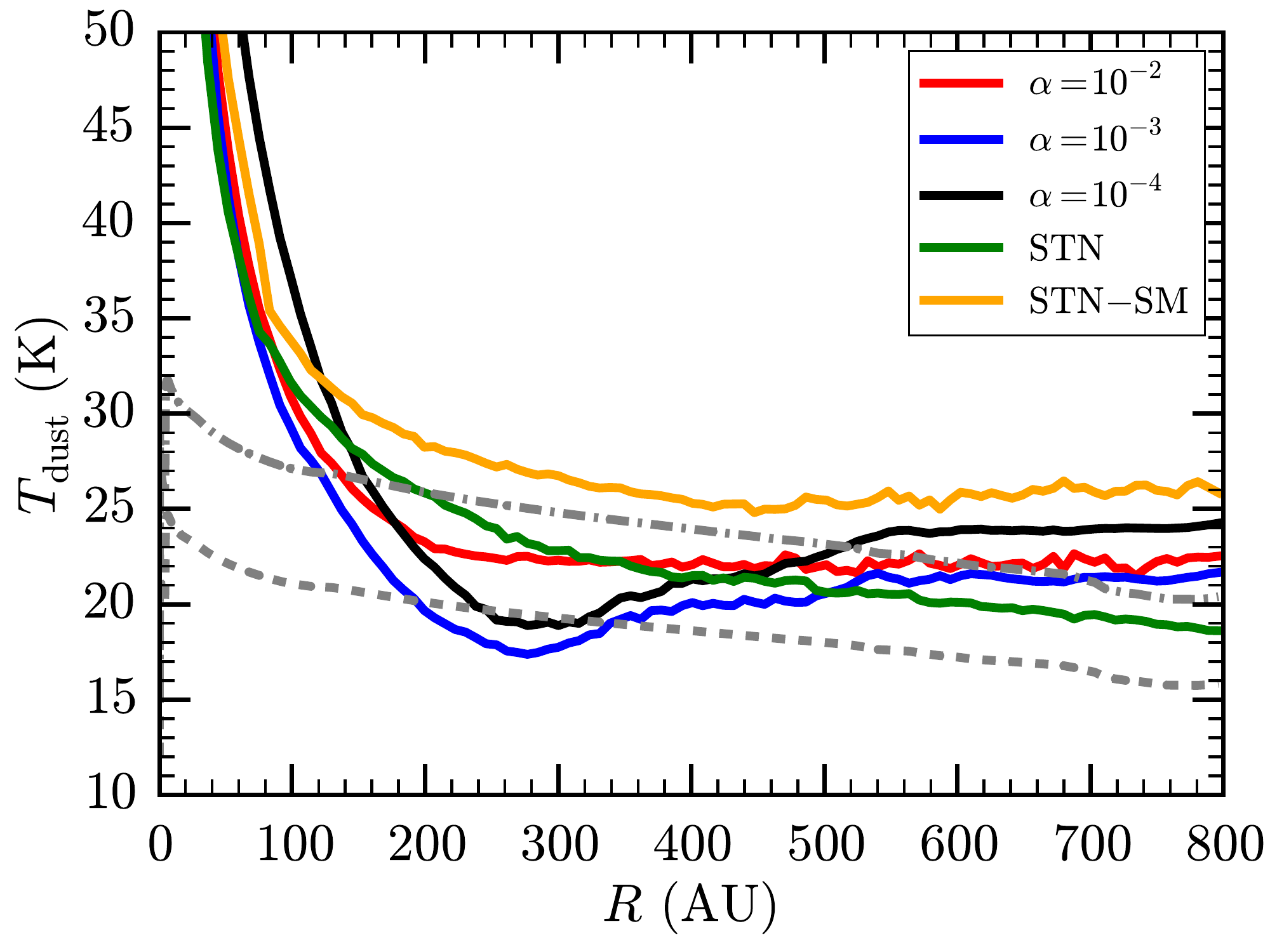}
\caption{Mid-plane dust temperatures of the models using the gas density parameters by \citet{2013A&A...557A.133D}. The grey lines show the sublimation temperature for CO in the $\alpha=10^{-3}$ model; the dashed line uses a binding energy of $855\,$K, the dotted-dashed line of $1100\,$K.}
\label{fig:t_dust_midplane}
\end{figure}

The results of the combination of grain growth, radial drift, and vertical settling in the dust density distribution are apparent in the left panels of Fig.~\ref{fig:vert_settling}, where the grain size evolution models are compared to the STN one. First, the dust density structure in the most viscous case is similar to that of the STN model. As turbulence gets lower, the vertical settling becomes more severe, with differences in the dust density in the upper layers of the disk of more than four orders of magnitude between the $\alpha=10^{-2}$ and $10^{-4}$ cases. This is due to a combination of grain growth and vertical settling: for low viscosities, \rev{larger size bins are more populated because the fragmentation limit is very close to the central star}, and the low turbulence is very inefficient in maintaining the massive dust grains at significant altitudes in the disk.

The average grain size $\bar{a}$ and the total dust surface area per volume $n_{\rm gr}\langle \sigma\rangle$ are two other important quantities that are directly related to the dust density structure (see Sec.~\ref{sec:method_aver_a}). From the combination of grain growth and vertical settling, lower viscosities lead to average grain sizes that are a steep function of both radius $R$ and height $z$, as can be seen from the central panels of Fig.~\ref{fig:vert_settling}. In the $\alpha=10^{-4}$ case, $\bar{a}$ can be as high as $>10\,\mu$m in the disk mid-plane. However, in this case $\bar{a}$ steeply decreases to average grain sizes of the order of a few nm at relatively low scale heights. The dust-to-gas ratio and the average grain size jointly determine the total dust surface area per unit volume (see Eq.~\ref{eq:n_gr_sigma} and \ref{eq:sigma_d}), which at any given location clearly decreases with turbulence. The difference in the upper layers of the disk can be higher than four orders of magnitude for $\alpha$ varying between $10^{-4}$ and $10^{-2}$. Interestingly, the total dust surface area that the STN model assumes is quite similar to the $\alpha=10^{-2}$ case, indicating that the parameters chosen for the STN model, with a settling parameter $\chi=0.2$, are representative of a quite turbulent disk. 

\begin{figure}
\center
\includegraphics[width=\columnwidth]{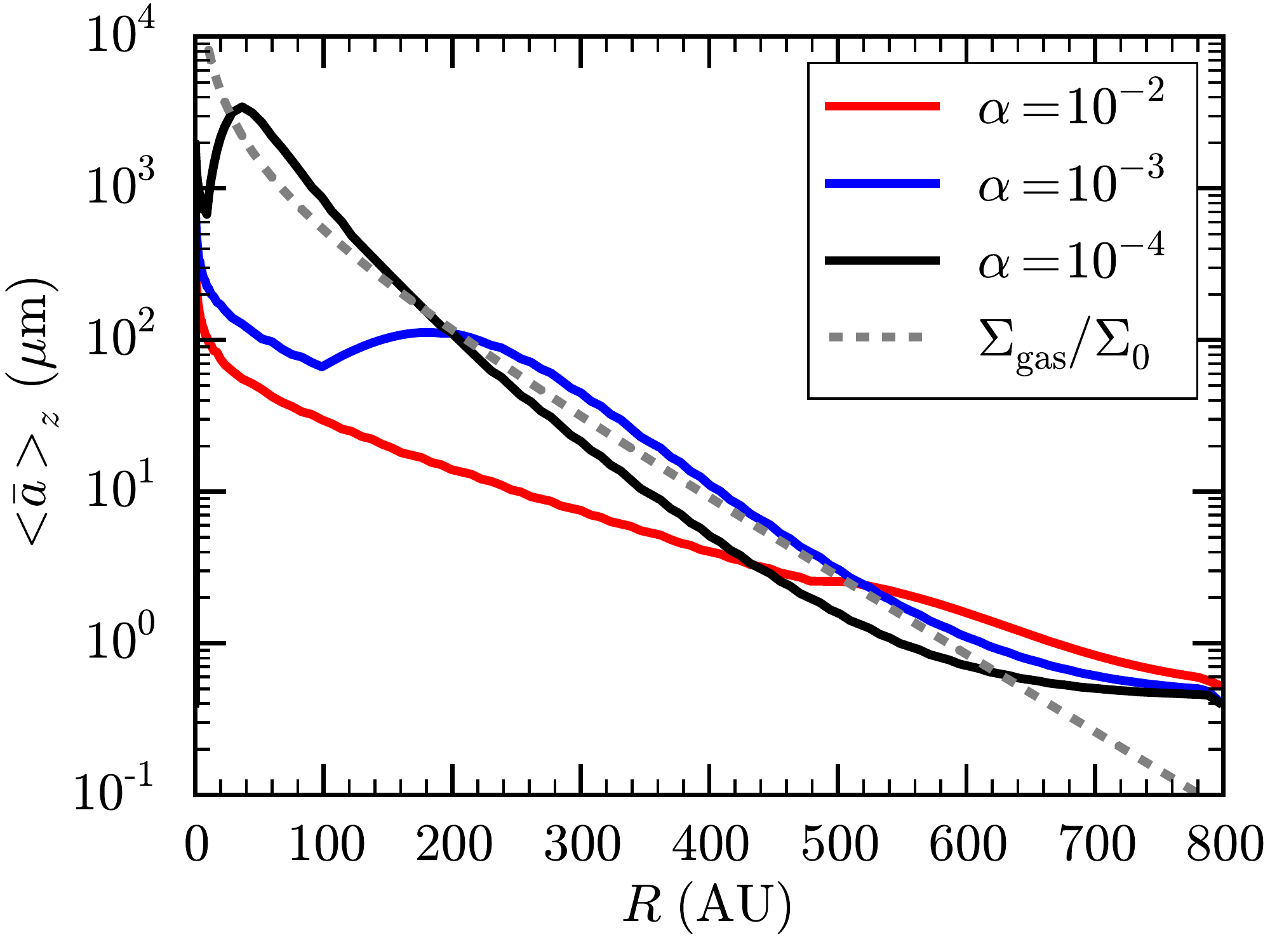}
\caption{\rev{Vertically averaged grain size ($<\bar{a}>_z$) as a function of radius for the turbulent models, compared to the gas surface density profile rescaled to some value $\Sigma_0$. For both $\alpha=10^{-3}$ and $\alpha=10^{-4}$ there is an intermediate region where the radial gradient of $<\bar{a}>_z$ is steeper than the radial gradient of $\Sigma_{\rm gas}$, thus leading to disk self-shadowing.}
}
\label{fig:aaver_vs_sigma}
\end{figure}

\begin{figure*}
\center
\includegraphics[width=.33\textwidth]{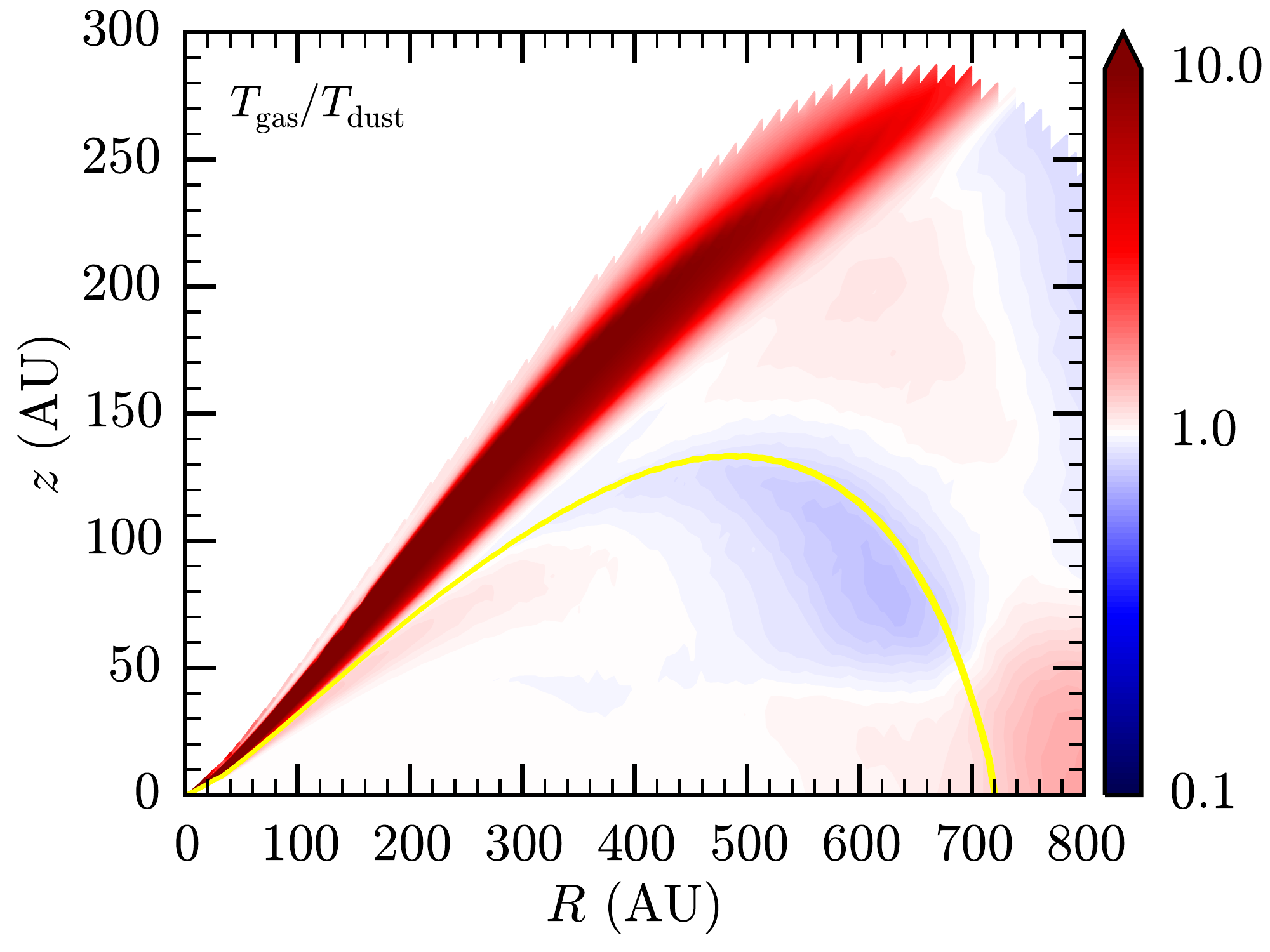}
\includegraphics[width=.33\textwidth]{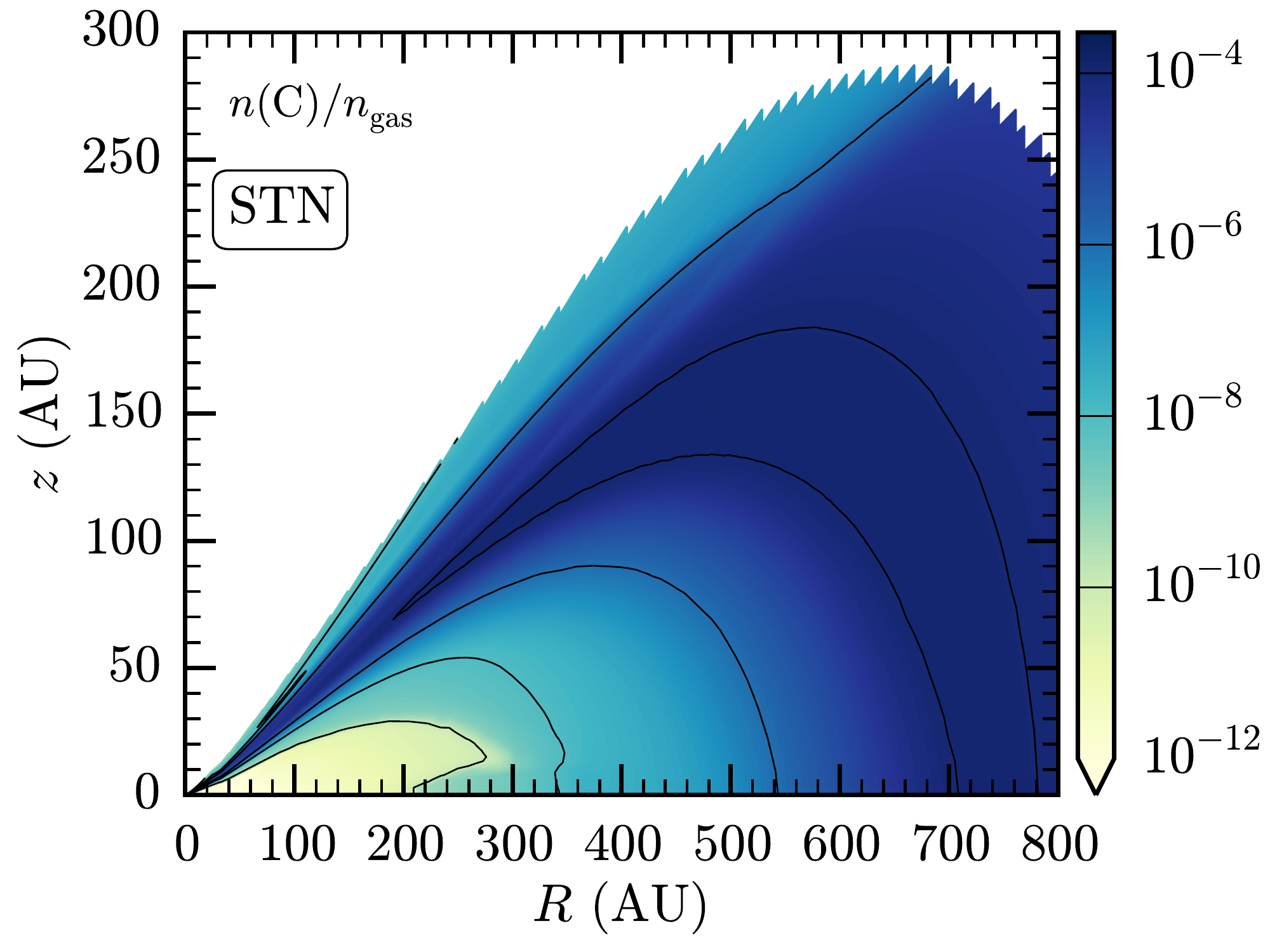}
\includegraphics[width=.33\textwidth]{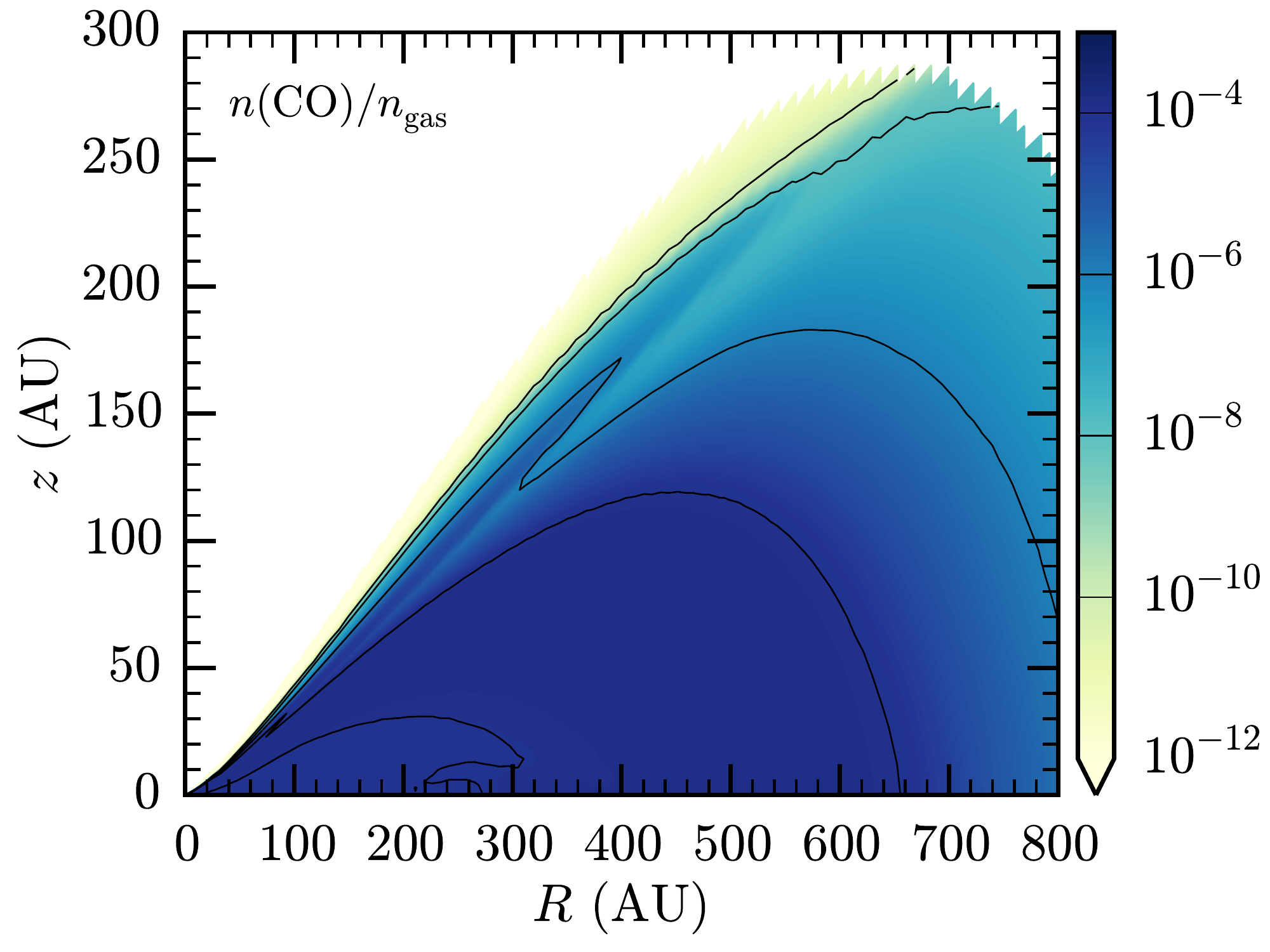}\\
\includegraphics[width=.33\textwidth]{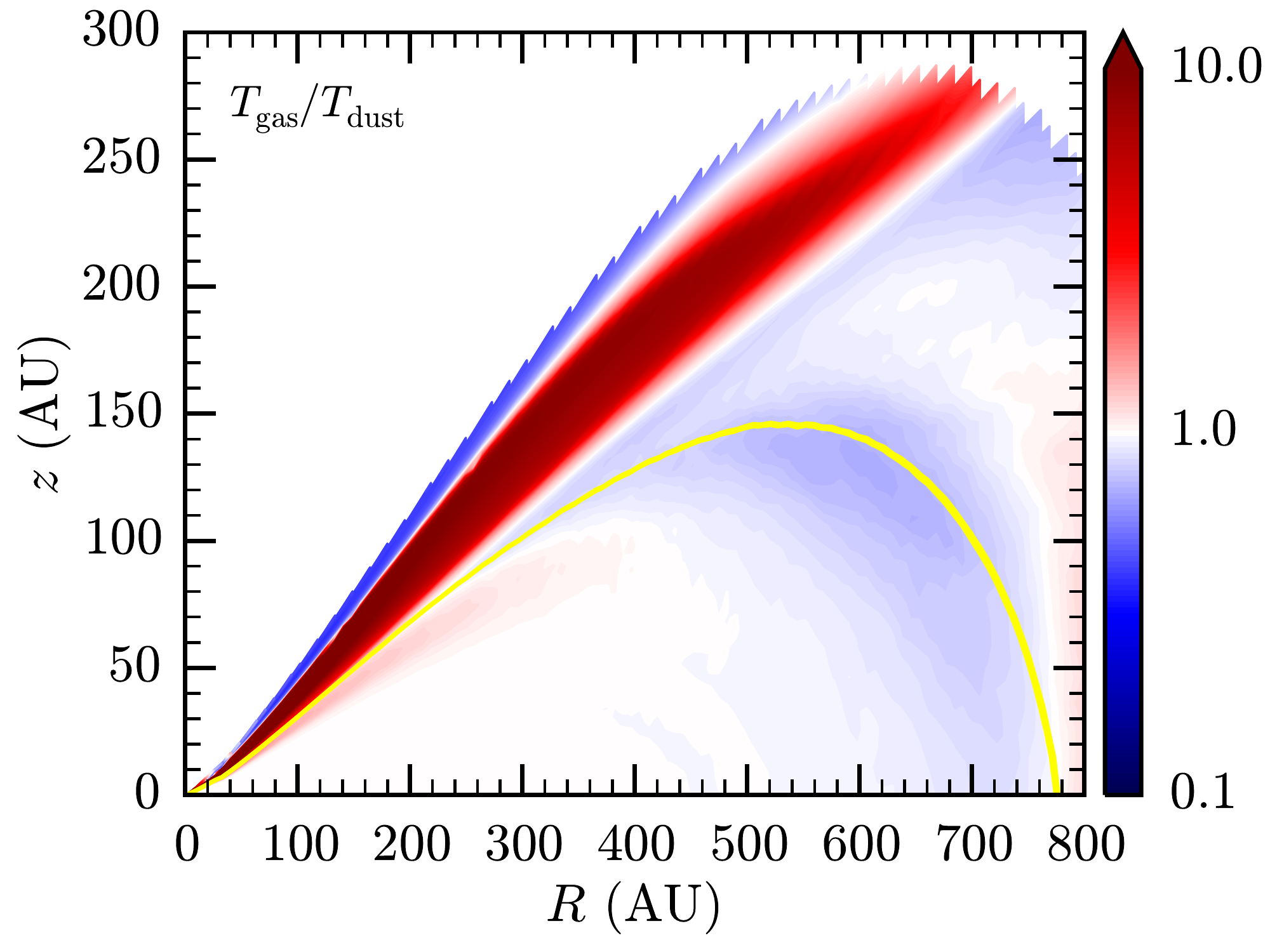}
\includegraphics[width=.33\textwidth]{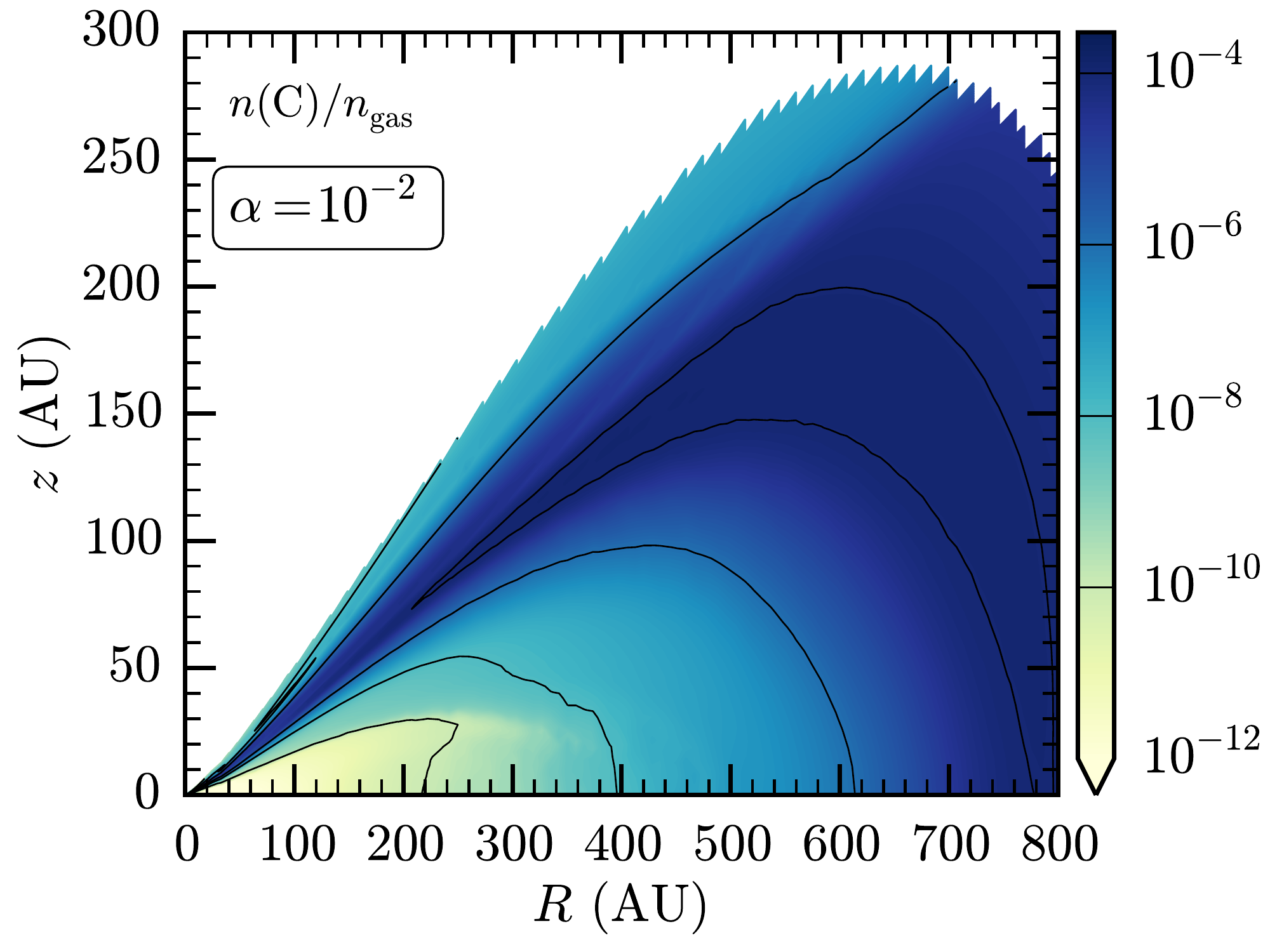}
\includegraphics[width=.33\textwidth]{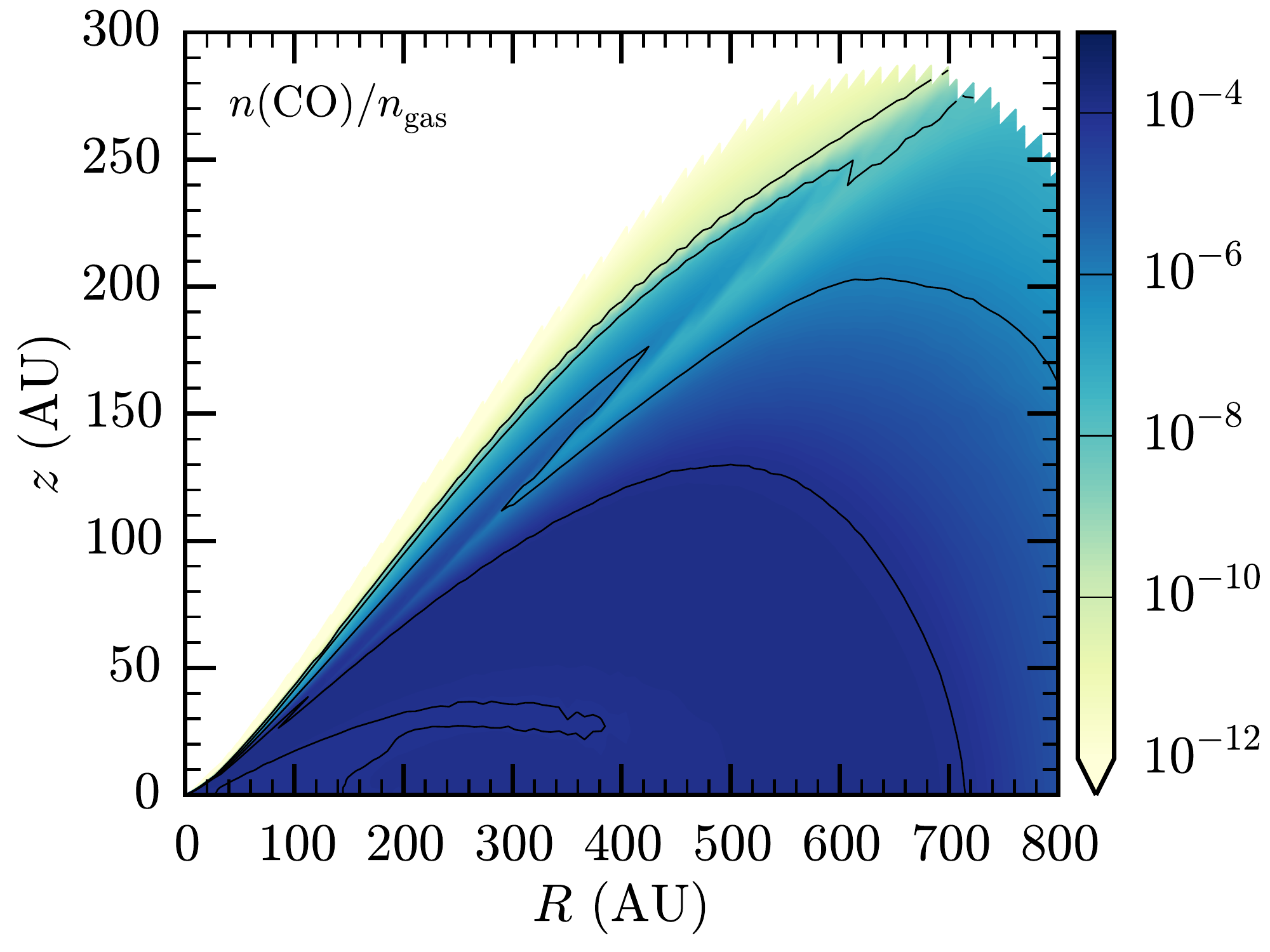}\\
\includegraphics[width=.33\textwidth]{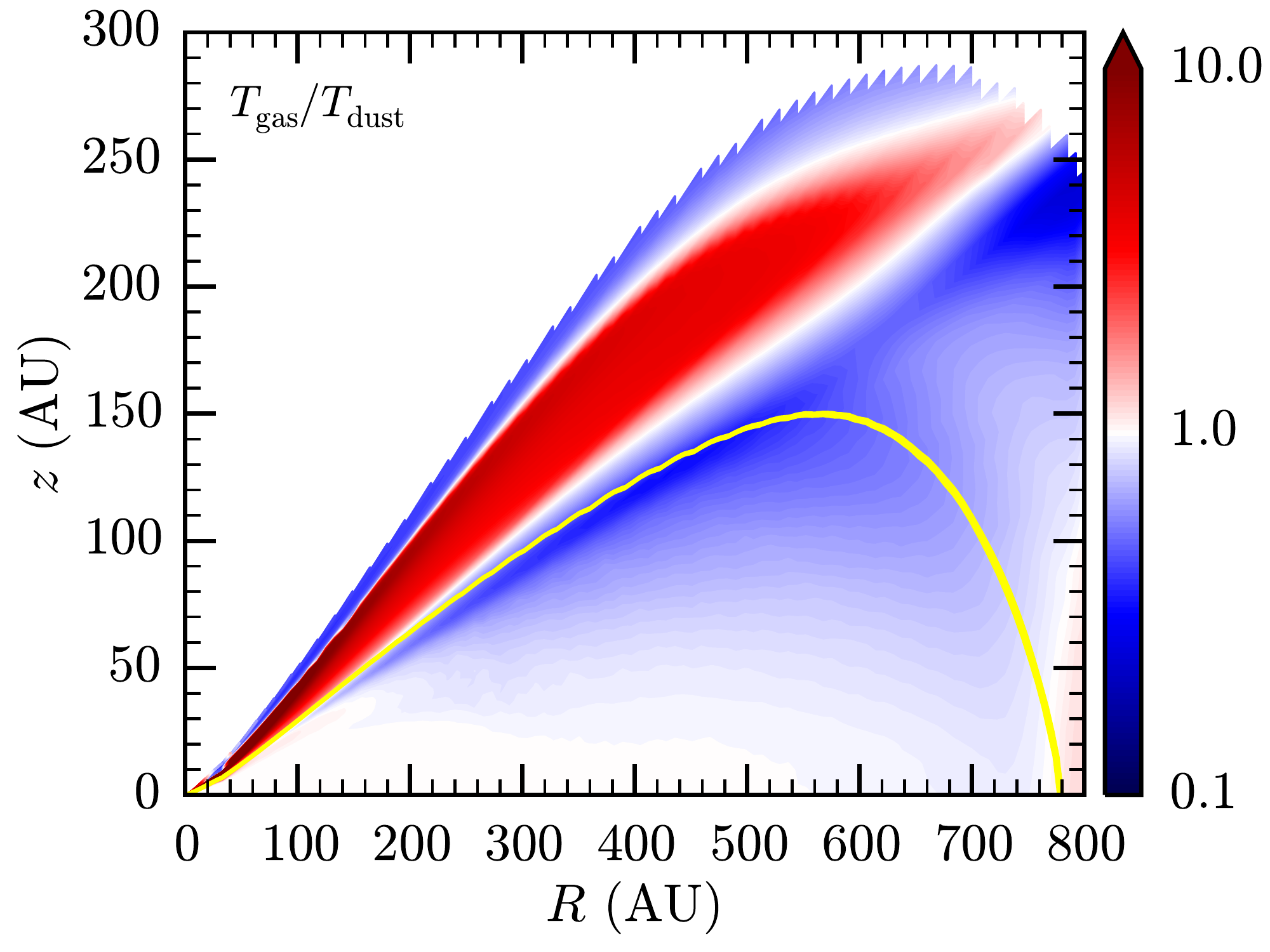}
\includegraphics[width=.33\textwidth]{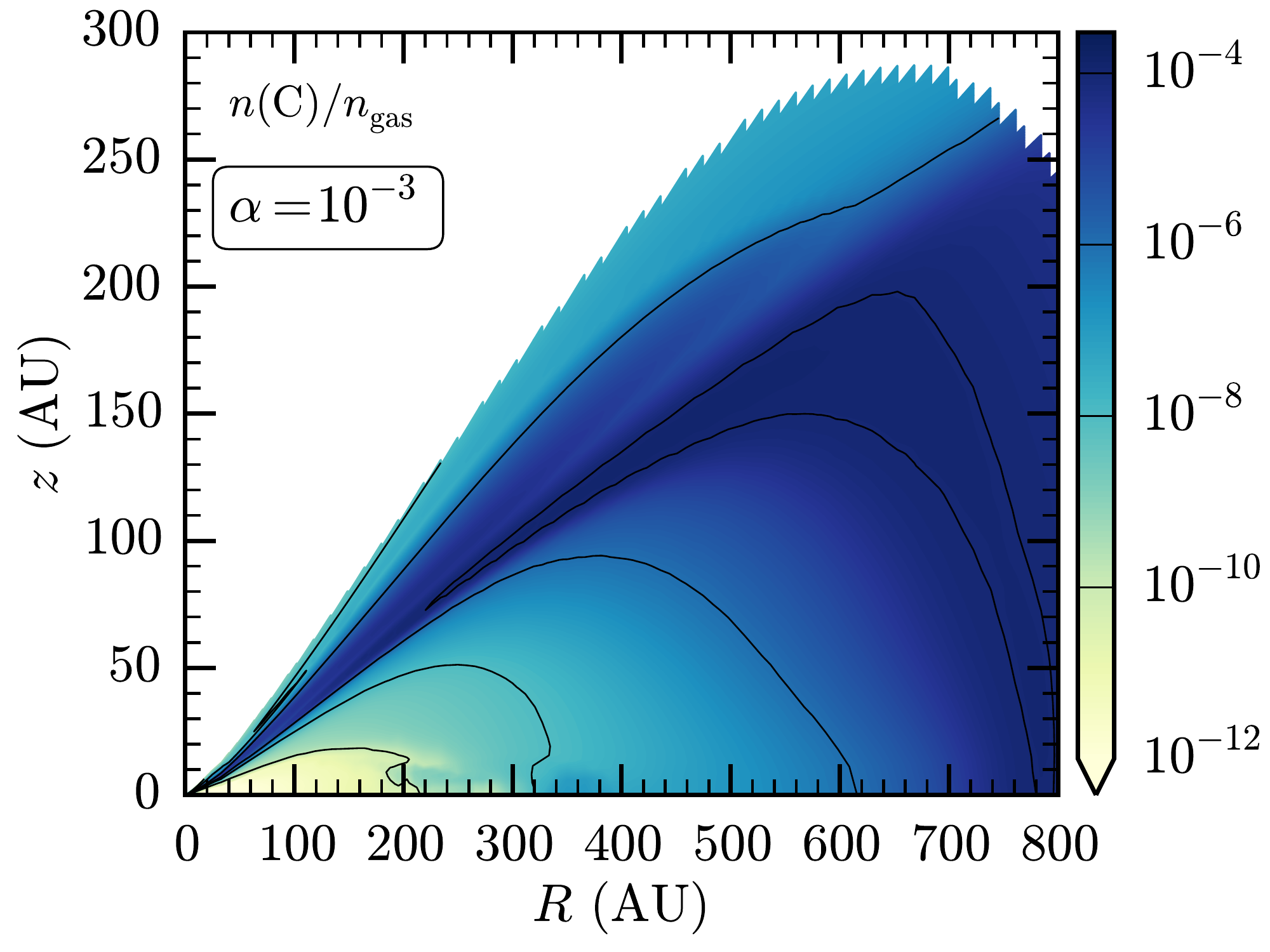}
\includegraphics[width=.33\textwidth]{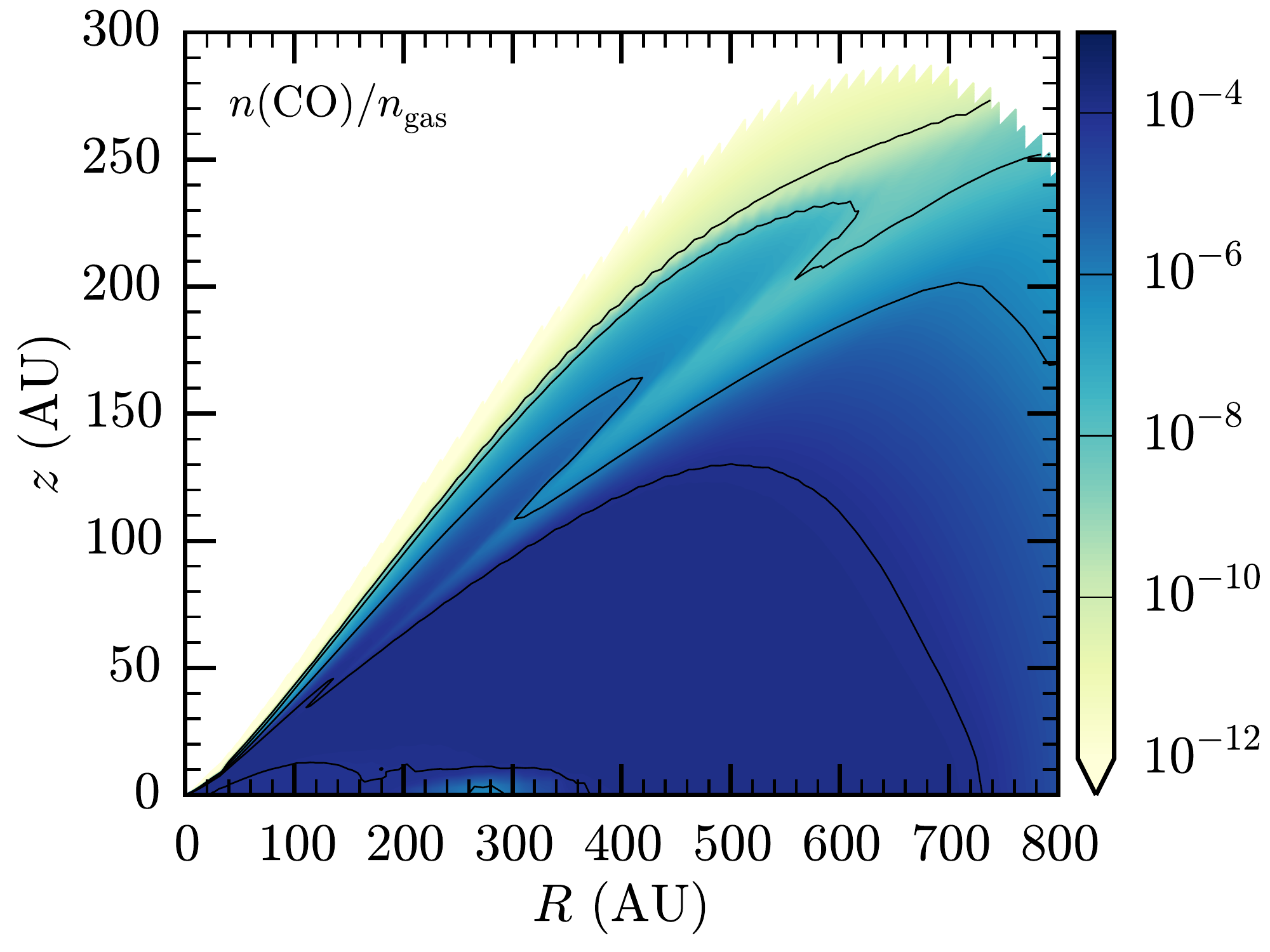}\\
\includegraphics[width=.33\textwidth]{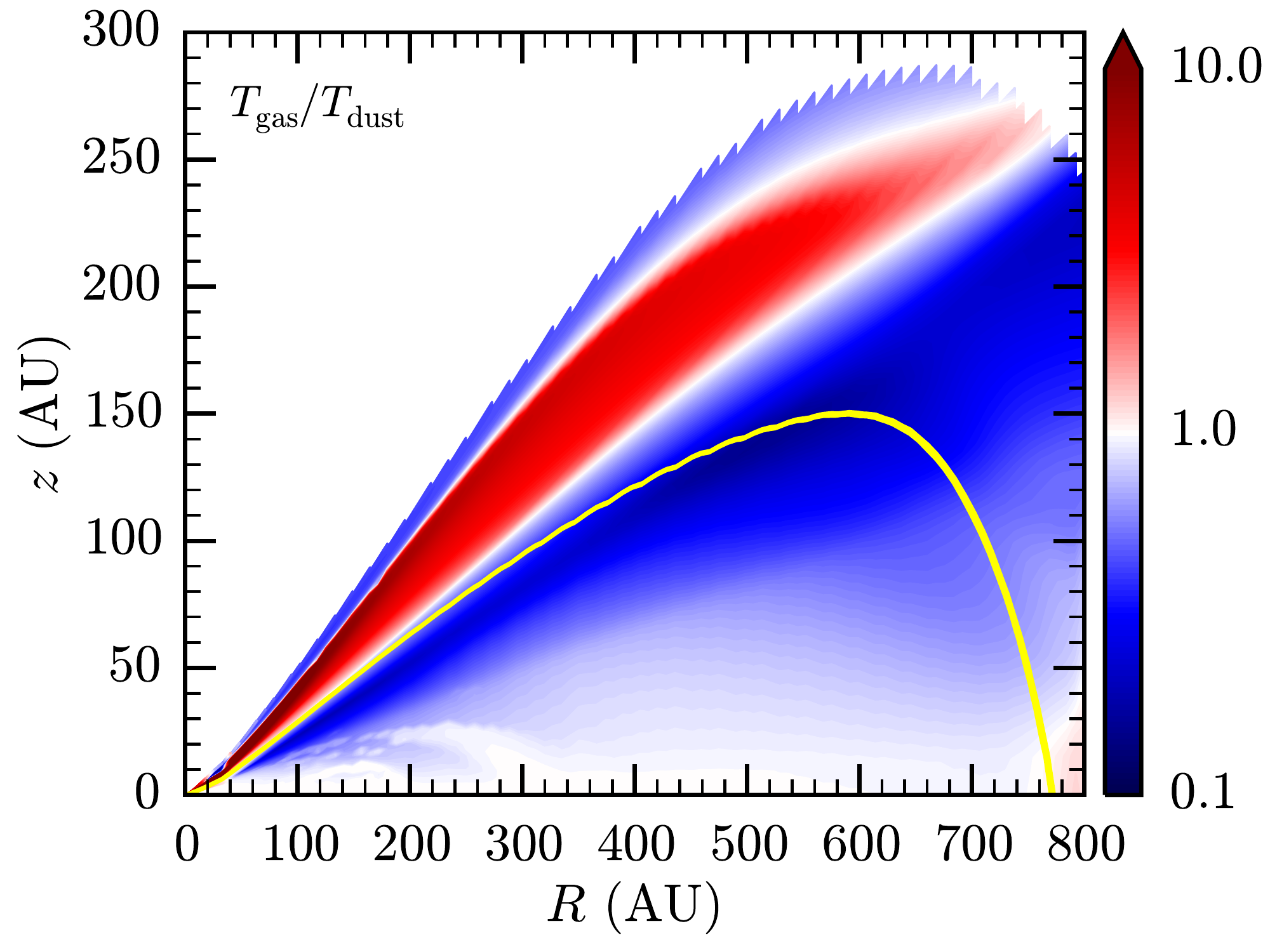}
\includegraphics[width=.33\textwidth]{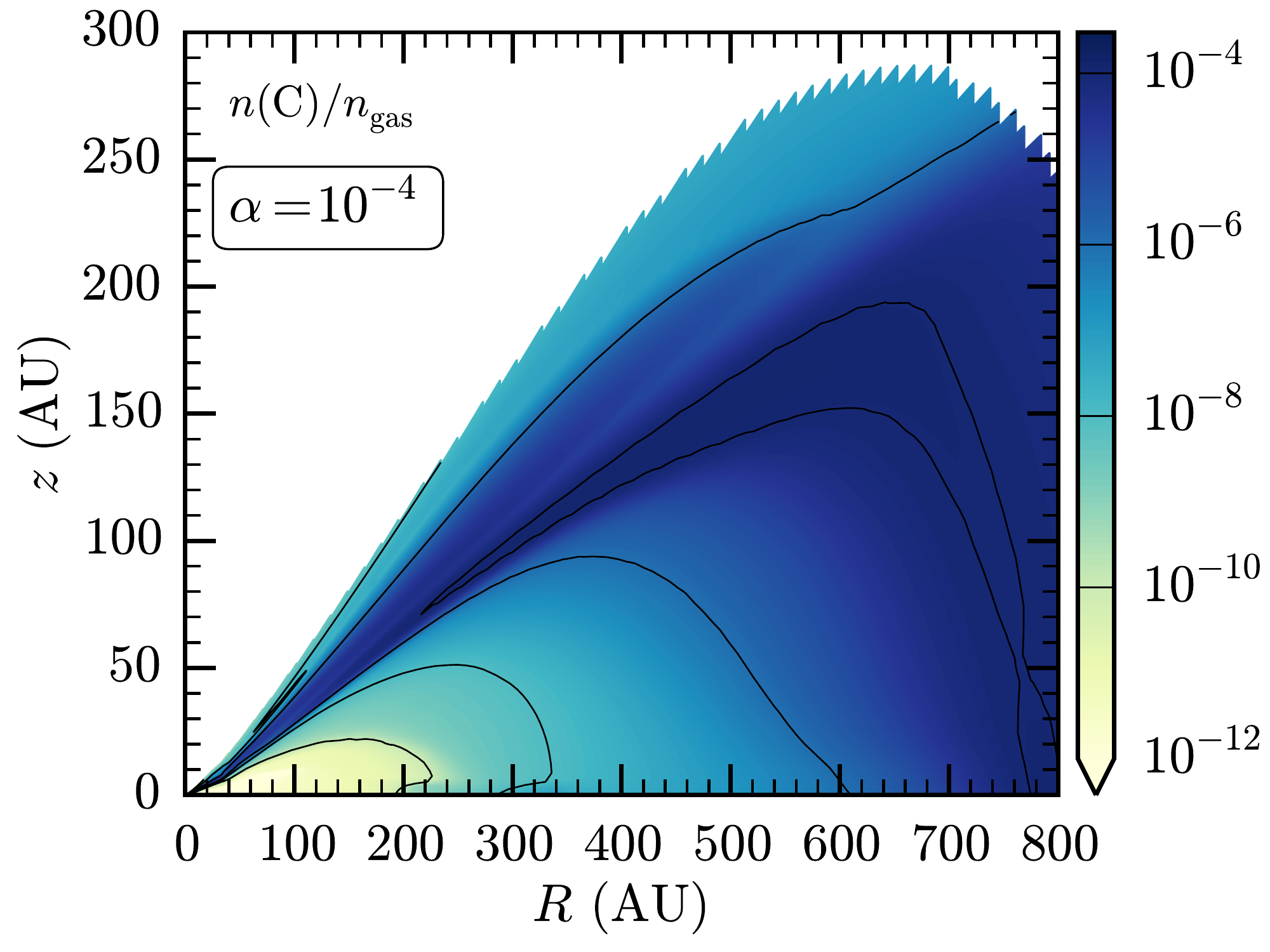}
\includegraphics[width=.33\textwidth]{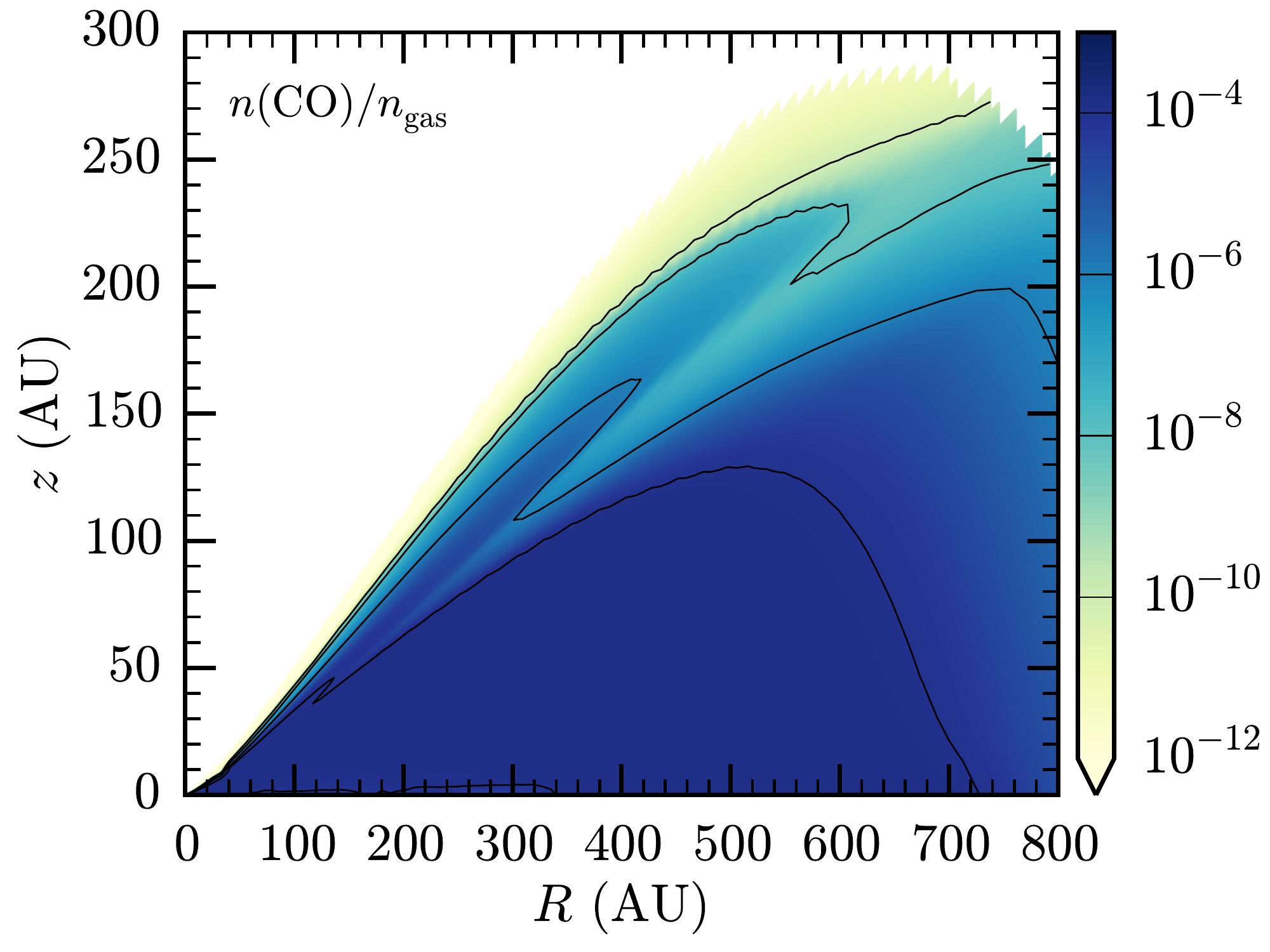}\\
\caption{From top to bottom: STN, $\alpha=10^{-2}$, $10^{-3}$ and $10^{-4}$ models. From left to right: ratio $T_{\rm gas}/T_{\rm dust}$, C abundance, and gas phase CO abundance. The yellow and red solid lines in the left and right panels indicate the vertical $\tau=1$ layer for the $^{12}$CO $J$=3-2 transition.}
\label{fig:abun-CO_lin}
\end{figure*}

\subsection{The effects of grain growth, radial drift, and dust settling on gas properties}

The distribution of the dust particles in the disk has multiple effects, which are addressed below. In particular, their concurrence can significantly affect the radial distribution of the CO emission and it is thus important to take them into account when modelling CO emission in a protoplanetary disk. The focus of this section is on the outer disk, which usually dominates the emission both in continuum and in the low-lying CO rotational lines.

\subsubsection{Dust and gas thermal structures}
\label{subsec:thermal_structure}

The opacities in a disk do depend on the size distribution of dust particles within the disk (see Sec.~\ref{sec:method_opacities}). The UV wavelength range is particularly important because it provides one of the most important heating sources for the gas phase via photoelectric heating of both PAHs and small particles \citep[e.g.][]{2001ApJ...548..296W,2012ApJ...747...81C}. At low viscosities, the smallest grains become the dominant mass carriers in the outer disk \rev{(see discussion in Section \ref{sec:results4_1})} and are generally more abundant than in the STN model, or in the $\alpha=10^{-2}$ case (which again look very similar). This has the important effect that for low viscosities, the small grains are very effective in screening the FUV (far-UV) photons (see Appendix~\ref{app:thermal} for more details). In these cases, the transition between optically thick and optically thin disk regions at FUV wavelengths becomes very sharp in the vertical direction. However, at longer wavelengths, the exact opposite happens. At lower viscosities, the opacities in the IR (infrared) at intermediate scale heights are much lower than in the more turbulent cases, simply due to the fact that the vertical settling for low $\alpha$ values is more severe. Thus, the IR photons, which are the most important in determining the dust temperature, can penetrate further into the disk, heating the dust to higher temperatures. At intermediate scale heights in the outer disk, close to the $^{12}$CO emitting layer, the dust temperatures can be as high as $\sim100\,$K in the $\alpha=10^{-4}$ case, whereas they reach at most $\sim50\,$K in the $\alpha=10^{-2}$ case.

\begin{figure*}
\center
\includegraphics[width=.32\textwidth]{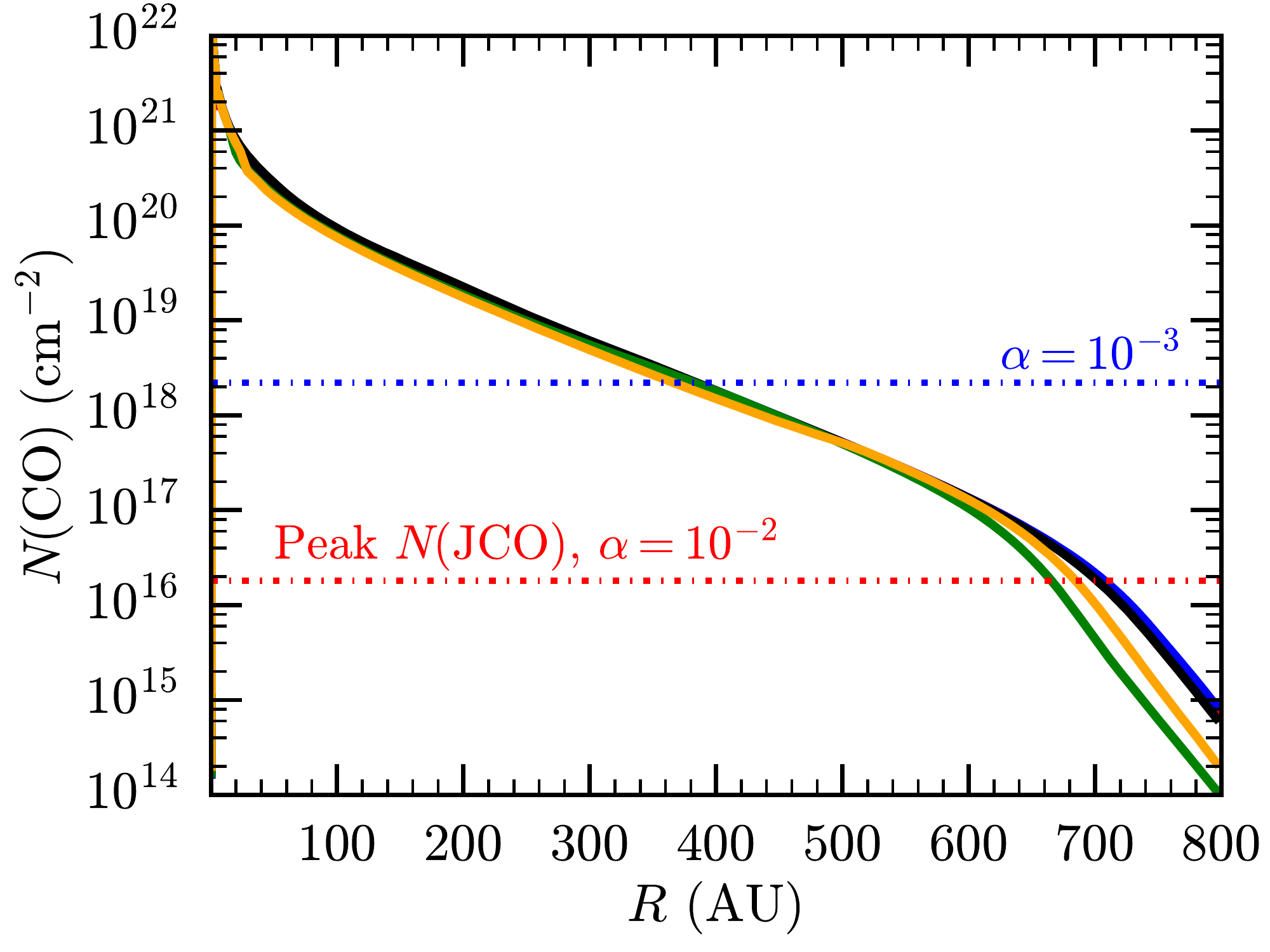}
\includegraphics[width=.32\textwidth]{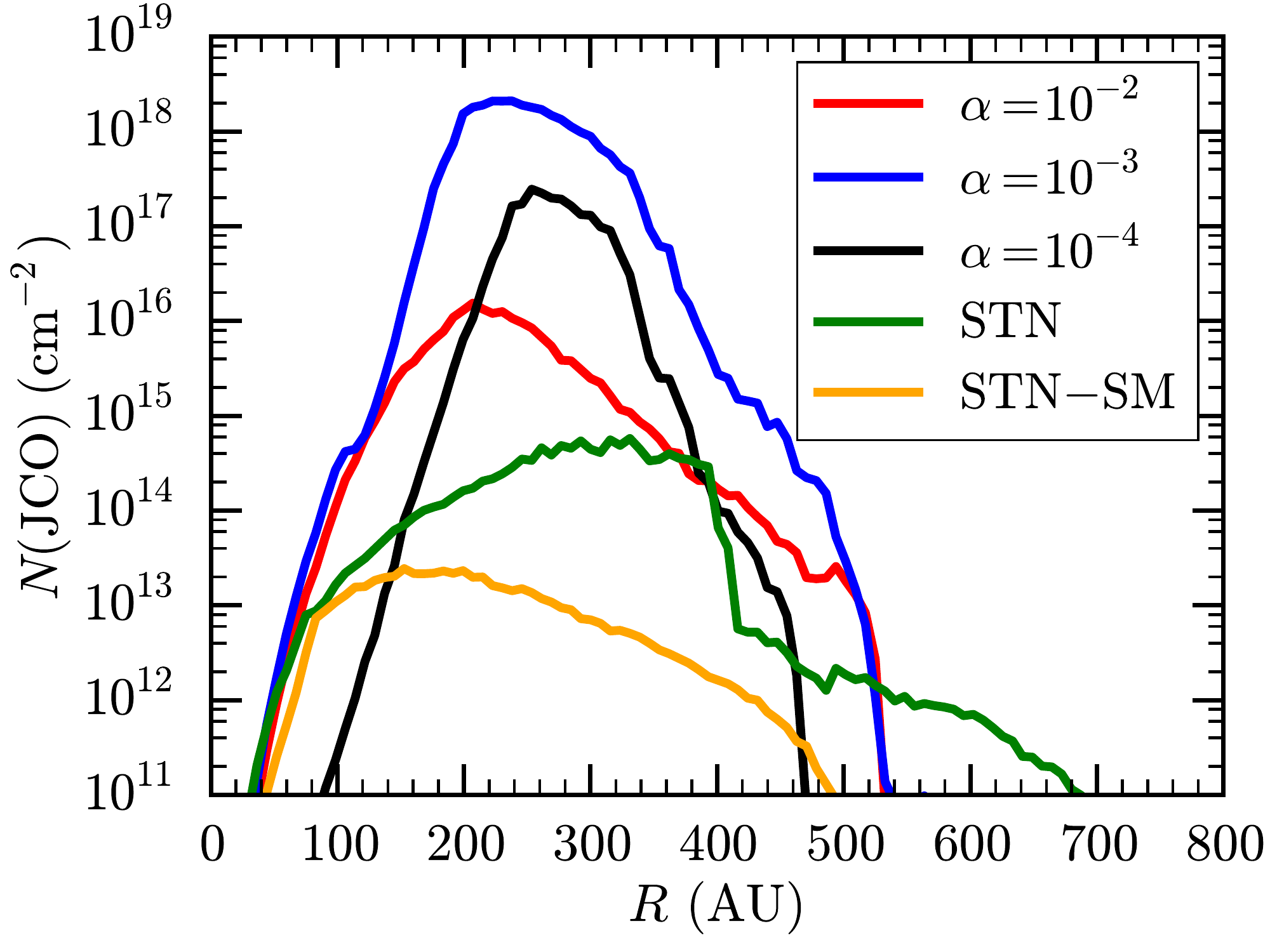}
\includegraphics[width=.31\textwidth]{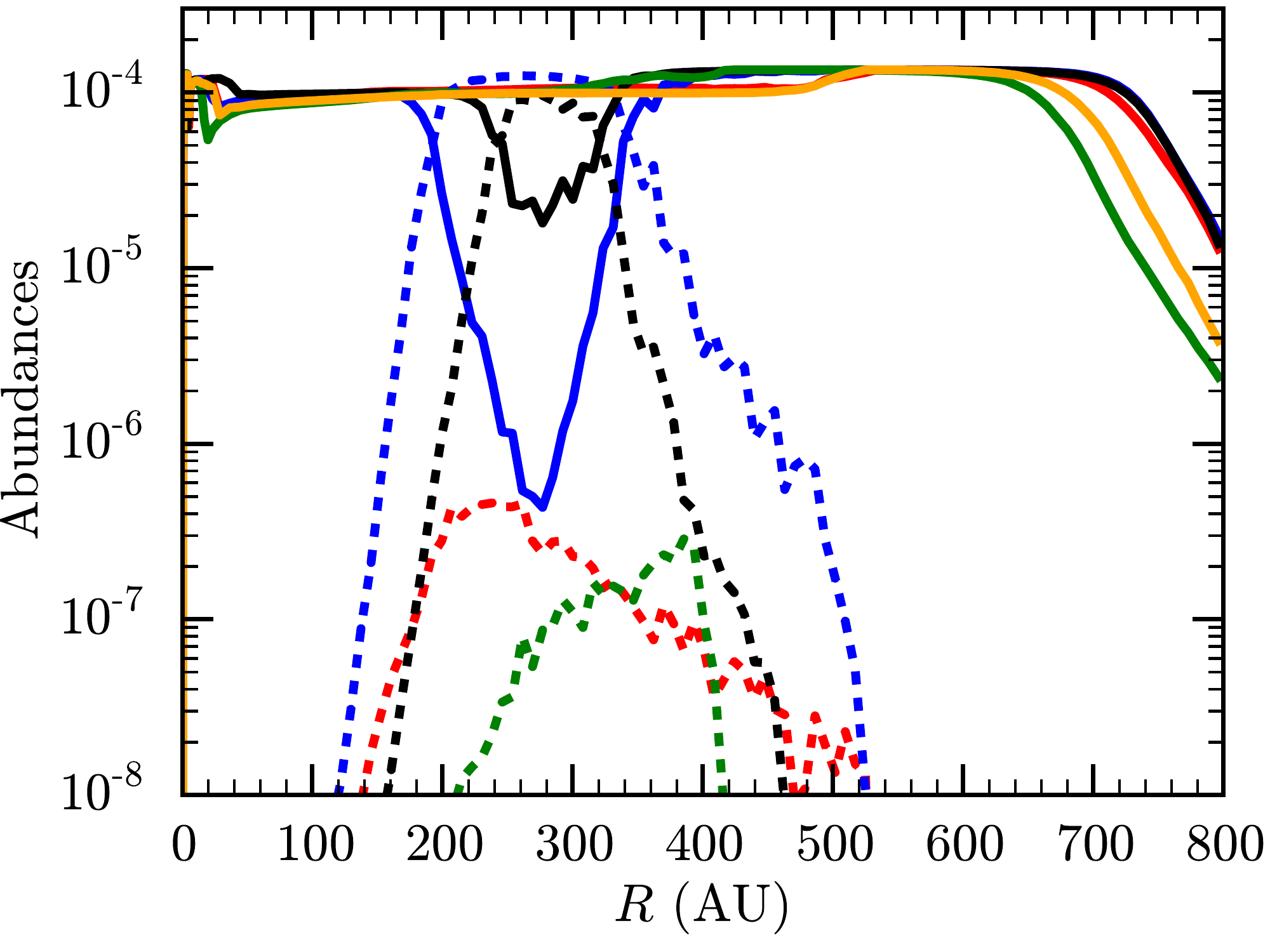}\\
\caption{Left and central panel: column density of CO gas and CO ice (JCO) of all models. We note the different scale on the $y$-axis in the two panels\rev{: $N$(JCO) is always much lower than $N$(CO) in the gas phase at all radii, even for $\alpha=10^{-3}$}. Right panel: abundance of CO (solid lines) and CO ice (dashed lines) along the disk mid-plane.}
\label{fig:co_col}
\end{figure*}

\begin{figure*}
\center
\includegraphics[width=.43\textwidth]{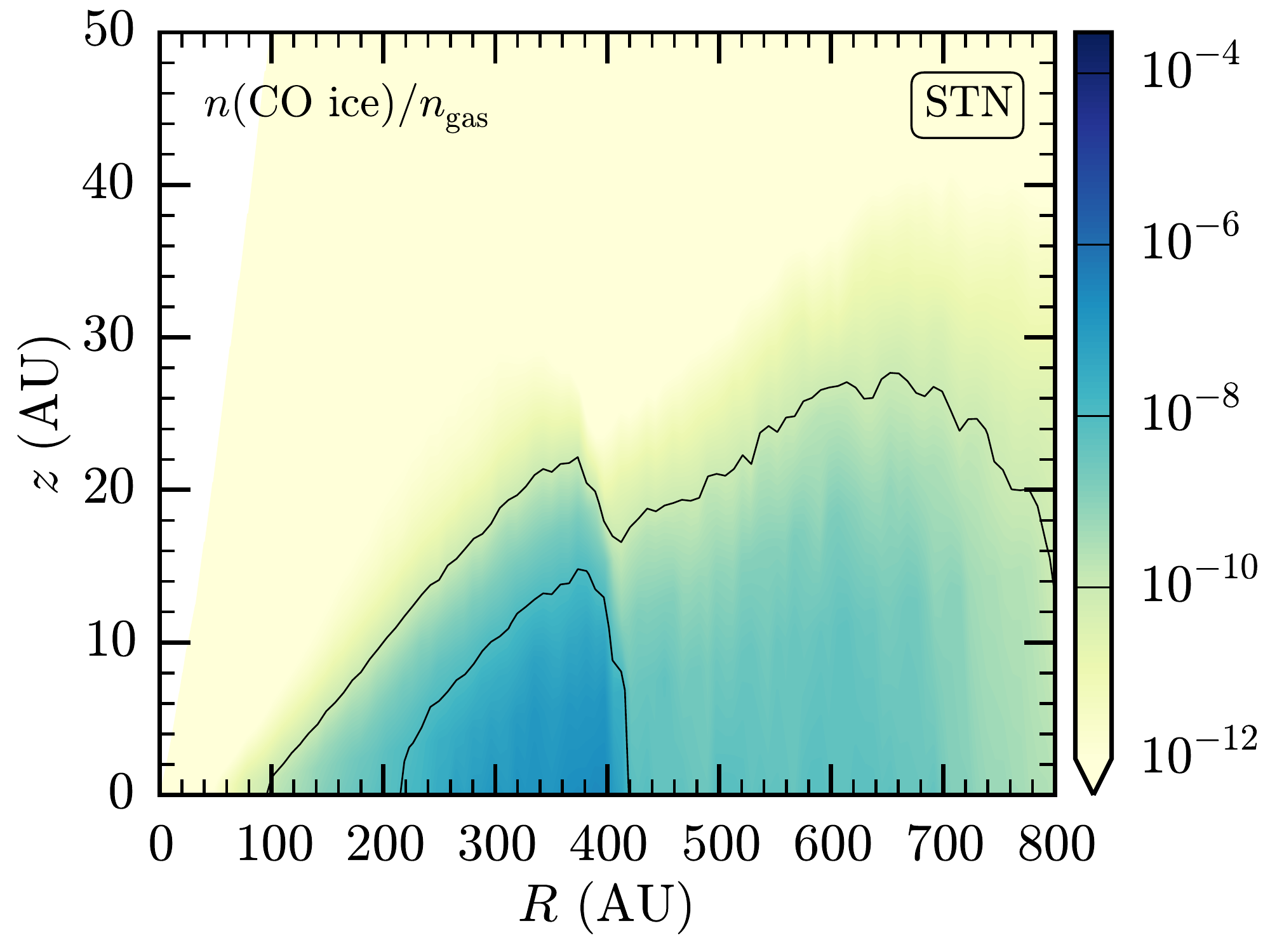}
\includegraphics[width=.43\textwidth]{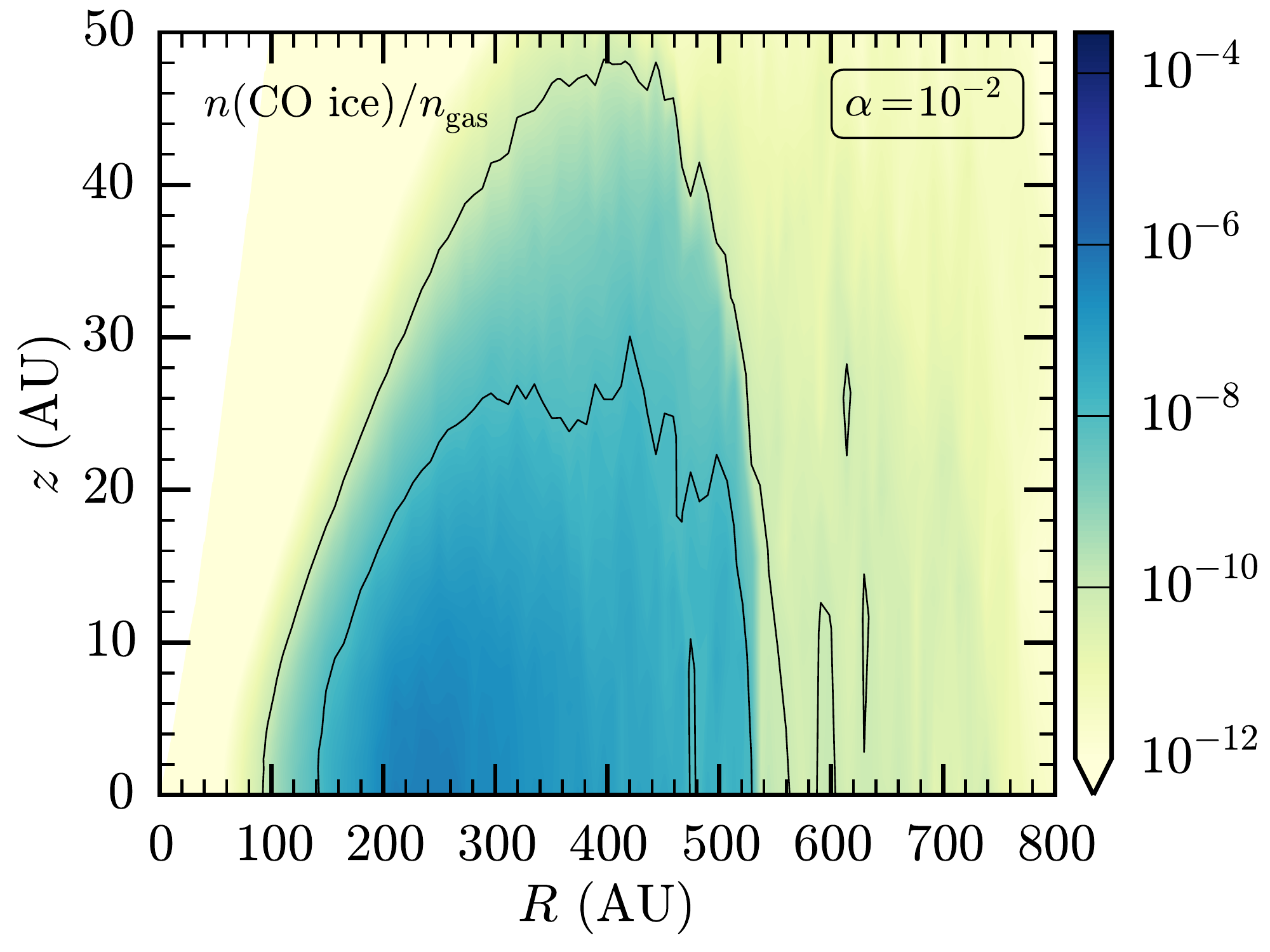}\\
\includegraphics[width=.43\textwidth]{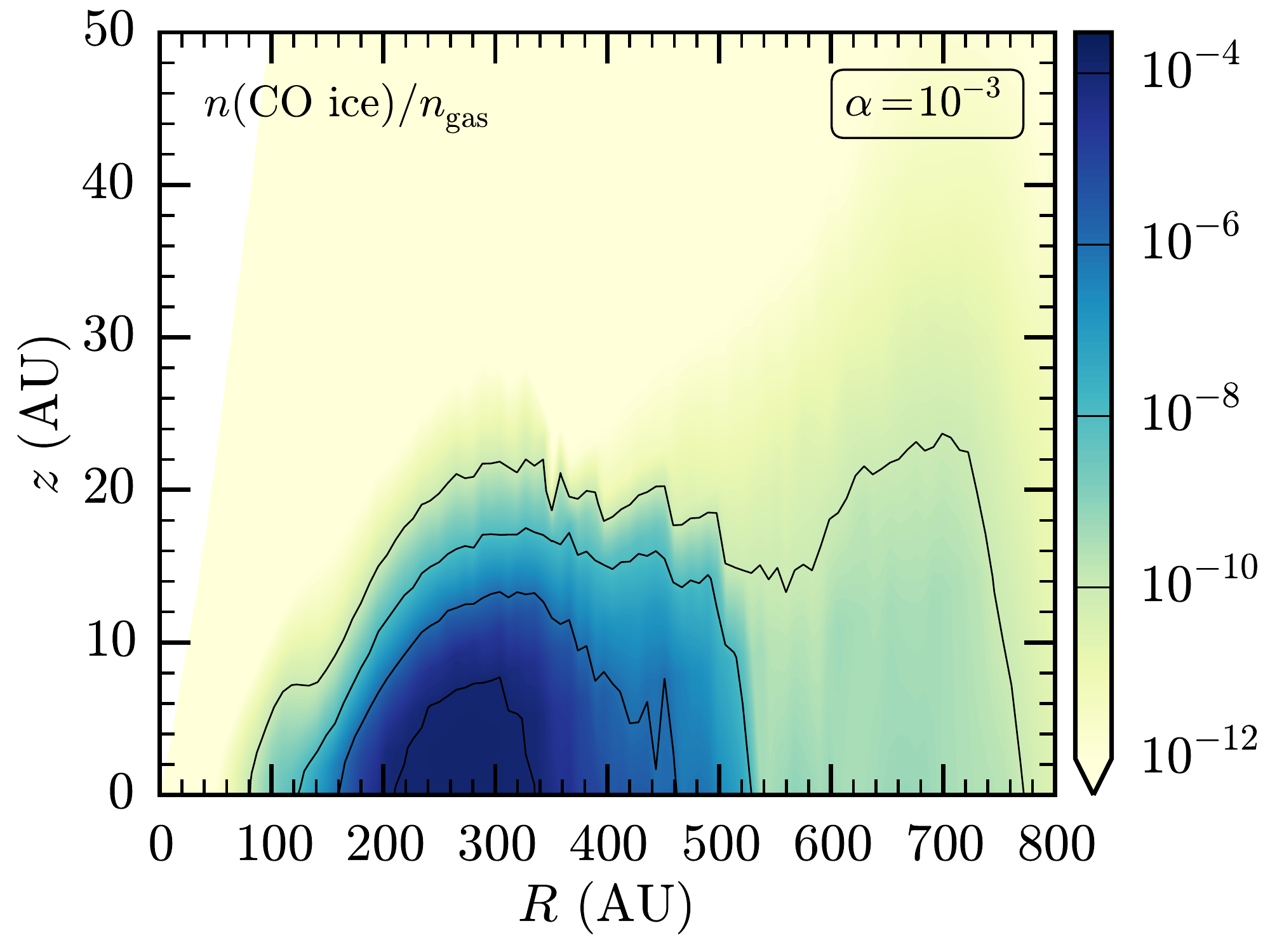}
\includegraphics[width=.43\textwidth]{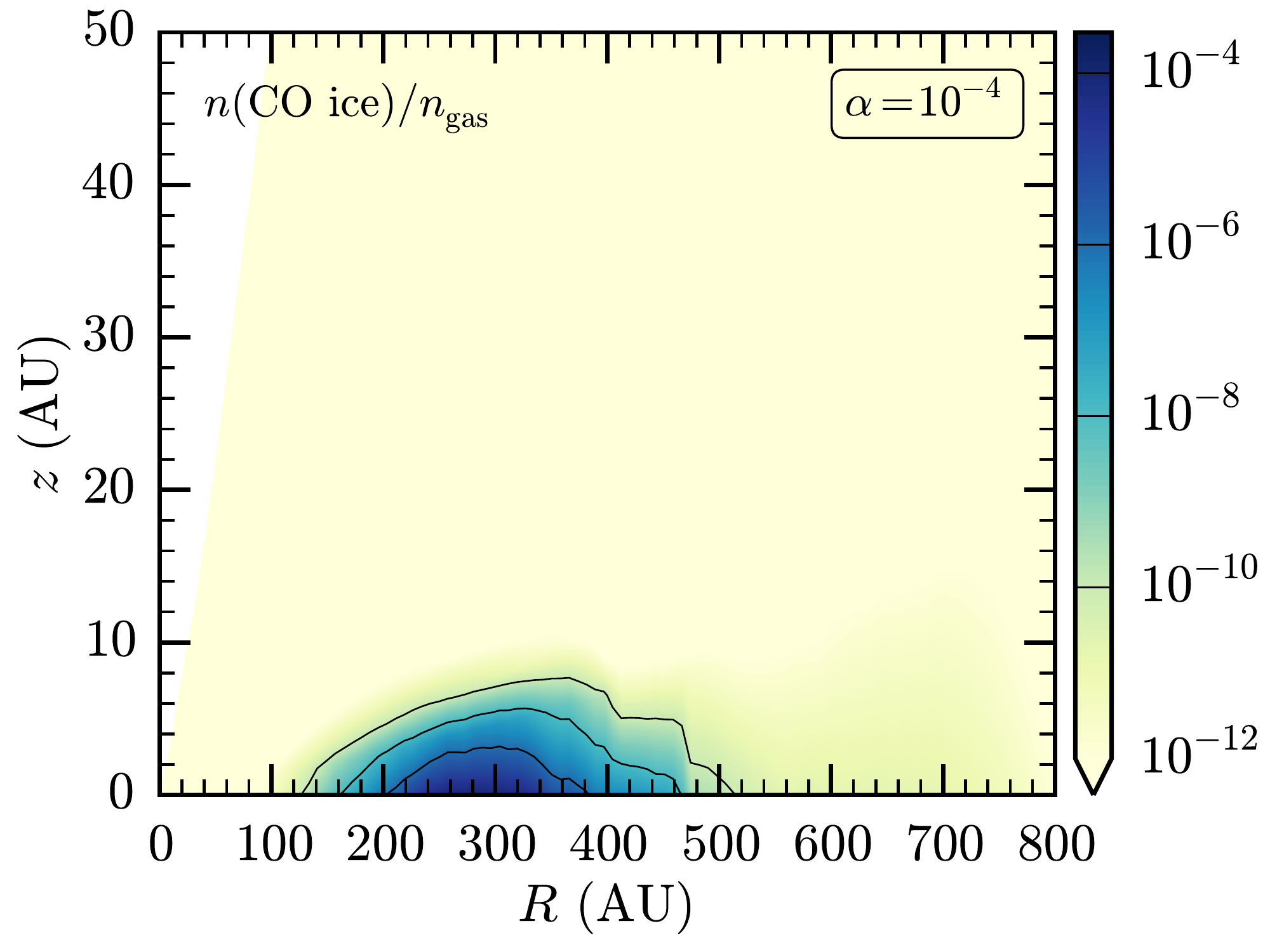}\\
\caption{From top left to bottom right: CO ice abundance in the STN, $\alpha=10^{-2}$, $10^{-3}$ , and $10^{-4}$ models. We note the different vertical scale from other similar plots.}
\label{fig:abun-JCO_lin}
\end{figure*}

\begin{figure*}
\center
\includegraphics[width=.9\textwidth]{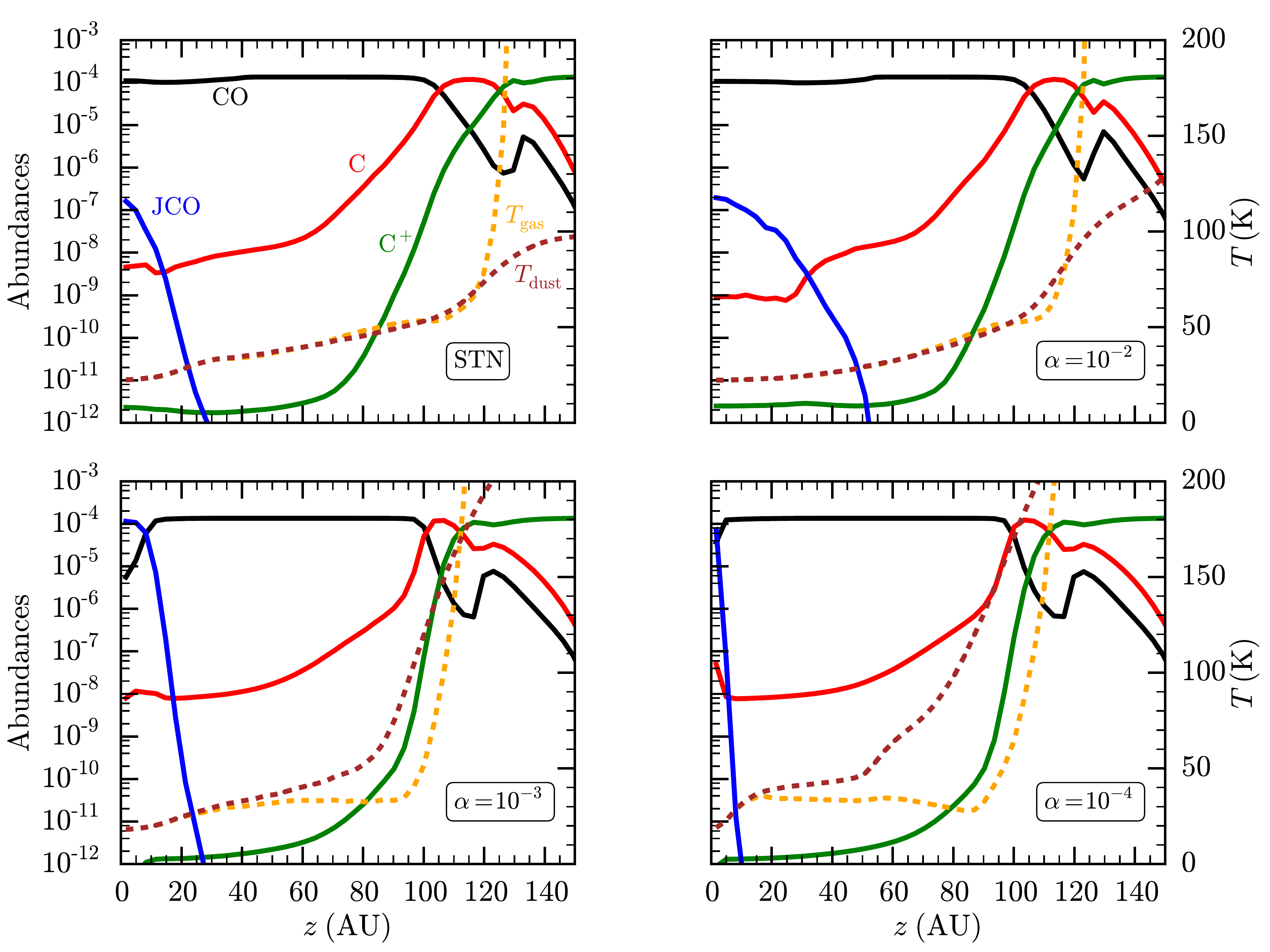}\\
\caption{From top left to bottom right: Vertical cuts at $\sim300\,$AU of the STN, $\alpha=10^{-2}$, $10^{-3}$ , and $10^{-4}$ models. Legend: solid lines in black, red, green, and blue represent abundances of CO, C, C$^+$ , and CO ice, respectively. Dashed lines in orange and brown are $T_{\rm gas}$ and $T_{\rm dust}$, respectively. \rev{We note that $T_{\rm gas}$ falls below $T_{\rm dust}$ at low values of $\alpha$ and that this thermal de-coupling starts at the C--CO transition.}}
\label{fig:vert_cut_jco}
\end{figure*}

We now focus on the disk mid-plane, since this is where most of the mass lies, and where important gas-grain chemistry dependent on the dust temperature occurs. The dust temperatures in the disk mid-plane are shown in Fig.~\ref{fig:t_dust_midplane}. While the STN and STN-SM models show a monotonically decreasing temperature (which is usually assumed in most models), grain growth and vertical settling concur in yielding a non monotonic dust temperature. The effect becomes more severe for lower values of turbulence. The reason can be understood as follows. At intermediate radii ($100-300\,$AU), the average size of the dust particles has grown considerably, in particular for low turbulence (see Fig.~\ref{fig:d2g_vs_r}). For such large particles, the aerodynamic drag of the turbulent eddies is quite inefficient and dust particles can easily settle close to the disk mid-plane. However, in the inner regions the gas densities are high enough that the dust particles are stirred to high altitudes (see left panels of Fig.~\ref{fig:vert_settling}). The net effect is that due to settling there is an intermediate region where the disk is self-shadowing \citep{2004A&A...421.1075D,2015ApJ...813L..14B}. At larger radii the average grain size decreases again, due to radial drift. The \rev{abundant} small particles are again stirred up easily, making them intercept \rev{significant} stellar radiation. Thus, the dust temperature increases at large radii ($R\gtrsim300\,$AU). The result is that there is a region in the disk mid-plane where, for low turbulent models, the dust temperatures can be lower by $\gtrsim20\%$ with respect to the STN model (Fig.~\ref{fig:t_dust_midplane}). \rev{Another way to understand this mechanism is by looking at the radial gradient of the vertically averaged grain size, which in intermediate regions decreases more steeply than the gas surface density profile (Fig. ~\ref{fig:aaver_vs_sigma}). Thus, Eq.~\ref{eq:stokes} implies that there is a region where the average Stokes number decreases with radius, with less dust particles being stirred up towards the upper layers.} We note that the disk analysed here is a relatively warm disk, with most of the disk mid-plane too warm for CO freeze-out. Interestingly, the difference between the models is just around the CO freeze-out temperature (see Fig.~\ref{fig:t_dust_midplane}), which is computed as the dust temperature where $k_{\rm ads}({\rm CO})=k_{\rm thdes}({\rm CO})$.

The gas temperature has somehow the opposite dependence on disk turbulence. The total dust surface area per volume at intermediate scale heights is much lower for lower turbulence values than in the STN and $\alpha=10^{-2}$ models, and this reduces the efficiency of photoelectric heating \citep[][]{2004A&A...428..511J,2007A&A...463..203J}. The low dust surface area in the low turbulence models also implies that the gas-grain collisions are very ineffective in yielding thermal coupling between the dust and gas components. The net effect is that the gas can get significantly colder in low viscosity disks, with gas temperatures $<15-20\,$K in large regions of the disk, even though in those same regions the dust temperature is warmer. In the left panels of Fig.~\ref{fig:abun-CO_lin}, it is apparent that the region of thermal coupling close to the disk mid-plane \rev{(highlighted in white in the left panels of Fig.~\ref{fig:abun-CO_lin})} becomes vertically thicker as viscosity increases. More plots on the resulting disk thermal structure are shown in Appendix \ref{app:thermal}.

By comparing the left panels of Fig.~\ref{fig:abun-CO_lin} with Fig.~\ref{fig:g0}, it emerges that the region where the gas temperature is lower than the dust temperatures occurs just below the $\tau_{\rm FUV}=1$ layer, where the heating via photoelectric effect is quenched. Moreover, line cooling becomes very significant in the same region. This is apparent if we look at the abundances plots of atomic carbon C and CO, shown in Fig.~\ref{fig:abun-CO_lin}. The region where $T_{\rm gas}/T_{\rm dust}<1$ in the low turbulence cases \rev{(highlighted in blue in the left panels of Fig.~\ref{fig:abun-CO_lin})} coincides with the transition between atomic C and CO, that is when the cooling becomes significant via the lines of these two species.

The same mechanisms can be observed in the vertical cuts in the outer disk (at $\sim550\,$AU) shown in Fig.~\ref{fig:temp_abun}. In these regions of the disk, while the mid-plane temperature is the same for all the models, $T_{\rm dust}$ gets warmer at $z>0$ in the low turbulence cases, as described above. At the same time, the gas temperature becomes considerably colder, due to the lack of photoelectric heating, to the poor thermal coupling, and to the effective cooling in the region of the C$^+$ -- C -- CO transition. Finally, we confirm the general results of the simpler models by \citet{2004A&A...428..511J,2007A&A...463..203J} and \citet{2011ApJ...727...76V}, where they suggest that settling causes the maximum of the abundances in the vertical direction to shift closer to the disk mid-plane. In particular, we note that the difference of this work from \citet{2011ApJ...727...76V} is that in the latter the dust evolution considered the fragmentation limit only, and more importantly the gas thermal balance was not computed.

\subsubsection{CO abundance and snowline}

The vertical column density of CO in the gas phase is almost the same in all models, as shown in the left panel of Fig.~\ref{fig:co_col}. The only differences appear at very large radii, where the STN and STN-SM models have higher photodissociation rates due to the deeper UV penetration in the disk (see Fig.~\ref{fig:g0}), which leads to slightly lower column densities. Settling does not affect the gas CO abundances significantly, as already suggested by \citet{2007A&A...463..203J}.

The thermal structure of both dust and gas phases in the disks, and the total dust surface area available for freeze-out and desorption, have a significant impact on the amount of CO ice (see column density of CO ice in the central panel of Fig. \ref{fig:co_col}) and the location of the CO snow surface. The snow surface is defined as the location (in both $R$ and $z$) where $50\%$ of a species is in the gas phase and $50\%$ is frozen out onto dust grains. Similarly, the snowline is defined as the radius where the same happens along the disk mid-plane. The CO ice abundance of most models is shown in Fig.~\ref{fig:abun-JCO_lin}. There are at least two effects that can be identified. The first one is that in the STN model, some CO ice is present out to very large radii ($>600\,$AU), since the dust is colder in the outer regions than in other models. The second one is that the CO ice abundance gets to rather high values ($\gtrsim10^{-4}$) for $\alpha\leq10^{-3}$. This is mainly due to vertical settling self-shadowing these regions of the disk, as explained in the previous section. In fact there is a clear correlation between the column density of CO ice (Fig.~\ref{fig:co_col}) and the dust temperature in the disk mid-plane (Fig.~\ref{fig:t_dust_midplane}). The dust temperatures in these models fall to $\sim17-20\,$K in the disk mid-plane, at $R\sim200-350\,$AU, and the large amount of dust surface area leads to significant CO freeze-out. This is shown more quantitatively in Fig.~\ref{fig:vert_cut_jco}, where the abundances of the main C carriers and the gas and dust temperature are shown in a vertical cut at $\sim300\,$AU. The amount of CO ice in the STN and $\alpha=10^{-2}$ case are again rather similar. Moreover, the abundance of CO ice clearly reflects the vertical settling of the dust particles, with the abundance being a steep function of $z$ for the low turbulence cases. \rev{Finally, we note that for the warm disk considered in this work, the column density of the gas phase CO is very similar in all models, since $N$(CO)$\gg N$(JCO) even in the lowest turbulence case (Fig.~\ref{fig:co_col}), since the gas phase CO column densities are dominated by the layers above the CO snow surface in all cases.}

The low turbulence models predict that CO freezes out in the disk mid-plane only in a very specific radial range. At larger radii, where the growth of dust particles has been limited by radial drift, the higher dust temperatures in the mid-plane (see Fig.~\ref{fig:t_dust_midplane}) enhance thermal desorption, leading to very low abundances of CO ice. This suggests that radial drift, grain growth, and settling can concur in having a second desorption front of CO at large radii, where the average grain size decreases substantially \citep[][]{2016ApJ...816L..21C}. We thus confirm that radial drift can indeed lead to a thermal inversion in the outer disk, thus leading to a second CO desorption front. Differently from the models by \citet{2016ApJ...816L..21C}, where the second CO desorption front is due to photodesorption, due to a low dust-to-gas ratio in the outer disk, in our models it is caused by thermal desorption, since in the outer regions $T_{\rm dust}$ becomes higher than the CO sublimation temperature. Interestingly, a few observations suggest that such a second CO desorption front could indeed be present in IM Lup, AS 209, and TW Hya, just outside the submm radius \citep[][respectively]{2015ApJ...810..112O,2016ApJ...823L..18H,2016ApJ...823...91S}. Whether these observations can be explained by the models presented here, or whether non-thermal desorption is important in these outer regions, is difficult to say at this stage, in particular given the very different properties of both disk and stellar mass of these three objects from the parameters used in this work.

\subsection{Emission}

Hitherto we have discussed the differences that grain evolution has on the density and temperature structure, and on CO abundances, in a representative model of a protoplanetary disk. We now investigate how these properties affect observable quantities, in particular the intensity profiles of both (sub-)mm continuum and CO isotopologues.

\subsubsection{Continuum}

\begin{figure*}
\center
\includegraphics[width=\textwidth]{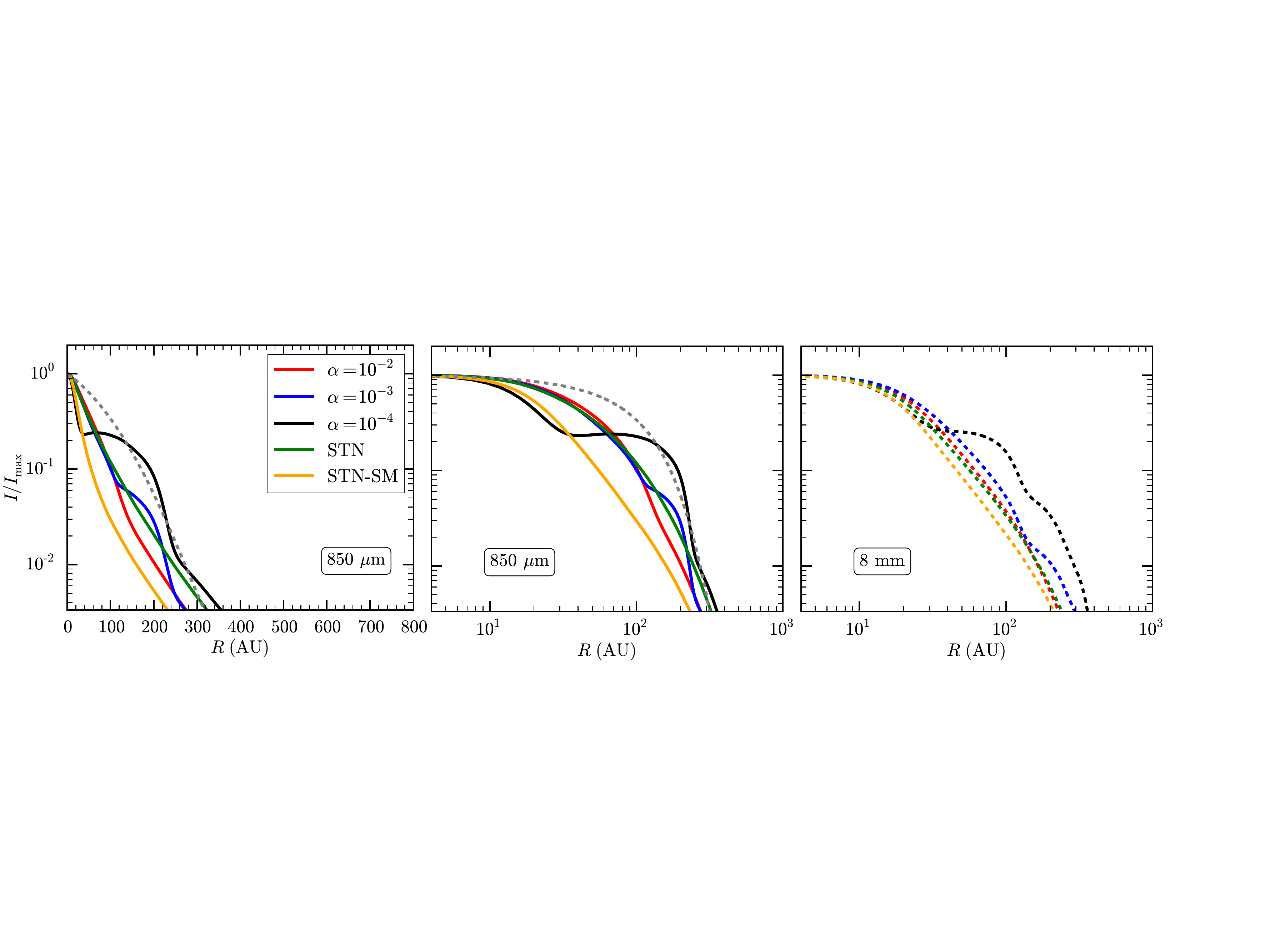}
\caption{Peak-normalised continuum intensity profiles of all models. \rev{In all panels the solid lines show the profiles at $850\,\mu$m. The difference between the two left panels is the radial scale. The dashed grey line in the two left panels indicates the normalised input surface density profile. In the right panel, the coloured dashed lines show the peak-normalised continuum intensity profiles at $8\,$mm, which look more compact than the $850\,\mu$m profiles, apart from the STN-SM model.}
}
\label{fig:cont_prof}
\end{figure*}

The peak-normalised continuum emission at $850\,\mu$m of all models is shown in Fig.~\ref{fig:cont_prof}. The STN-SM model is clearly different from all the others, since it lacks grains larger than $1\,\mu$m. Its mm emission is thus the most compact in radial extent. The radial extent of the STN, $\alpha=10^{-2}$ , and $10^{-3}$ cases is quite similar. The STN model does not have a radially varying grain size distribution, that is, radial drift is not considered in this parametric model. However, the sharpness of the continuum emission in the very outer regions does depend on the radial gradient of the grain size distribution \citep[or on the radial gradient of the dust-to-gas ratio, which we did not consider in this paper, see][]{birnstiel_14}. As the turbulence gets lower (and thus the steepness of the radial grain size distribution becomes more significant), the outer edge of the continuum emission becomes sharper. Thus, this sharp structure can be naturally explained by a combination of grain growth and radial drift. Interestingly, these models predict that the sharpness of the edge is correlated to a large scale substructure in the continuum emission that is clearly visible in the $\alpha=10^{-4}$ case in Fig.~\ref{fig:cont_prof}, but also present in the $\alpha=10^{-3}$ model. We recall that this substructure is due to the accumulation of large grains (and depletion of small particles) outside the fragmentation radius \citep{2015ApJ...813L..14B}. For this radial substructure to arise, there is no need of any planet or hydrodynamical instability, it is a pure consequence of radial drift and fragmentation of dust particles.

\rev{From our dust evolution models, we also recover the well-known result that the size of the emitting surface in continuum depends on the wavelength, as predicted by grain models considering radial drift. As an example, the right panel of Fig.~\ref{fig:cont_prof} compares the normalised synthetic emission profiles at $850\,\mu$m and $8\,$mm. Apart from the STN-SM model, which does not consider either grain growth or radial drift, all other cases show more compact emission at longer wavelengths.}

\begin{figure*}
\center
\includegraphics[width=\textwidth]{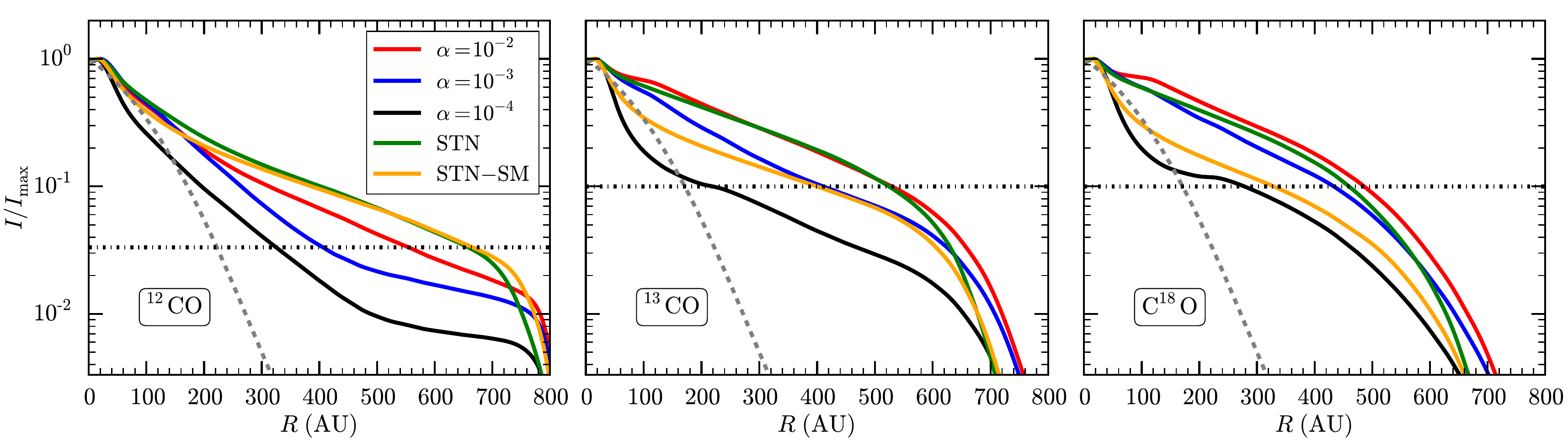}
\caption{From left to right: intensity profiles of the $^{12}$CO, $^{13}$CO, and C$^{18}$O $J$=3-2 line. The grey dashed line depicts the input surface density, convolved by the same beam size as the synthetic line intensity profiles. The horizontal dashed-dotted line represents an arbitrary intensity cut, with a dynamic range of $30$ for $^{12}$CO, and a dynamic range of $10$ for $^{13}$CO and C$^{18}$O.
}
\label{fig:profiles_gas}
\end{figure*}

\subsubsection{CO versus continuum radial profiles}
\label{sec:profiles_gas}
The emission of CO isotopologues is often used to determine the temperature and density structure of protoplanetary disks. The most abundant $^{12}$CO is used to constrain the thermal profile, whereas the more optically thin $^{13}$CO and C$^{18}$O are studied to determine the gas surface density profile \citep[e.g.][]{2013A&A...557A.133D,2016A&A...585A..58V,2016ApJ...823L..18H,2016ApJ...832..110C,2016ApJ...823...91S}.

\begin{figure*}
\center
\includegraphics[width=.33\textwidth]{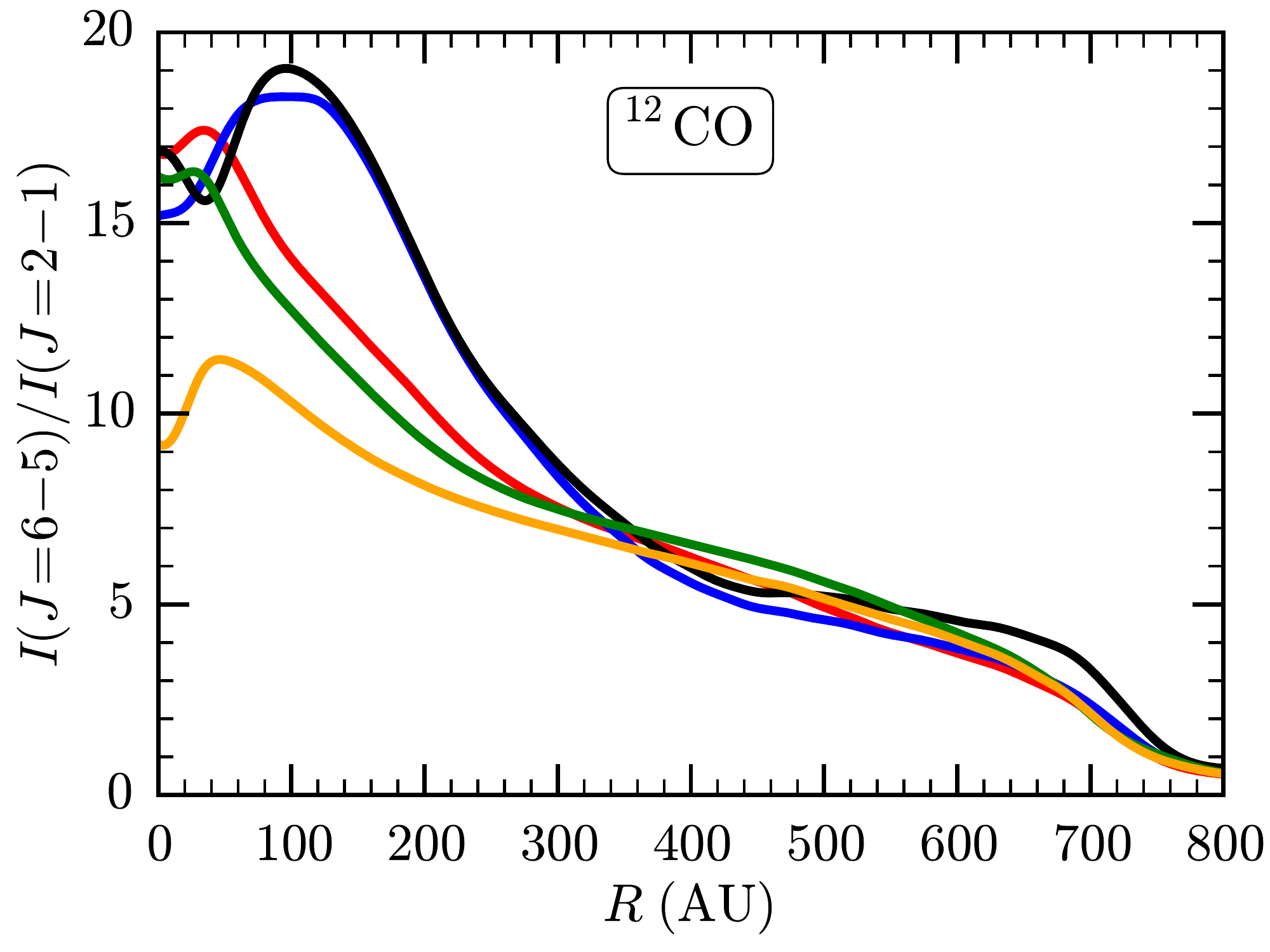}
\includegraphics[width=.33\textwidth]{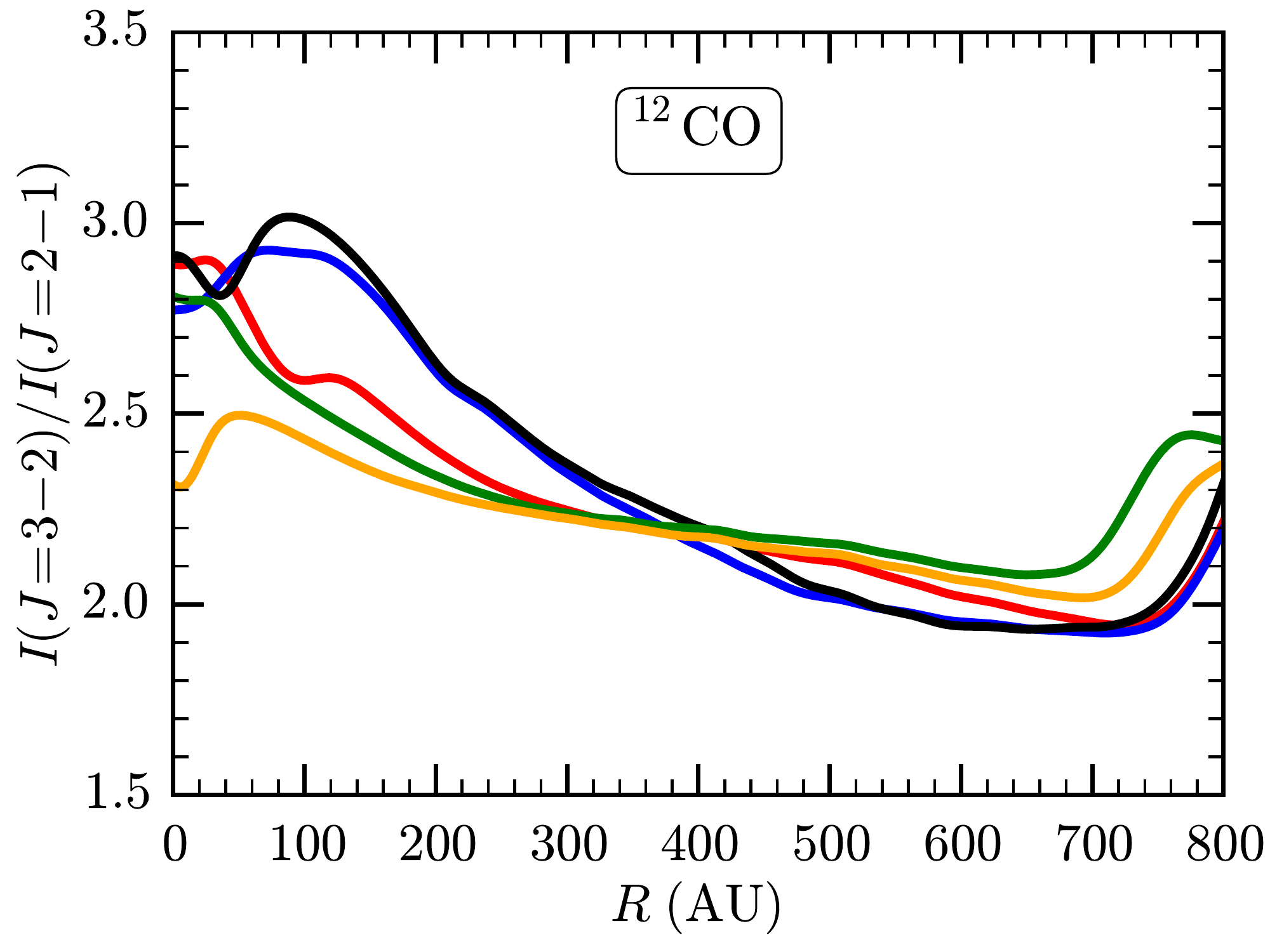}
\includegraphics[width=.33\textwidth]{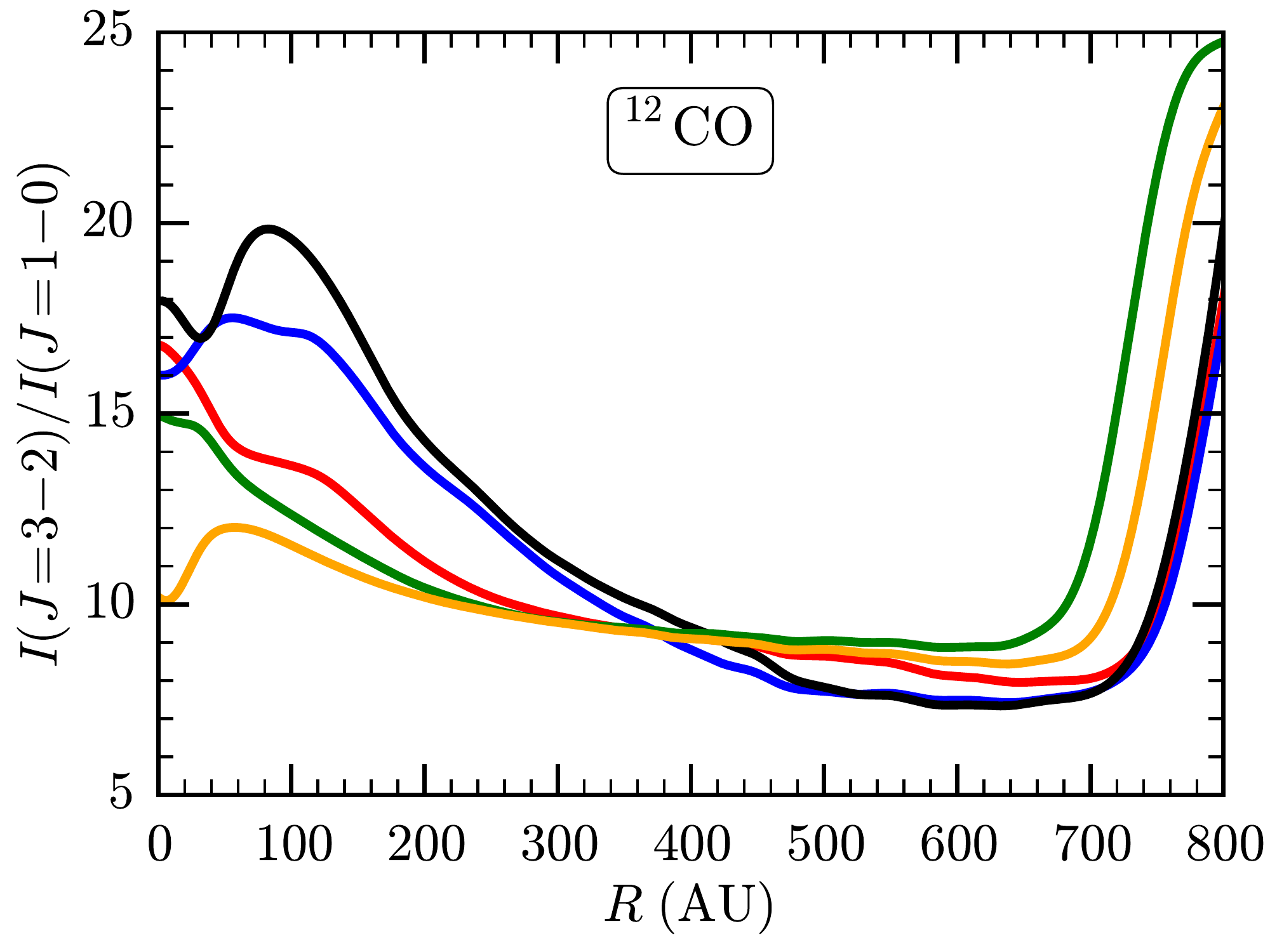}\\
\includegraphics[width=.33\textwidth]{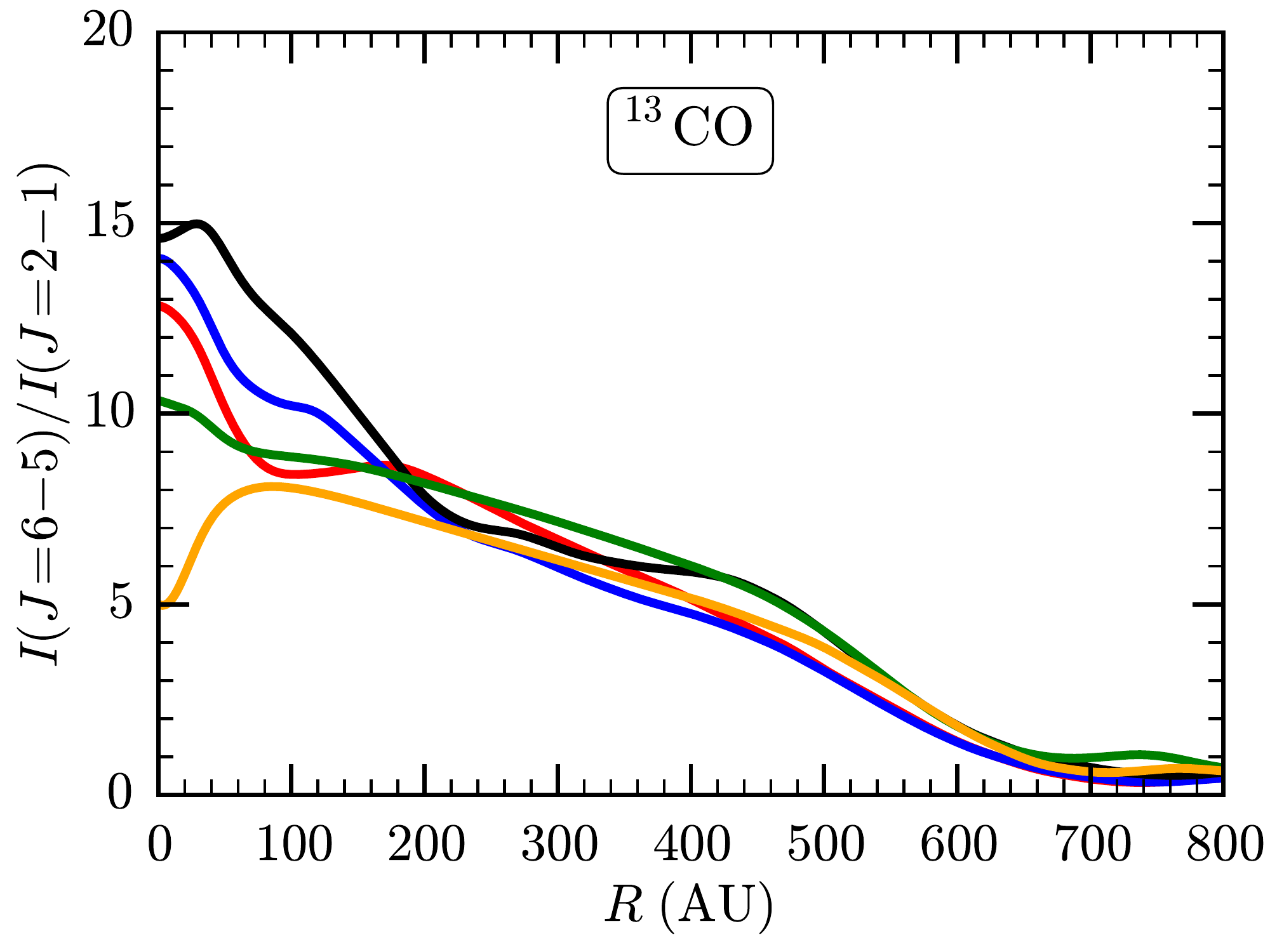}
\includegraphics[width=.33\textwidth]{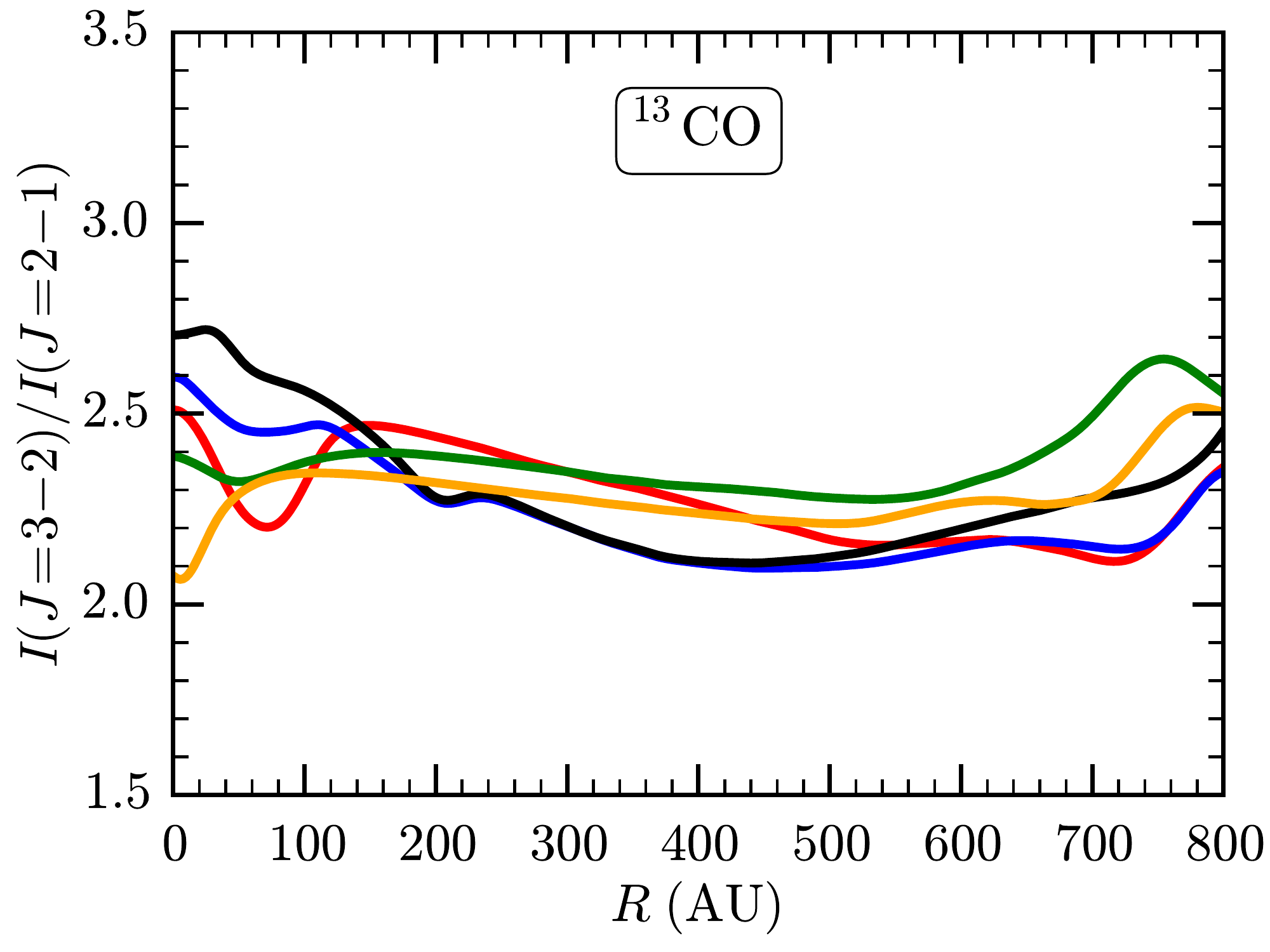}
\includegraphics[width=.33\textwidth]{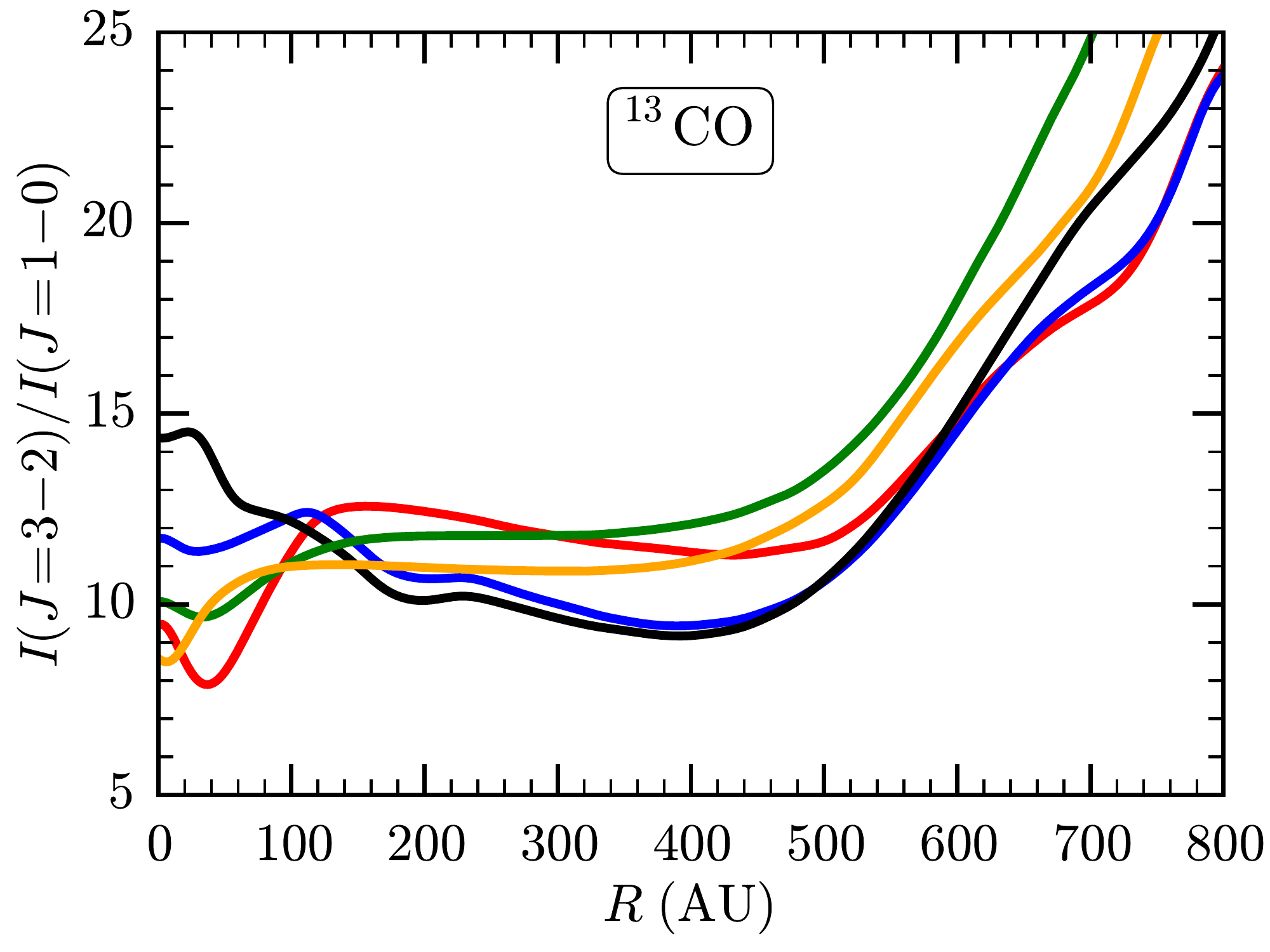}
\caption{Line ratios of $J$=6-5/$J$=2-1 (left column), $J$=3-2/$J$=2-1 (middle column), and $J$=3-2/$J$=1-0 (right column) for $^{12}$CO and $^{13}$CO (top and bottom row, respectively). The legend is the same as in Fig.~\ref{fig:profiles_gas}.}
\label{fig:line_ratio}
\end{figure*}

The radial intensity profiles of the CO $J$=3-2 line are shown in Fig.~\ref{fig:profiles_gas}. While the STN and $\alpha=10^{-2}$ models are very similar as usual, the $\alpha=10^{-3}$ and $10^{-4}$ profiles become dimmer in the outer regions of the disks, in particular for $^{12}$CO. This can be well understood from what has been outlined in Sec.~\ref{subsec:thermal_structure}. For lower turbulence values, the gas becomes colder, in particular in the regions where $^{12}$CO is emitting most, not far from the C -- CO transition. The lower temperature implies that the $J$=3 state is less populated, leading to lower emission in the $J$=3-2 line. The $\tau=1$ surface of the $J=$3-2 $^{12}$CO line crosses the region of thermal de-coupling of dust and gas in the low turbulent cases (see Fig.~\ref{fig:abun-CO_lin}), where the gas can be much colder than the dust. This is confirmed by looking at the line ratios, in particular at the ratio of the $J$=6-5/$J$=2-1 lines, the $J$=3-2/$J$=2-1 lines, and the $J$=3-2/$J$=1-0 lines, for both $^{12}$CO and $^{13}$CO (Fig.~\ref{fig:line_ratio}). The $^{12}$CO line ratios clearly show that there is a correlation between the line ratio being steeper and the steeper intensity profile in the $J$=3-2 line. This confirms that the difference seen between the models in the profiles of $^{12}$CO is due to a different thermal structure, in particular due to the vertical settling (and weak thermal dust-gas coupling) leading to lower temperatures in the outer disk. Such temperature effects are still seen in the $^{13}$CO line ratios, even though the trend with turbulence is less clear, due to the lower optical thickness of the lines. This confirms that the discrepancy is not caused by a difference in the column density, which is almost the same in all models (see Fig.~\ref{fig:co_col}), but by an excitation effect.

Figure~\ref{fig:profiles_gas} shows another significant result. Given the same gas surface density profile structure (the input gas density structure is the same for all models), the radial intensity profile of all CO isotopologues can be quite different depending on the dust properties of the disk, that is, depending on the radial grain size distribution and vertical settling. In particular, observations performed with ALMA at high sensitivity usually lead to a dynamic range in $^{12}$CO of the order of $\sim30$, whereas in $^{13}$CO and C$^{18}$O this is of the order of $\lesssim10$. In all panels of Fig.~\ref{fig:profiles_gas} the horizontal dashed-dotted line shows an arbitrary intensity cut with such dynamic ranges. These plots indicate that an observed disk with the same gas surface density profile would lead to very different estimates of the disk gas outer radius ($R_{\rm CO}^{\rm out}$), since this would be observationally determined by the sensitivity cut. In particular, from the $^{12}$CO line, the outer radius would vary between $\sim350 - 550\,$AU for $\alpha$ ranging between $10^{-2}$ and $10^{-4}$, whereas the same estimate based on the $^{13}$CO line would yield disk outer radii varying between $\sim250 - 550\,$AU for the same range of turbulence. Such an effect is less prominent for the C$^{18}$O isotope, since its lower optical depth makes it depend more strongly on the CO column density. For optically thinner isotopologues, the line emission would look smaller in size, as observed by for example. \citet{isella_16}. In general, for the same gas surface density profile, disks with lower turbulence would look smaller in CO intensity maps.

\rev{Both in the turbulent models and in the STN model}, the $^{12}$CO emission is more extended than the continuum (compare Figs.~\ref{fig:cont_prof}-\ref{fig:profiles_gas}), in most cases by a factor $\gtrsim$ two, considering a dynamic range of $30$ for both intensity profiles. This confirms the early suggestion by \citet{1998A&A...338L..63D} and \citet{1998A&A...339..467G} that the difference in radial extents between gas and dust is (mostly) due to optical depth effects. However, the ratio of $R_{\rm CO}^{\rm out}$ and $R_{\rm mm}^{\rm out}$ (taken as the radius where the intensity profiles reach a dynamic range of $30$ at $850\,\mu$m), depends strongly on turbulence. In particular, for the turbulent models shown in Figs.~\ref{fig:cont_prof}-\ref{fig:profiles_gas}, the ratio $R_{\rm CO}^{\rm out}/R_{\rm mm}^{\rm out}$ scales from $\sim1.4$ for $\alpha=10^{-4}$ to $\sim4$ for $\alpha=10^{-2}$, indicating that the ratio increases with turbulence.

Finally, we note that in this paper we have not considered how the CO emission is affected by potential C and O depletion in the disk, which has been suggested by thermo-chemical models of recent observations \citep[e.g.][]{2016A&A...592A..83K,2016ApJ...831..101B,2016ApJ...831..167M,2017A&A...599A.101V,2017A&A...599A.113M}.

\section{A closer view of HD 163296}
\label{sec:results_hd16}

As a representative case for our models, we consider the HD 163296 system. As mentioned in the Introduction, this is one of the first systems where the need for grain growth and radial drift has been advocated to explain the different radial intensity profiles in (sub)mm continuum and CO rotational lines, and in particular to explain the steep radial decline of the $850\,\mu$m emission. This system has been analysed in continuum and CO emission by a few different groups \citep{2013A&A...557A.133D,2013ApJ...774...16R,2015ApJ...813...99F,2016A&A...588A.112G,2016MNRAS.461..385B,2016ApJ...830...32W}, who have all exploited the ALMA Science Verification data, program 2011.0.00010.SV. In particular, \citet{2013A&A...557A.133D} and \citet{2013ApJ...774...16R} derived a density structure of the disk from the continuum and $^{12}$CO data, whereas more recently \citet{2016ApJ...830...32W} fitted the more optically thin $^{13}$CO and C$^{18}$O to obtain directly the gas surface density profile. They all used a surface density parametrisation given by Eq.~\ref{eq:surf_dens}, with very similar stellar mass, stellar luminosity, and flaring angle. We thus run models with the best fit parameters determining the gas surface density structure from these three papers. The actual values are reported in Table \ref{tab:hd163296}. We run both STN and STN-SM models of the three parametrisations, together with models with turbulence $\alpha=10^{-2}$, $10^{-3}$ , and $10^{-4}$. All parameters not specified in Table~\ref{tab:hd163296} are kept fixed. All models presented in the previous sections have the gas surface density parameters obtained by \citet[][see Sec.~\ref{sec:setup}]{2013A&A...557A.133D}. We stress that we do not aim to find a best fit model for HD 163296. The aim is to show how the properties of dust grains affect the emission of CO isotopologues, which are the most used probe for the gas surface density, and whether we can recover the continuum and lines intensity profiles of actual observations within a single framework.

\begin{figure}
\center
\includegraphics[width=\columnwidth]{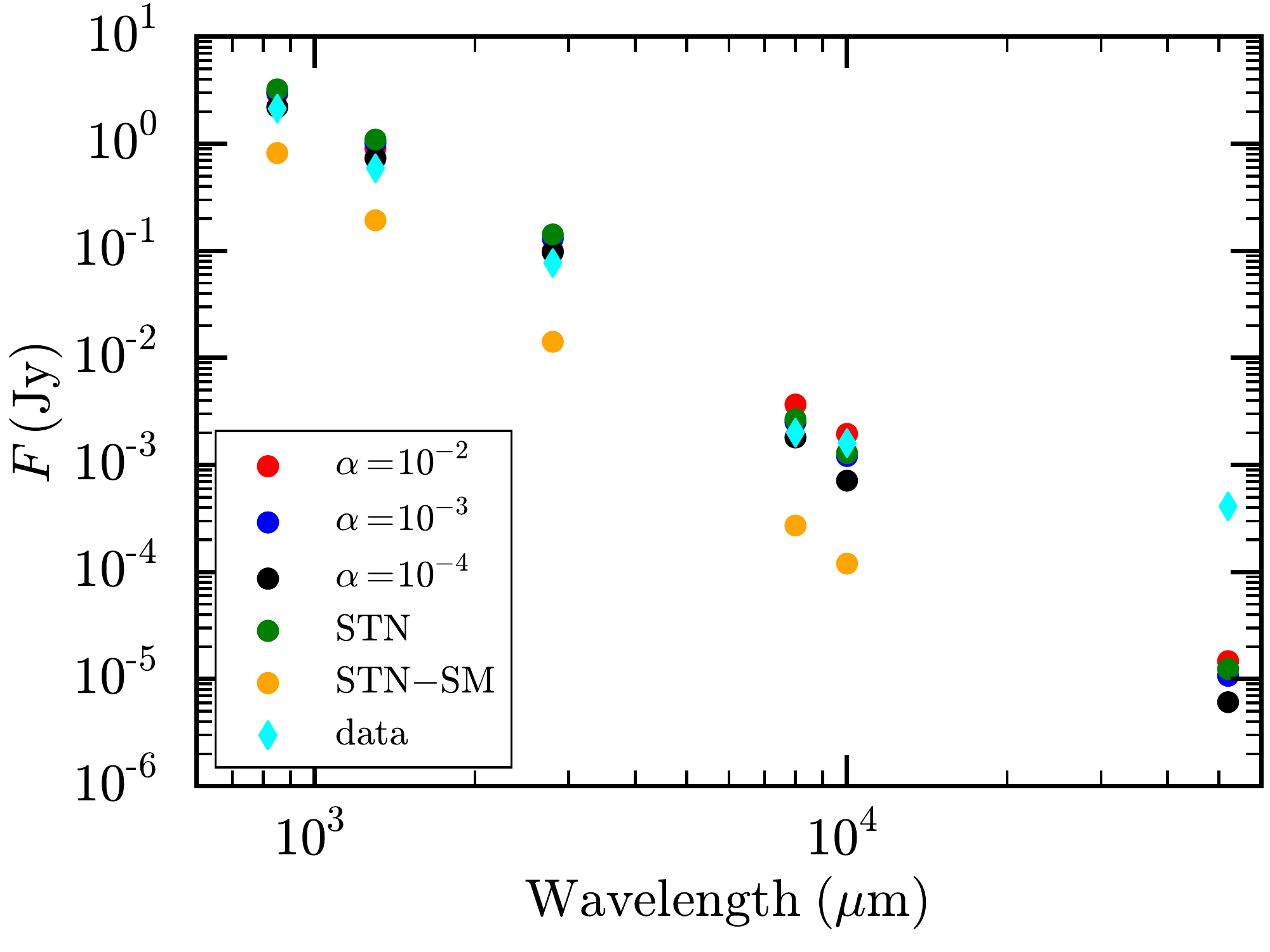}
\caption{Comparison of continuum fluxes predicted by all models using the gas density parameters from \citet{2013A&A...557A.133D} with the data from \citet[][at $2.8\,$mm]{2007A&A...469..213I} and \citet{2016A&A...588A.112G}.
}
\label{fig:fluxes}
\end{figure}

\begin{table}
\centering
\begin{tabular}{cccl}
\hline
\hline
$M_{\rm disk}\ (M_\odot)$ & $R_{\rm c}$ (AU) & $\gamma$ & Reference\\
\hline
$7\times10^{-2}$ & $125$ & $0.9$ & A \\
$9\times10^{-2}$ & $115$ & $0.8$ & B \\
$4.8\times10^{-2}$ & $213$ & $0.39$ & C \\
\hline
\end{tabular}
\caption{\rev{Main parameters of HD163296 used to produce the models in Fig.~\ref{fig:image2_de_greg_lin}, as derived in the following references}: [A] \citet{2013A&A...557A.133D}; [B] \citet{2013ApJ...774...16R}; [C] \citet{2016ApJ...830...32W}. }
\label{tab:hd163296}
\end{table}

First, we check that the models do recover the integrated continuum fluxes observed at (sub)mm wavelengths. In Fig.~\ref{fig:fluxes} we show a comparison of the continuum fluxes at $0.85$, $1.3$, $2.8$, $8$, $10,$ and $52\,$mm of the models using the parametrisation by \citet{2013A&A...557A.133D} with the continuum fluxes reported in \citet[][at $2.8\,$mm]{2007A&A...469..213I} and \citet{2016A&A...588A.112G}. The STN-SM model under-predicts the flux at all wavelengths, confirming that grain growth is needed to explain the observed emission. The other models are all roughly consistent with the data. All models under-predict the flux at $5.2\,$cm, but free-free emission is likely dominating at these long wavelengths and probably contributes also at $1\,$cm \citep{2016A&A...588A.112G}.

\begin{figure}
\center
\includegraphics[width=\columnwidth]{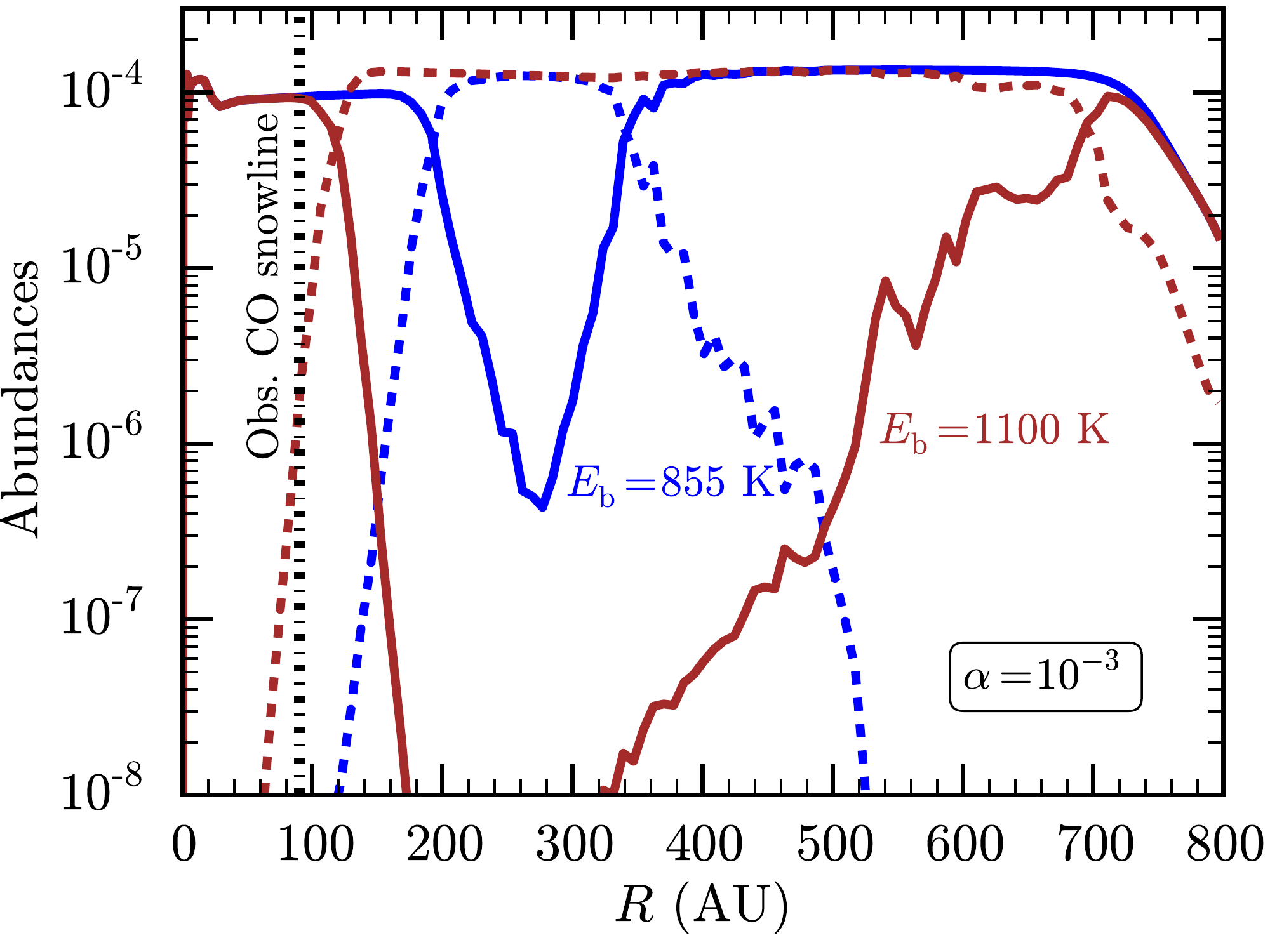}
\caption{ Abundance of CO (solid lines) and CO ice (dashed lines) along the disk mid-plane for the $\alpha=10^{-3}$ model, with CO binding energy of $855\,$K (blue lines) and $1100\,$K (brown lines). The vertical dashed-dotted black line shows the location of the CO snowline as determined by \citet{2015ApJ...813..128Q}. The parametrisation of the model is by \citet{2013A&A...557A.133D}. \rev{To reproduce the observed snowline location, the CO binding energy needs to be higher than that of pure CO ice, with $E_{\rm b}=1100\,$K leading to a much better agreement than $E_{\rm b}=855\,$K.}
}
\label{fig:co_col_eb1100}
\end{figure}

\citet{2015ApJ...813..128Q} determined a radial location for the CO snowline at $90\pm10\,$AU by modelling the observed N$_2$H$^+$ and C$^{18}$O emission \citep[see also the earlier work by][on DCO$^+$]{2013A&A...557A.132M}. Even though the recent work by \citet{2017A&A...599A.101V} has shown that the interpretation of N$_2$H$^+$ emission as a CO snowline tracer is more subtle than previously assumed, the C$^{18}$O emission still suggests that the snowline is at about $90-100\,$AU. In the models shown in Fig.~\ref{fig:co_col}, we obtain a CO snowline around $250\,$AU, larger by a factor of two in the most optimistic case ($\alpha=10^{-3}$). However, all those models assume a CO binding energy of $855\,$K, typical for pure CO ice (see Sec.~\ref{sec:freeze_out} for a more detailed discussion). By running a turbulent model ($\alpha=10^{-3}$) with a CO binding energy more appropriate for ice mixtures ($1100\,$K), we obtain a location of the snowline of $\sim115\,$AU, which is more consistent with the data (see Fig.~\ref{fig:co_col_eb1100}), as is also suggested by \citet{2015ApJ...813..128Q}. The location of the snowline corresponds to the radius at which the dust temperature reaches the CO sublimation temperature (see Fig.~\ref{fig:t_dust_midplane}). The column density of gaseous CO is not significantly affected. Interestingly, the least turbulent models predict a second thermal desorption front in the outer regions of the disk, around $\gtrsim500\,$AU. Deeper ALMA observations should be able to determine whether this prediction is correct, for example by looking for DCO$^+$ emission in these outer regions.

\begin{figure*}
\center
\includegraphics[width=.33\textwidth]{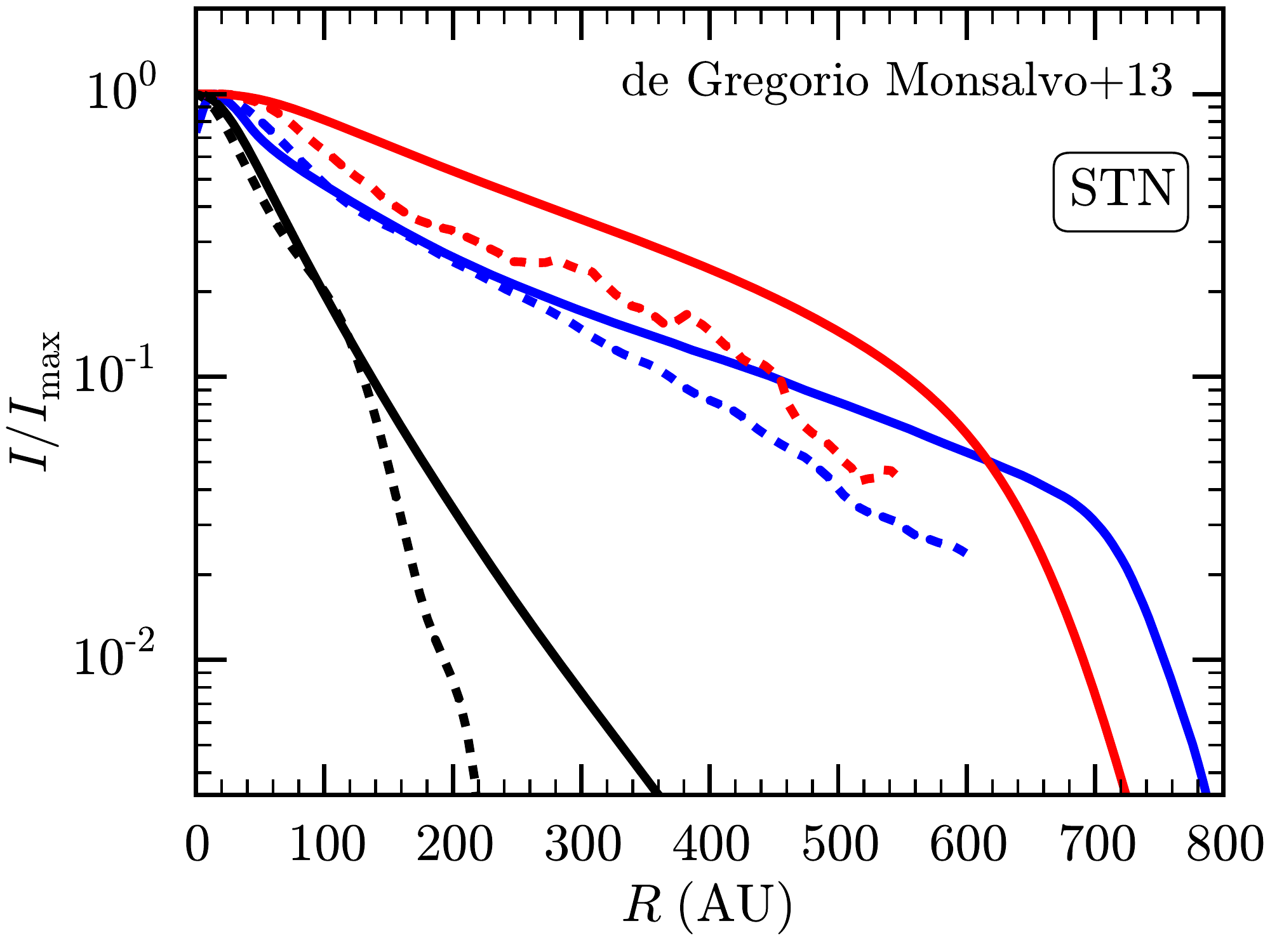}
\includegraphics[width=.33\textwidth]{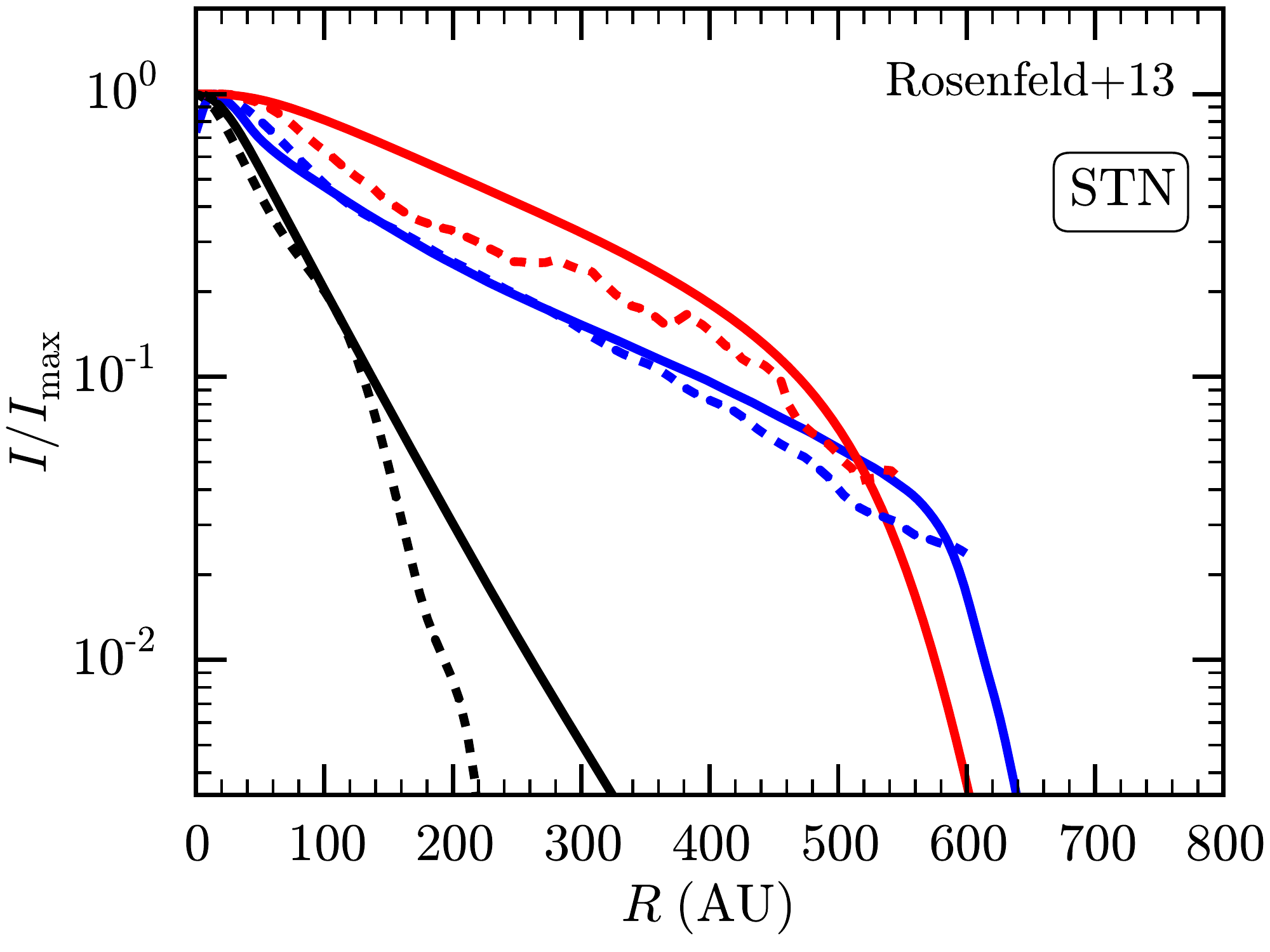}
\includegraphics[width=.33\textwidth]{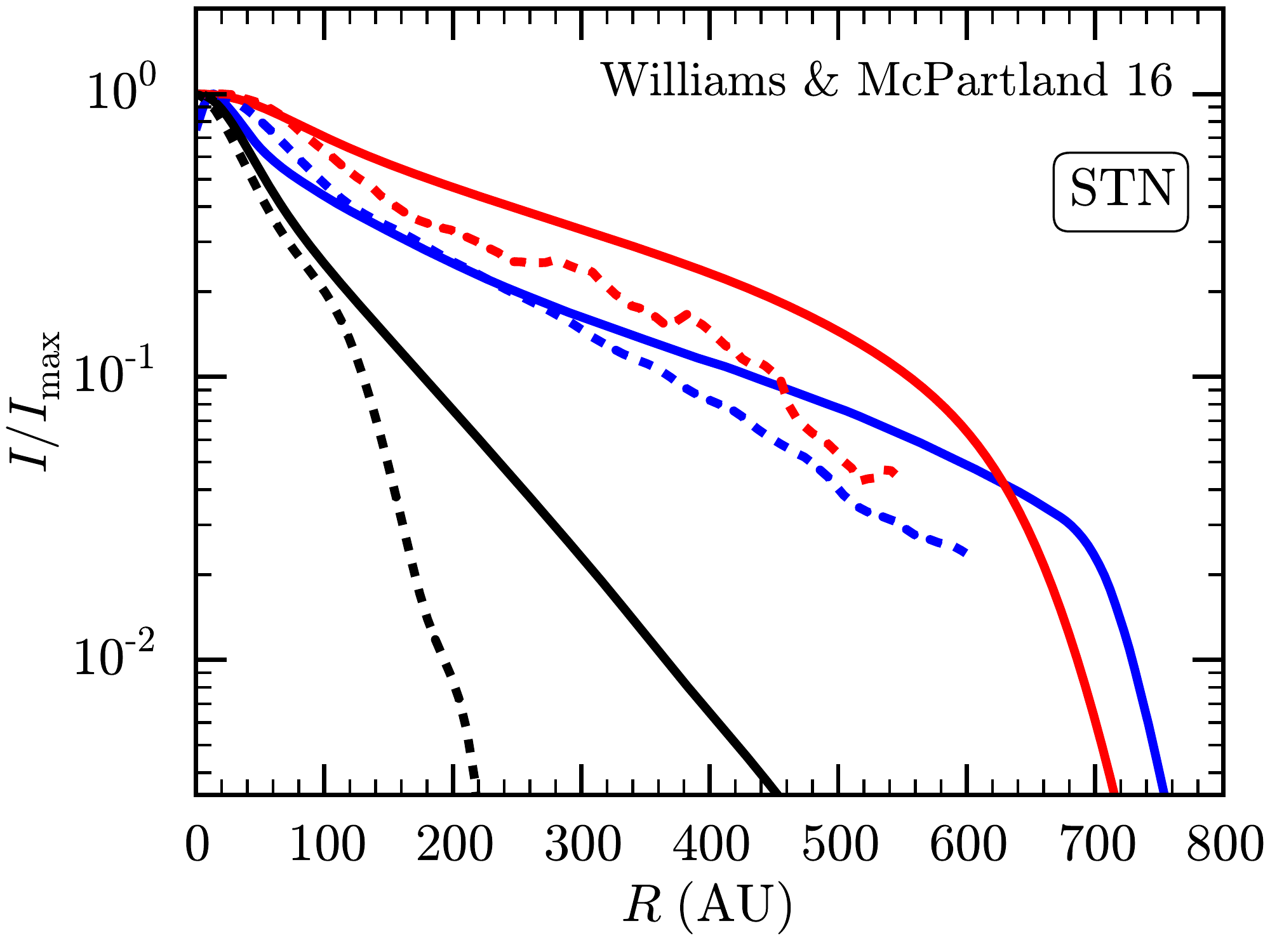}\\
\includegraphics[width=.33\textwidth]{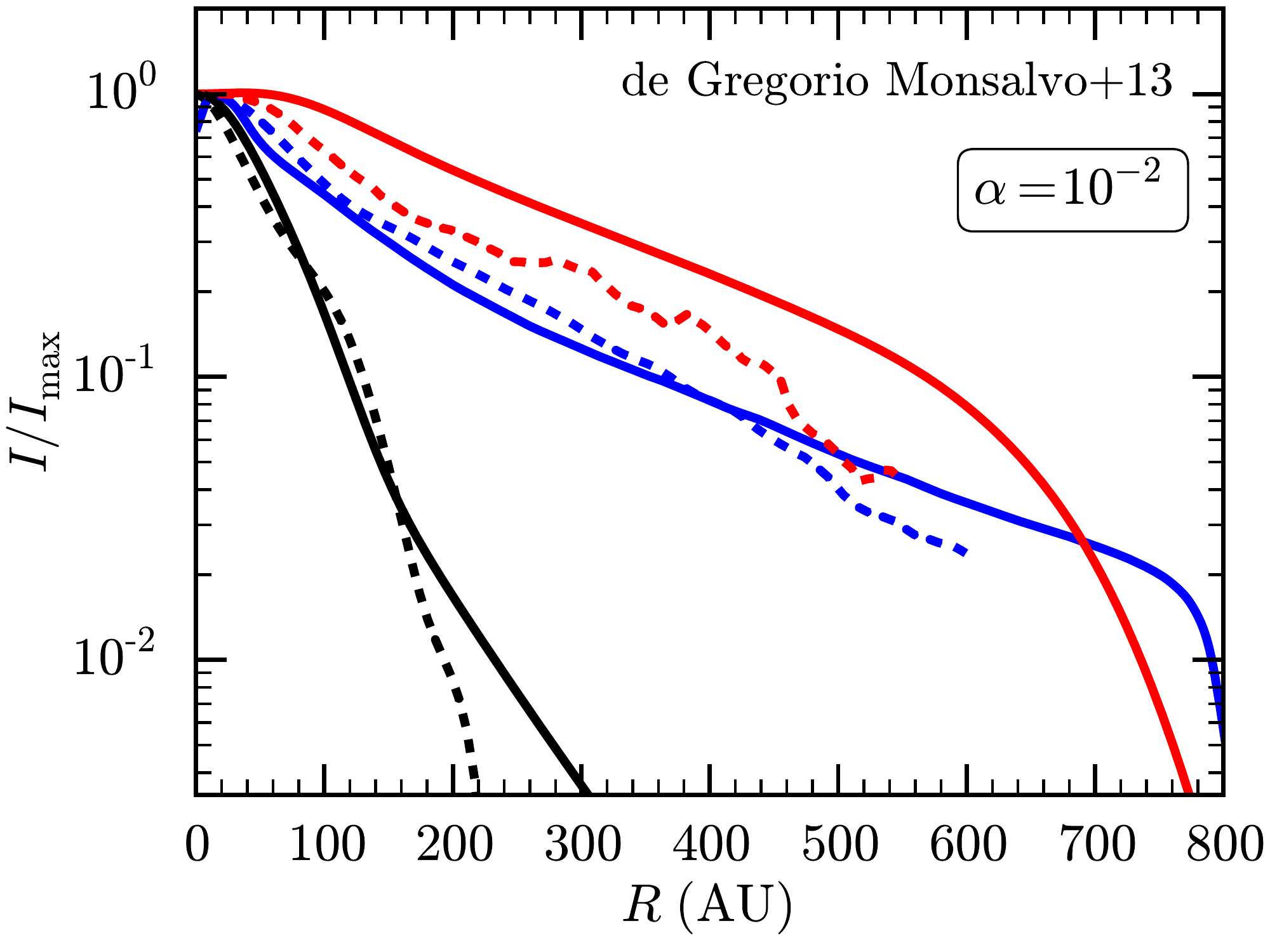}
\includegraphics[width=.33\textwidth]{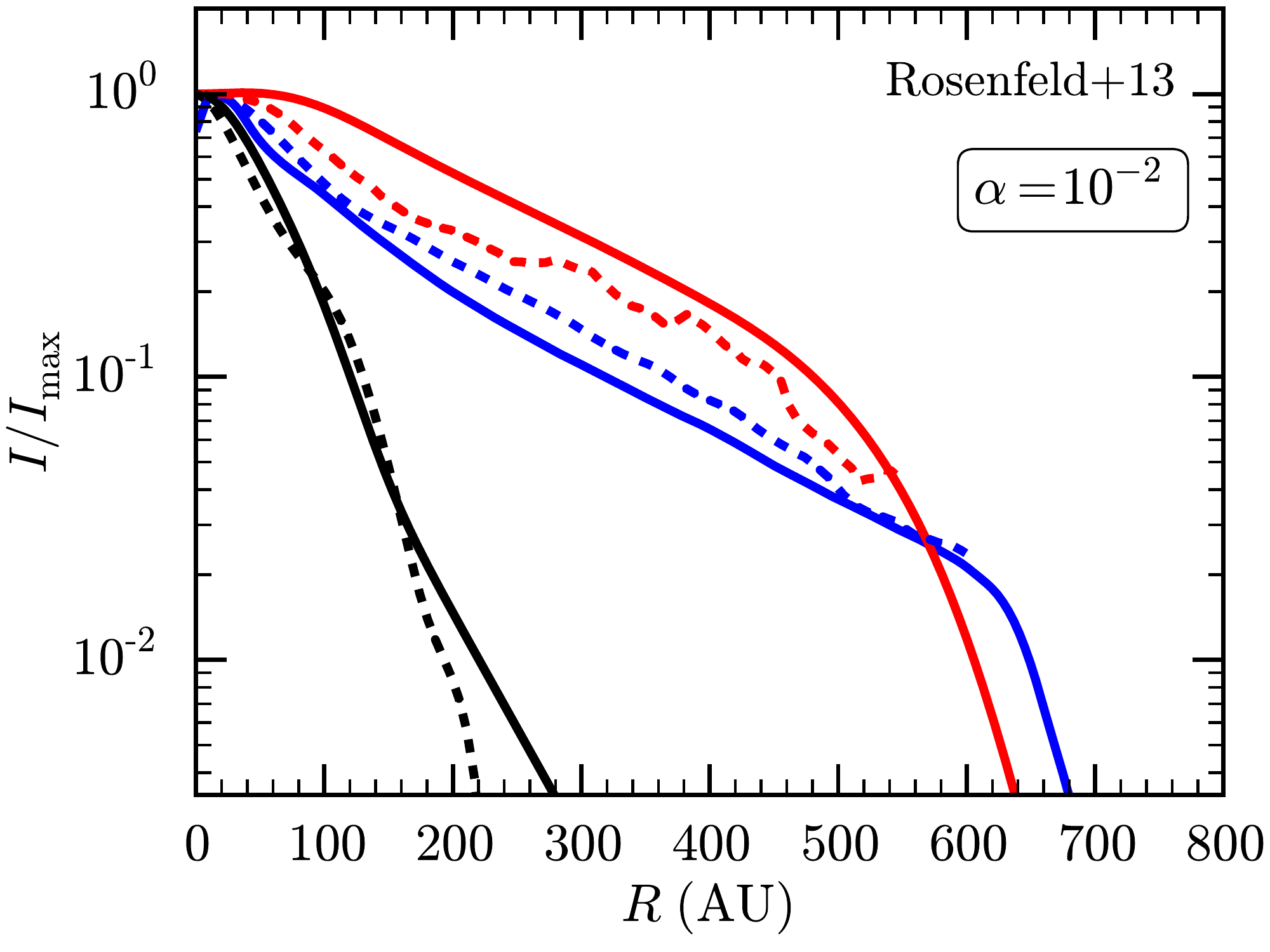}
\includegraphics[width=.33\textwidth]{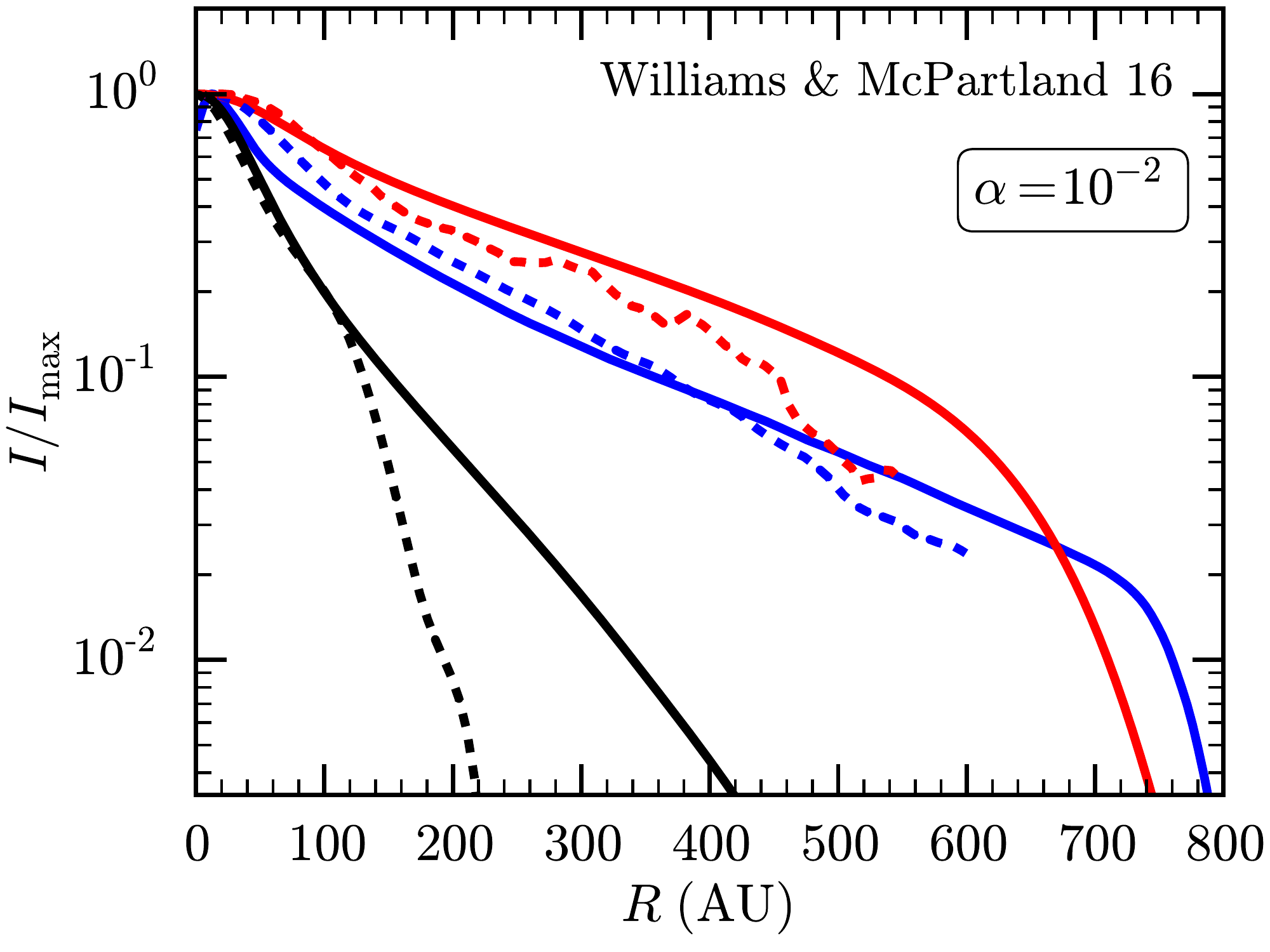}\\
\includegraphics[width=.33\textwidth]{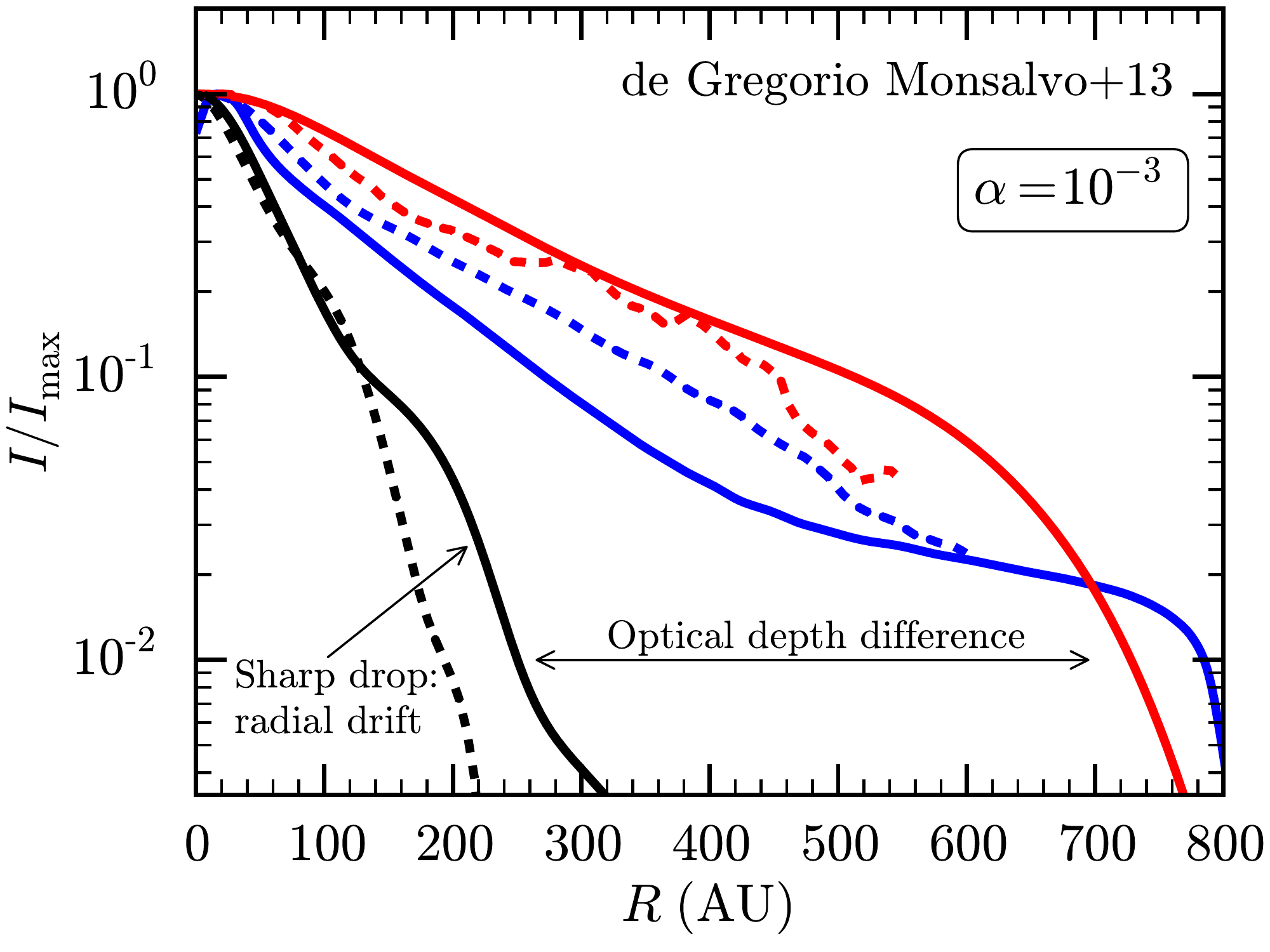}
\includegraphics[width=.33\textwidth]{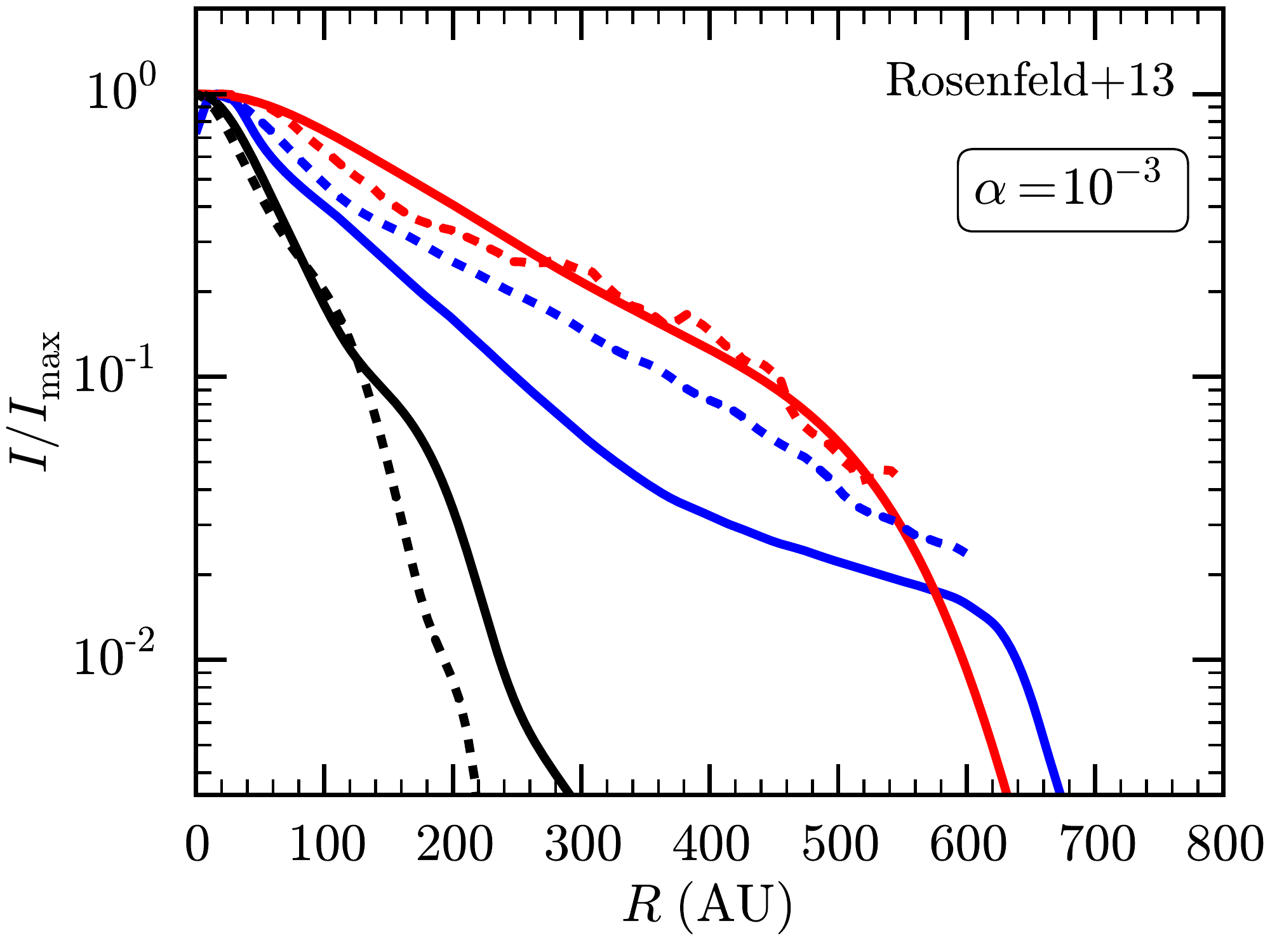}
\includegraphics[width=.33\textwidth]{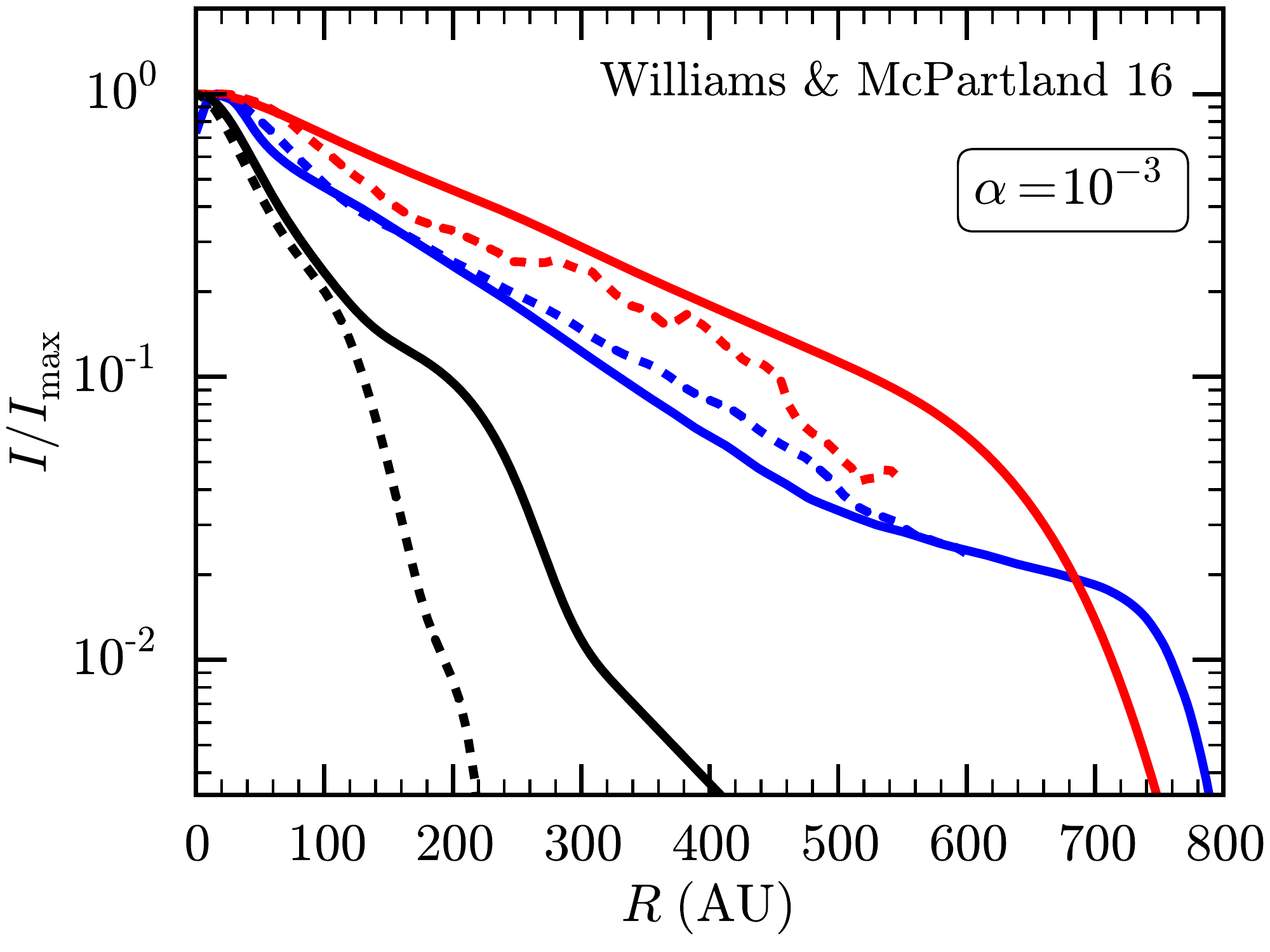}\\
\includegraphics[width=.33\textwidth]{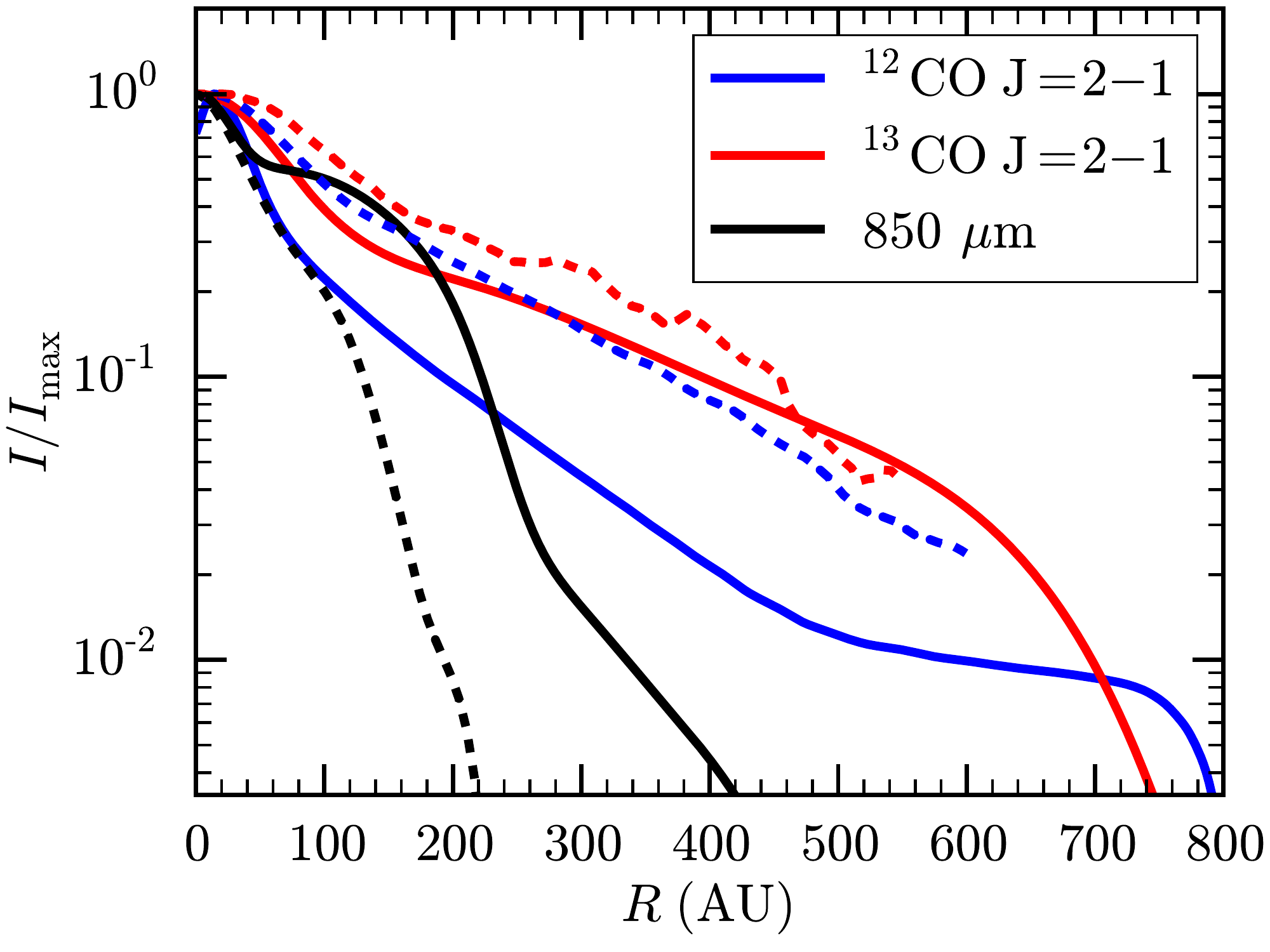}
\includegraphics[width=.33\textwidth]{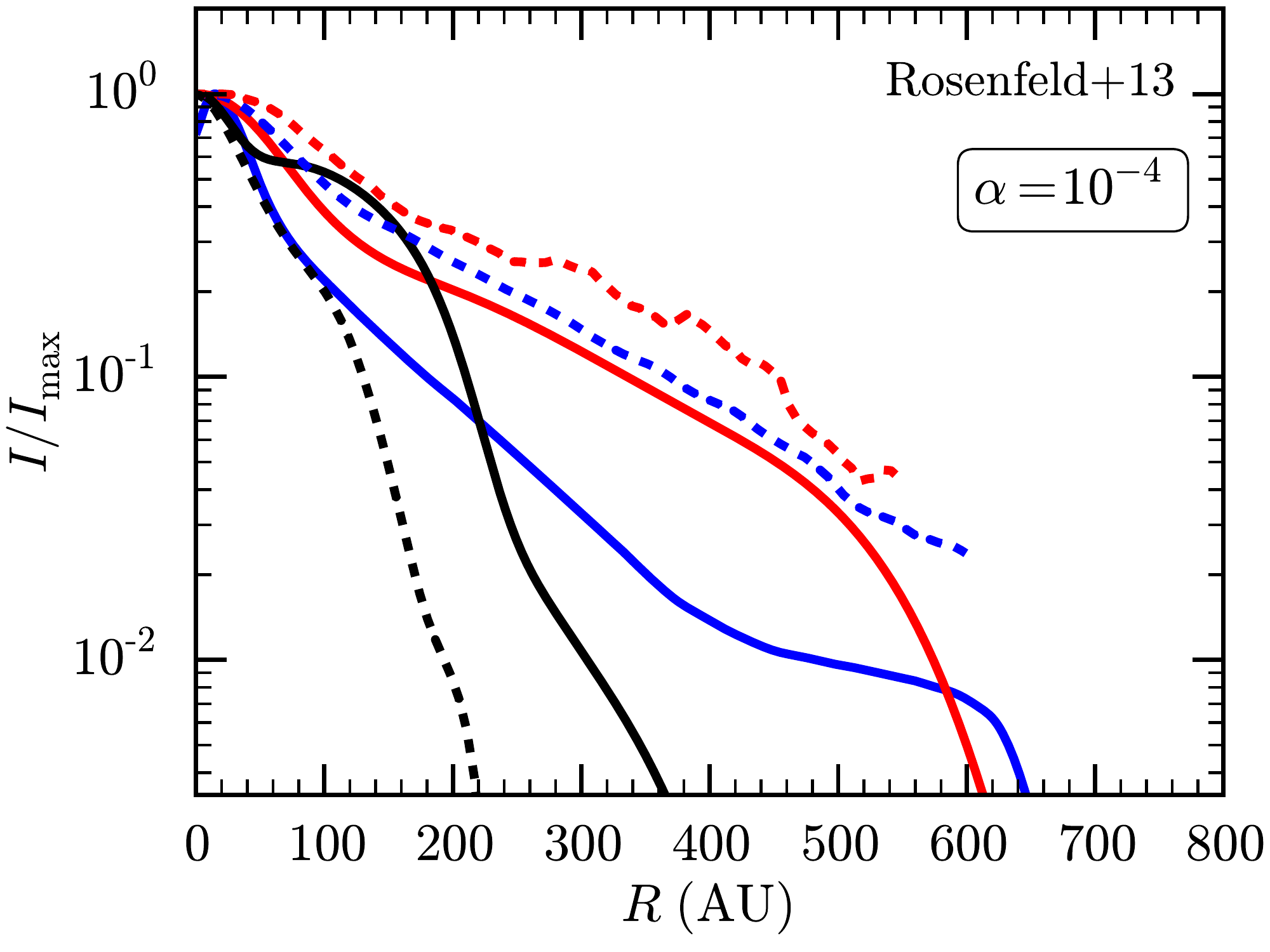}
\includegraphics[width=.33\textwidth]{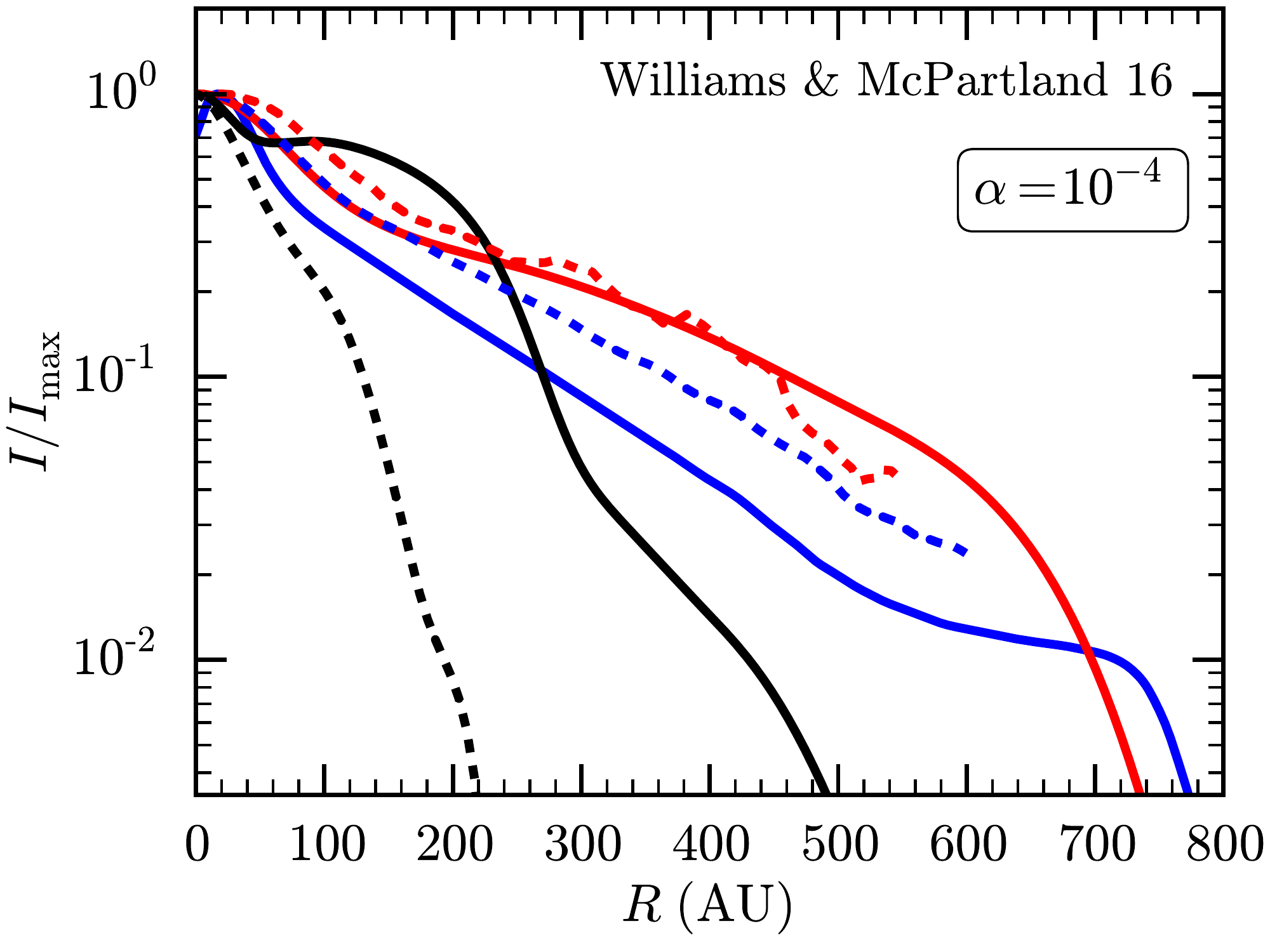}\\
\caption{Comparison between the Science Verification ALMA data and models of HD 163296. The dashed lines represent the ALMA data, the solid lines the models. From top to bottom: STN, $\alpha=10^{-2}$, $10^{-3}$ , and $10^{-4}$ models. From left to right: model parameters taken from \citet{2013A&A...557A.133D}, \citet{2013ApJ...774...16R}, and \citet{2016ApJ...830...32W}. \rev{The $\alpha=10^{-2}$ case in the middle column is the best representation of the data among our models.}
}
\label{fig:image2_de_greg_lin}
\end{figure*}

Figure~\ref{fig:image2_de_greg_lin} shows the observed and modelled peak-normalised intensity profiles of both continuum at $850\,\mu$m and of the $^{12}$CO and $^{13}$CO $J$=2-1 lines. The $^{12}$CO and $^{13}$CO synthetic images are obtained with a beam convolution of $0.68\arcsec\times 0.55\arcsec$ and $0.72\arcsec\times 0.57\arcsec$, respectively, with a PA$=72^\circ$. The vertical scale is used to mimic a peak value of $\sim300\sigma$. With the parameters used by \citet{2016ApJ...830...32W}, we always obtain a dust disk outer radius that is too large compared to the ALMA data. More interestingly, the continuum observations of the STN models all fail to reproduce the sharp edge in the outer disk, whereas models including dust evolution are more consistent with the data. For low turbulence values, the sharp edge is associated with substructure in the disk continuum emission, with the level of substructure increasing with decreasing turbulence (i.e. with $R_{\rm frag}$ shifting towards the inner regions of the disk). The $^{12}$CO data are well recovered by many models, in particular with $\alpha=10^{-2}-10^{-3}$. When $\alpha=10^{-4}$, the temperature gradient along the CO emitting layer becomes too steep, leading to a fast decline in CO emission, which is not observed in the ALMA data.

\section{Discussion}
\label{sec:discussion}

\subsection{Gas versus dust radial extent}

So far, the quantity that has been mostly derived for protoplanetary
disks is disk mass, since the observations have been limited by
angular resolution and sensitivity. This fact has led the community to
mostly look at models of disk integrated CO emission, in particular at
the fluxes of different CO isotopologues which can be used to
constrain the gaseous disk mass
\citep[e.g.][]{williams_14,2016A&A...594A..85M}. However, recent and
upcoming ALMA observations allow us to investigate spatially resolved
chemical and physical properties of the gaseous component of disks, in
particular the surface density profile and the disk outer radius
\citep[e.g.][]{2016ApJ...830...32W}.

\begin{figure}
\center
\includegraphics[width=0.92\columnwidth]{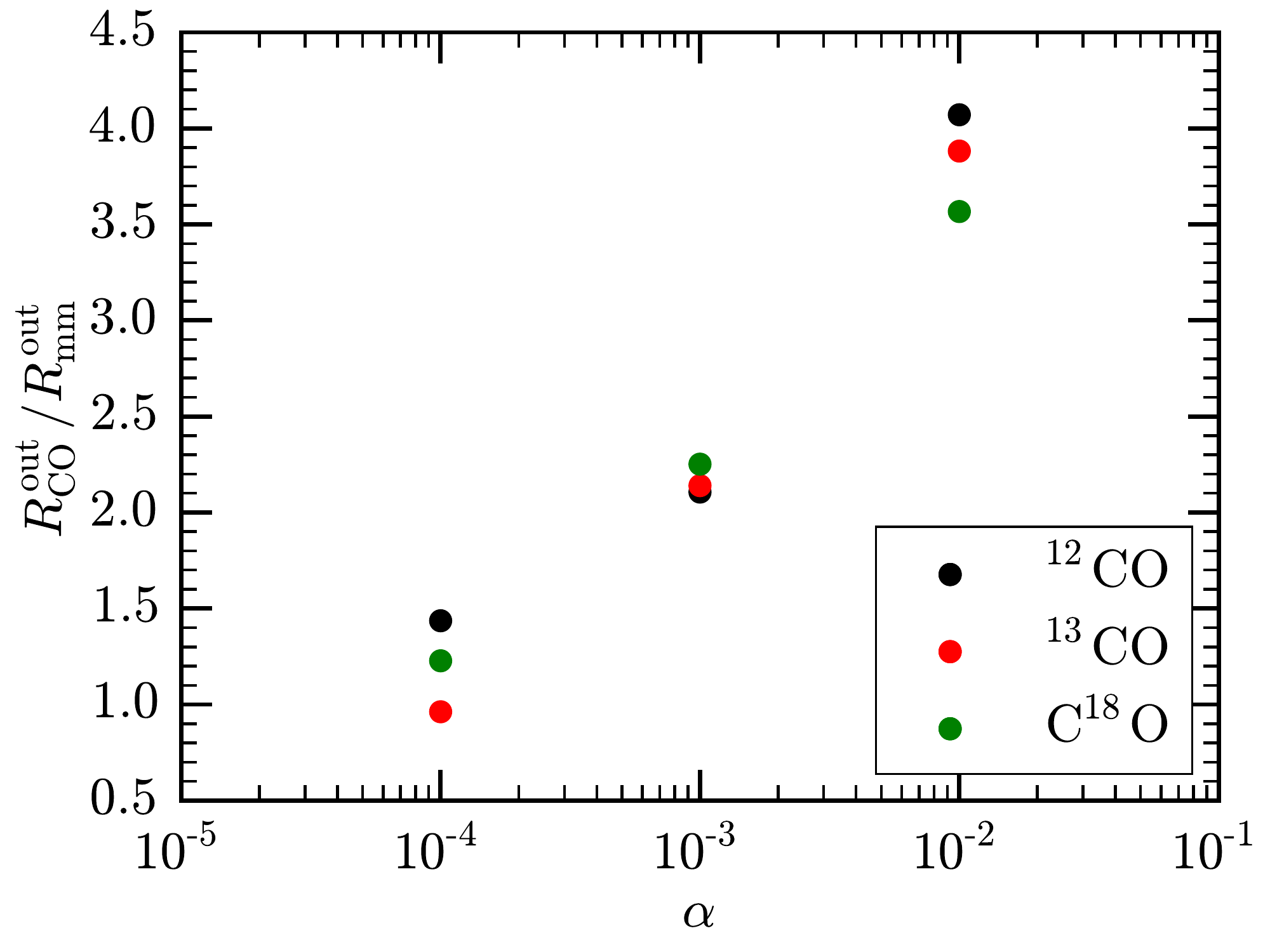}
\caption{\rev{Ratio $R_{\rm CO}^{\rm out}/R_{\rm mm}^{\rm out}$ derived for the models using the disk parametrisation of \citet{2013A&A...557A.133D}, with a $0.52\arcsec$ synthetic beam. The quantity $R_{\rm CO}^{\rm out}$ is derived assuming a dynamic range of 30 for $^{12}$CO and a dynamic range of 10 for $^{13}$CO and C$^{18}$O. For the continuum at $850\,\mu$m, the assumed dynamic range is 30. For all isotopologues, the ratio $R_{\rm CO}^{\rm out}/R_{\rm mm}^{\rm out}$ is a steep increasing function of disk turbulence.}
}
\label{fig:rco}
\end{figure}

Our models including physically motivated grain properties confirm the
early suggestion by \citet{1998A&A...338L..63D} and \citet{1998A&A...339..467G}
that the observed different radial extents in gas and dust can be
largely explained by the difference in the optical depth of gas versus
dust. This feature is retrieved by all models mimicking grain growth
and vertical settling (i.e. the STN models), without the need to invoke radial
drift. However, a simple quantity such as the disk gas outer radius as
derived from $^{12}$CO and $^{13}$CO radial intensity profiles can
depend significantly on the properties of the dust grains. In
particular, the radial extent of the (sub)mm continuum does not depend
strongly on turbulence (considering all the assumptions in our
models), whereas the radial extent of the CO emission does. In our
models, the ratio of the CO and mm continuum radius
$R_{\rm CO}^{\rm out}/R_{\rm mm}^{\rm out}$ is directly related to the amount of
turbulence in the disk, with the ratio increasing with $\alpha$ \rev{(Fig.~\ref{fig:rco}). The exact values of the ratio will depend on the dynamic range of the CO emission used to determine $R_{\rm CO}^{\rm out}$. The value of the ratio will also depend on the wavelength considered to extract $R_{\rm mm}^{\rm out}$, since the size of the dust disk can depend on wavelength (Fig.~\ref{fig:cont_prof}), with disk size slowly decreasing with increasing wavelength}. In this preliminary study we did not explore the
dependence on other important parameters, such as stellar mass (thus
stellar spectrum), disk mass, and surface density profile, although
qualitatively the trend should be robust.

In order to interpret measurements of the disk outer radius and line
emission, a proper treatment of the dust properties is needed.  In the
future, observations at multiple (sub)mm frequencies will allow us to
determine some fundamental properties of the large dust grains as a
function of radius, in particular meaningful constraints on the radial
dependence of the grain size distribution
\citep[e.g.][]{2011A&A...529A.105G,2012ApJ...760L..17P,2015ApJ...813...41P,2014A&A...564A..93M,2016A&A...588A..53T}. Such inferred dust properties could then be
used as a parametric input for the thermo-chemical models, in order to
interpret the emission of important molecules (as CO) more
consistently with the continuum observations.
 
In this work we have mostly looked at the large scale structure of
disks, which are assigned a smooth gas surface density profile. We
have not explored the potential effects of small substructures in the
surface density \citep[as observed
by][]{2015ApJ...808L...3A,2016ApJ...820L..40A,2016ApJ...829L..35T,isella_16}
and we have not looked at how substructures in the dust properties can
affect the gas emission. However, from the results shown in this
paper, it is reasonable to assume that radial substructures in dust
properties will affect the CO emission, by changing the gas thermal
variations and possibly abundances on the same scales \citep[e.g.][]{2013A&A...559A..46B,2016ApJ...820L..25Y,isella_16,2017ApJ...835..228T}.

\subsection{Effects of grain growth, vertical settling, and radial drift on chemistry and line emission}

The combination of dust evolution processes, in particular grain
growth, vertical settling, and radial drift, largely determine the dust
density and thermal structure in protoplanetary disks. Moreover, they
also affect the continuum mean intensity throughout the disk, which is
fundamental in the thermo-chemical processes of the gaseous
material. In particular, severe vertical settling and substantial
grain growth decrease the efficiency of physical and chemical
processes dependent on the dust surface area.  In these cases, the gas
can be largely colder than the dust component at intermediate scale
heights due to the low rate of gas-grain collisions. Optically thick
molecular lines emitting from the disk intermediate layers, as low $J$
CO transitions, can show the imprint of such cold gas temperatures by
being less excited than expected from simple models. Thus, the largest
effect of dust evolution and radial drift is on the gas temperature
structure; enhanced UV penetration heating the gas and dissociating or
desorbing CO has a minor effect. As a result, the radial CO emission
profile can be significantly steeper than the actual gas surface density
profile.  In this paper we have focused on CO emission lines, but the
same can indeed occur for other molecular lines emitted from the same
regions of disks.

The dust properties also affect the abundances of ices in the disk
mid-plane. The dust density structure and grain size distribution
determine the dust temperature in the disk mid-plane and the grain
size distribution sets the total dust surface area available for
adsorption and desorption processes. We have shown how these two
mechanisms determine the amount of CO ice in the disk mid-plane for
the disk parametrisation analysed in this paper, which is typical of a
Herbig star. The combination of radial drift and vertical settling in
particular can induce a thermal inversion in the disk mid-plane in the
outer regions of the disk, without the need to reduce the dust-to-gas
ratio in these outer regions. Since the ice abundances are so
sensitive to the thermal structure and dust total surface area
available, including more realistic dust properties in the
thermo-chemical models is important to follow the chemical evolution
of some species, in particular for processes depending on the dust
surface area, such as grain surface reactions in the disk mid-plane. The
formation of some precursors of complex molecules, such as CH$_3$OH and
H$_2$CO \rev{\citep[the latter has been detected in HD163296 by][]{2016ApJ...832..204Y}}, is indeed coupled to the ice composition and abundances on
the mantles of dust grains.

\subsection{Dust growth versus freeze-out timescales}

\label{sec:discussion_timescales}

In the results shown above, the main implicit assumption is that the growth timescale of dust particles is shorter than the main chemical timescales for the reactions considered in our chemical network, such that the thermal balance can be computed on static dust properties. This assumption is made only for numerical reasons, that is, there is still no code capable of coupling dust evolution and thermo-chemistry in a 2D disk for a viscous timescale. We can however check {\it a posteriori} whether the methodology presented here should be improved in the near future \citep[see][for a more generic discussion]{2016PASA...33...53H}. We can do so by comparing the growth timescale of dust particles of a given size with the CO freeze-out timescale on particles of the same size, in order to focus in particular on the CO snow surface. The model that is best suited to make such comparison is the STN-SM one, where the grains are still small ($\bar{a}\lesssim0.1\,\mu$m). In the comparison, the gas thermal structure computed by the model is used.

The growth timescale $t_{\rm growth}$ of a dust particle of size $a$ due to turbulent motions can be written as $t_{\rm growth}=a/\dot{a}$, where $\dot{a}$ is \citep{2008A&A...480..859B}:

\begin{equation}
\dot{a} = \frac{\rho_{\rm dust}}{\rho_{\rm gr}} \Delta v.
\end{equation}
The relative velocities due to turbulence can be expressed as \citep{2007A&A...466..413O}:

\begin{equation}
\Delta v = c_{\rm s} \sqrt{\frac{3}{2} \alpha\, {\rm St} } ,
\end{equation}
where the local Stokes number can be written as:

\begin{figure}
\center
\includegraphics[width=0.92\columnwidth]{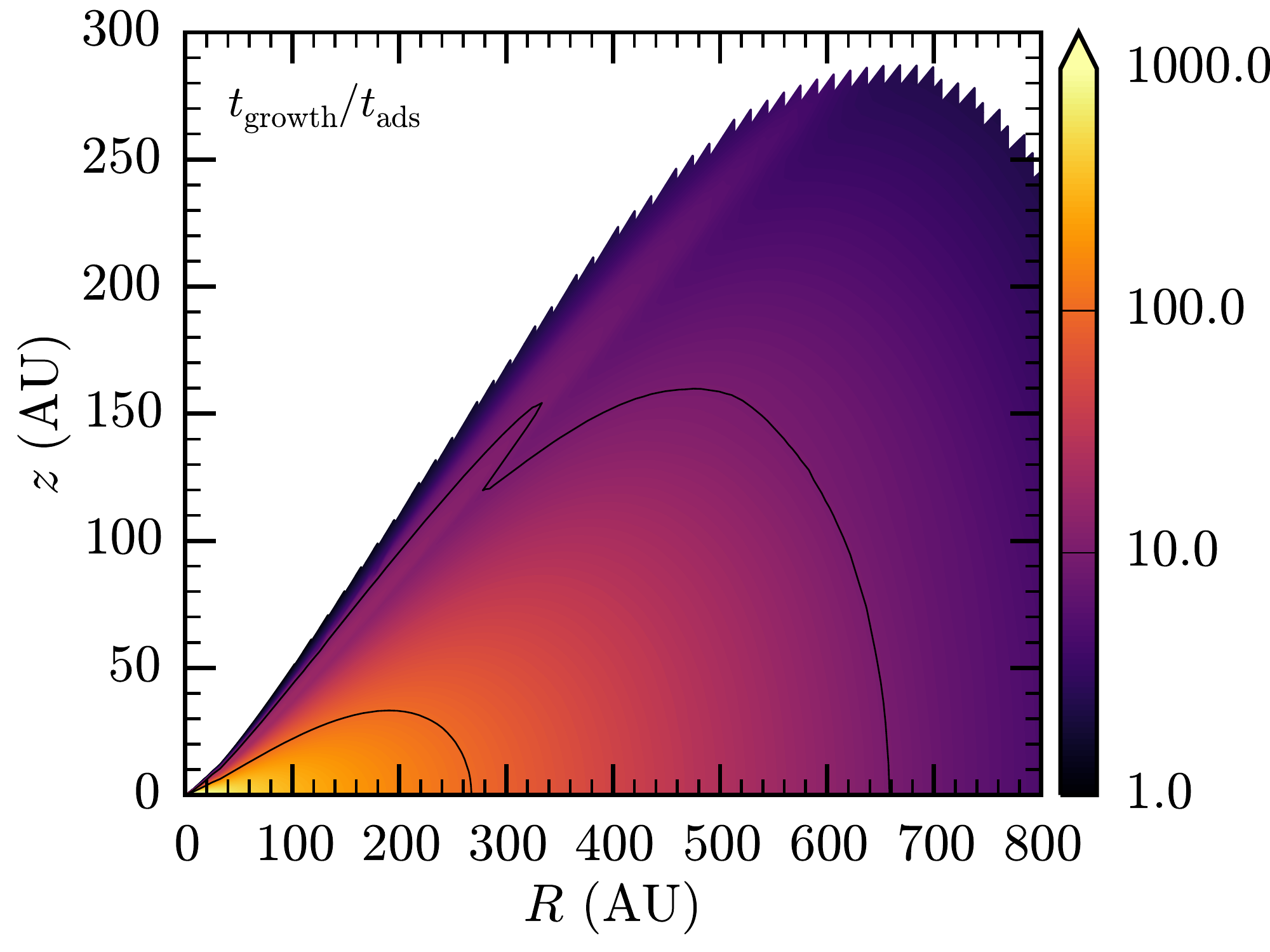}
\caption{Ratio of the growth timescale of dust particles and the freeze-out timescale of $^{12}$CO. For both timescales, particles with sizes of $0.1\,\mu$m have been considered. The thermal structure is taken from the STN-SM model. A turbulent $\alpha=10^{-3}$ has been used in the calculations (Eq.~\ref{eq:timescales_ratio}).
}
\label{fig:timescales}
\end{figure}

\begin{equation}
{\rm St} = \frac{a\rho_{\rm gr}}{\rho_{\rm dust}v_{\rm th}}\Omega.
\end{equation}
We thus obtain that:

\begin{equation}
t_{\rm growth} = \sqrt{\frac{2}{3} \sqrt{\frac{8}{\pi}} \frac{a}{\alpha c_{\rm s}\Omega} \frac{\rho_{\rm gr}}{\rho_{\rm dust}}  }.
\end{equation}
The quantity $c_{\rm s}$ is obtained from the STN-SM model, where $c_{\rm s}=\sqrt{P/\rho_{\rm gas}}$, and the contribution of chemical species to both $P$ and $\rho_{\rm gas}$ is taken into account.

Similarly, the freeze-out timescale can be expressed as $t_{\rm ads}\sim1/k_{\rm ads}$ (see Eq.~\ref{eq:ads}), where $k_{\rm ads}$ is the adsorption rate. If we consider an ensemble of grains with size $a$ for the freeze-out timescale, the ratio of the two timescales can be written as:

\begin{equation}
\frac{t_{\rm growth}}{t_{\rm ads}} \sim \sqrt{ \left( \frac{72}{\pi^{3}}\right)^{1/2}\frac{\rho_{\rm dust}}{\rho_{\rm gr}} \frac{1}{\sqrt{M(X)}} \frac{c_{\rm s}}{\alpha\Omega} \frac{1}{a} }.
\label{eq:timescales_ratio}
\end{equation}
For the most abundant $^{12}$CO, $M({\rm CO})=28$.

Figure~\ref{fig:timescales} shows the ratio of the two timescales for particles of size $a=0.1\,\mu$m, an assumed turbulent $\alpha=10^{-3}$, and $\rho_{\rm gr}=2.5\,$g\,cm$^{-3}$. All the other quantities are taken from the STN-SM model. In the whole disk, $t_{\rm growth}$ is larger than $t_{\rm ads}$, and in particular in the mid-plane of the inner disk, $t_{\rm growth}\gg t_{\rm ads}$. This implies that grains should be growing once the CO gas has already frozen onto dust grains, where the physical conditions of the disk allow this to happen. This fact strongly supports the hypothesis that in order to determine the abundance of ices in a protoplanetary disk, more sophisticated models, with coupled dynamical grain growth and thermo-chemistry, are needed. Some recent papers are indeed going in this direction, coupling different aspects of dust evolution (e.g. radial drift, grain growth, or vertical settling) with the chemical evolution of a disk \citep[e.g.][]{2015ApJ...815..109P,2016ApJ...833..203P,2016ApJ...833..285K,2016ApJ...831..101B,2016A&A...592A..83K}.

\section{Conclusions}
\label{sec:conclusion}

In this work we have coupled dust evolution models to thermo-chemical
models of protoplanetary disks, focusing in particular on the radial
properties of continuum and CO rotational lines emission. Detailed
results are presented for a model representative of the HD 163296 disk
and compared with observations of that disk. Our main conclusions can
be summarised as follows:

\begin{enumerate}
\item Differences in gas and dust radial extents of protoplanetary disks are readily found in our models and can be largely explained by the difference in optical depth of gas versus dust, without the need to invoke radial drift\rev{ (the STN model leads to a similar size in continuum as the turbulent models)}.
\item Different turbulence values can dramatically affect the estimate of the disk gas outer radius. The gas outer radius probed by $^{12}$CO emission can differ by a factor of $\sim$ three for a turbulent $\alpha$ ranging between $10^{-4}$ and $10^{-2}$, with the ratio of the CO and  mm radius $R_{\rm CO}^{\rm out}/R_{\rm mm}^{\rm out}$ increasing with turbulence \rev{(Fig.~\ref{fig:rco})}.
\item For the massive disk considered in this paper, $^{13}$CO exhibits the same trend as $^{12}$CO, with the outer radius increasing with turbulence. The intensity profile of the more optically thin C$^{18}$O depends more shallowly on turbulence.
\item The effect of radial drift is primarily visible in the sharp outer edge of the (sub)mm continuum intensity profile, in particular for low values of turbulence, when the maximum attained dust grain size is a steep function of radius. This may also lead to radial substructure in the continuum emission at large radii.
\item Proper treatment of dust evolution via grain growth, radial drift, and vertical settling leads to gas outer disks that are significantly colder than in models with simpler parametrised dust treatments. This is caused mostly by settled large grains coupling poorly with the gas, with the gas temperature even falling below the dust temperature at intermediate disk heights. This low gas temperature should be reflected in the CO line intensity ratios. Enhanced penetration of UV radiation has a smaller effect on CO intensities.
\item Lower turbulence allows particles to grow to larger sizes. Together with low vertical stirring, this leads to severe settling. As a consequence, steeper CO intensity profiles than the actual surface density profile are obtained for low $\alpha$, due to a steeper gas temperature gradient in the intermediate disk layers.
\item Radial drift and dust settling concur in causing a thermal inversion of the dust, potentially leading to a second mid-plane CO thermal (rather than photo) desorption front at large radii. 
\item Using HD 163296 as a test case, we are able to obtain good agreement between our models and the (sub)mm continuum and $^{12}$CO - $^{13}$CO lines intensity profiles for fairly high values of $\alpha$.
\item The CO snowline location of HD 163296 at $\sim90\,$AU, as determined by \citet{2015ApJ...813..128Q}, can be retrieved by using a binding energy for CO typical of ice mixtures ($E_{\rm b}\sim1100\,$K). Lower binding energies, in particular for pure CO ice ($E_{\rm b}\sim855\,$K) fail to reproduce the CO snowline location.
\end{enumerate}

In summary, models mimicking grain growth and vertical settling with a
parametrised dust treatment but no radial drift reproduce several of
the observed features well, but fail to explain sharp dust edges and
result in different radial and vertical gas temperature
structures. Using such models to infer the underlying gas surface
density structure from observed radial profiles could therefore also
lead to incorrect conclusions.

\newpage

\begin{acknowledgements}
We thank Ruud Visser, Paola Pinilla, Arthur Bosman, Merel van' t Hoff , and Paolo Cazzoletti for useful discussions, and Michiel Hogerheijde \rev{and the Allegro team, and especially Nico Salinas,} for providing the cleaned ALMA data. \rev{We thank the anonymous referee for providing comments that helped improve the clarity of the paper.} This paper makes use of the following ALMA data: ADS/JAO.ALMA\#2011.0.00010.SV. ALMA is a partnership of ESO (representing its member states), NSF (USA), and NINS (Japan), together with NRC (Canada) and NSC and ASIAA (Taiwan), in cooperation with the Republic of Chile. The Joint ALMA Observatory is operated by ESO, AUI/NRAO, and NAOJ. \rev{T.B. acknowledges funding from the European Research Council (ERC) under the European Union's Horizon 2020 research and innovation programme under grant agreement No 714769.} Astrochemistry in Leiden is supported by the European Union A-ERC grant 291141 CHEMPLAN, by the Netherlands Research School for Astronomy (NOVA), and by a Royal Netherlands Academy of Arts and Sciences (KNAW) professor prize. All the figures were generated with the \textsc{python}-based package \textsc{matplotlib} \citep{2007CSE.....9...90H}.
\end{acknowledgements}

%-------------------------------------------------------------------

\bibliographystyle{aa}
\bibliography{references}

\begin{appendix}

\section{Disk thermal structure}
\label{app:thermal}

\begin{figure*}
\center
\includegraphics[width=.45\textwidth]{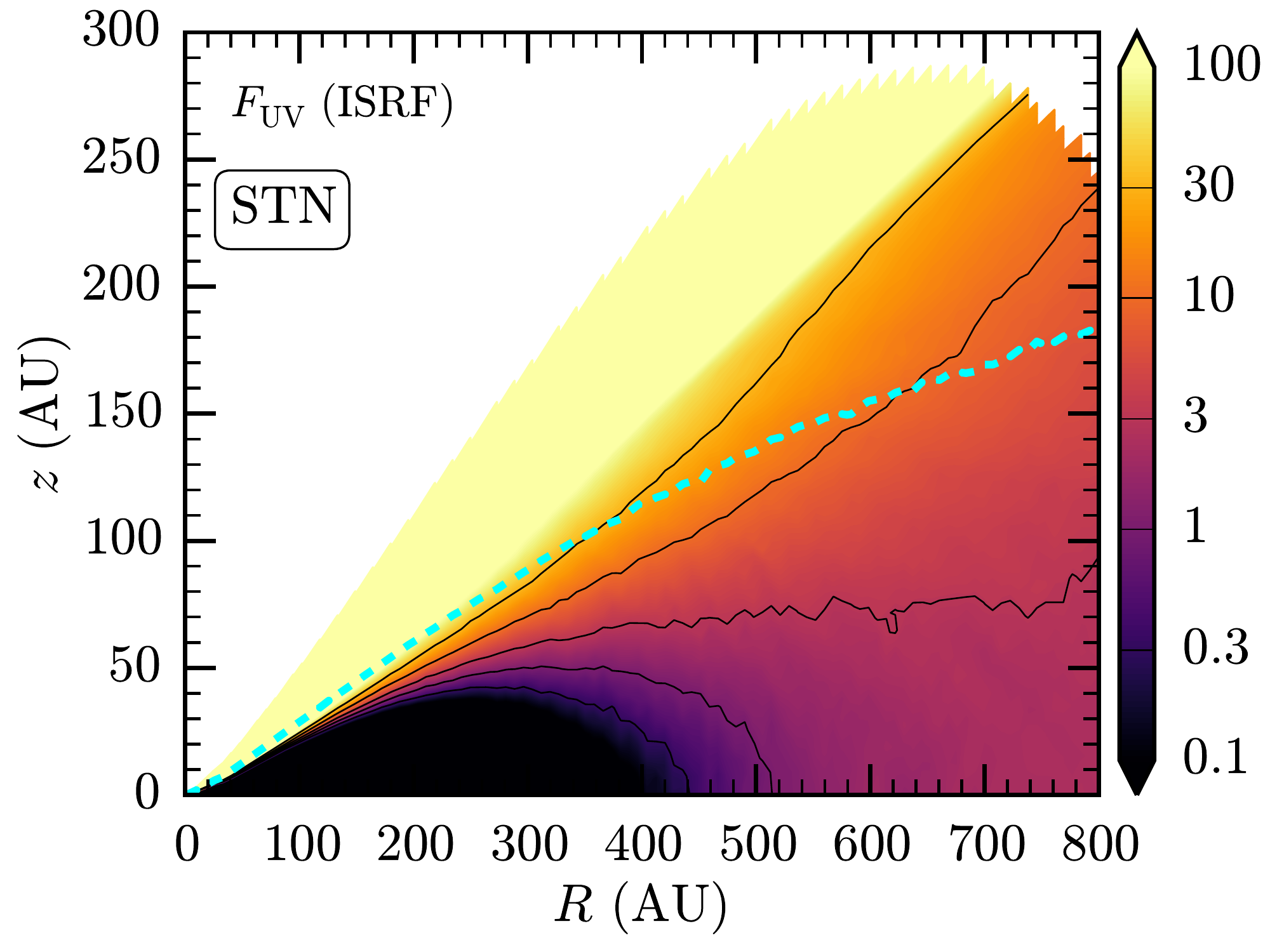}
\includegraphics[width=.45\textwidth]{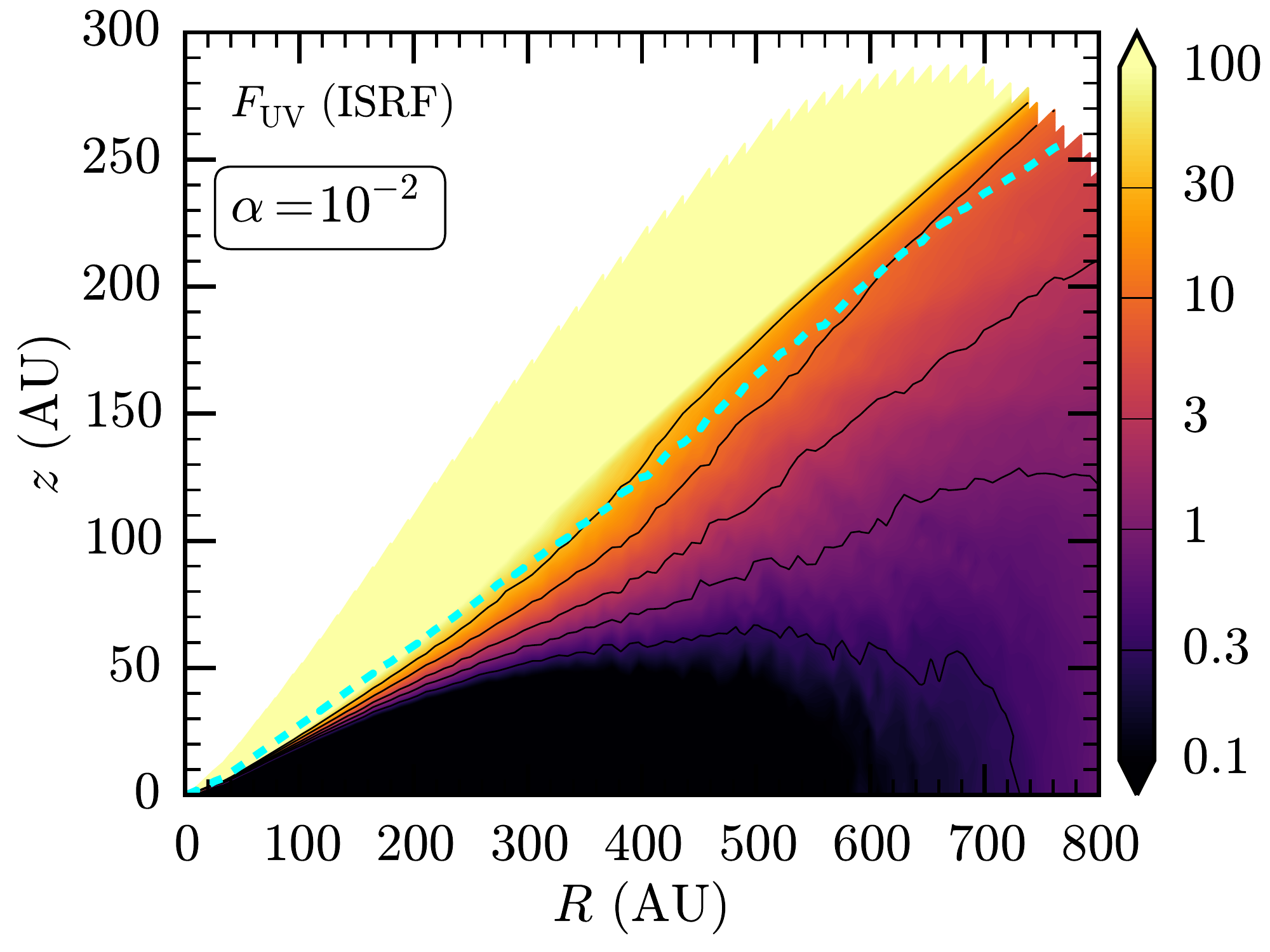}\\
\includegraphics[width=.45\textwidth]{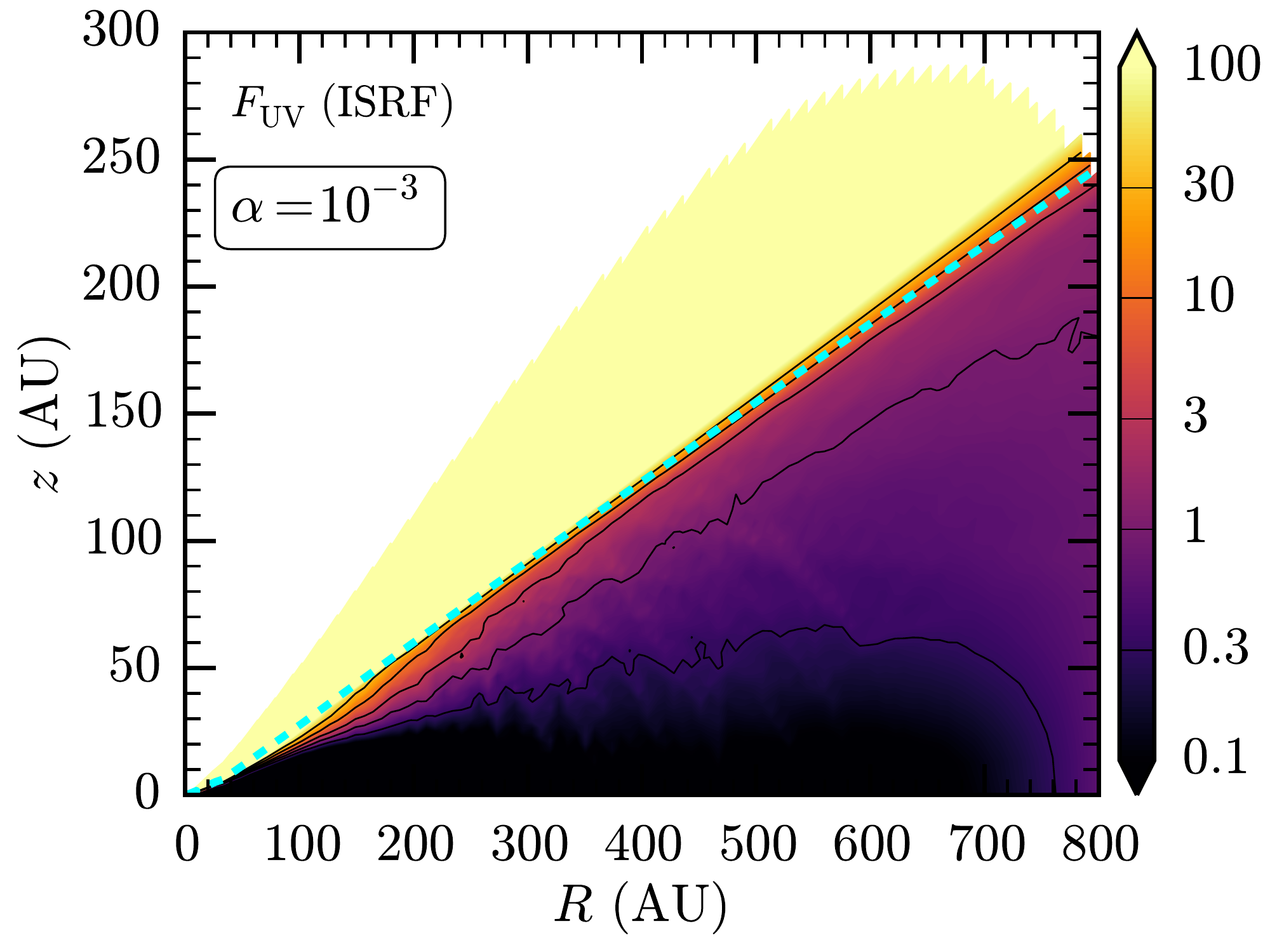}
\includegraphics[width=.45\textwidth]{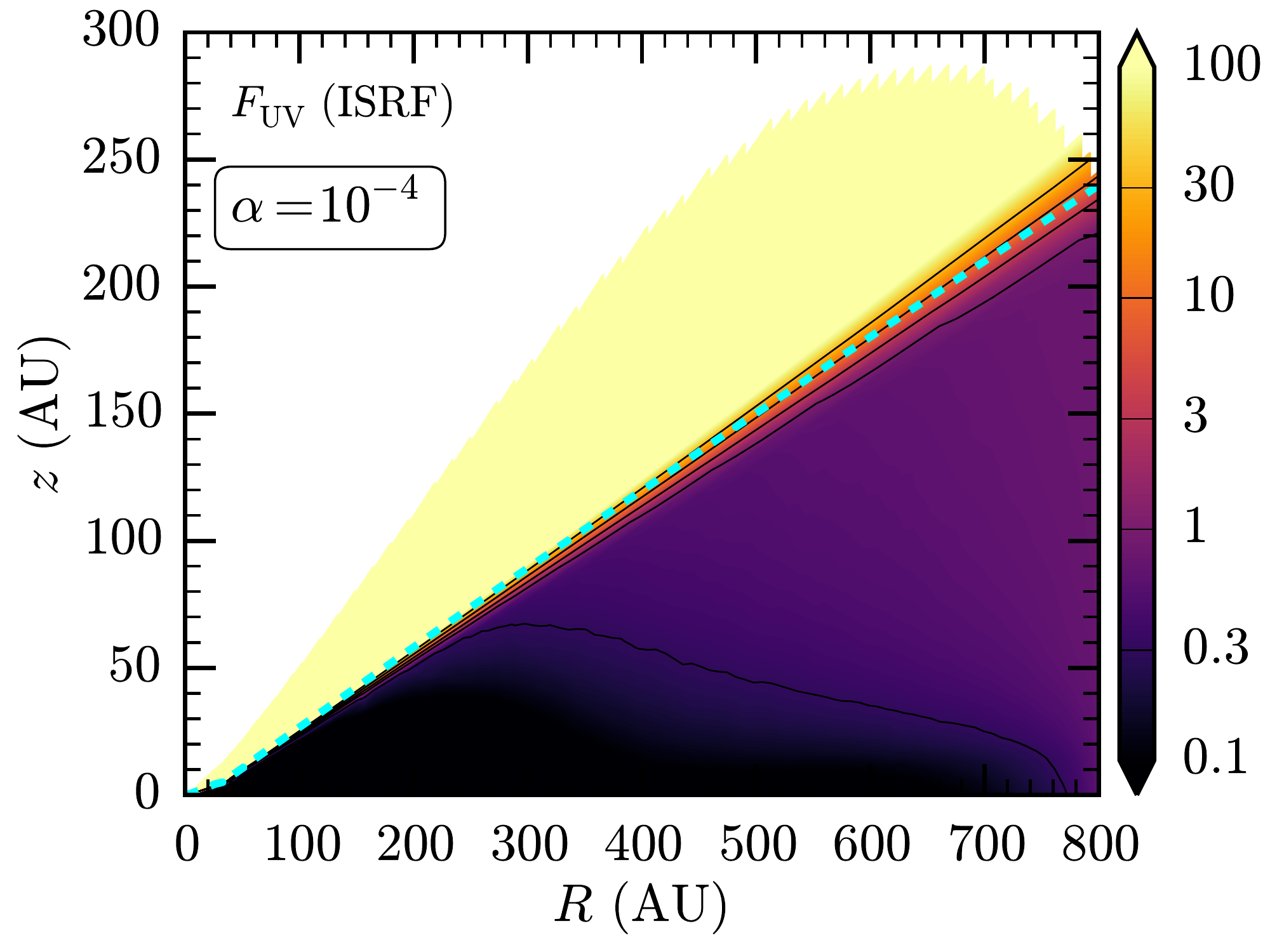}\\
\caption{From top left to bottom right: STN, $\alpha=10^{-2}$, $10^{-3}$ , and $10^{-4}$ model. Colours show the UV field in $G_0$ units, where $G_0\sim2.7\times10^{-3}\,$erg\,s$^{-1}$\,cm$^{-2}$ is the UV interstellar radiation field (ISRF) between $911\,\AA$ and $2067\,\AA$ \citep{1978ApJS...36..595D}. \rev{The cyan dashed lines indicate the $\tau_{\rm FUV}=1$ surface.}
}
\label{fig:g0}
\end{figure*}

\begin{figure*}
\center
\includegraphics[width=.33\textwidth]{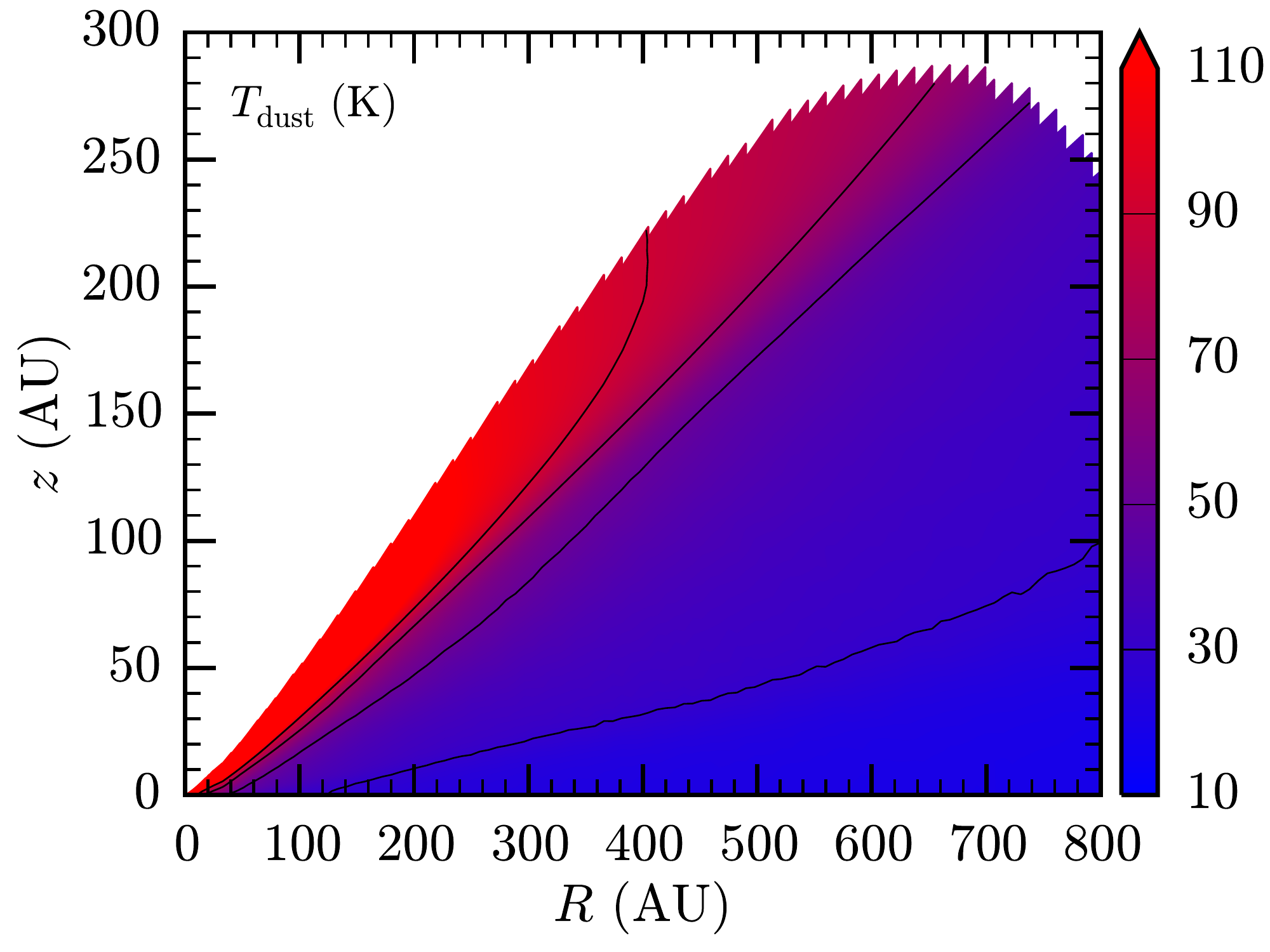}
\includegraphics[width=.33\textwidth]{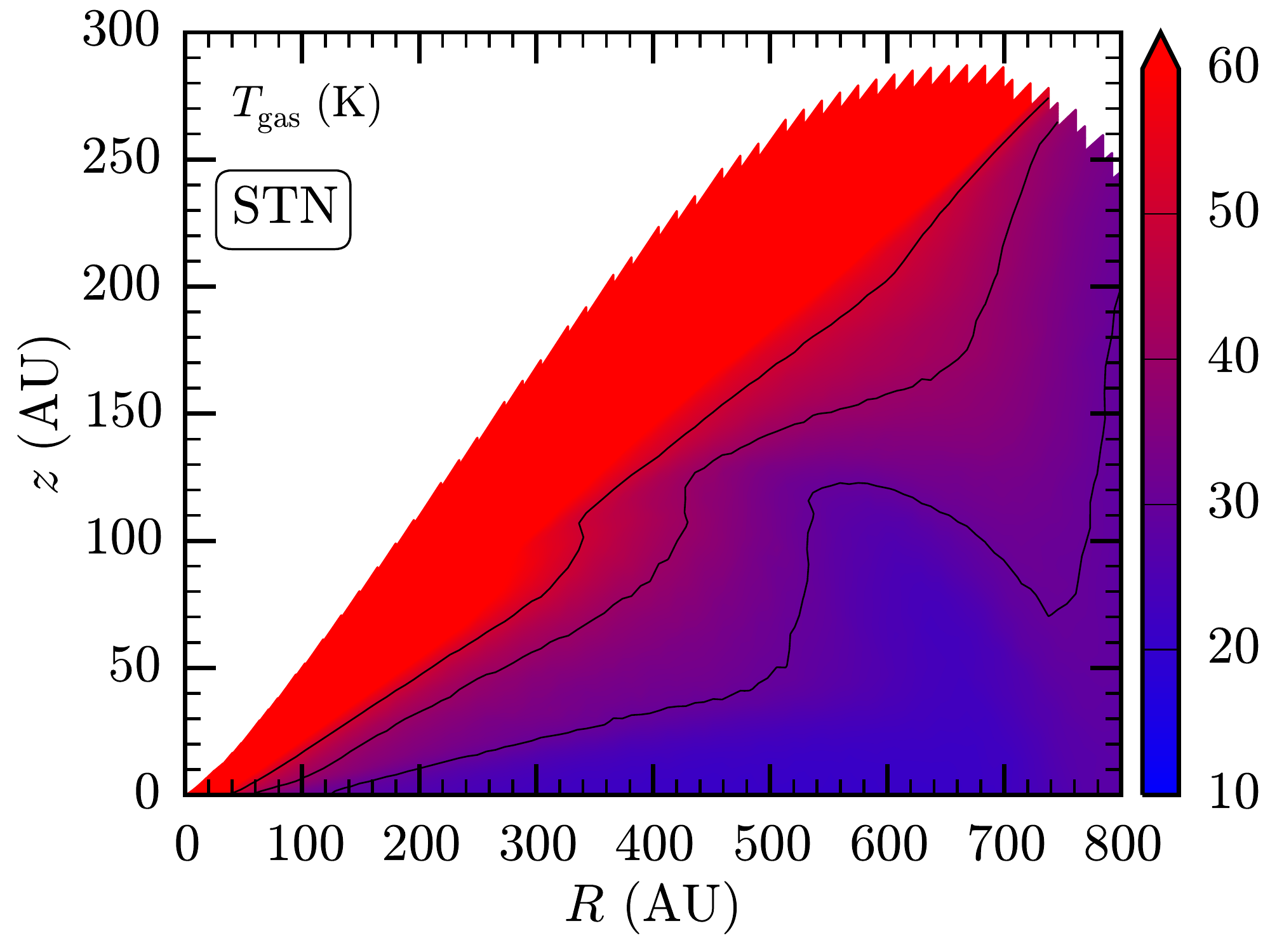}
\includegraphics[width=.33\textwidth]{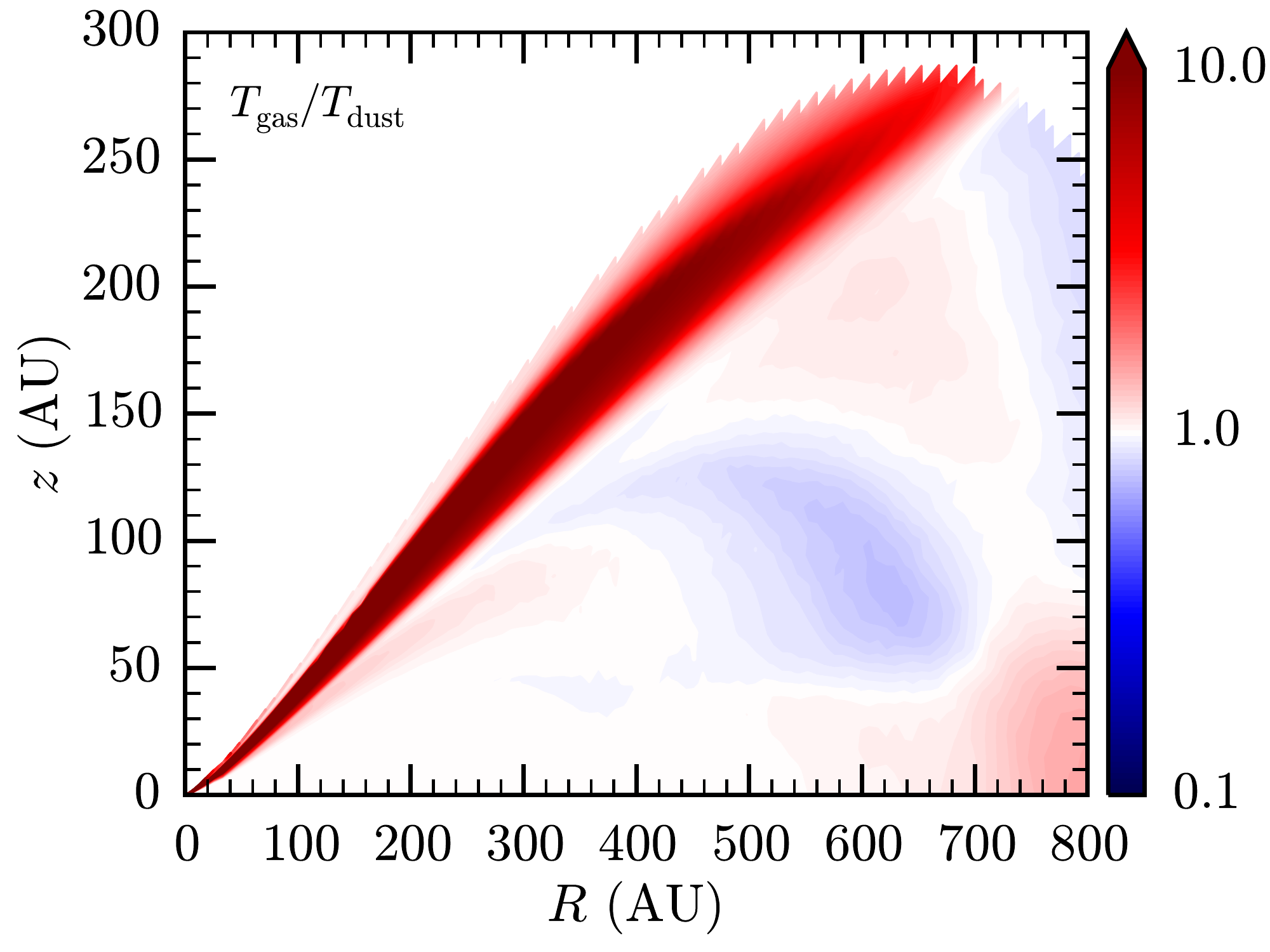}\\
\includegraphics[width=.33\textwidth]{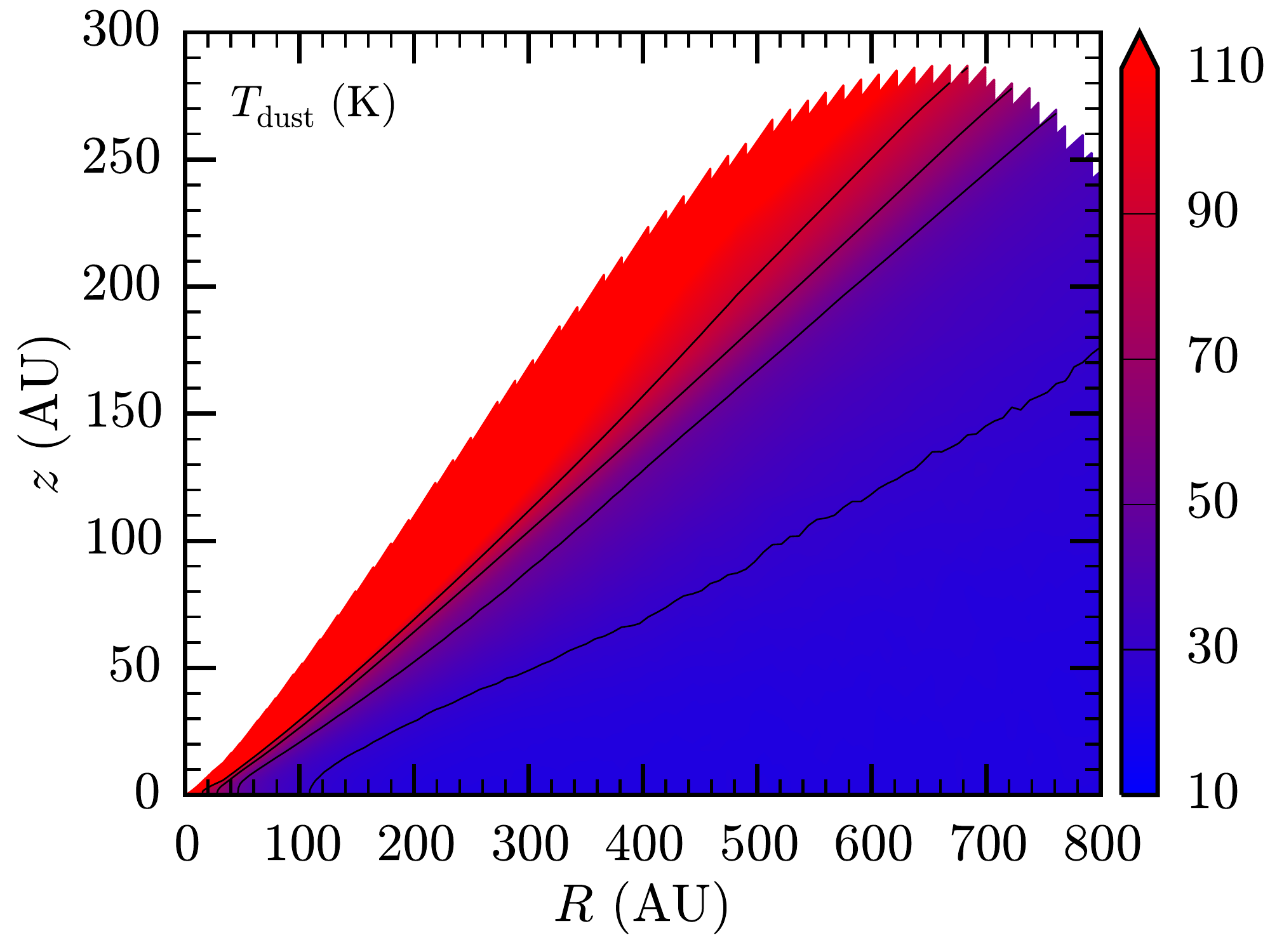}
\includegraphics[width=.33\textwidth]{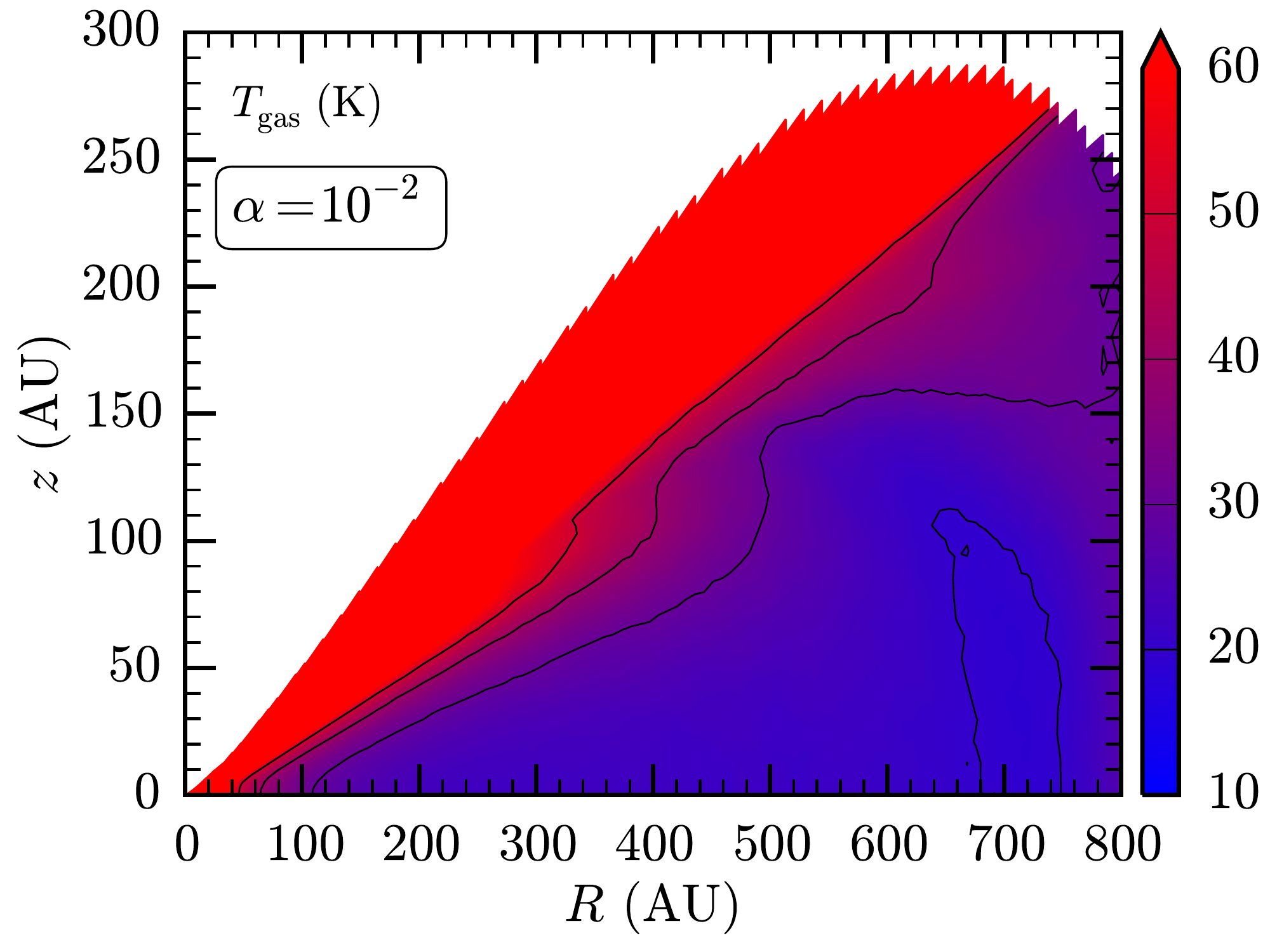}
\includegraphics[width=.33\textwidth]{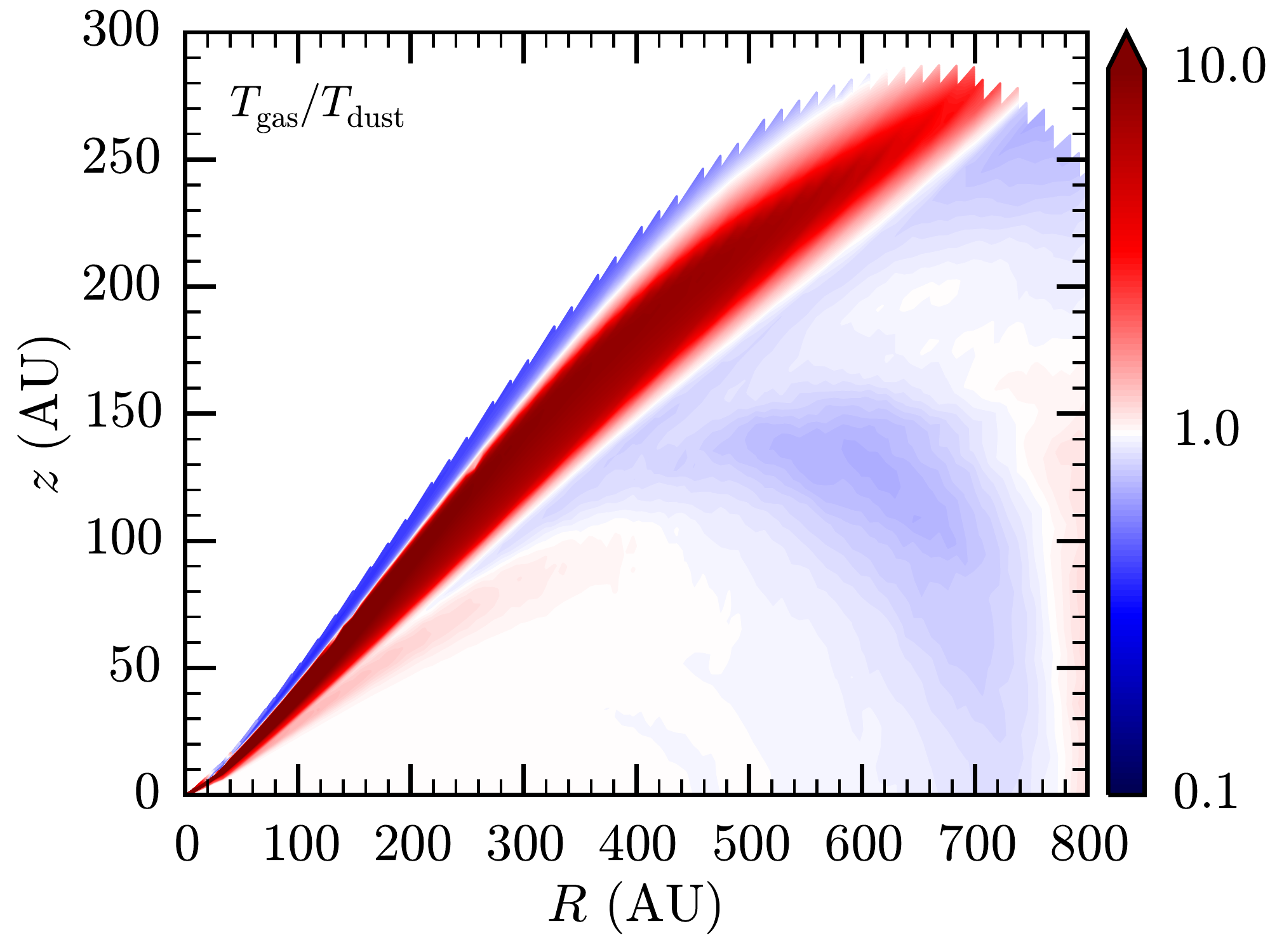}\\
\includegraphics[width=.33\textwidth]{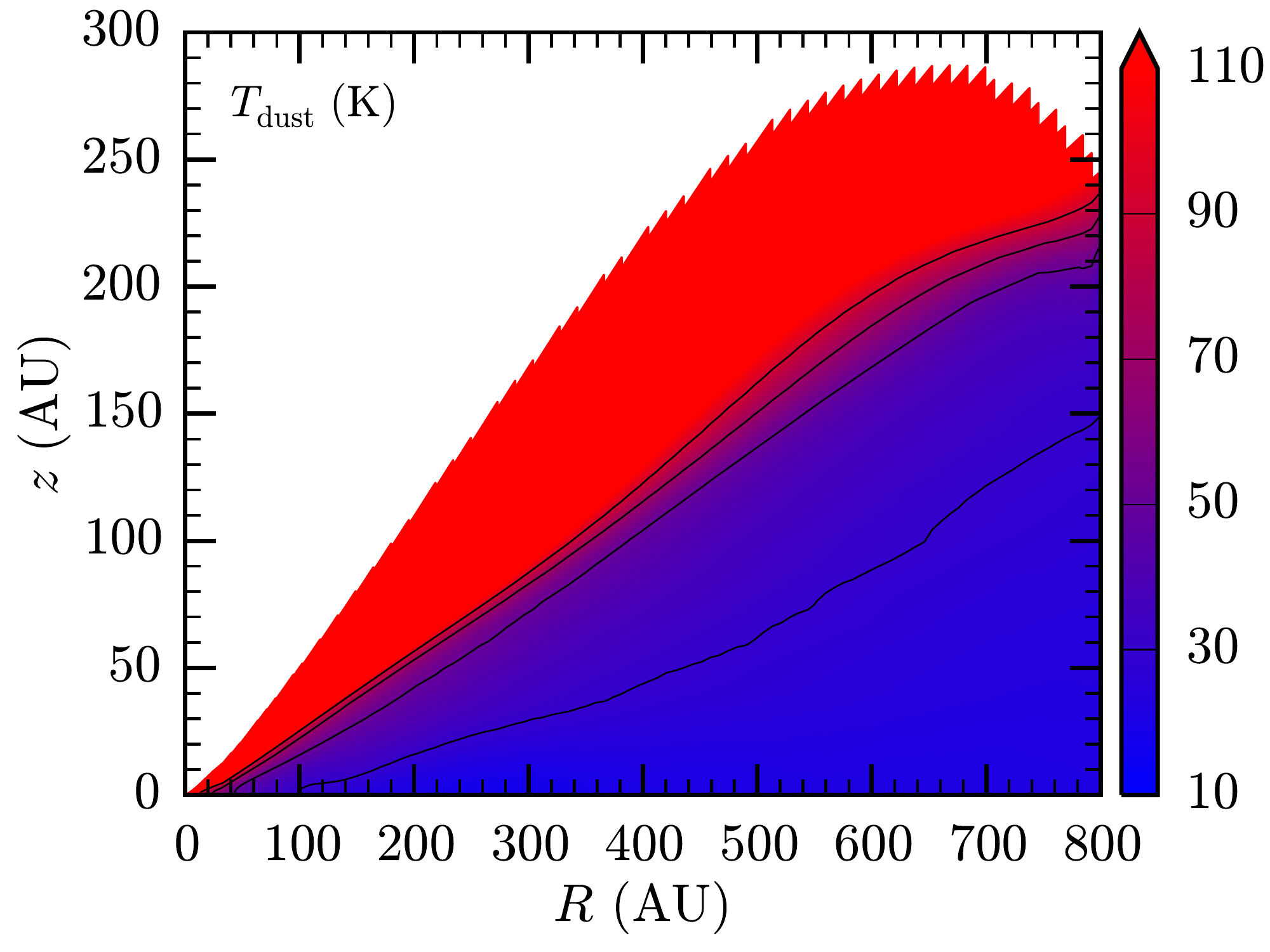}
\includegraphics[width=.33\textwidth]{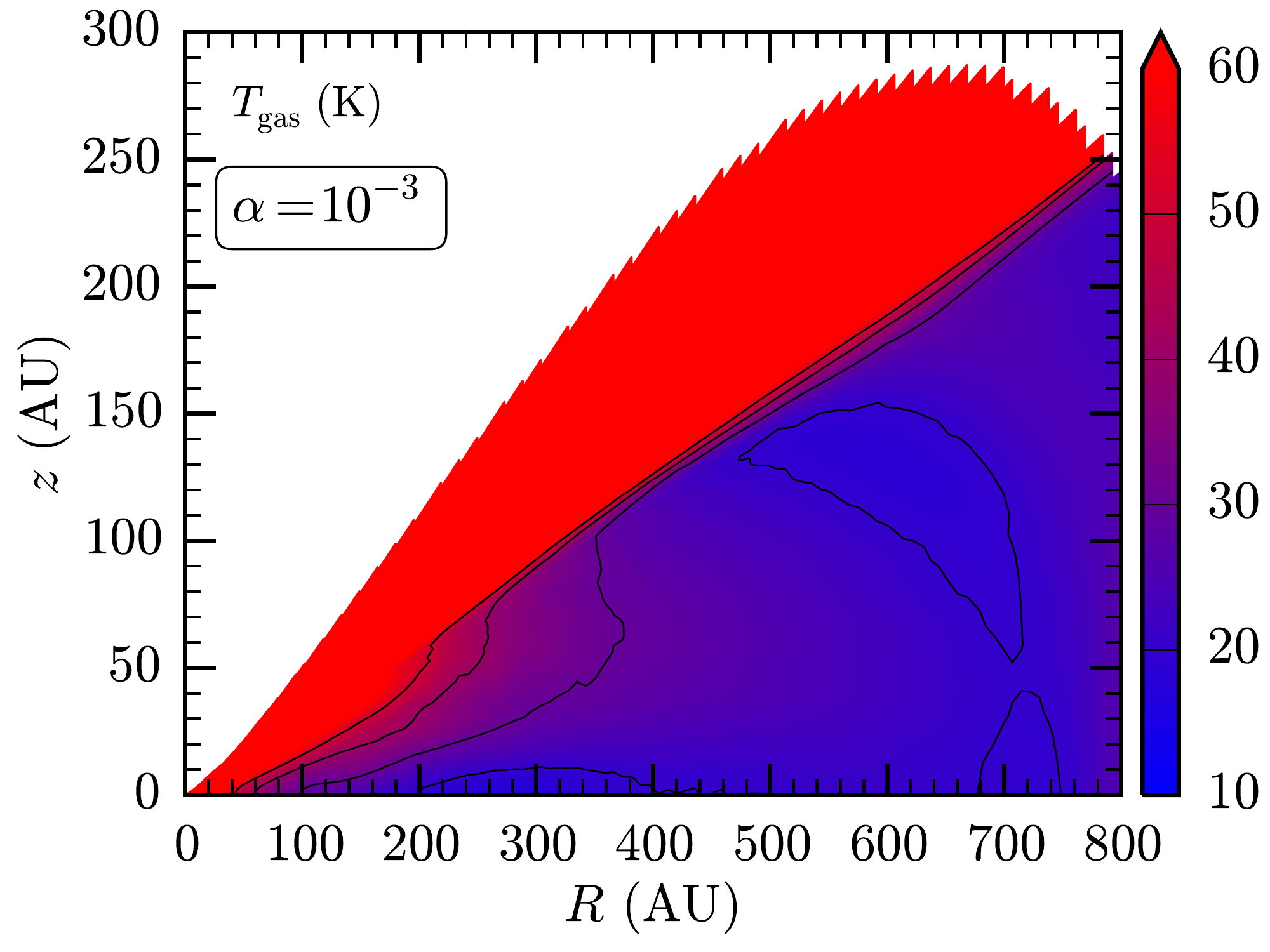}
\includegraphics[width=.33\textwidth]{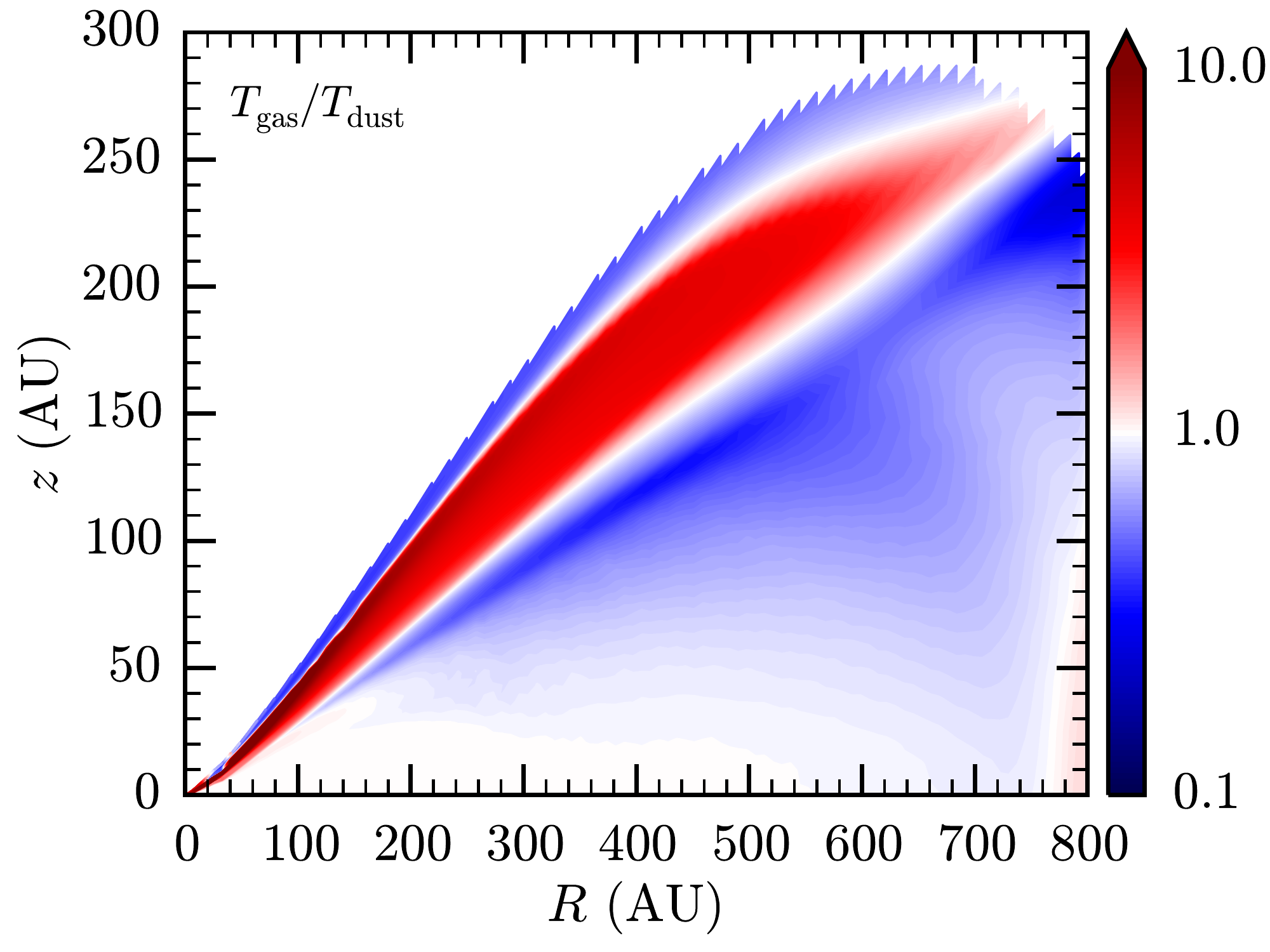}\\
\includegraphics[width=.33\textwidth]{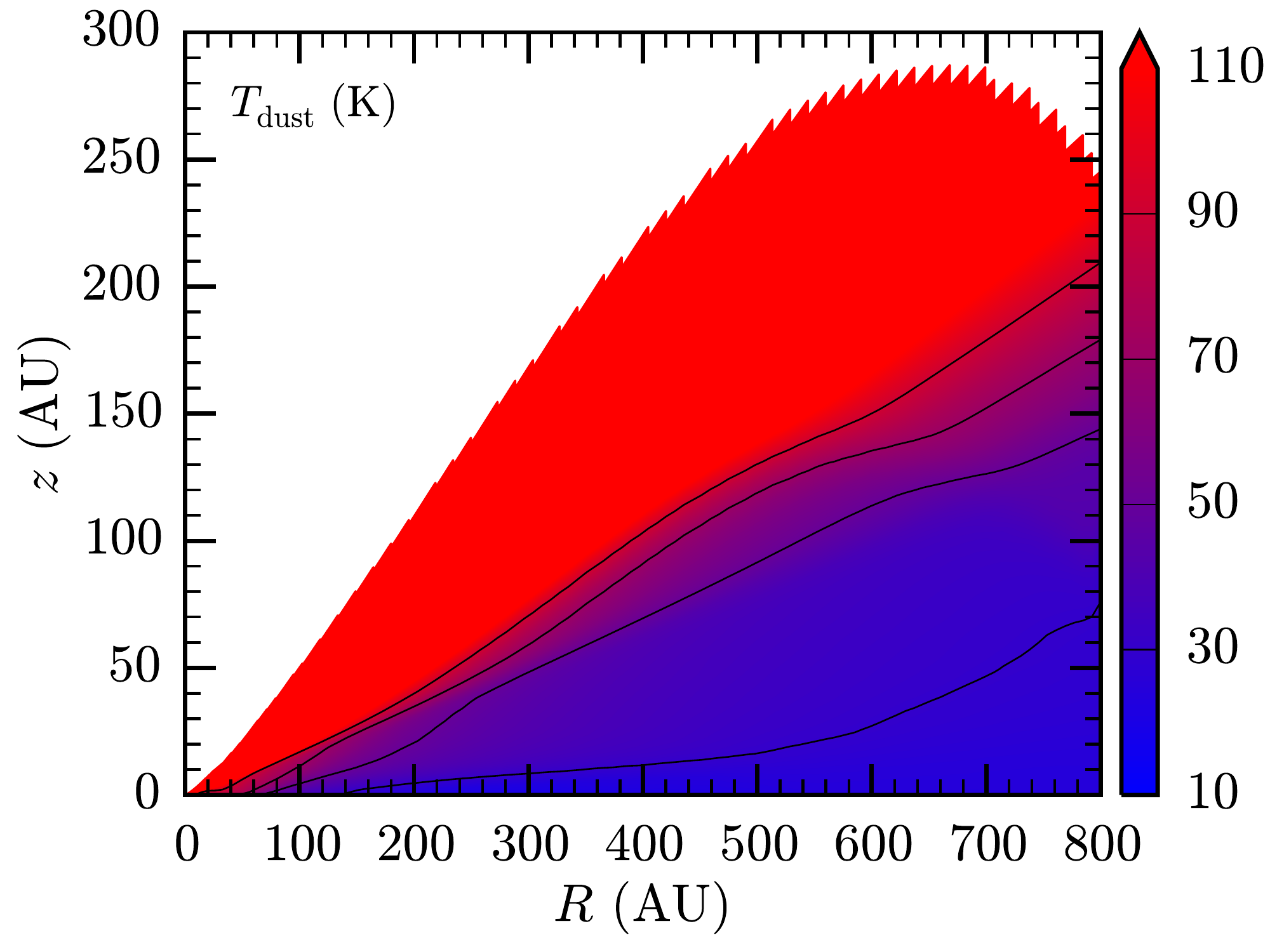}
\includegraphics[width=.33\textwidth]{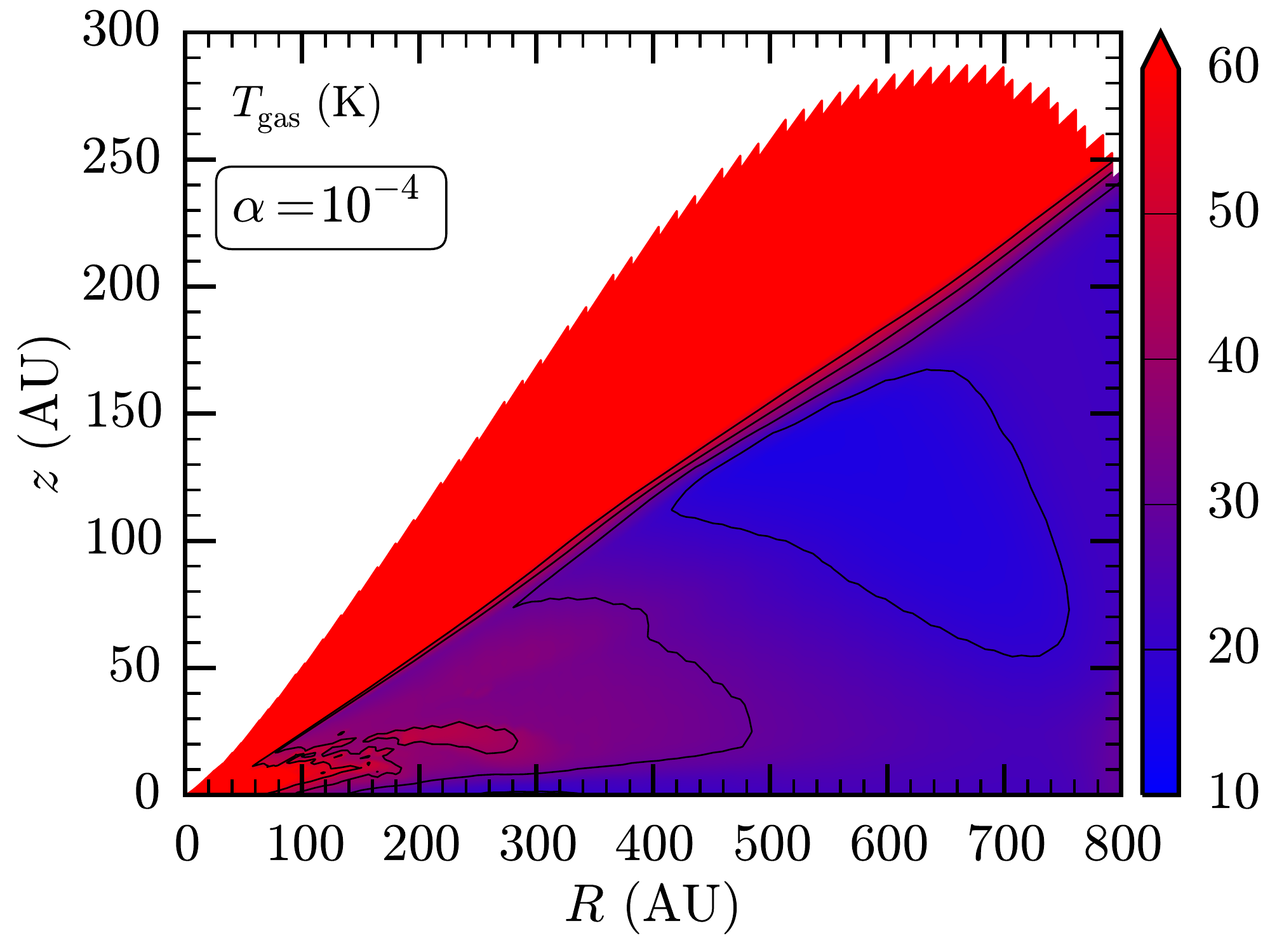}
\includegraphics[width=.33\textwidth]{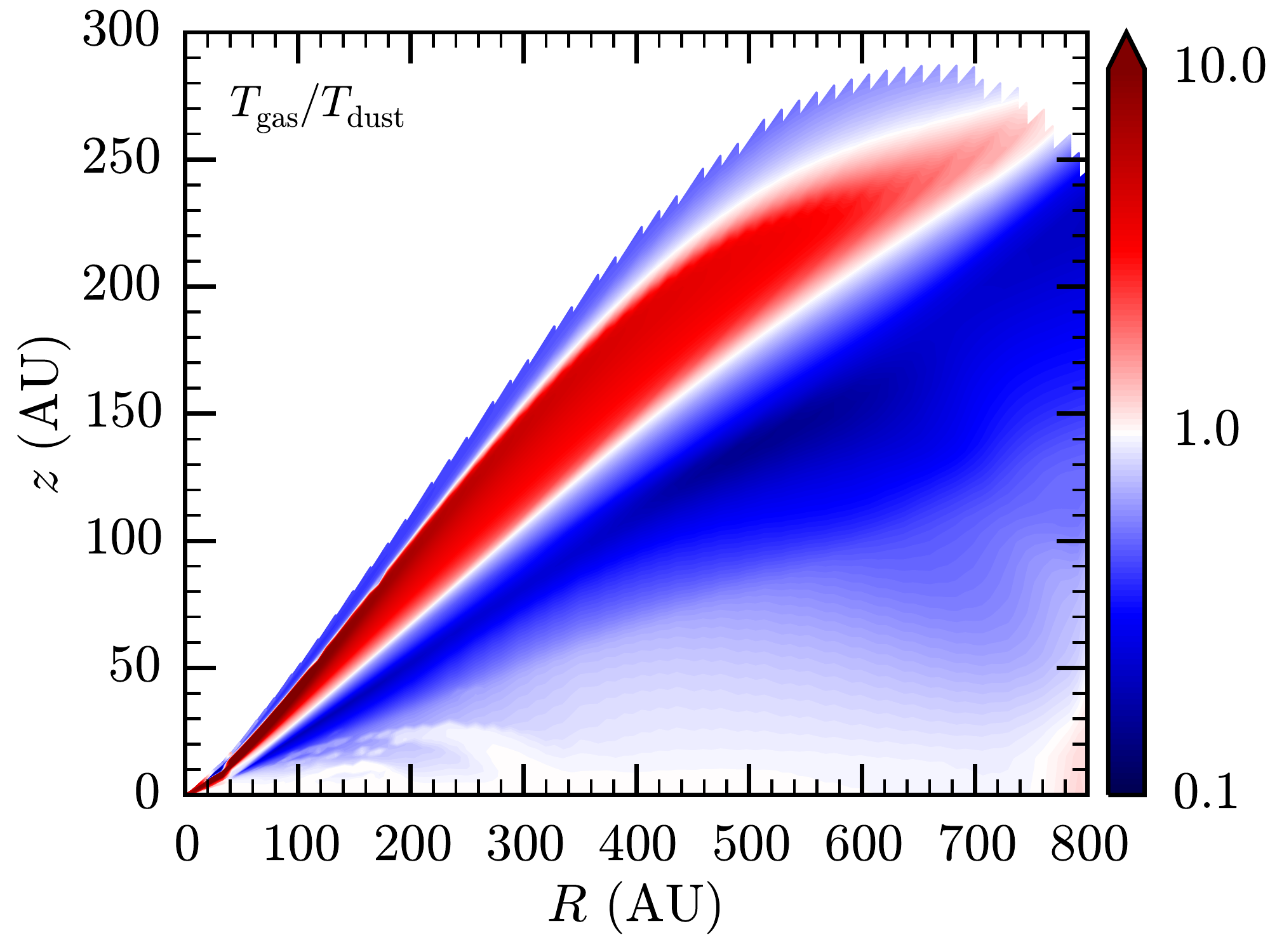}\\
\caption{From top to bottom: STN, $\alpha=10^{-2}$, $10^{-3}$ , and $10^{-4}$ models. From left to right: dust temperature $T_{\rm dust}$,  gas temperature $T_{\rm gas}$, and ratio of the two.}
\label{fig:temp}
\end{figure*}

\begin{figure*}
\center
\includegraphics[width=.9\textwidth]{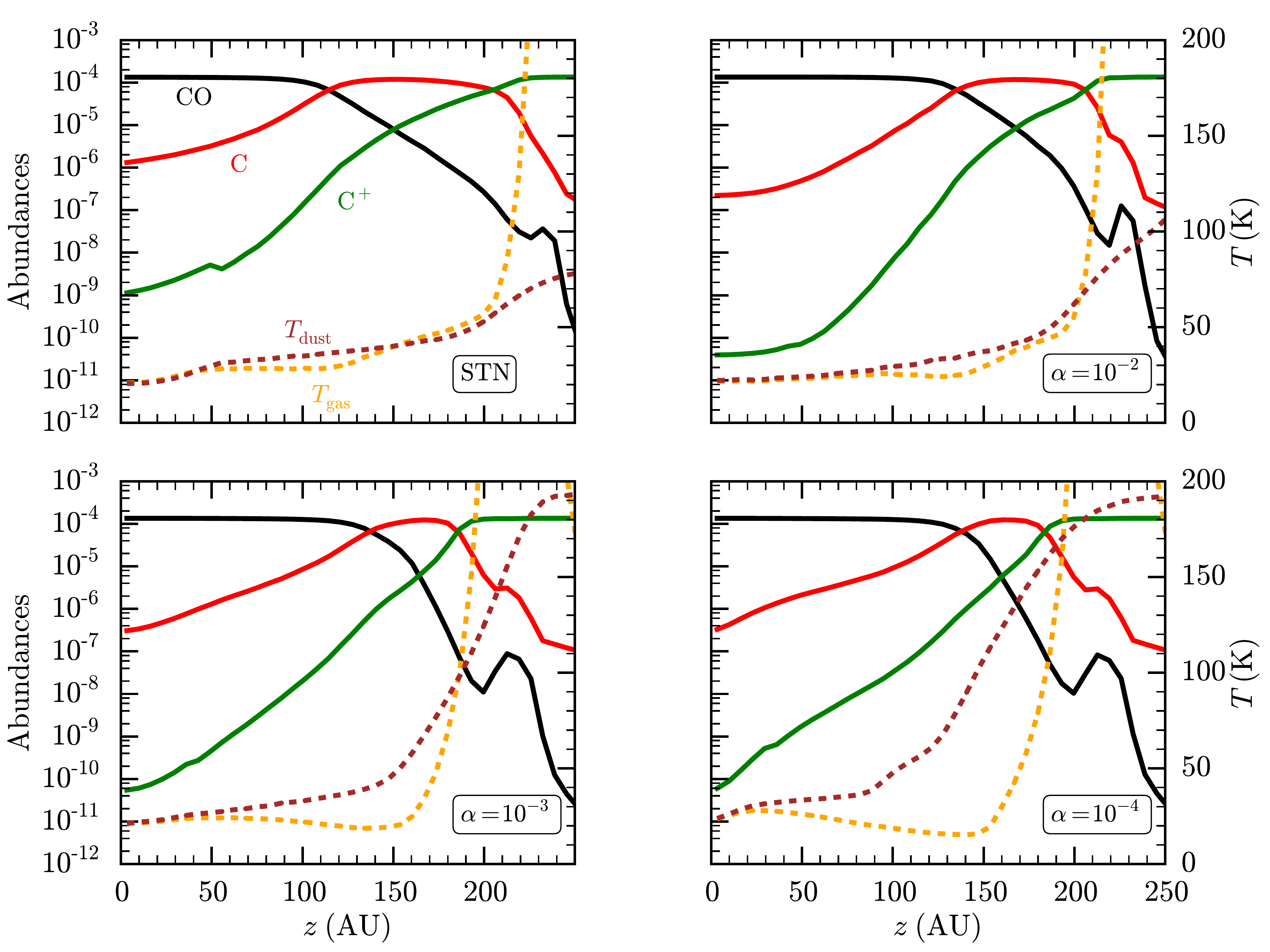}
\caption{From top left to bottom right: vertical cuts at $\sim550\,$AU of the STN, $\alpha=10^{-2}$, $10^{-3}$ , and $10^{-4}$ models. Legend: solid lines in black, red, and green represent abundances of CO, C, and C$^+$, respectively. Dashed lines in orange and brown are $T_{\rm gas}$ and $T_{\rm dust}$, respectively.}
\label{fig:temp_abun}
\end{figure*}

Figure~\ref{fig:g0} shows the attenuated UV field in STN and turbulent models.  Figure~\ref{fig:temp} portrays the dust and gas thermal structure of the same models. Finally, Fig.~\ref{fig:temp_abun} shows the gas and dust temperatures, and the abundances of CO, C, and C$^+$, in a vertical cut at $550\,$AU.

\section{Line profiles of CO isotopologues}

\begin{figure*}
\center
\includegraphics[width=.33\textwidth]{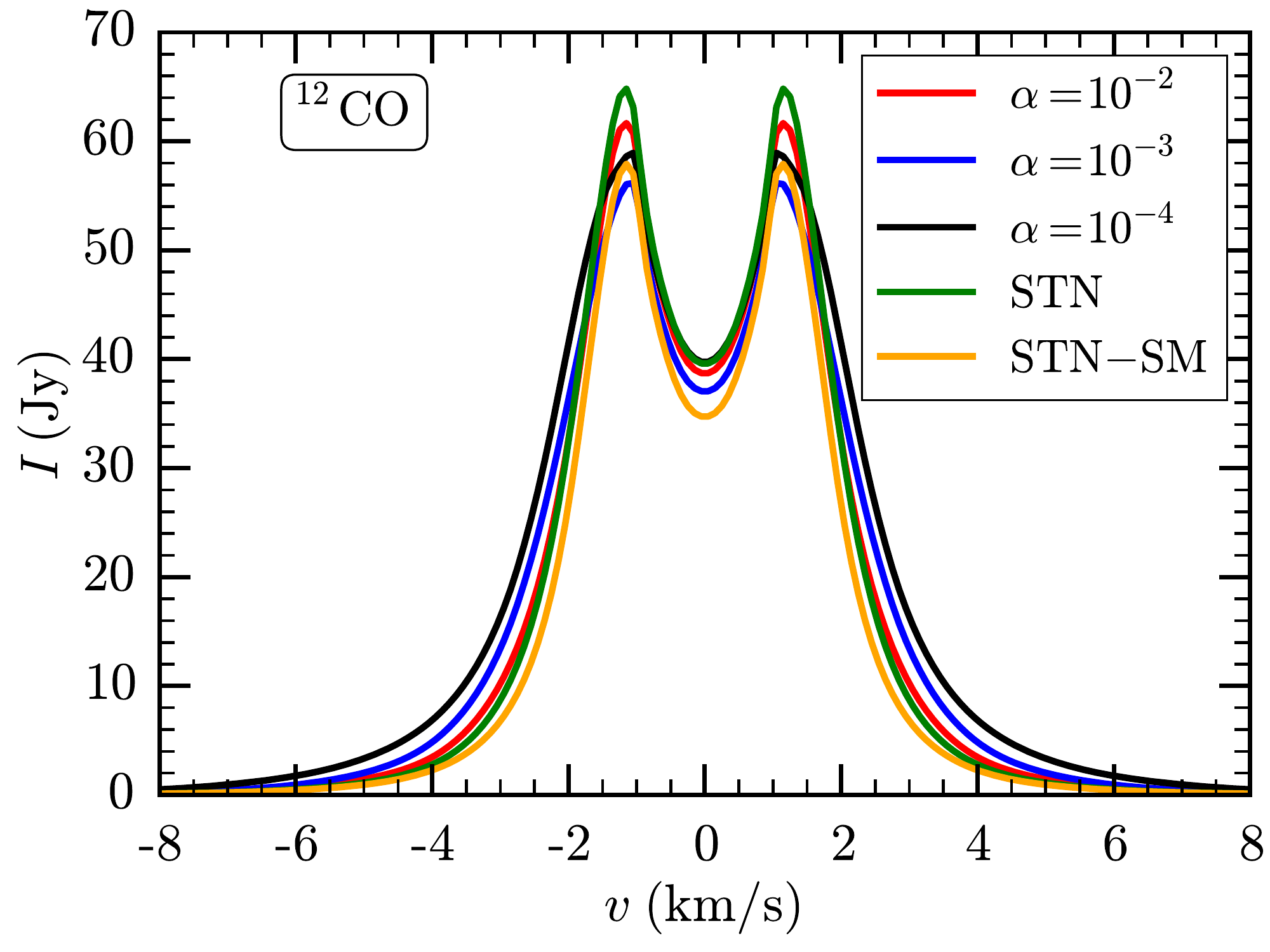}
\includegraphics[width=.33\textwidth]{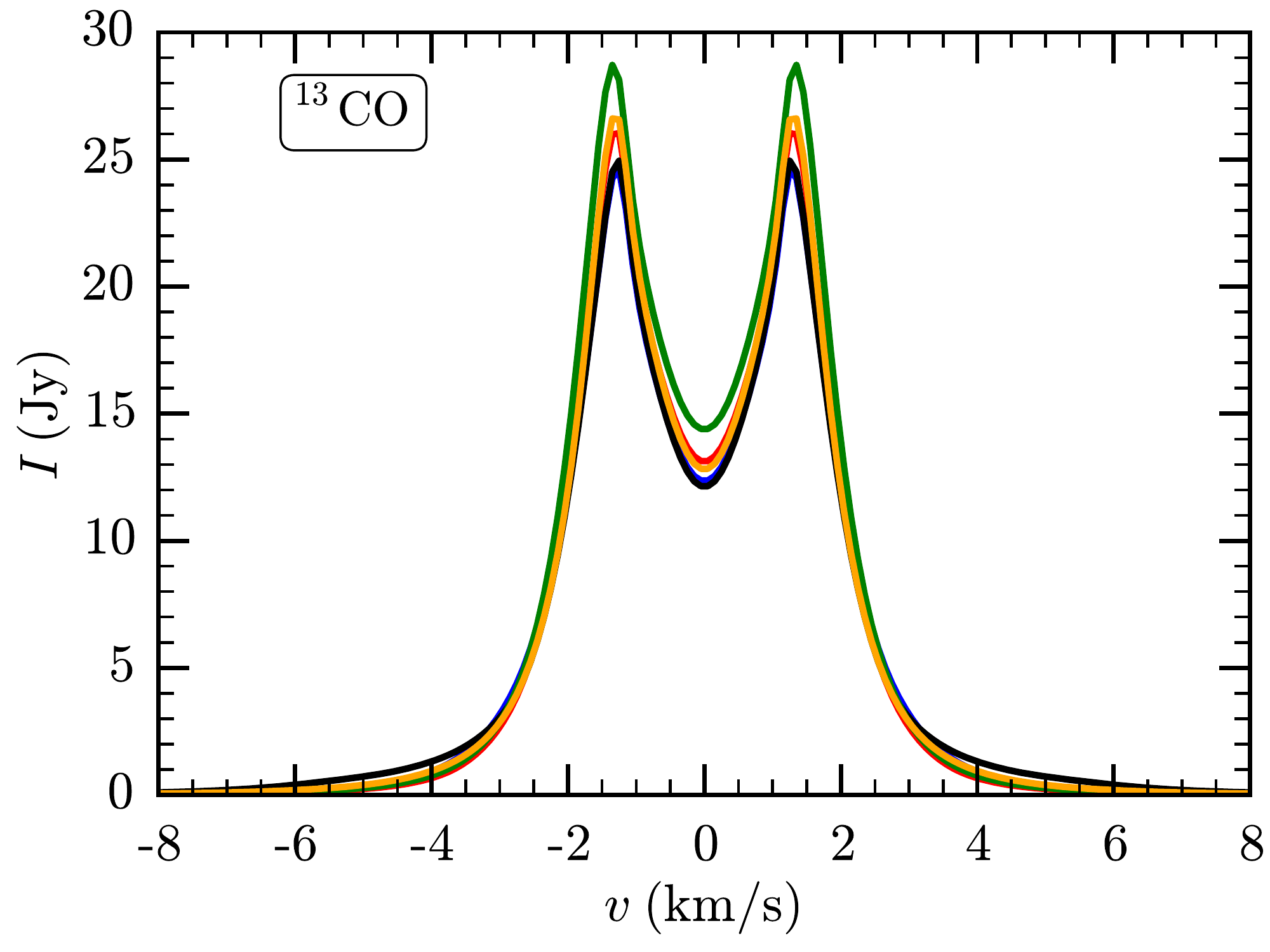}
\includegraphics[width=.33\textwidth]{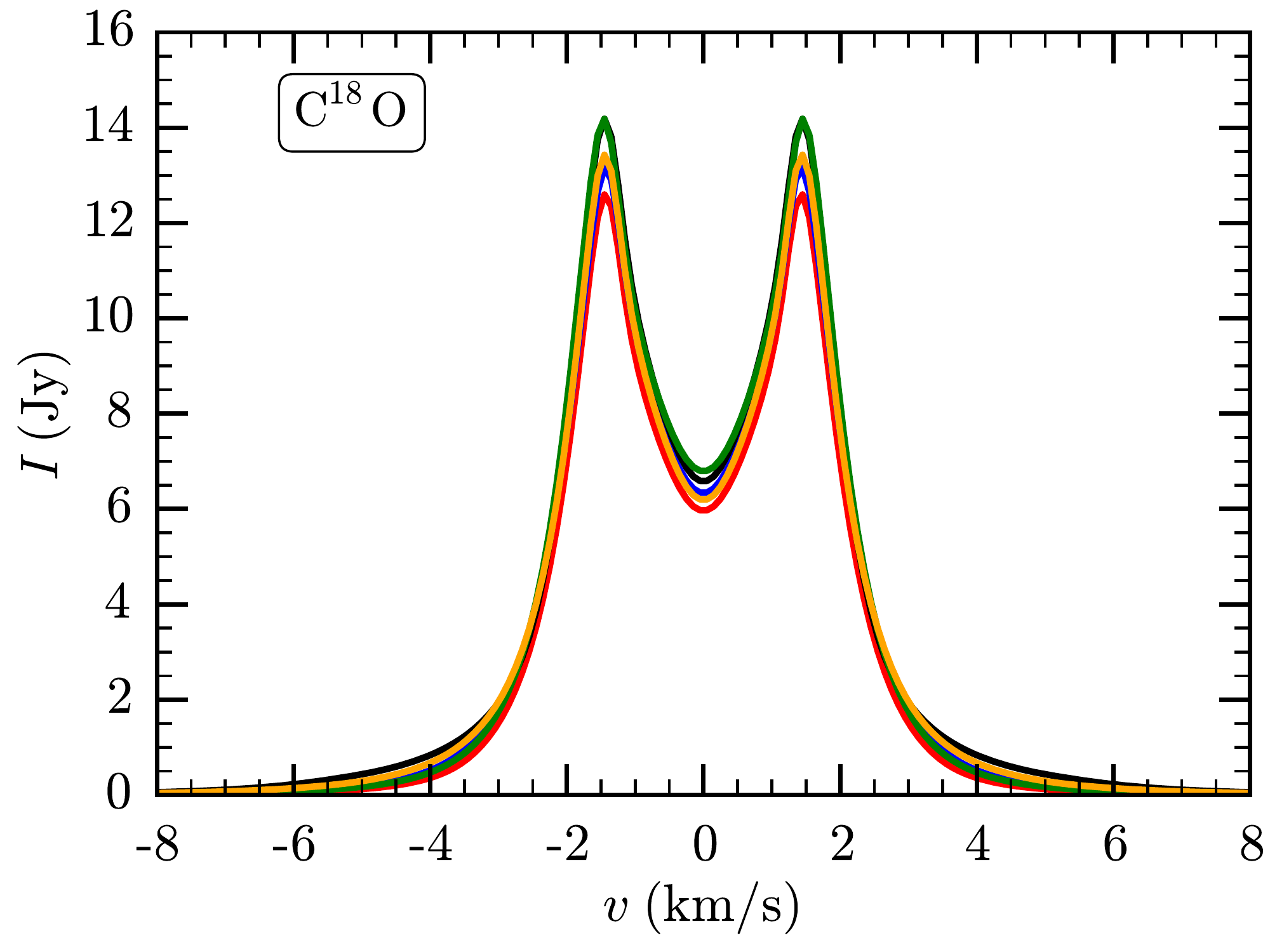}\\
\caption{From left to right: integrated line profile of the $^{12}$CO, $^{13}$CO, and C$^{18}$O $J$=3-2 transition of the different models shown in Fig.~\ref{fig:profiles_gas}.
}
\label{fig:line_profiles}
\end{figure*}

The spatially unresolved line profiles of three CO isotopologues discussed and shown in Section~\ref{sec:profiles_gas} already contain important information. The profiles of the $J$=3-2 transition are reported in Fig.~\ref{fig:line_profiles}. The first result is that the total fluxes do not change significantly between different models, that is, settling  and grain growth do not affect the total CO flux \citep[e.g.][]{2014A&A...572A..96M,2016A&A...586A.103W}. Moreover, the optically thinner isotopologues (in particular C$^{18}$O) show almost perfect agreement between all models. For optically thicker CO isotopes, the differences between the models become more prominent. This suggests that the discrepancy is not caused by a difference in the column density, which is almost the same in all models (see Fig.~\ref{fig:co_col}), but by an excitation effect, that is, by a difference  in the gas temperature. In particular, the emission from the outer regions (thus at low velocities) is clearly correlated to the disk turbulence, with the peaks of the double-horned line profile being brighter for higher viscosity. Moreover, the $^{12}$CO line profile shows brighter high velocity wings for lower viscosities, suggesting higher temperatures from the very inner regions.

\section{High resolution synthetic images}

\begin{figure*}
\center
\includegraphics[width=\columnwidth]{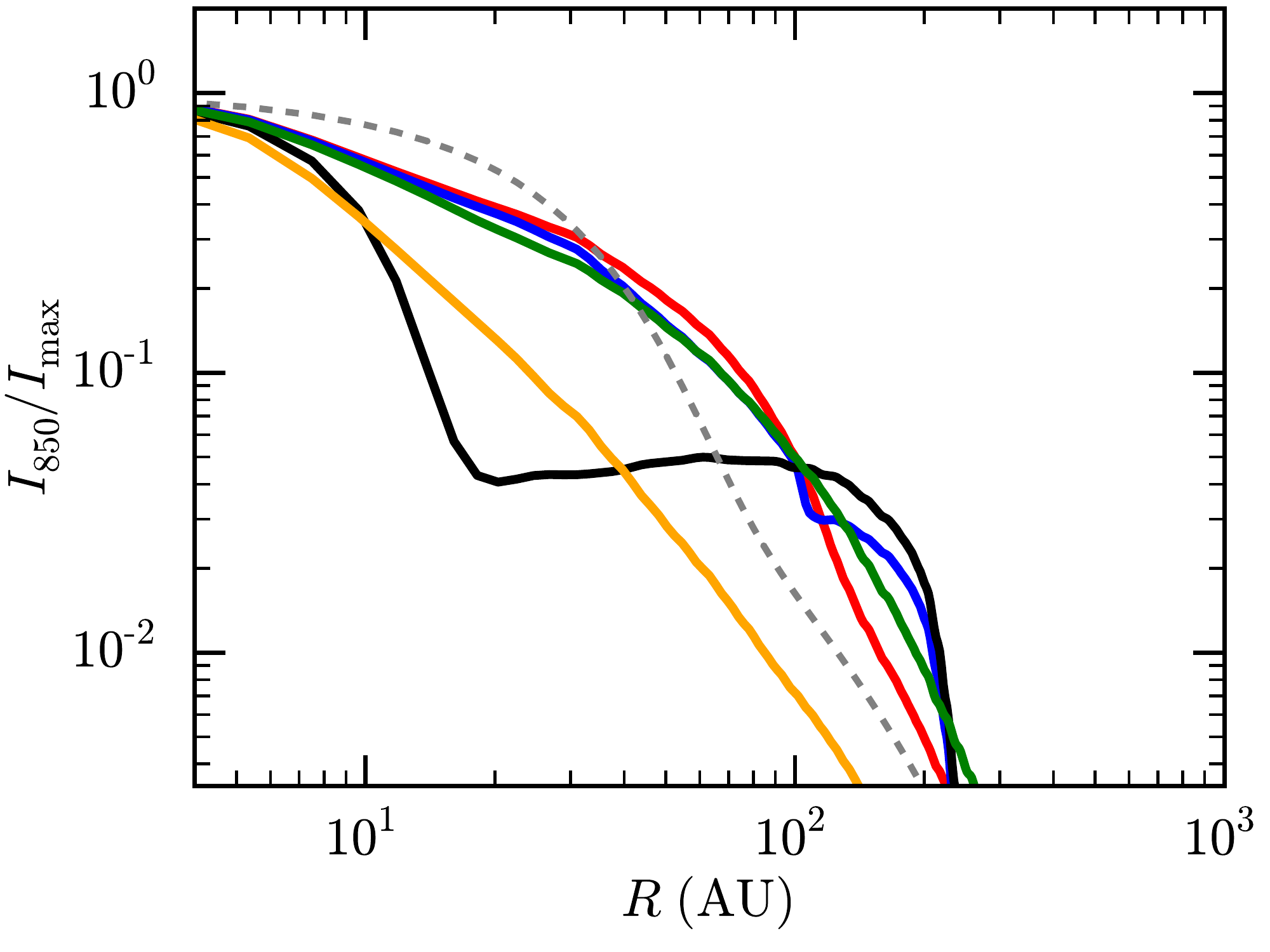}
\includegraphics[width=\columnwidth]{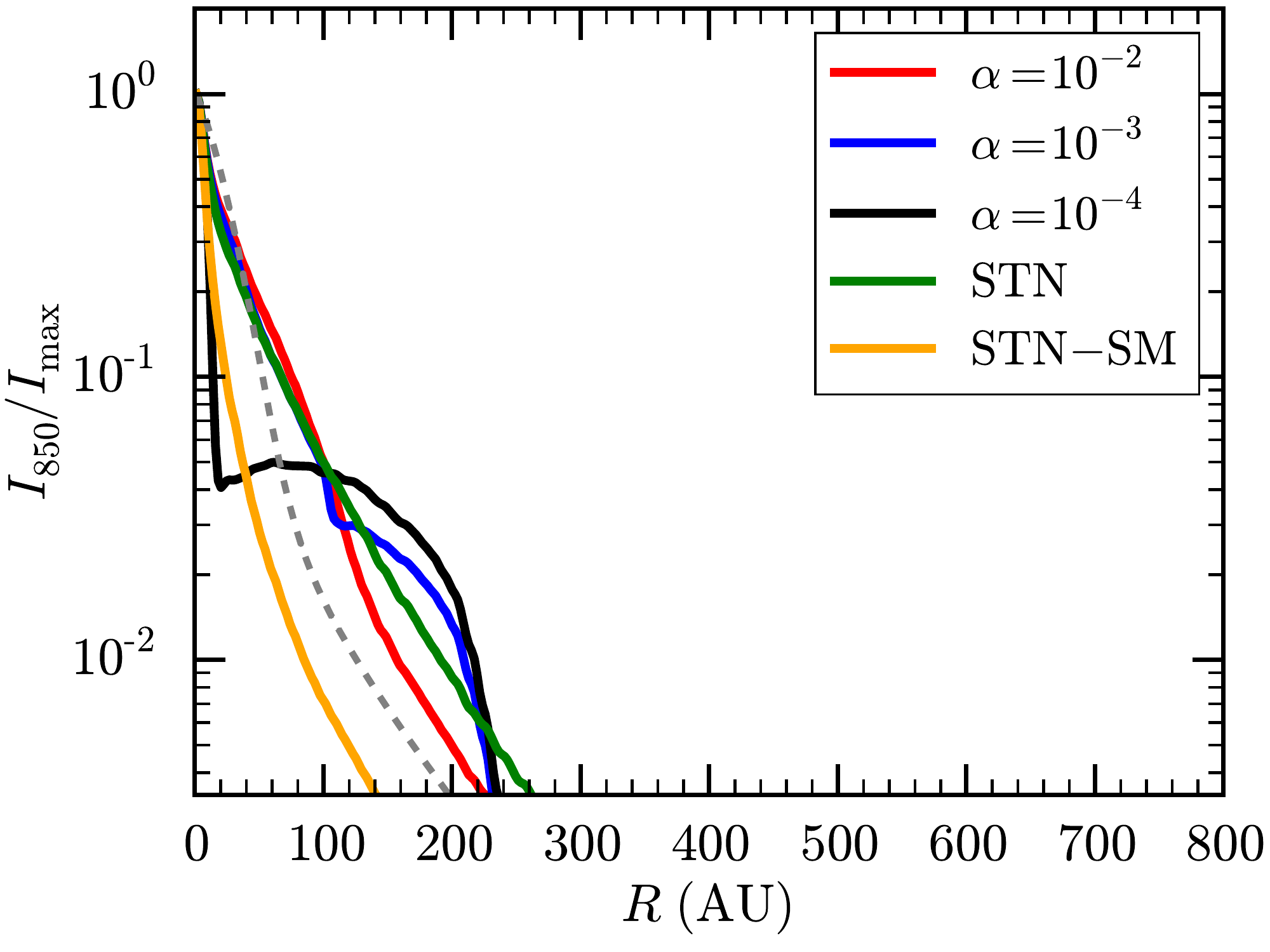}\\
\caption{As in Fig.~\ref{fig:cont_prof}, with a $0.1\arcsec$ beam convolution.}
\label{fig:cont_prof_high_res}
\end{figure*}

\begin{figure*}
\center
\includegraphics[width=.33\textwidth]{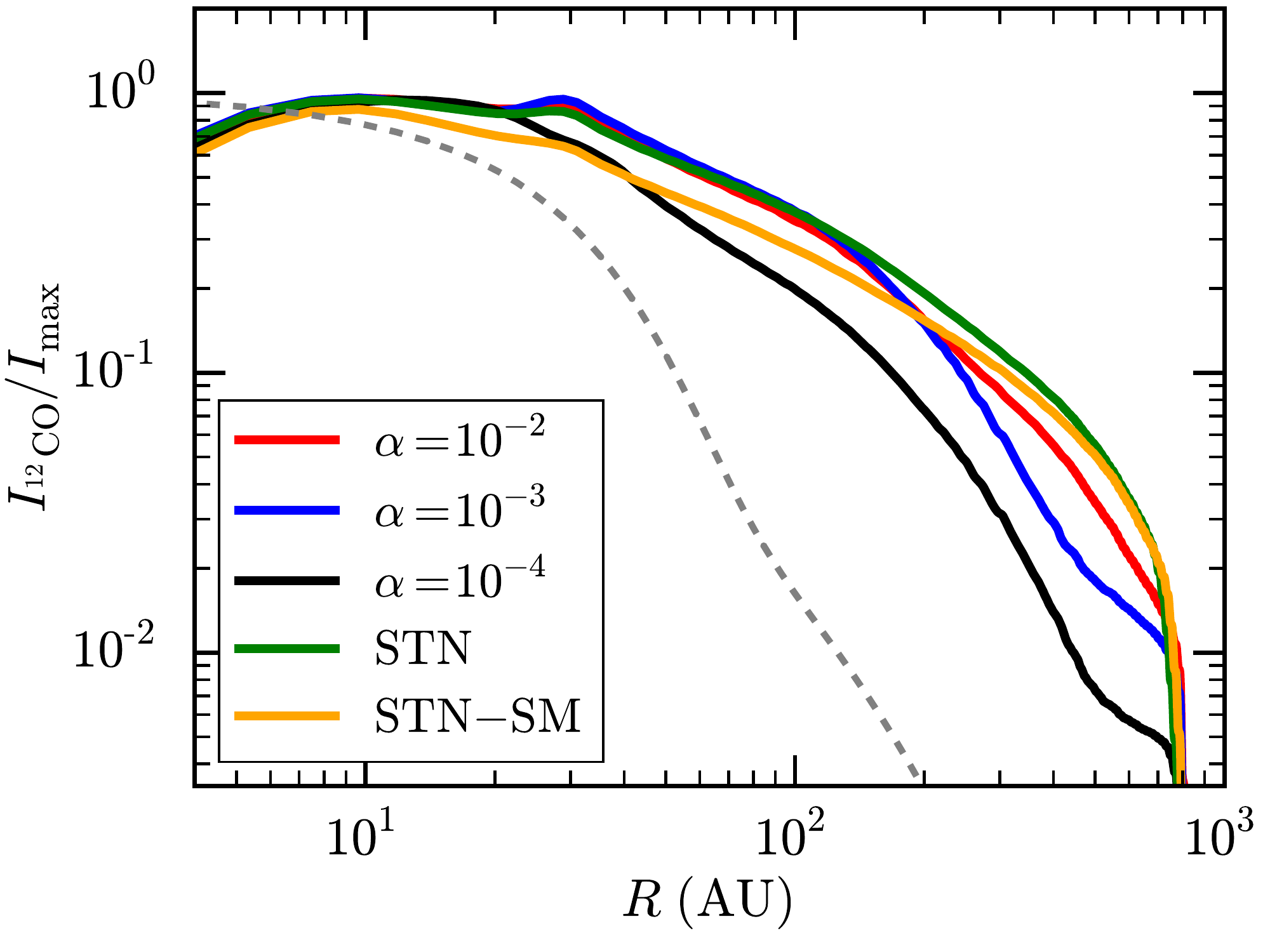}
\includegraphics[width=.33\textwidth]{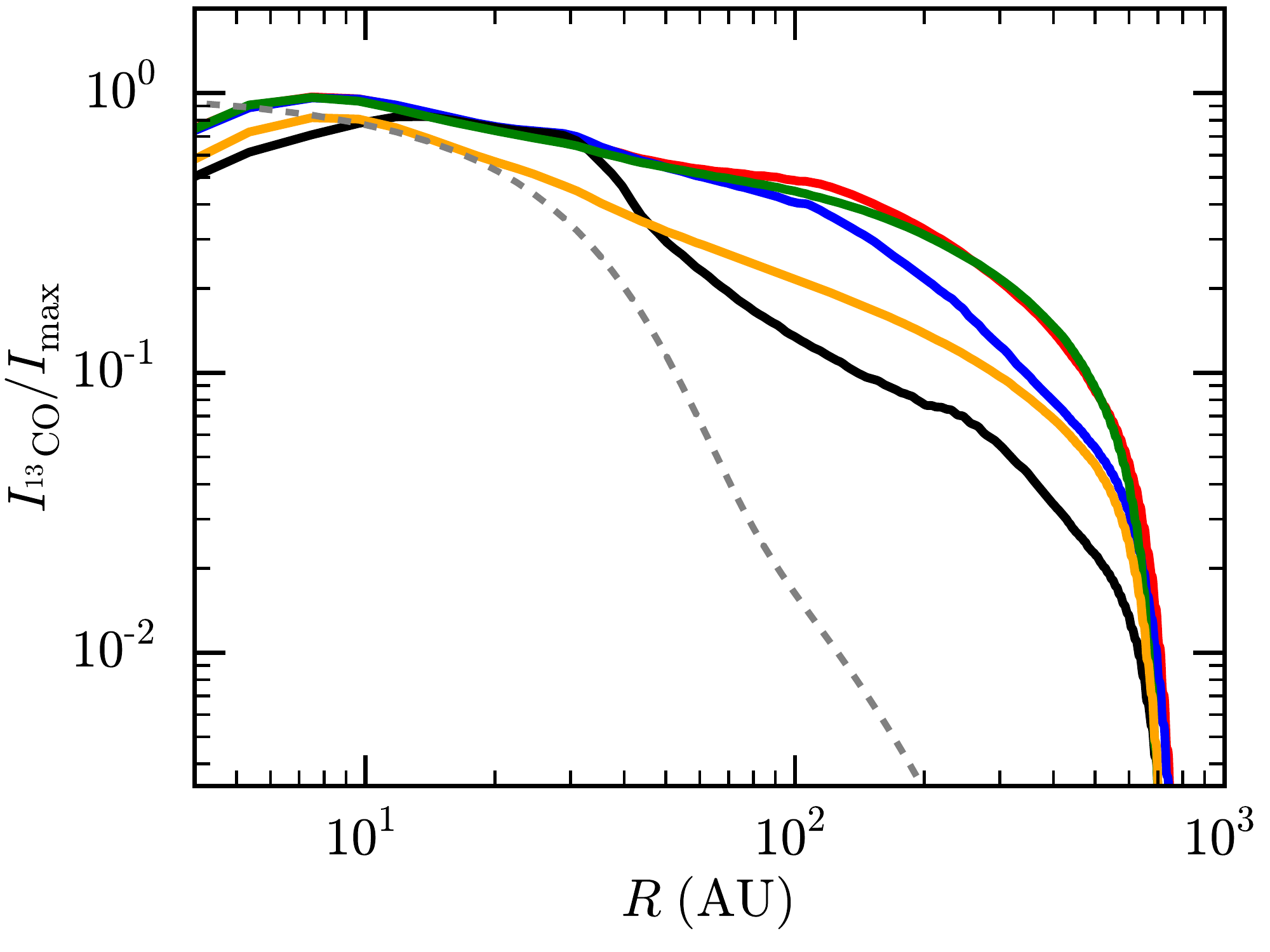}
\includegraphics[width=.33\textwidth]{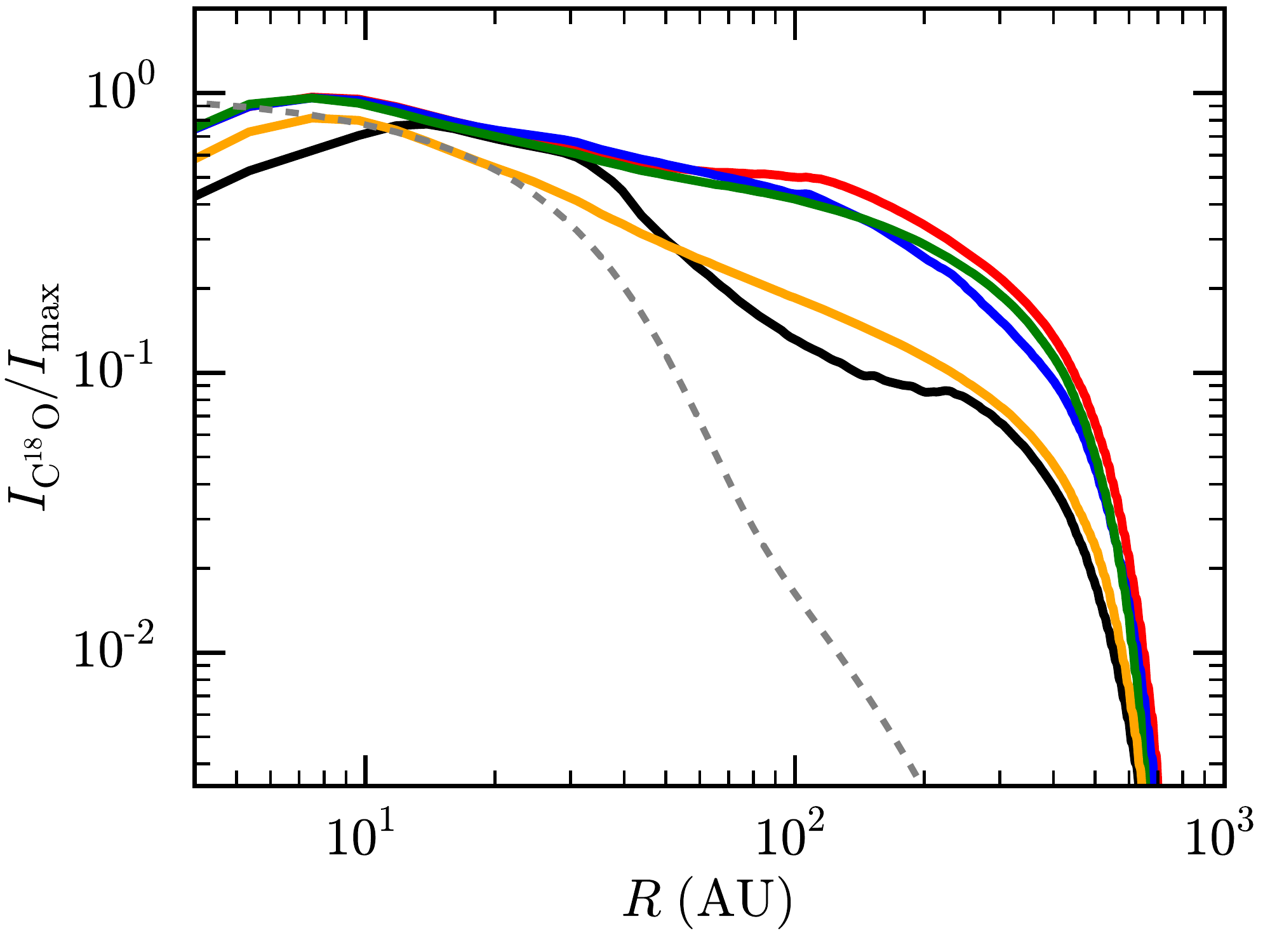}\\
\includegraphics[width=.33\textwidth]{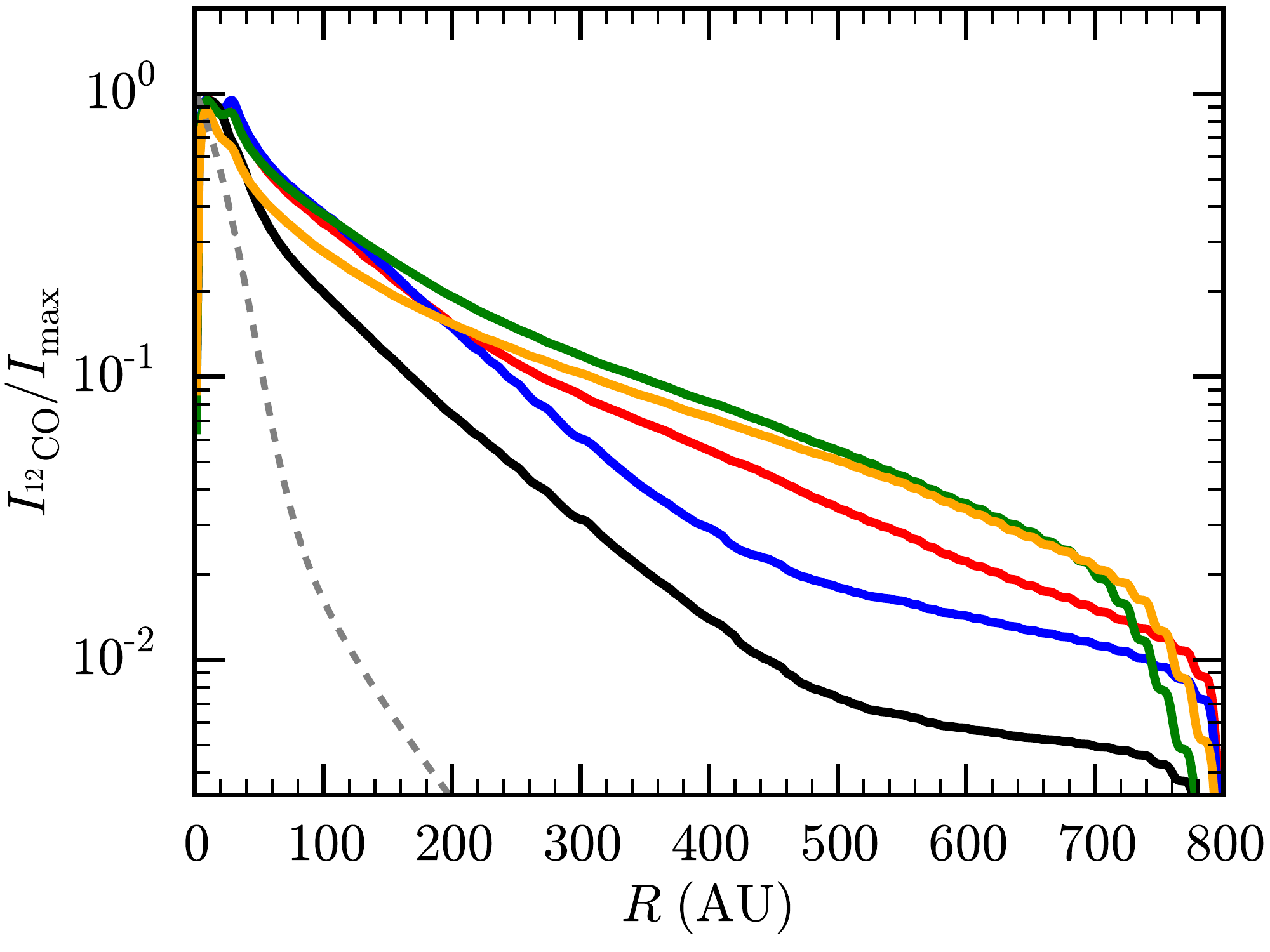}
\includegraphics[width=.33\textwidth]{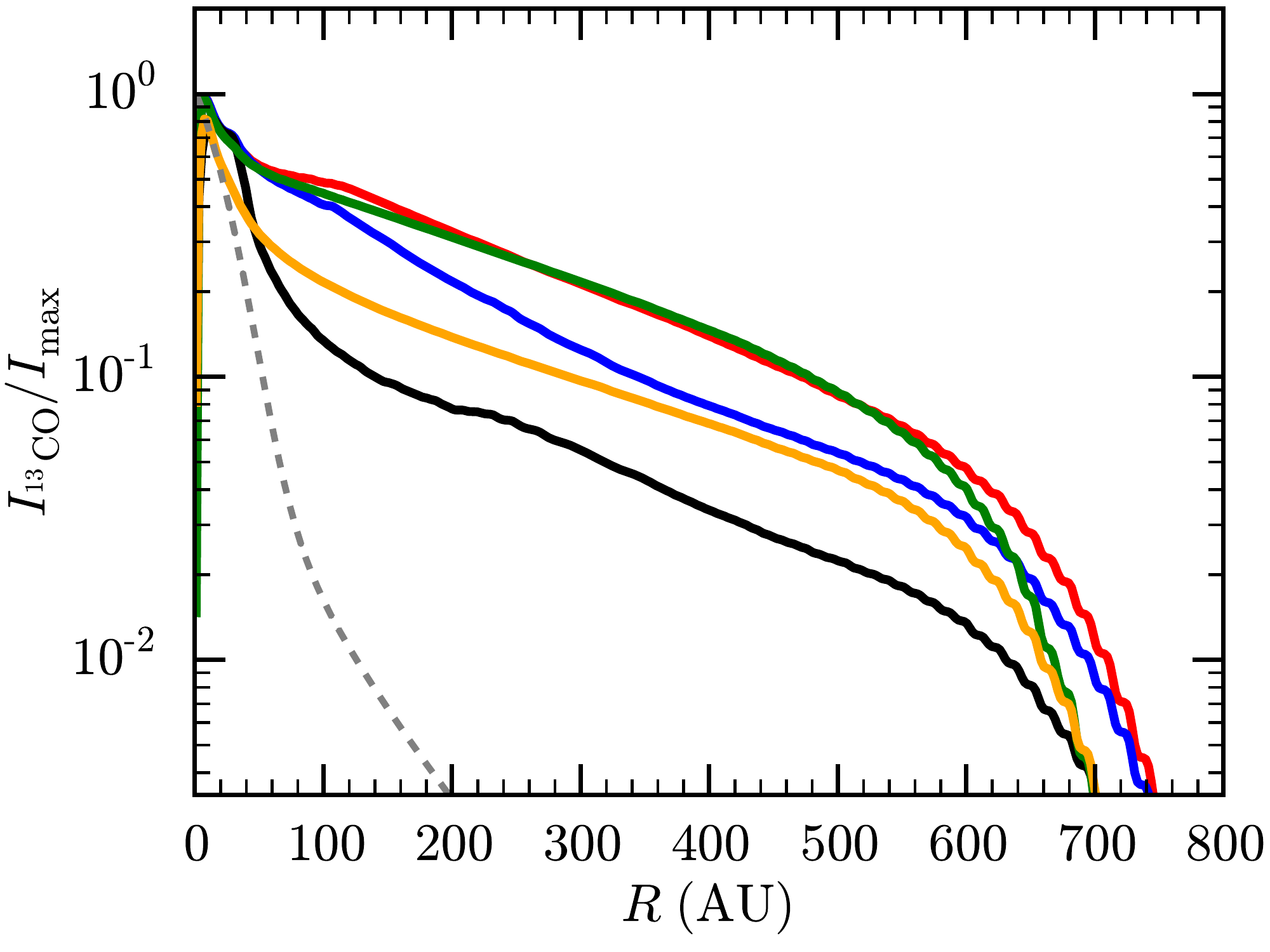}
\includegraphics[width=.33\textwidth]{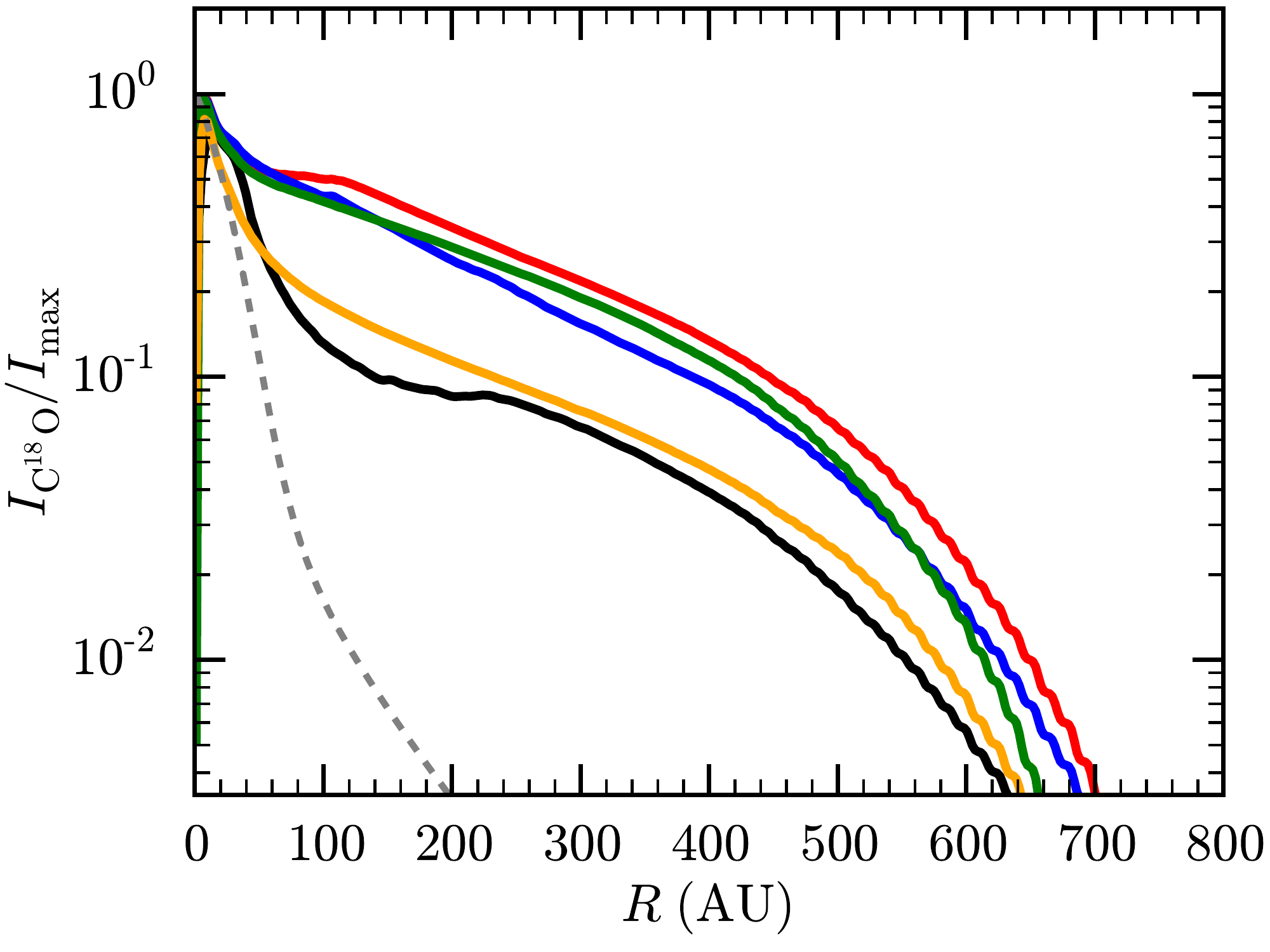}\\
\caption{As in Fig.~\ref{fig:profiles_gas}, with a $0.1\arcsec$ beam convolution. Top panels show the same plots as in the bottom panels, but with a different radial scale.}
\label{fig:profiles_gas_high_res}
\end{figure*}

All the synthetic intensity profiles shown in the paper use a convolution beam of $0.52\arcsec\times0.38\arcsec$ (Section~\ref{sec:setup}). However, ALMA has and is going to provide higher angular resolution observations of protoplanetary disks. We thus show the peak-normalised synthetic intensity profiles of both continuum and CO lines of our main models convolved with a smaller circular beam ($0.1\arcsec$ resolution) in Figs.~\ref{fig:cont_prof_high_res}-~\ref{fig:profiles_gas_high_res}. With higher angular resolution, the substructure outside the fragmentation radius is more prominent. Moreover, the difference of the sharpness of the outer edge between the STN model and models including dust evolution is even more significant. The models with $\alpha\leq10^{-3}$ show a rather sharp edge in the $850\,\mu$m intensity profiles.

The intensity profiles of the line emission are similar to the low angular resolution ones. The main difference appears in the inner disk (see top panels of Fig.~\ref{fig:profiles_gas_high_res}), where part of the line flux is lost due to high optical thickness of the continuum. This is apparent in the $\alpha=10^{-4}$ case, where the inner disk is heavily populated by large grains, due to low turbulent velocities (see Fig.~\ref{fig:d2g_vs_r}).

\begin{figure*}
\center
\includegraphics[width=.49\textwidth]{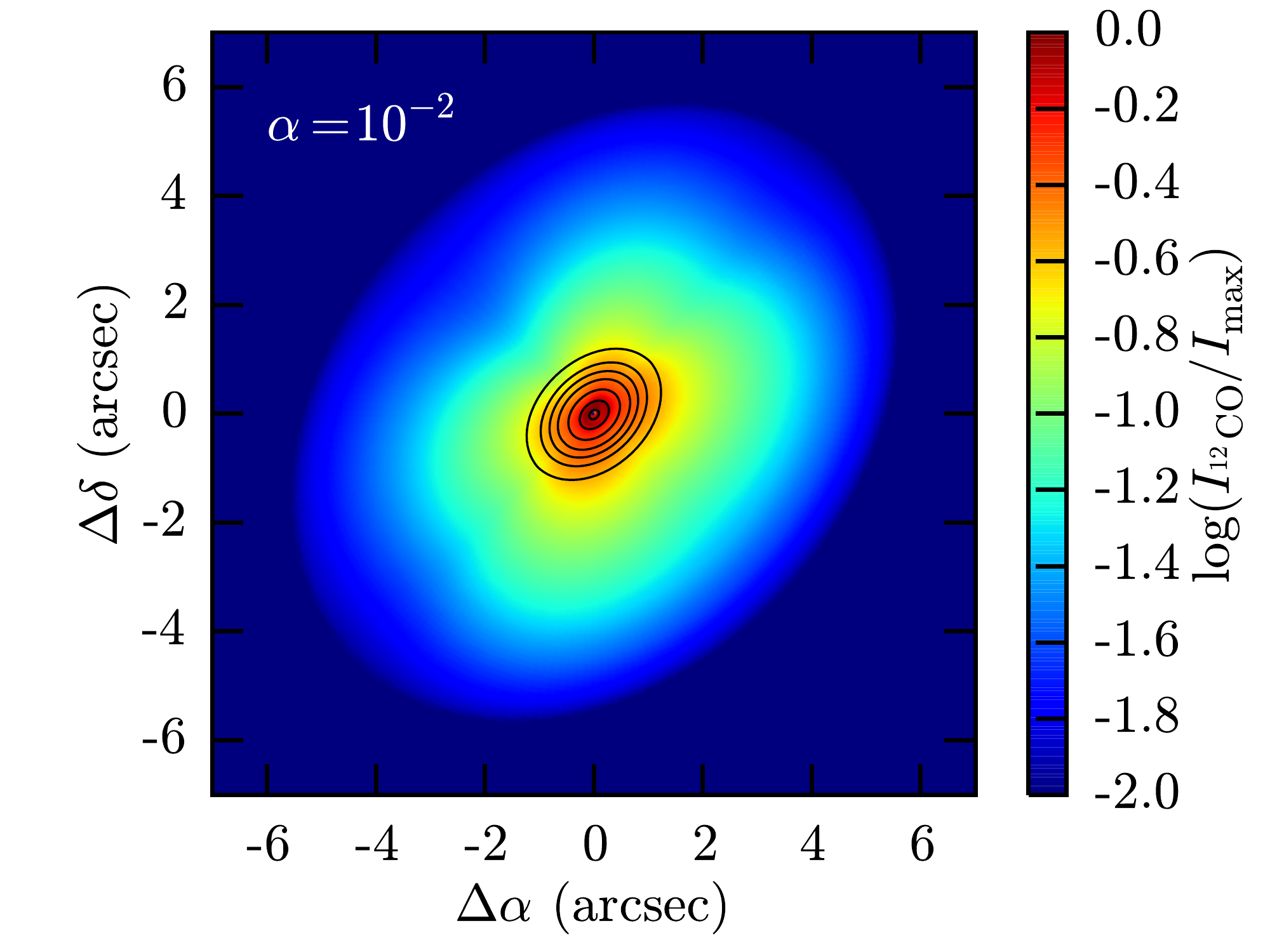}
\includegraphics[width=.49\textwidth]{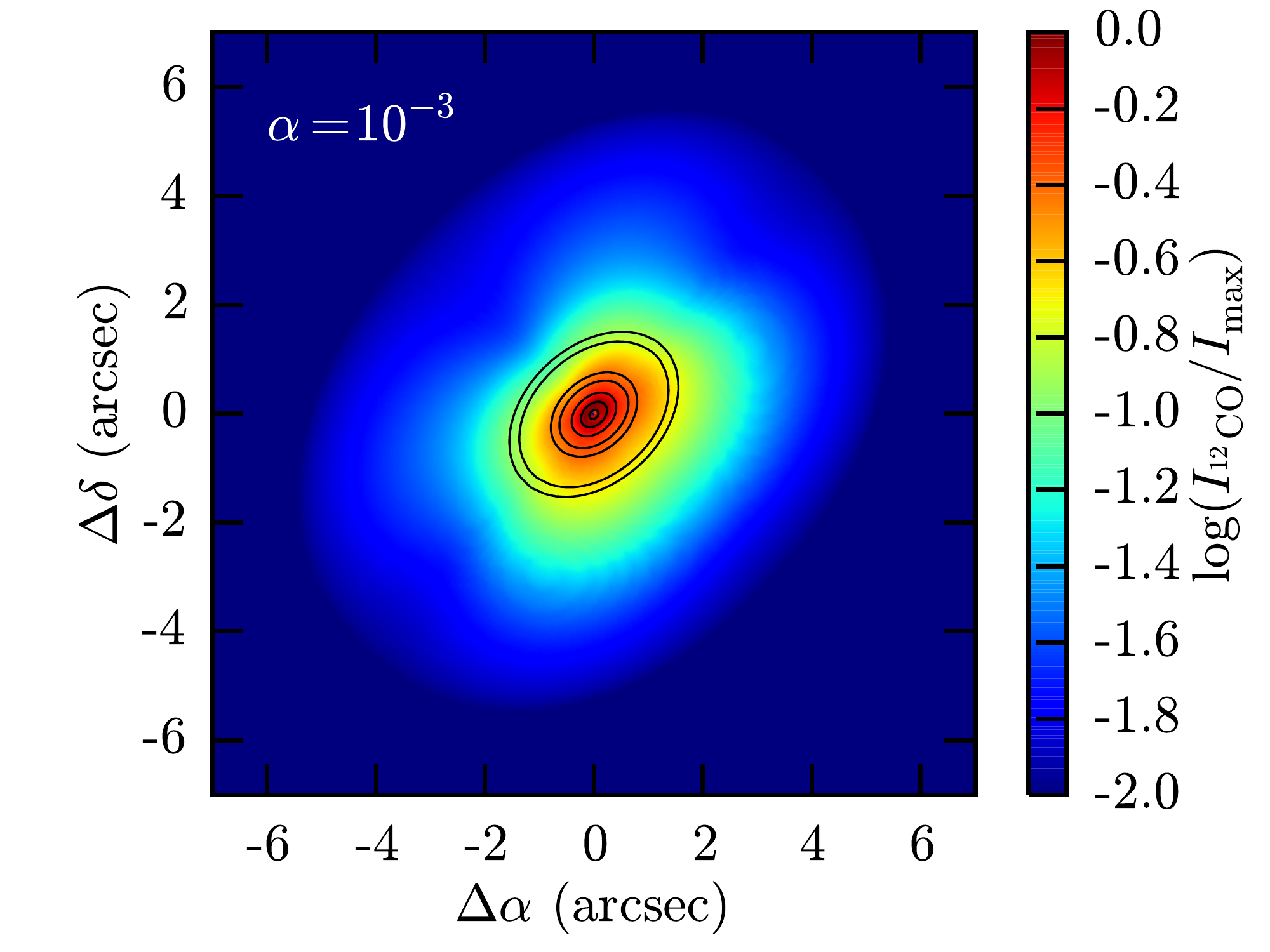}\\
\includegraphics[width=.49\textwidth]{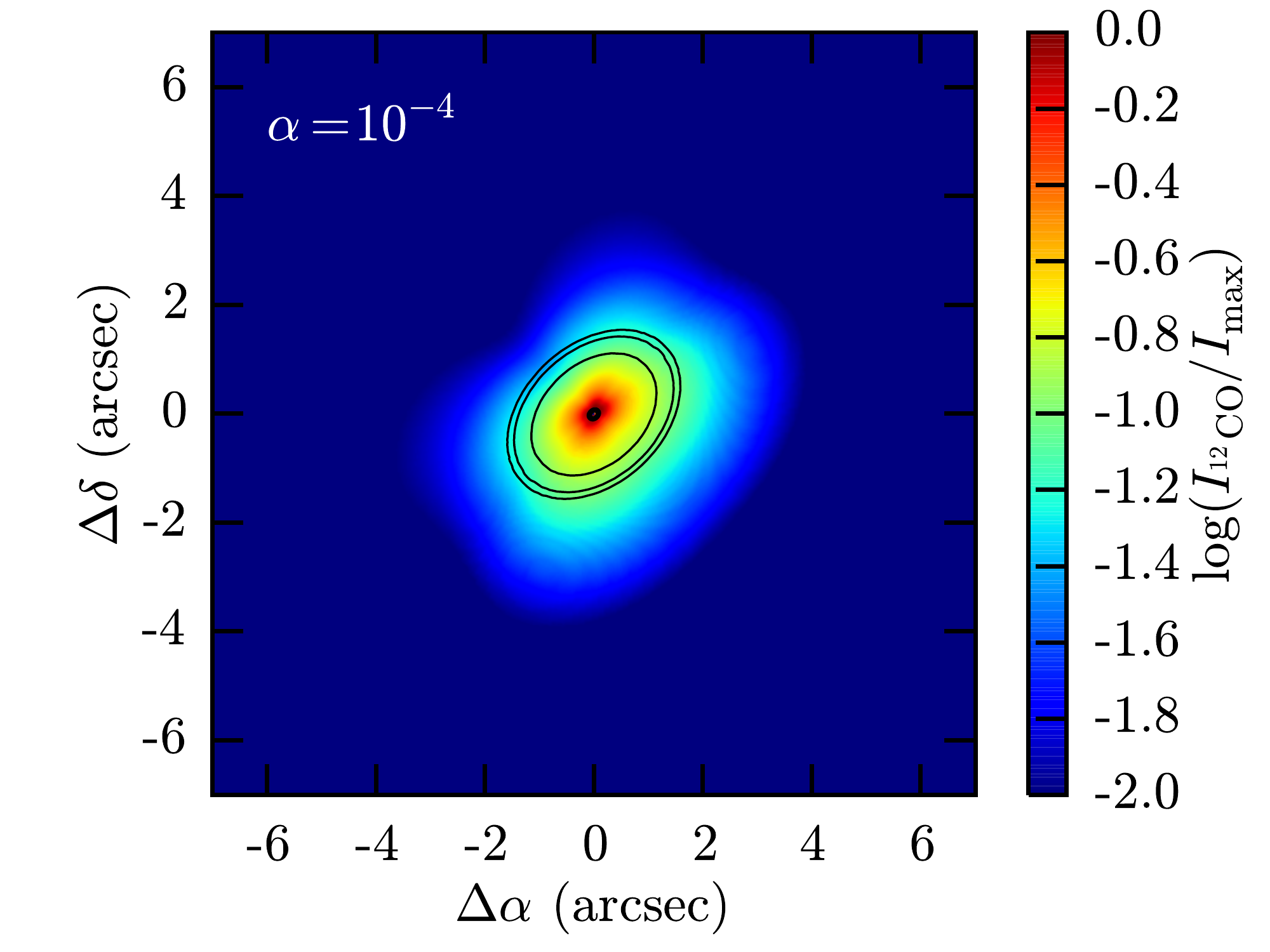}
\includegraphics[width=.49\textwidth]{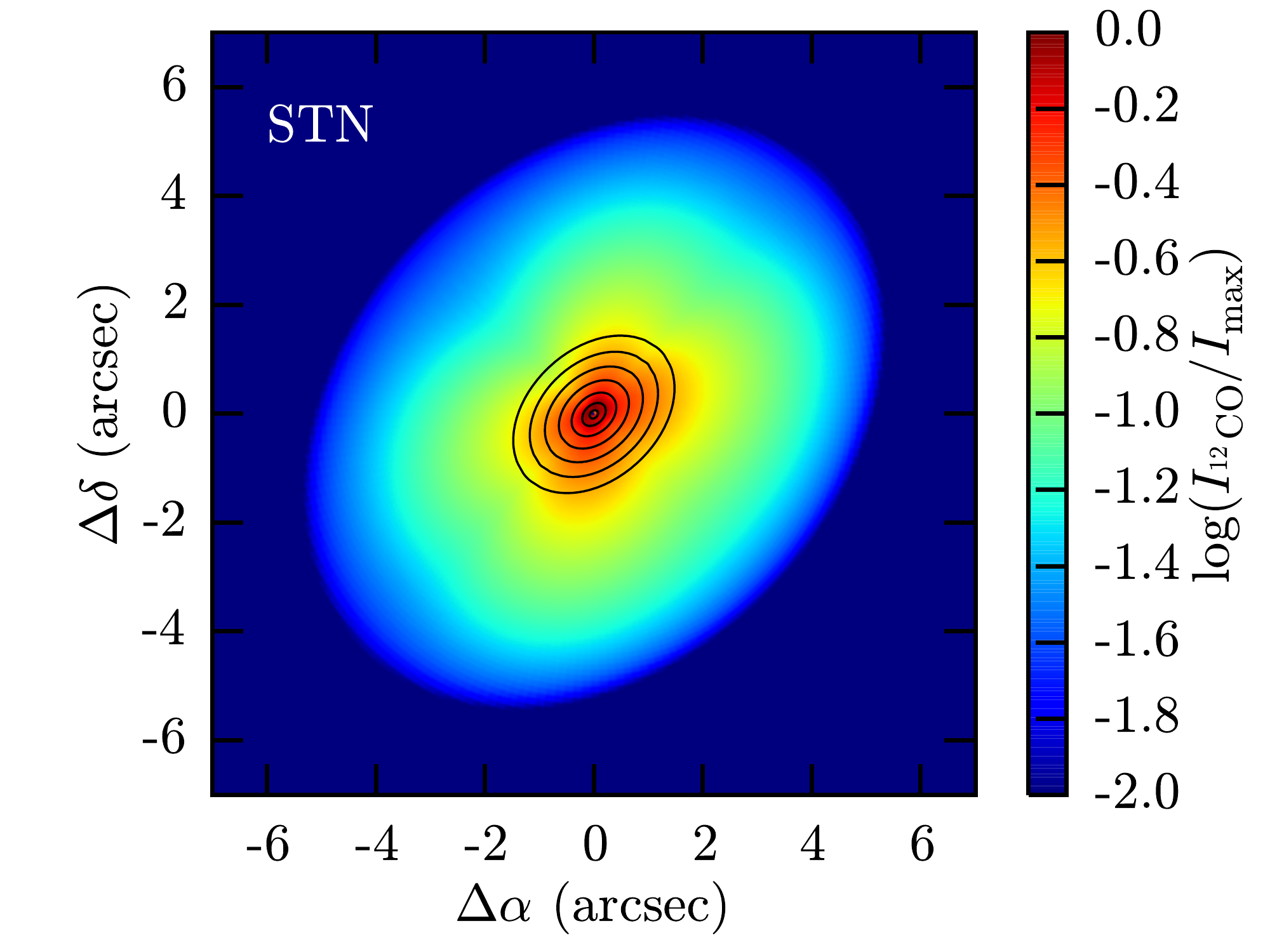}\\
\includegraphics[width=.49\textwidth]{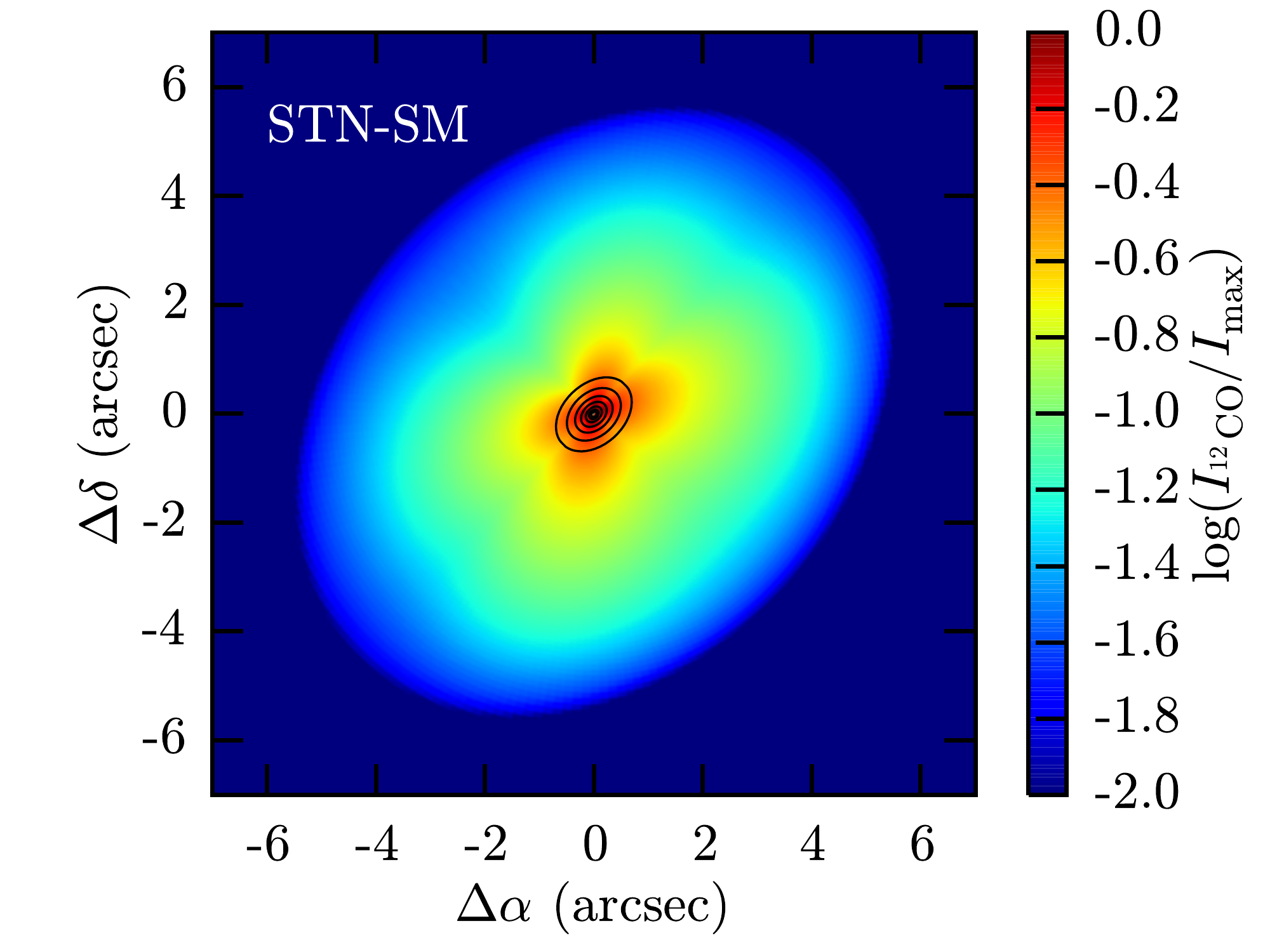}
\caption{Moment 0 map of the $^{12}$CO $J$=3-2 line of HD 163296 parametrised with the prescription by \citet{2013A&A...557A.133D}. The moment 0 map is shown in log scale, normalised to peak value, with dynamic range equalling 100 in the emission. Contours show peak normalised continuum levels at $850\,\mu$m, at peak value over $2$, $4$, $8$, $16$, $32$, $64,$ and $128$. The images are shown with an angular resolution $0.1\arcsec$.}
\label{fig:images_hd16}
\end{figure*}

In Fig.~\ref{fig:images_hd16} we show the peak normalised moment 0 maps of the $^{12}$CO $J$=3-2 line of the models of HD 163296, where the parameters by \citet{2013A&A...557A.133D} are used for the gas density structure. While the dust radial extent seems similar for all models, expect the STN-SM one, the gas radial extent depends significantly on $\alpha$, with the disk appearing smaller for lower turbulent parameters. The gas extents are derived for a fixed dynamic range of $100$ in the line emission.

\end{appendix}

\end{document}